\documentclass[epj-spec]{svjour}
\usepackage{graphics}
\newcommand{\be}{\begin{equation}}
\newcommand{\ee}{\end{equation}}
\newcommand{\bea}{\begin{eqnarray}}
\newcommand{\eea}{\end{eqnarray}}
\newcommand{\nn}{\nonumber}
\newcommand{\nnl}{\nonumber\\}

 \def\l{\left}
 \def\r{\right}
 \def\Im{{\rm Im}}
 \def\Re{{\rm Re}}
 \def\bm#1{\mbox{\boldmath$#1$}}
 \def\gsim{\mathrel{\rlap{\lower0.2em\hbox{$\sim$}}\raise0.2em\hbox{$>$}}}
 \def\gtrless{\mathrel{\rlap{\lower0.0em\hbox{$>$}}\raise0.41em\hbox{$<$}}}
 \def\lessgtr{\mathrel{\rlap{\lower0.0em\hbox{$<$}}\raise0.41em\hbox{$>$}}}

\begin{document}

\title{From Kadanoff-Baym dynamics to off-shell parton transport}

\author{W. Cassing\inst{1}\fnmsep\thanks{\email{Wolfgang.Cassing@theo.physik.uni-giessen.de}}}
%
%
\institute{Institut f\"ur Theoretische Physik, Universit\"at
Giessen, 35392 Giessen, Germany}

\abstract{The  description of strongly interacting quantum fields
is based on two-particle irreducible (2PI) approaches that allow
for a consistent treatment of quantum systems out-of-equilibrium
as well as in thermal equilibrium.  As a theoretical test case the
quantum time evolution of $\Phi^4$-field theory in 2+1 space-time
dimensions is investigated numerically for out-of-equilibrium
initial conditions on the basis of the Kadanoff-Baym-equations
including the tadpole and sunset self energies.  In particular we
address the dynamics of the spectral (`off-shell') distributions
of the excited quantum modes and the different phases in the
approach to equilibrium described by Kubo-Martin-Schwinger
relations for thermal equilibrium states. A detailed comparison of
the full quantum dynamics to approximate schemes  like that of a
standard kinetic (on-shell) Boltzmann equation is performed.  We
find that far off-shell $1\leftrightarrow 3$ processes are
responsible for chemical equilibration, which is not included in
the Boltzmann limit. Furthermore, we derive generalized transport
equations for the same theory in a first order gradient expansion
in phase space thus explicitly retaining the off-shell dynamics as
inherent in the time-dependent spectral functions. The solutions
of these equations compare very well with the exact solutions of
the full Kadanoff-Baym equations with respect to the occupation
numbers of the individual modes, the spectral evolution as well as
the chemical equilibration process. Furthermore, the proper
equilibrium off-shell distribution is reached for large times
contrary to the quasiparticle Boltzmann limit. We additionally
present a direct comparison of the solution of the generalized
transport equations in the Kadanoff-Baym (KB) and
Botermans-Malfliet (BM) form; both solutions are found to agree
very well with each other. The off-shell transport equation in the
BM scheme allows for an explicit solution within an extended
dynamical quasiparticle Ansatz. This leads to generalized
equations of motion for dynamical quasiparticles that exceed the
classical Hamilton equations and allow for the description of
dynamical spectral functions. The dynamical quasiparticle model
(DQPM) is used to extract partonic spectral functions from lattice
QCD in the temperature range 0.8 $\leq T/T_c \leq$ 10. By
consideration of time-like and space-like sectors of 'observables'
such as number densities, energy densities etc. mean-field
potentials as well as effective interactions are extracted at
different temperature $T$ and quark chemical potentials $\mu_q$.
The latter determine the off-shell dynamics in the
Parton-Hadron-String Dynamics (PHSD) transport approach.
Illustrative examples for off-shell dynamics in the hadron and
parton sector are presented for $e^+e^-$ production from
nucleus-nucleus collisions from SIS to RHIC energies. The
generalized transport approach, furthermore, qualifies for the
description of hadronization without leading to the problem of
entropy reduction as in conventional coalescence models.}

\maketitle

\section*{Introduction}

Non-equilibrium many-body theory or quantum field theory has
become a major topic of research for   transport processes in
nuclear physics, in cosmological particle physics as well as
condensed matter physics. The multidisciplinary aspect arises due
to a common interest to understand the various relaxation
phenomena of quantum dissipative systems. Recent progress in
cosmological observations has also intensified the research on
quantum fields out of equilibrium. Important questions in
high-energy nuclear or particle phyics at the highest energy
densities are: i) how do nonequilibrium systems in extreme
environments  evolve, ii) how do they eventually thermalize, iii)
how phase transitions do occur in real time with possibly
nonequilibrium remnants, and iv) how do such systems evolve for
unprecedented short and nonadiabatic timescales?

 The very early
history of the universe provides important scenarios, where
non-equili\-brium effects might have played an important role,
like in the (post-) inflationary epoque (see e.g.
\cite{inflation,boya6,Linde,Pietroni}), for the understanding of
baryogenesis (see e.g. \cite{inflation,Janka}) and also for the
general phenomena of cosmological decoherence \cite{GMH93}.
Referring to modern nuclear physics the dynammics of heavy-ion
collisions at various bombarding energies has always been a major
motivation for research on nonequilbrium quantum many-body physics
and relativistic quantum-field theories, since the initial state
of a collision resembles an extreme nonequilibrium situation while
the final state might even exhibit a certain degree of
thermalization. Indeed, at the presently highest energy heavy-ion
collider experiments at RHIC, where one expects to create
experimentally a transient deconfined state of matter denoted as
quark-gluon plasma (QGP) \cite{Mul85}, there are experimental
indications - like the build up of collective flow - for an early
thermalization accompanied with the build up of a very large
pressure.  These examples demonstrate that one needs an {\it
ab-initio} understanding of the dynamics of out-of-equilibrium
quantum-field theory.

Especially the powerful method of the `Schwinger-Keldysh'
\cite{Boy95,Sc61,BM63,Ke64,Cr68} or `closed time path' (CTP)
(non-equilibrium) real-time Greens functions has been shown to
provide an appropriate basis for  the formulation of the  complex
problems in the various  areas of nonequilibrium quantum many-body
physics. Within this framework one can derive  valid
approximations - depending, of course, on the problem under
consideration - by preserving  overall consistency relations.
Originally, the resulting causal Dyson-Schwinger equation of
motion for the one-particle Greens functions (or two-point
functions), i.e. the Kadanoff-Baym (KB) equations \cite{KB}, have
served as the underlying scheme for deriving various transport
phenomena and generalized transport equations. For review articles
on the Kadanoff-Baym equations in the various areas of
nonequilibrium quantum physics we refer  the reader to Refs.
\cite{DuBois,dan84,Ch85,RS86,calhu,Haug}. We note in passing, that
also the 'influence functional formalism' has been shown to be
directly related to the KB equations \cite{GL98a}. Such a relation
allows to address inherent stochastic aspects of the latter and
also to provide a rather intuitive interpretation of the various
self-energy parts that enter the KB equations.

Furthermore, kinetic transport theory is a convenient tool to
study many-body nonequilibrium systems, nonrelativistic or
relativistic. Kinetic equations, which do play the central role in
more or less all practical  simulations, can be derived by means
of appropriate KB equations within suitable approximations. Hence,
a major impetus in the past has been to derive semi-classical
Boltzmann-like transport equations within the standard
quasi-particle approximation. Additionally, off-shell extensions
by means of a gradient expansion in the space-time inhomogenities
- as already introduced by Kadanoff and Baym \cite{KB} - have been
formulated: for an relativistic electron-photon plasma
\cite{BB72}, for transport of electrons in a metal with external
electrical field \cite{Schoeller,LSV86}, for transport of nucleons
at intermediate heavy-ion reactions \cite{botmal}, for transport
of particles in $\Phi^4$-theory \cite{calhu,danmrow}, for
transport of electrons in semiconductors \cite{Haug,SL94}, for
transport of partons or fields in high-energy heavy-ion reactions
\cite{Ma95,Ge96,BD98,BI99}, or for a trapped Bose system described
by effective Hartree-Fock-Bogolyubov kinetic equations
\cite{Gri99}. We recall that on the formal level of the
KB-equations the various forms assumed for the self-energy have to
fulfill consistency relations in order to preserve symmetries of
the fundamental Lagrangian \cite{KB,knoll1,knoll2}. This allows
also for a unified treatment of stable and unstable (resonance)
particles.

In  non-equilibrium quantum-field theory typically the
nonperturbative description of (second-order) phase transitions
has been in the foreground of interest by means of mean-field
(Hartree) descriptions \cite{boya6,Boy95,Co94,boya3,boya2}, with
applications for the evolution of disoriented chiral condensates
or the decay of the (oscillating) inflaton in the early reheating
era of the universe. `Effective' mean-field dissipation (and
decoherence) - solving the so called `backreaction' problem - was
incorporated by particle production through order parameters
explicitly varying in time. However, it had been then realized
that such a dissipation mechanism, i.e. transferring collective
energy from the time-dependent order parameter to particle degrees
of freedom, can not lead to true dissipation and thermalization.
Such a conclusion has already been known for quite some time
within the effective description of heavy-ion collisions at low
energy. Full time-dependent Hartree or Hartree-Fock descriptions
\cite{negele} were insufficient to describe the reactions with
increasing collision energy; additional Boltzmann-like collision
terms had to be incorporated in order to provide a more adequate
description of the collision processes.

The incorporation of true collisions then has been formulated also
for  various quantum-field theories
\cite{Peter2,Peter3,boya1,boya4,boya5,CH02}. Here, a systematic
$1/N$ expansion of the "2PI effective action" is conventionally
invoked \cite{calhu,Co94,CH02} serving as a nonperturbative
expansion parameter. Of course, only for  large $N$ this might be
a controlled expansion. In any case, the understanding and the
influence of dissipation with the chance for true thermalization -
by incorporating collisions - has become a major focus of recent
investigations. The resulting equations of motion always do
resemble the KB equations; in their general form (beyond the mean
field or Hartree(-Fock) approximation) they do break time
invariance and thus lead to irreversibility. This macroscopic
irreversibility arises from the truncations of the full theory to
obtain the self-energy operators in a specific limit. As an
example we mention the truncation of the (exact) Martin-Schwinger
hierarchy in the derivation of the collisional operator in Ref.
\cite{botmal} or the truncation of the (exact) BBGKY hierarchy in
terms of $n$-point functions \cite{botmal,Peter2,Peter3,SKK03}.

In principle, the non-equilibrium quantum dynamics is
nonperturbative in nature. Unphysical singularities only appear in
a limited truncation scheme, e.g. ill-defined pinch singularities
\cite{AS94}, which do arise at higher order in a perturbative
expansion in out of equilibrium quantum-field theory, are
regularized by a consistent nonperturbative description (of
Schwinger-Dyson type) of the non-equilibrium evolution, since the
resummed propagators obtain a finite width \cite{GL99}. Such a
regularization is also observed by other resummation schemes like
the dynamical renormalization-group technique \cite{boya5,Pawlo}.

Although the analogy of KB-type equations to a Boltzmann-like
process is quite obvious,  this analogy is far from being trivial.
The full quantum formulation contains much more information than a
 semi-classical (generally) on-shell Boltzmann equation. The
dynamics of the spectral (i.e. `off-shell') information is fully
incorporated in the quantum dynamics while it is missing in the
Bolzmann limit. A full answer to the question of quantum
equilibration can thus only be obtained by detailed numerical
solutions of the quantum description itself.

We briefly address  previous works that have investigated
numerically approximate or full solutions of KB-type equations. A
seminal work has been carried out by Danielewicz \cite{dan84a},
who investigated for the first time the full KB equations for a
spatially homogenous system with a deformed Fermi sphere in
momentum space for the initial distribution of occupied momentum
states in order to model the initial condition of a heavy-ion
collision in the nonrelativistic domain. In comparison to a
standard on-shell semi-classical Boltzmann equation the full
quantum Boltzmann equation showed quantitative differences, i.e. a
larger collective relaxation time for complete equilibration of
the  momentum distribution $f(\vec{p},t)$. This slowing down of
the quantum dynamics was attributed to quantum interference and
off-shell effects. Similar quantum modifications in the
equilibration and momentum relaxation have been found for a
relativistic situation in Ref. \cite{CGreiner}.  In the following,
full and more detailed solutions of nonrelativistic KB equations
have been performed by K\"ohler \cite{koe1,koe2} with special
emphasis on the build up of initial many-body correlations on
short time scales. The role of memory effects has been clearly
shown experimentally by femtosecond laser spectroscopy in
semiconductors \cite{Haug95} in the relaxation of excitons.
Solutions of quantum transport equations for semiconductors
\cite{Haug,WJ99} - to explore relaxation phenomena  on short time
and distance scales - has become also a very active field of
research \cite{Schoeller,morabuch}.

In the last decade the numerical treatment of general
out-of-equilibrium quantum dynamics, as described within 2PI
effective approaches at higher order,  have become more frequent.
The subsequent equations of motion are very similar to the KB
equations, although more involved expressions for a non-local
vertex function - as obtained by the 2PI scheme - are
incorporated. The numerical solutions are also obtained only for
homogenous systems, so far (cf.
\cite{berges1,berges2,berges3,CDM02,berges4,Lindner}).

Apart from complete thermalization of all single-particle degrees
of freedom  the quantum dynamics of the spectral function is also
a lively discussed issue in the microscopic modeling of hadronic
resonances with a broad mass distribution. This is of particular
relevance for simulations of heavy-ion reactions, where e.g. the
$\Delta $-resonance or the $\rho $-meson already show a large
decay witdh in vacuum
\cite{caju1,caju2,caju3,Leupold,HHab,knoll3}. Especially the $\rho
$ vector meson  represents the most promising hadronic particle
for showing possible in-medium modifications in hot and compressed
nuclear matter  since the leptonic decay products  are of only
weakly interacting electromagnetic nature. Hence a consistent
formulation for the transport of extremely short-lived particles
beyond the standard quasi-particle approximation is needed. On the
one side, there exist purely formal developments starting from a
first-order gradient expansion of the underlying KB equations
\cite{Leupold,HHab,knoll3}, while on the other side already
practical realisations for various questions have emerged
\cite{caju1,caju2,caju3,cas03}. The general idea is to obtain a
description for the propagation of dynamical spectral functions,
i.e. a propagation in the off-shell mass squared $M^2$.

The aim of these lectures is to provide a short survey on
(nonperturbative) nonequilibrium quantum field theory with actual
applications (and numerical illustrations) for the model case of
scalar $\Phi^4$-theory. This includes the derivation of off-shell
as well as on-shell (Boltzmann) transport equations and their
comparison to the exact solution for a variety of
out-of-equilibrium initial conditions. Furthermore, a generalized
testparticle Ansatz is introduced that allows for a convenient
solution of off-shell transport equations which involve only the
complex (retarded) selfenergies for the relevant degrees of
freedom. In case of hadrons these selfenergies may be determined
from effective Lagrangian models whereas in case of partons
suitable selfenergies can be fixed by the dynamical quasiparticle
model in comparison to lattice QCD at finite temperature $T$.
Applications for dilepton production in nucleus-nucleus collisions
provide a practical case for the spectral off-shell dynamics and
bridge the gap to current spectra taken by various experimenal
collaborations \cite{Stroth}. On the partonic side we present an
illustrative example for the build up of collective flow,
strangeness equilibration as well as the dynamics of
hadronization.

\section{Quantum field dynamics and thermalization}

\subsection{Non-equilibrium dynamics}

\subsubsection{The Kadanoff-Baym equations}
\begin{figure}[t]
\vspace{0.5cm}
\begin{center}
\resizebox{0.7\columnwidth}{!}{\includegraphics{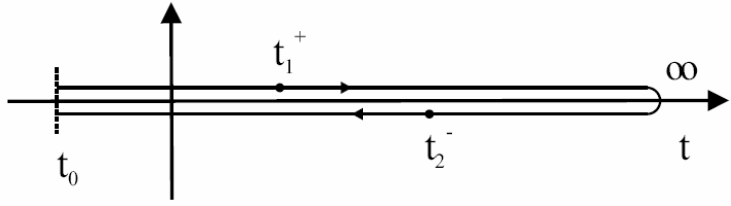} }
\end{center}
\vspace{0.5cm} \caption[]{The closed time contour in the Schwinger-Keldysh formalism}
\label{Plot0} \vspace{0.3cm}
\end{figure}

As mentioned in the Introduction, a natural starting point for
non-equilibrium theory is provided by the closed-time-path (CTP)
method. Here all quantities are given on a special real-time
contour with the time argument running from $-\infty$ to $\infty$
on the chronological branch $(+)$ and returning from $\infty$ to
$-\infty$ on the antichronological branch $(-)$. In cases of
systems prepared at a time $t_0$ this value is (instead of
$-\infty$) the start and end point of the real-time contour (cf.
Fig. \ref{Plot0}). In particular the path ordered Green functions
(in case of real scalar fields $\phi(x)$) are defined as
\bea G(x,y) & = & \langle \, T^P \, \{ \, \phi(x) \: \phi(y) \, \}
\, \rangle
\\[0.4cm]
       & = & \Theta^P(x_0-y_0) \langle \, \phi(x) \: \phi(y) \,
       \rangle
       \;+\; \Theta^P(y_0-x_0) \langle \, \phi(y) \: \phi(x) \, \rangle \nn
\label{pathgreendef} \eea
where the operator $T^P$ orders the field operators according to
the position of their arguments on the real-time path as
accomplished by the path step-functions $\Theta^P$. The
expectation value in (\ref{pathgreendef}) is taken with respect to
some given density matrix $\rho_0$, which is constant in time,
while the operators in the Heisenberg picture contain the entire
time dependence of the non-equilibrium system, i.e. $O(t) = \exp(i
H(t-t_0)) \ O \ \exp(-i H(t-t_0))$.

Self-consistent equations of motion for these Green functions can
be obtained with help of the two-particle irreducible (2PI)
effective action $\Gamma[G]$. It is given by the Legendre
transform of the generating functional of the connected Green
functions $W$ as
\bea \Gamma[G] \; = \; \Gamma^o \: + \: \frac{i}{2} \: \left[ \;
ln ( 1 - \odot_p \, G_o \odot_p \Sigma ) \: + \: \odot_p \, G
\odot_p \Sigma \;\right] \; + \; \Phi[G] \:  \label{effaction}
\eea
in case of vanishing vacuum expectation value
$<\!\!0|\phi(x)|0\!\!>= 0$. In (\ref{effaction}) $\Gamma^o$
depends only on free Green functions and is treated as a constant,
while the symbols $\odot_p$ represent convolution integrals over
the closed time path in Fig. \ref{Plot0}. The functional $\Phi$ is
the sum of all closed $2PI$ diagrams built up by {\it full
propagators} $G$; it determines the {\it self-energies} by
functional variation as
\bea \Sigma(x,y) \; = \; 2 i \, \frac{\delta \Phi}{\delta G(y,x)}
\: . \label{effactionsigma} \eea
From the effective action $\Gamma(G)$ the {\it equations of motion
for the Green function} are determined by the stationarity
condition
\bea \delta \Gamma / \delta G = 0. \label{effactioneom} \eea

\subsubsection{$\Phi^4$-theory}
The scalar $\phi^4$-theory is an example for a fully relativistic
field theory of interacting scalar particles that allows to test
theoretical approximations \cite{Peter2,Peter3} without coming to
the problems of gauge-invariant truncation schemes  as encountered
for gauge (vector) fields \cite{Peter2,berges1,berges2,berges3}.
Its Lagrangian density is given by $(x=(t,\vec{x}))$
\bea \label{lagrangian} {\cal L}(x) \; = \;
  \frac{1}{2} \, \partial_{\mu} \phi(x) \, \partial^{\mu} \phi(x)
\: - \: \frac{1}{2} \, m^2 \, \phi^2(x) \: - \: \frac{\lambda}{4
!} \, \phi^4(x) \; , \eea
where $m$ denotes the 'bare' mass and $\lambda$ is the coupling
constant determining the interaction strength of the scalar
fields.

\begin{figure}[t]
\vspace{0.5cm}
\begin{center}
\resizebox{0.5\columnwidth}{!}{\includegraphics{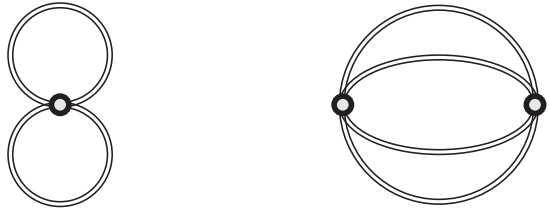} }
\end{center}
\vspace{0.5cm} \caption[]{Contributions to the $\Phi$-functional
for the Kadanoff-Baym equation: two-loop contribution (l.h.s.)
giving the tadpole self-energy and three-loop contribution
(r.h.s.) generating the sunset self-energy. The $\Phi$-functional
is built-up by full Green functions (double lines) while open dots
symbolize the integration over the inner coordinates.
\label{diagram_phi}} \vspace{0.3cm}
\end{figure}

In the present calculation we take into account contributions up
to the 3-loop order for the $\Phi$-functional (cf. Fig.
\ref{diagram_phi}) which reads explicitly
\bea i \Phi \; = \; \frac{i \lambda}{8} \int_{\cal C} d^{d+1\!}x
\; \: G(x,x)^2 \; - \; \frac{\lambda^2}{48} \int_{\cal C}
d^{d+1\!}x \int_{\cal C} d^{d+1\!}y \; \: G(x,y)^4 ,
\label{phifunctional} \eea
where $d$ denotes the spatial dimension of the problem.

\begin{figure}[hbtp]
\begin{center}
\resizebox{0.6\columnwidth}{!}{\includegraphics{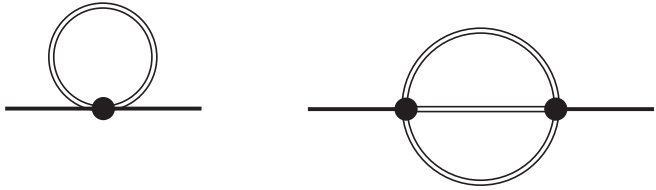} }
\end{center}
\vspace{0.5cm}
\caption[]{Self-energies of the Kadanoff-Baym equation: tadpole
self-energy (l.h.s.) and sunset self-energy (r.h.s.). Since the
lines represent full Green functions the self-energies are
self-consistent (see text) with the external coordinates indicated by
full dots.
\label{diagram_self}}
\vspace{0.3cm}
\end{figure}

This approximation corresponds to a weak coupling expansion such
that we consider contributions up to the second superficial order
in the coupling constant $\lambda$ (cf. Fig. \ref{diagram_self}).
For the superficial coupling constant order we count the explicit
coupling factors $\lambda$ associated with the visible vertices.
The hidden dependence on the coupling strength - which is
implicitly incorporated in the self-consistent Green-functions
that built up the $\Phi$-functional and the self-energies  - is
ignored on that level. For our present purpose this approximation
is sufficient since we include the leading mean-field effects as
well as the leading order scattering processes that pave the way
to thermalization.

For the actual calculation it is advantageous to change to a
single-time representation for the Green functions and
self-energies defined on the closed-time-path. In line with
 the position of the coordinates on the contour there exist
four different two-point functions
\bea i \, G^{c}(x,y) & = & i \, G^{++}(x,y) \;\; = \;\; < \, T^c
\, \{ \, \phi(x) \: \phi(y) \, \} \, > \, , \\[0.2cm] i \,
G^{<}(x,y) & = & i \, G^{+-}(x,y) \;\; = \;\;\;\, \;\;\;\: < \{ \,
\phi(y) \: \phi(x) \, \} > \, , \nnl[0.2cm] i \, G^{>}(x,y) & = &
i \, G^{-+}(x,y) \;\; = \;\;\;\, \;\;\;\: < \{ \, \phi(x) \:
\phi(y) \, \} > \, , \nnl[0.2cm] i \, G^{a}(x,y) & = & i \,
G^{--}(x,y) \;\; = \;\; < \, T^a \, \{ \, \phi(x) \: \phi(y) \, \}
\, > . \nn \label{green_def} \eea
Here $T^c \, (T^a)$ represent the (anti-)time-ordering operators
in case of both arguments lying on the (anti-)chronological branch
of the real-time contour. These four functions are not independent
of each other. In particular the non-continuous functions $G^c$
and $G^a$ are built up by the Wightman functions $G^>$ and $G^<$
and the usual $\Theta$-functions in the time coordinates. Since
for the real boson theory (\ref{lagrangian}) the relation
$G^{>}(x,y) = G^{<}(y,x)$ holds, the knowledge of the Green
functions $G^{<}(x,y)$ for all $x, y$ characterizes the system
completely. Nevertheless, we will give the equations for $G^{<}$
and $G^{>}$ explicitly since this is the familiar representation
for general field theories \cite{danmrow}.

By using the stationarity condition for the action
(\ref{effactioneom}) and resolving the time structure of the path
ordered quantities we obtain the Kadanoff-Baym equations for the
time evolution of the Wightman functions \cite{berges1,Juchem03}:
\bea
\label{kabaeqcs}
- \left[
\partial_{\mu}^{x} \partial_{x}^{\mu} \!+ m^2
\right] \, G^{\gtrless}(x,y) & = & \Sigma^{\delta}(x) \;
G^{\gtrless}(x,y) \\[0.3cm] & + & \! \int_{t_0}^{x_0} \!\!\!\!\!
dz_0 \int \!\!d^{d}\!z \;\; \left[\,\Sigma^{>}(x,z) -
\Sigma^{<}(x,z) \,\right] \: G^{\gtrless}(z,y) \nnl[0.2cm] & - &
\! \int_{t_0}^{y_0} \!\!\!\!\! dz_0 \int \!\!d^{d}\!z \;\;\,
\Sigma^{\gtrless}(x,z) \: \left[\,G^{>}(z,y) - G^{<}(z,y)
\,\right] \!, \nnl[1.5cm]
- \left[
\partial_{\mu}^{y} \partial_{y}^{\mu} \!+ m^2
\right] \, G^{\gtrless}(x,y) & = & \Sigma^{\delta}(y) \;
G^{\gtrless}(x,y) \nnl[0.3cm] & + & \! \int_{t_0}^{x_0} \!\!\!\!\!
dz_0 \int \!\!d^{d}\!z \;\; \left[\,G^{>}(x,z) - G^{<}(x,z)
\,\right] \: \Sigma^{\gtrless}(z,y) \nnl[0.2cm] & - & \!
\int_{t_0}^{y_0} \!\!\!\!\! dz_0 \int \!\!d^{d}\!z \;\;\,
G^{\gtrless}(x,z) \: \left[\,\Sigma^{>}(z,y) - \Sigma^{<}(z,y)
\,\right] \!, \nn \eea

Within the 3-loop approximation for the 2PI effective action (i.e.
the $\Phi$-functional (\ref{phifunctional})) we get two different
self-energies: In leading order of the coupling constant only the
tadpole diagram (l.h.s. of Fig. \ref{diagram_self}) contributes
and leads to the generation of an effective mass for the field
quanta. This self-energy (in coordinate space) is given by
\bea
\label{tadpole_cs}
\Sigma^{\delta}(x) \; = \; \frac{\lambda}{2} \; i \: G^{<}(x,x) \; ,
\eea\\
and is local in space and time. In next order in the coupling
constant (i.e. $\lambda^2$) the non-local sunset self-energy
(r.h.s. of Fig. \ref{diagram_self}) enters the time evolution as
\bea \label{sunset_cs} \Sigma^{\gtrless}(x,y) \; = \; -
\frac{\lambda^2}{6} \; G^{\gtrless}(x,y) \; G^{\gtrless}(x,y) \;
G^{\lessgtr}(y,x) \\[0.3cm] \longrightarrow \quad
\Sigma^{\gtrless}(x,y) \; = \; - \frac{\lambda^2}{6} \; \left[ \,
G^{\gtrless}(x,y) \, \right]^3 \, . \eea
Thus the Kadanoff-Baym equation (\ref{kabaeqcs}) in our case
includes the influence of a mean-field on the particle propagation
-- generated by the tadpole diagram -- as well as  scattering
processes as inherent in the sunset diagram.

The Kadanoff-Baym equation describes the full quantum
nonequilibrium time evolution on the two-point level for a system
prepared at an initial time $t_0$, i.e. when higher order
correlations are discarded. The causal structure of this initial
value problem is obvious since the time integrations are performed
over the past up to the actual time $x_0$ (or $y_0$, respectively)
and do not extend to the future.

Furthermore, also linear combinations of the Green functions in
single-time representation are of interest and will be exploited
for the spectral properties of the system later on. The retarded
Green function $G^R$ and the advanced Green function $G^A$ are
given as
\bea \label{defret} G^{R}(x_1,x_2) & = & \phantom{-} \, \Theta(t_1
- t_2) \; \left[ \, G^{>}(x_1,x_2) - G^{<}(x_1,x_2) \, \right]
\\[0.3cm]
& = & \phantom{-} \, \Theta(t_1 - t_2) \; \langle \, \left[ \,
\phi(x_1) \, , \, \phi(x_2) \, \right]_{-} \, \rangle \nnl[0.3cm]
& = & \phantom{-} \, G^{c}(x_1,x_2) \:-\: G^{<}(x_1,x_2)
         \;\: = \;\: G^{>}(x_1,x_2) \:-\: G^{a}(x_1,x_2) \, ,
\nnl[0.5cm] \label{defadv} G^{A}(x_1,x_2) & = & - \, \Theta(t_2 -
t_1) \; \left[ \, G^{>}(x_1,x_2) - G^{<}(x_1,x_2) \, \right]
\\[0.3cm]
& = & - \, \Theta(t_2 - t_1) \; \langle \, \left[ \, \phi(x_1) \,
, \, \phi(x_2) \, \right]_{-} \, \rangle \nnl[0.3cm] & = &
\phantom{-} \, G^{c}(x_1,x_2) \:-\: G^{>}(x_1,x_2) \;\: = \;\:
G^{<}(x_1,x_2) \:-\: G^{a}(x_1,x_2) \, . \nn \eea
These Green functions contain exclusively spectral, but no
statistical information of the system. Their time evolution is
determined by  Dyson-Schwinger equations  and given by (cf. Ref.
\cite{Juchem03})
\bea \label{dseqretcs} && - \left[
\partial_{\mu}^{x_1} \partial_{x_1}^{\mu} \!+ m^2 + \Sigma^{\delta}(x_1)
\right] \, G^{R}(x_1,x_2) \;=\; \\[0.3cm] && \qquad \qquad \qquad
\qquad \qquad \delta^{(d+1)}(x_1\!-\!x_2) \:+\: \int \!d^{d+1}\!z
\;\; \Sigma^{R}(x_1,z) \;\; G^{R}(z,x_2) \, , \nnl[0.6cm]
\label{dseqadvcs} && - \left[
\partial_{\mu}^{x_1} \partial_{x_1}^{\mu} \!+ m^2 + \Sigma^{\delta}(x_1)
\right] \, G^{A}(x_1,x_2) \:=\:\\[0.3cm] && \qquad \qquad \qquad
\qquad \qquad \delta^{(d+1)}(x_1\!-\!x_2) \:+\: \int \!d^{d+1}\!z
\;\; \Sigma^{A}(x_1,z) \;\; G^{A}(z,x_2) \, , \nn \eea
where the retarded and advanced self-energies $\Sigma^{R}$,
$\Sigma^{A}$ are defined via $\Sigma^{>}$, $\Sigma^{<}$ similar to
the Green functions. Thus the retarded (advanced) Green functions
are determined by retarded (advanced) quantities, only.

\subsubsection{Homogeneous systems in space}
In the following we will consider homogeneous systems in space. To
obtain a numerical solution the Kadanoff-Baym equation
(\ref{kabaeqcs}) is transformed to momentum space:
\bea
\label{kabaeqms}
\partial^2_{t_1} \, G^{<}(\vec{p},t_1,t_2)
& = & - [ \, \vec{p}^{\,2} + m^2 + \bar{\Sigma}^{\delta}(t_1) \, ]
\; G^{<}(\vec{p},t_1,t_2) \\[0.5cm] & - & \int_{t_0}^{t_1} \!\!\!
dt^{\prime} \; \left[ \, \Sigma^{>}(\vec{p},t_1,t^{\prime}) -
\Sigma^{<}(\vec{p},t_1,t^{\prime}) \, \right] \;
G^{<}(\vec{p},t^{\prime},t_2) \nnl[0.1cm] & + & \int_{t_0}^{t_2}
\!\!\! dt^{\prime} \; \Sigma^{<}(\vec{p},t_1,t^{\prime}) \; \left[
\, G^{>}(\vec{p},t^{\prime},t_2) - G^{<}(\vec{p},t^{\prime},t_2)
\, \right] \nnl[0.4cm] & = & - [ \, \vec{p}^{\,2} + m^2 +
\bar{\Sigma}^{\delta}(t_1) \, ] \; G^{<}(\vec{p},t_1,t_2) \; + \;
I_1^{<}(\vec{p},t_1,t_2) \nn \eea
where both memory integrals are summarized in the function
$I_1^{<}$. The equation of motion in the second time direction
$t_2$ is given analogously. In two-time and momentum space
($\vec{p},t,t'$) representation the self-energies read
\bea
\label{sems}
\bar{\Sigma}^{\delta}(t)
& = &
\frac{\lambda}{2} \, \int \!\! \frac{d^{d\!}p}{(2\pi)^d} \; \;
i \, G^{<}\!(\vec{p},t,t) \; ,
\\[0.5cm]
\Sigma^{\gtrless}\!(\vec{p},t,t^{\prime}) & = & -
\frac{\lambda^2}{6} \int \!\! \frac{d^{d\!}q}{(2\pi)^{d}} \! \int
\!\! \frac{d^{d\!}r}{(2\pi)^{d}} \;\; G^{\gtrless}\!(\vec{q}
,t,t^{\prime}) \;\; G^{\gtrless}\!(\vec{r} ,t,t^{\prime}) \;\;
G^{\lessgtr}\!(\vec{q} \!+\! \vec{r} \!-\! \vec{p}
,t^{\prime}\!,t) \, . \nnl[0.2cm] & = & - \frac{\lambda^2}{6} \int
\!\! \frac{d^{d\!}q}{(2\pi)^{d}} \! \int \!\!
\frac{d^{d\!}r}{(2\pi)^{d}} \;\; G^{\gtrless}\!(\vec{q}
,t,t^{\prime}) \;\; G^{\gtrless}\!(\vec{r} ,t,t^{\prime}) \;\;
G^{\gtrless}\!(\vec{p} \!-\! \vec{q} \!-\! \vec{r} ,t,t^{\prime})
\, . \nn \eea

\subsection{Numerical studies on equilibration}

\subsubsection{Numerical implementation}
For the solution of the Kadanoff-Baym equations (\ref{kabaeqms}) a
flexible and accurate algorithm has been developed: Instead of
solving the second order differential equation (\ref{kabaeqms})
one can generate a set of first order differential equations for
the Green functions in the Heisenberg picture,
\bea \label{gfdefall} i \, G_{\phi \phi}^{<}(x_1,x_2) & = & <
\phi(x_2) \, \phi(x_1) > \: = \: i \, G^{<}(x_1,x_2) \; ,
\\[0.3cm]
i \, G_{\pi \phi}^{<}(x_1,x_2) & = & < \phi(x_2) \, \pi(x_1) > \:
= \:
\partial_{t_1} \, i \, G_{\phi \phi}^{<}(x_1,x_2) \; ,
\nnl[0.3cm]
i \, G_{\phi \pi}^{<}(x_1,x_2) & = & < \pi(x_2) \, \phi(x_1) > \:
= \: \partial_{t_2} \, i \, G_{\phi \phi}^{<}(x_1,x_2) \; ,
\nnl[0.3cm]
i \, G_{\pi \pi}^{<}(x_1,x_2) & = & < \pi(x_2) \, \pi(x_1) > \: =
\: \partial_{t_1} \, \partial_{t_2} \, i \, G_{\phi
\phi}^{<}(x_1,x_2) \; , \nn \eea
with the canonical field momentum $\pi(x) = \partial_{x_0}
\phi(x)$. The first index $\pi$ or $\phi$ is always related to the
first space-time argument. Exploiting the time-reflection symmetry
of the Green functions some of the differential equations are
redundant. The required equations of motion are given as
\bea \label{eomall}
\partial_{t_1} \, G_{\phi \phi}^{<}(\vec{p},t_1,t_2)
& = & G_{\pi \phi}^{<}(\vec{p},t_1,t_2) \; ,
\\[0.4cm]
\partial_{\bar{t}} \: G_{\phi \phi}^{<}(\vec{p},\bar{t},\bar{t})
& = & 2 \, i \; Im \, \{ \, G_{\pi
\phi}^{<}(\vec{p},\bar{t},\bar{t}) \, \} \; , \nnl[0.6cm]
\partial_{t_1} \, G_{\pi \phi}^{<}(\vec{p},t_1,t_2)
& = & - \, \Omega^2(t_1) \; G_{\phi \phi}^{<}(\vec{p},t_1,t_2) \;
+ \; I_1^{<}(\vec{p},t_1,t_2) \; , \nnl[0.4cm]
\partial_{t_2} \, G_{\pi \phi}^{<}(\vec{p},t_1,t_2)
& = & G_{\pi \pi}^{<}(\vec{p},t_1,t_2) \; , \nnl[0.4cm]
\partial_{\bar{t}} \: G_{\pi \phi}^{<}(\vec{p},\bar{t},\bar{t})
& = & - \, \Omega^2(\bar{t}) \; G_{\phi
\phi}^{<}(\vec{p},\bar{t},\bar{t}) \; + \; G_{\pi
\pi}^{<}(\vec{p},\bar{t},\bar{t}) \; + \;
I_1^{<}(\vec{p},\bar{t},\bar{t}) \; , \nnl[0.6cm]
\partial_{t_1} \, G_{\pi \pi}^{<}(\vec{p},t_1,t_2)
& = & - \, \Omega^2(t_1) \; G_{\phi \pi}^{<}(\vec{p},t_1,t_2) \; +
\; I_{1,2}^{<}(\vec{p},t_1,t_2) \; , \nnl[0.4cm]
\partial_{\bar{t}} \: G_{\pi \pi}^{<}(\vec{p},\bar{t},\bar{t})
& = & - \, \Omega^2(\bar{t}) \; 2 \, i \; Im \, \{ \, G_{\pi
\phi}^{<}(\vec{p},\bar{t},\bar{t}) \, \} \; + \; 2 \, i \; Im \,
\{ \, I_{1,2}^{<}(\vec{p},\bar{t},\bar{t}) \, \} \; , \nn \eea
where $\bar{t} = (t_1 + t_2)/2$ is the mean time variable. Thus
one explicitly considers the propagation in the time diagonal (cf.
Ref. \cite{koe1}). In the equations of motion (\ref{eomall}) the
current (renormalized) effective energy including the time
dependent tadpole contribution enters,
\bea \Omega^2(t) \; = \; \vec{p}^{\,2} \; + \; m^2 \; + \; \delta
m^2_{tad} \; + \; \delta m^2_{sun} \; + \;
\bar{\Sigma}^{\delta}(t) , \eea with $\delta m^2_{tad}$ and
$\delta m^2_{sun}$ specified in the next Subsection. The evolution
in the $t_2$ direction has not be taken into account for
$G^<_{\phi \phi}$ and $G^<_{\pi \pi}$ since the Green functions
beyond the time diagonal ($t_2 > t_1$) are determined via the time
reflection symmetry $G^<_{\cdot \cdot}(\vec{p},t_1,t_2) = - [ \,
G^<_{\cdot \cdot}(\vec{p},t_2,t_1) \, ]^{*}$ from the known values
for the lower time triangle in both cases. Since there is no time
reflection symmetry for the $G_{\pi \phi}$ functions, they have to
be calculated (and stored) in the whole $t_1$, $t_2$ range.
However, we can ignore the evolution of $G_{\phi \pi}$ since it is
obtained by the relation $G^{<}_{\phi \pi}(\vec{p},t_1,t_2) = - [
\, G^{<}_{\pi \phi}(\vec{p},t_2,t_1) \,]^{*}$. The correlation
integrals in (\ref{eomall}) are given by
\bea \label{corrint} I_1^{<}(\vec{p},t_1,t_2) \: = \: & - &
\!\!\!\! \int_{0}^{t_1} \!\!\! dt^{\prime} \; \; \left[ \,
\Sigma^{>}(\vec{p},t_1,t^{\prime}) -
\Sigma^{<}(\vec{p},t_1,t^{\prime}) \, \right] \; \; G_{\phi
\phi}^{<}(\vec{p},t^{\prime},t_2) \\[0.0cm]
& + & \!\!\!\! \int_{0}^{t_2} \!\!\! dt^{\prime} \; \;
\Sigma^{<}(\vec{p},t_1,t^{\prime}) \; \; \left[ \, G_{\phi
\phi}^{<}(-\vec{p},t_2,t^{\prime}) - G_{\phi \phi}^{<}(
\vec{p},t^{\prime},t_2) \, \right] \; , \nnl[0.6cm]
\label{numreali12} I_{1,2}^{<}(\vec{p},t_1,t_2) \: \equiv \: && \!
\! \! \!
\partial_{t_2} I_{1}^{<}(\vec{p},t_1,t_2)
\\[0.2cm]
\: = \: & - & \!\!\!\! \int_{0}^{t_1} \!\!\! dt^{\prime} \; \;
\left[ \, \Sigma^{>}(\vec{p},t_1,t^{\prime}) -
\Sigma^{<}(\vec{p},t_1,t^{\prime}) \, \right] \; \; G_{\phi
\pi}^{<}(\vec{p},t^{\prime},t_2) \nnl[0.0cm]
& + & \!\!\!\! \int_{0}^{t_2} \!\!\! dt^{\prime} \; \;
\Sigma^{<}(\vec{p},t_1,t^{\prime}) \; \; \left[ \, G_{\pi
\phi}^{<}(-\vec{p},t_2,t^{\prime}) - G_{\phi \pi}^{<}(
\vec{p},t^{\prime},t_2) \, \right] \; .  \nn \eea
In (\ref{eomall}) and (\ref{numreali12}) one can replace
$G^{<}_{\phi \pi}(\vec{p},t_1,t_2) = - [ \, G^{<}_{\pi
\phi}(\vec{p},t_2,t_1) \,]^{*}$ such that the set of equations is
closed in the Green functions $G^{<}_{\phi \phi}$, $G^{<}_{\pi
\phi}$ and $G^{<}_{\pi  \pi }$.

The disadvantage, to integrate more Green functions in time in
this first-order scheme, is compensated by its good accuracy. As
mentioned before, we especially take into account the propagation
along the time diagonal which leads to an improved numerical
precision. The set of differential equations (\ref{eomall}) is
solved by means of a 4th order Runge-Kutta algorithm. For the
calculation of the self energies  a Fourier method similar to
Refs. \cite{dan84,koe1} is applied. The self energies
(\ref{sems}), furthermore, are calculated in coordinate space
where they are products of coordinate-space Green functions (that
are available by Fourier transformation) and finally transformed
to momentum space.

\subsubsection{Renormalization of $\phi^4$-theory in 2+1
dimensions}
In 2+1 space-time dimensions both self-energies
(\ref{tadpole_cs}), (\ref{sunset_cs})  incorporated in the present
case are ultraviolet divergent. Since the particles have a finite
mass no problems arise from the infrared momentum regime. The
ultraviolet regime, however, has to be treated explicitly.

For the renormalization of the divergences we only assume that the
time-dependent nonequilibrium distribution functions are
decreasing for large momenta comparable to the equilibrium
distribution functions, i.e exponentially. Thus one can apply the
conventional finite temperature renormalization scheme. By
separating the real time (equilibrium) Green functions into vacuum
($T=0$) and thermal parts it becomes apparent, that only the pure
vacuum contributions of the self-energies are divergent. For the
linear divergent tadpole diagram  a mass counter term (at the
renormalized mass $m$) can be introduced as
\bea \delta\!m^2_{tad} \; = \; \int \frac{d^2\!p}{(2\pi)^2} \;
\frac{1}{2 \omega_{\vec{p}}} \; , \qquad \qquad \omega_{\vec{p}} =
\sqrt{\vec{p}^2 + m^2} \; , \label{countermass_tadpole} \eea
which cancels the contribution from the momentum integration of
the vacuum part of the Green function.

In case of the sunset diagram only the logarithmically divergent
pure vacuum part requires a renormalization, while it remains
finite as long as at least one temperature line is involved.
Contrary to the case of 3+1 dimensions it is not necessary to
employ the involved techniques developed for the renormalization
of self-consistent theories (in equilibrium) in Refs.
\cite{knollren1} because in 2+1 dimensions $\Phi^4$ theory is
superrenormalizable. Since the divergence only appears (in
energy-momentum space) in the real part of the Feynman self-energy
$\Sigma^{c}$ at $T=0$ (and equivalently in the real part of the
retarded/advanced self-energies $\Sigma^{ret/adv}$), it can be
absorbed by another mass counterterm
\bea \delta\!m^2_{sun} & = & - Re\,\Sigma^{c}_{T=0}(p^2) \: = \: -
Re\,\Sigma^{ret/adv}_{T=0}(p^2) \\[0.2cm] & = &
\frac{\lambda^2}{6} \int \!\!\! \frac{d^2\!q}{(2 \pi)^2} \, \int
\!\!\! \frac{d^2\!r}{(2 \pi)^2} \; \, \frac{1}{4
\,\omega_{\vec{q}} \, \omega_{\vec{r}} \,
\omega_{\vec{q}+\vec{r}-\vec{p}}} \; \; \frac{
\omega_{\vec{q}}\!+\!\omega_{\vec{r}}\!+
\!\omega_{\vec{q}+\vec{r}-\vec{p}} } { [ \,
\omega_{\vec{q}}\!+\!\omega_{\vec{r}}\!+
\!\omega_{\vec{q}+\vec{r}-\vec{p}} \, ]^2 - p_0^2 } \nn
\label{countermass_sunset} \eea
at given four-momentum $p=(p_0,\vec{p})$ and renormalized mass
$m$.
\begin{figure}[t]
\begin{center}
\resizebox{1.0\columnwidth}{!}{\includegraphics{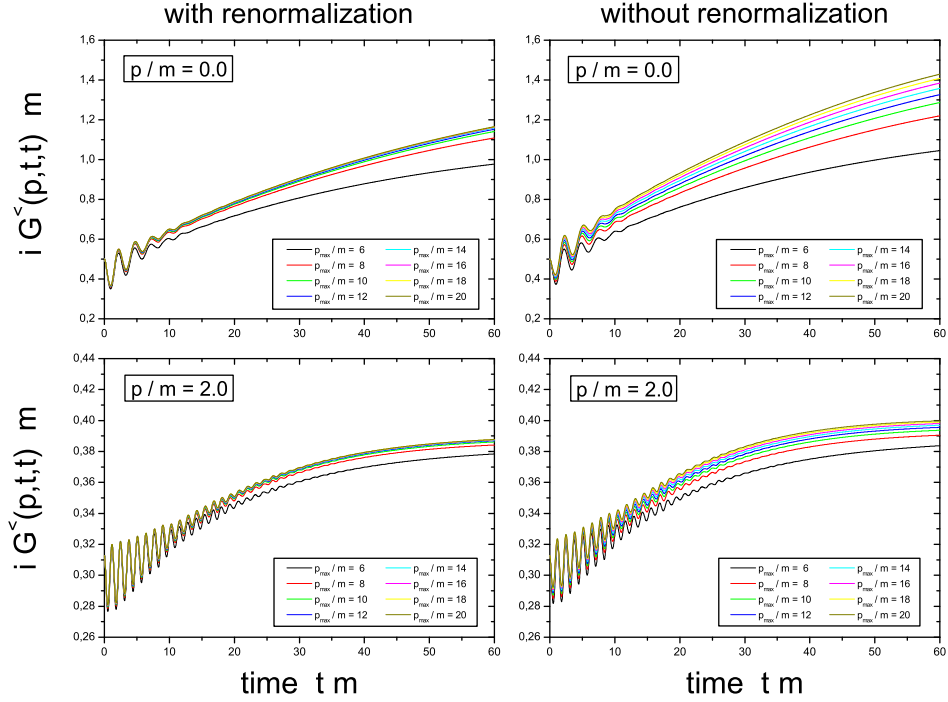} }
\end{center}
\vspace{-1.0cm} \caption[]{ Time evolution of two momentum modes
$|\,\vec{p}\,|/m = 0.0$, $|\,\vec{p}\,|/m = 2.0$ of the equal-time
Green function starting from the initial distribution D2 (as
specified in Section 1.2.3) with coupling constant $\lambda / m$ =
14. One observes, that with the renormalization of the sunset
diagram (left plots) a proper limit is obtained when increasing
the momentum cut-off $p_{max} / m = 6 (2) 24$, while  without
renormalization (right plots) the curves tend to infinity when
increasing the ultraviolet cut-off \label{plot_renorm}}.
\end{figure}

In Fig. \ref{plot_renorm} we demonstrate the applicability of the
renormalization prescription. To this aim we display two momentum
modes $|\,\vec{p}\,|/m = 0.0$ (upper plots) and $|\,\vec{p}\,|/m$
= 2.0 (lower plots) of the equal-time Green function $i
G^{<}(|\vec{p}|,t,t)$ for various momentum cut-offs $p_{max}/m =
6\,(2)\,20$ with (left plots) and without (right plots)
renormalization of the sunset self-energy. For both cases the
renormalization of the tadpole diagram has been used. We mention,
that a non-renormalization of the tadpole self-energy has even
more drastic consequences in accordance with the linear degree of
divergence.  For the non-renormalized calculations -- with respect
to the sunset diagram -- we observe that both momentum modes do
not converge with increasing momentum space cut-off. In fact, all
lines tend to infinity when the maximum momentum is enlarged
(since the gridsize of the momentum grid is kept constant).
Although the divergence as a function of the momentum cut-off is
rather weak - in accordance with the logarithmic divergence of the
sunset self-energy in 2+1 space-time dimensions - a proper
ultraviolet limit is not obtained.

This problem is cured by the  sunset mass counter term
(\ref{countermass_sunset}) as seen on the left  side of Figure
\ref{plot_renorm}. For the momentum mode $|\,\vec{p}\,|/m = 2.0$
the calculations converge to a limiting curve with increasing
momentum cut-off. Even for the more selective case of the
$|\,\vec{p}\,|/m = 0.0$ mode of the equal-time Green function the
convergence is established. We point out that this limit is
obtained for the unequal-time Green functions as well (not shown
here explicitly). In fact, it turns out that the equal-time
functions provide the most crucial test for the applicability of
the renormalization prescription, since the divergent behaviour
appears to be less pronounced for the propagation along a single
time direction $t_1$ or $t_2$. Thus we can conclude that the
 renormalization scheme introduced above, i.e. including mass counter
terms for the divergent tadpole and sunset self-energies, leads to
ultraviolet stable results.

In the following Subsections we will use the 'bare' mass $m$ = 1,
which implies that times are given in units of the inverse mass or
$t \cdot m$ is dimensionless. Accordingly, the bare coupling
$\lambda$ in (7) is given in units of the mass $m$ such that
$\lambda/m$ is dimensionless, too.

\subsubsection{Initial conditions}
\begin{figure}[hbt]
\resizebox{0.55\columnwidth}{!}{\includegraphics{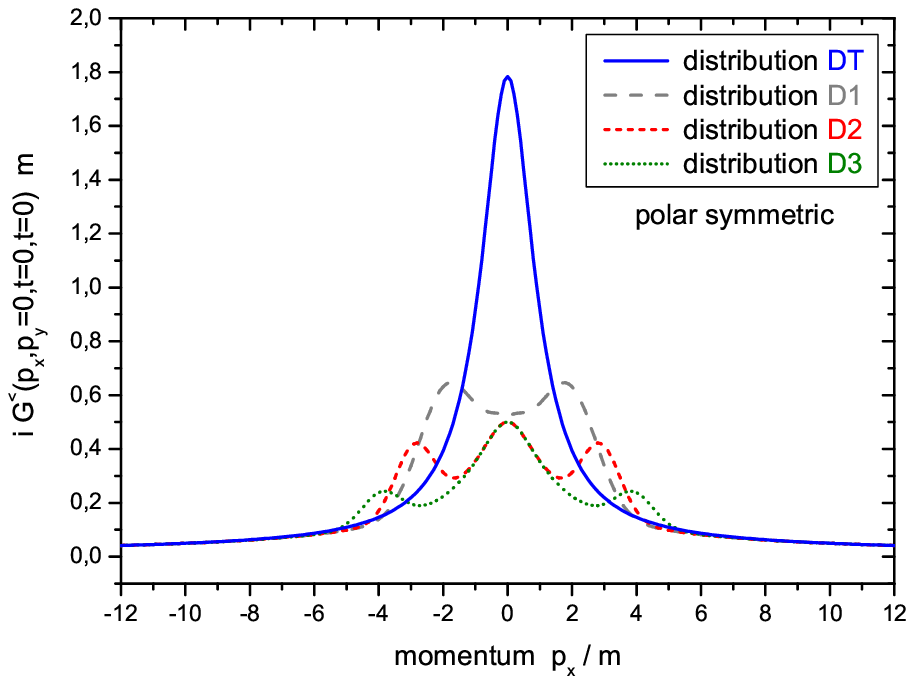} }
\hspace*{-10mm}
\resizebox{0.55\columnwidth}{!}{\includegraphics{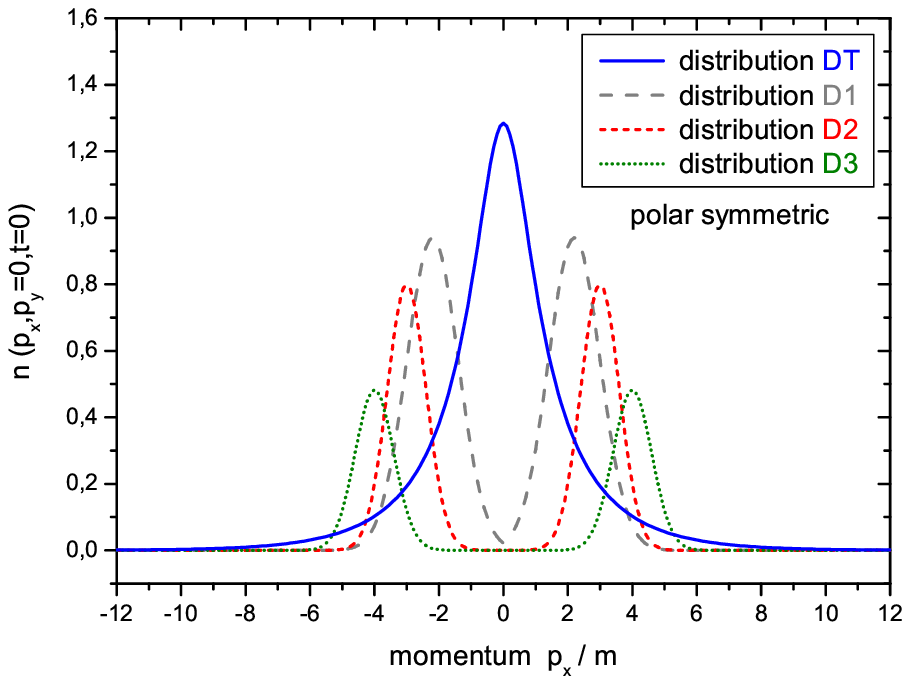} }
\caption{ Initial Green functions $i G^{<}(|\,\vec{p}\,|,t=0,t=0)$
(l.h.s.) and corresponding initial distribution functions
$n(|\,\vec{p}\,|,t=0)$ (r.h.s.) for the distributions D1, D2, D3
and DT in momentum space (for a cut of the polar symmetric
distribution in $p_x$ direction for $p_y = 0$).
\label{plot_ini01}}
\end{figure}

In order to investigate equilibration phenomena on the basis of
the Kadanoff-Baym equations for the 2+1 dimensional problem, one
first has to specify the initial conditions for the time
integration. To this aim we consider four different initial
distributions that are all characterized by the same energy
density. Consequently, for large times ($\rightarrow \infty$) all
initial value problems should lead to the same equilibrium final
state. The initial equal-time Green functions $i G^{<}({\bf
p},t=0,t=0)$ adopted are displayed in Fig. \ref{plot_ini01}
(l.h.s.) as a function of the momentum $p_z$. We first concentrate
on polar symmetric configurations due to the large numerical
expense for this first investigation\footnote{In Section 1.4  we
will present also calculations for non-symmetric systems.}. Since
the equal-time Green functions $G^{<}({\bf p},t,t,)$ are purely
imaginary we show only the real part of $i \, G^{<}$ in Fig.
\ref{plot_ini01}.

\phantom{a}\vspace*{-5mm}

Furthermore, the corresponding initial distribution functions in
the occupation density $n({\bf p},t=0)$, related to $i G^{<}({\bf
p},t=0,t=0)$ via
\begin{equation}
\label{ew}
 2 \omega_{{\bf p}} i G^{<}({\bf p},t=0,t=0) = 2
n({\bf p},t=0) +1 ,
\end{equation}
are shown in Fig. \ref{plot_ini01} on the r.h.s. While the initial
distributions D1, D2, D3 have the shape of (polar symmetric)
'tsunami' waves with maxima at different momenta in $p_z$, the
initial distribution DT corresponds to a free Bose gas at a given
initial temperature $T_0$ that is fixed by the initial energy
density. According to (\ref{ew}) the difference between the Green
functions and the distribution functions is basically given by the
vacuum contribution, which has its maximum at small momenta. Thus
even for the distributions D1, D2, D3 the corresponding Green
functions are non-vanishing for $\bf{p} \approx$ 0.

Since we consider a finite volume $V=a^2$ we work in a basis of
momentum modes characterized by the number of nodes in each
direction. The number of momentum modes is typically in the order
of 40 which is found to be sufficient for numerically stable
results. For times $t < 0$ we consider the systems to be
noninteracting and switch on the interaction ($\sim \lambda$) for
$t$=0 to explore the quantum dynamics of the interacting system
for $t> 0$.

\subsubsection{Equilibration in momentum space}
\begin{figure}[hbt]
 \vspace*{-1 cm} \hspace*{1.0 cm}
 \resizebox{0.8\columnwidth}{!}{\includegraphics{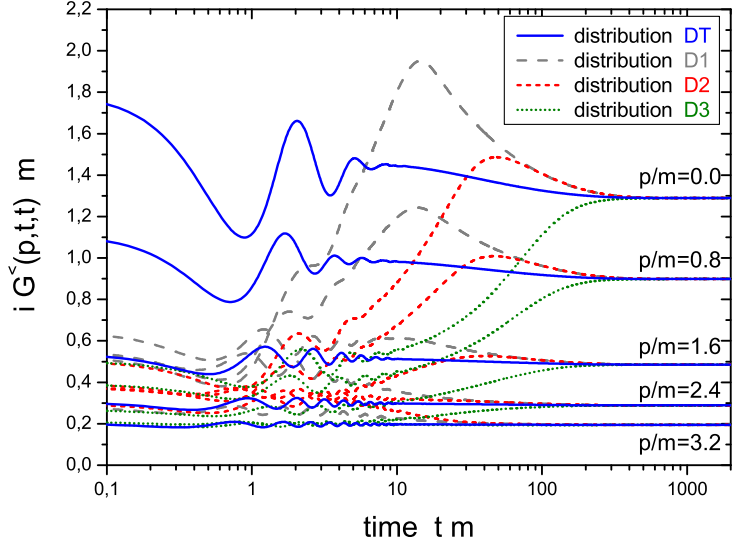} }
\vspace*{-5mm}
\caption{Time evolution of selected momentum modes of the
equal-time Green function $|\,\vec{p}\,|/m = 0.0, 0.8, 1.6, 2.4,
3.2$ (from top to bottom) for four different initial
configurations D1, D2, D3, and DT (characterized by the different
line types) with the same energy density. For the rather strong
coupling constant $\lambda/m = 18$ the initial oscillations - from
switching on the interaction at $t$=0 - are damped rapidly and
 disappear for $t \cdot m > 20$. Finally, all momentum modes
assume the same respective equilibrium value for long times
$t\cdot m > 500$) independent of the initial state.
\label{plot_equi01}}
\end{figure}

The time evolution of various (selected) momentum modes of the
equal-time Green function for the different initial states D1, D2,
D3 and DT is shown in Fig. \ref{plot_equi01}, where the
dimensionless time $t\cdot m$ is displayed on a logarithmic scale.
 We observe that starting from very different initial conditions
- as introduced above - the single momentum modes converge to the
same respective occupation numbers for large times as
characteristic for a system in equilibrium. As noted above, the
initial energy density is the same for all distributions and
energy conservation is fulfilled strictly in the time integration
of the Kadanoff-Baym equations. The different momentum modes in
Fig. \ref{plot_equi01} typically show a three-phase structure: For
small times ($t\cdot m < 10$) one finds damped oscillations that
can be identified with a typical switching-on effect at $t$=0,
where the system  is excited by a sudden increase of the coupling
constant to $\lambda/m = 18$. The damping of the initial
oscillations depends on the coupling strength $\lambda/m$ and is
more pronounced for strongly coupled systems. We mention that for
very strong interactions $\lambda/m > 30$ the initial oscillations
are even hard to recognize.

For 'intermediate' time scales ($10 < t\cdot m < 500$) one
observes a strong change of all momentum modes in the direction of
the final stationary state. We address this phase to 'kinetic'
equilibration and point out, that - depending on the initial
conditions and the coupling strength - the momentum modes can
temporarily even exceed their respective equilibrium value. This
can be seen explicitly for the lowest momentum modes ($p/m = 0$ or
$=0.8$) of the distribution D1 in Fig. \ref{plot_equi01}, which
possesses initially maxima at  small momentum. Especially the
momentum mode $|\vec{p}|=0$ of the equal-time Green function
$G^{<}$, which starts at around 0.52, is rising to a value of
$\sim$1.95 before decreasing again to its equilibrium value of
$\sim$ 1.29. Thus the time evolution towards the final equilibrium
value is -- after an initial phase with damped oscillations -- not
necessarily monotonic. For different initial conditions this
behaviour may be weakened significantly as seen for example in
case of the initial distribution D2 in Figure \ref{plot_equi01}.
Coincidently, both calculations D1 and D2 show approximately the
same equal-time Green function values for times  $t \cdot m \geq$
80. Note, that for the initial distribution D3 the non-monotonic
behaviour is not seen any more.

In general, one observes that only initial distributions (of the
well type) show this feature during their time evolution, if the
maximum is located at sufficiently small momenta. Initial
configurations like the distribution DT - where the system at
$t=0$ is given by a free gas of particles at a temperature $T_0$ -
do not show this property. Although the DT distribution is not the
equilibrium state of the interacting theory, the actual numbers
are much closer to the equilibrium state of the interacting system
than the initial distributions D1, D2 and D3. Therefore, the
evolution for DT proceeds less violently. We point out, that in
contrast to the calculations performed for $\phi^4$-theory in 1+1
space-time dimensions \cite{berges3} we find no power law
behaviour for intermediate time scales.

The third phase, i.e. the late time  evolution ($t\cdot m > 500$)
is characterized by a smooth approach of the single momentum modes
to their respective equilibrium values. As we will see in Section
1.4 this phase is adequately characterized by chemical
equilibration processes.

\subsection{The different phases of quantum
equilibration}

\subsubsection{Build-up of initial correlations}
\begin{figure}[ht]
 \vspace*{-1.0 cm} \hspace*{2.0 cm}
 \resizebox{0.8\columnwidth}{!}{\includegraphics{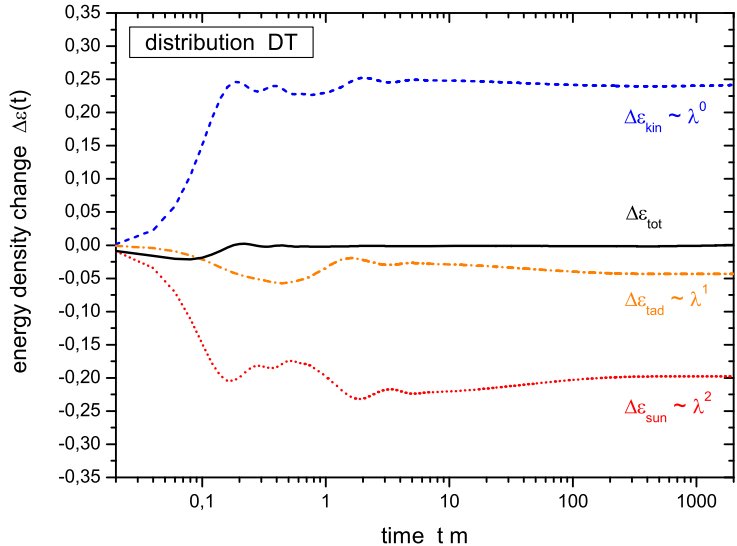} }
\vspace*{-5mm}
\caption{Change of the different contributions to the total energy
density in time. The sunset energy density $\varepsilon_{sun}$
decreases rapidly in time; this contribution  is approximately
compensated by an increase of the kinetic energy density
$\varepsilon_{kin}$. Together with the smaller tadpole
contribution $\varepsilon_{tad}$ the total energy
$\varepsilon_{tot}$ is conserved. \label{plot_energy01}}
\end{figure}
The time evolution of the interacting system within the
Kadanoff-Baym equations is characterized by the build-up of early
correlations. This can be seen from Fig. \ref{plot_energy01} where
all  contributions to the energy-density are displayed separately
as a function of time and the total energy density
$\varepsilon_{tot}(t=0)$ is subtracted. The kinetic energy density
$\varepsilon_{kin}$ is represented by all parts of
$\varepsilon_{tot}$ that are independent of the coupling constant
($\propto \lambda^0$). All terms proportional to $\lambda^1$ are
summarized by the tadpole energy density $\varepsilon_{tad}$
including the actual tadpole term as well as the corresponding
tadpole mass counterterm (cf. Section 1.2.2). The contributions
from the sunset diagram $(\propto \lambda^2)$ are represented by
the sunset energy density $\varepsilon_{sun}$.
\bea \varepsilon_{tot}(t) & = & \varepsilon_{kin}(t) \; + \;
\varepsilon_{tad}(t) \; + \; \varepsilon_{sun}(t) \; ,
\\[0.6cm]
\varepsilon_{kin}(t) & = & \phantom{-} \frac{1}{2} \, \int \!
\frac{d^{d}p}{(2\pi)^d} \; \; ( \, \vec{p}^{\,2} + m_0^2 \, ) \;
\;  i \, G^{<}_{\phi \phi}(\vec{p},t,t) \; + \; \frac{1}{2} \,
\int \! \frac{d^{d}p}{(2\pi)^d} \; \; i \, G^{<}_{\pi
\pi}(\vec{p},t,t) \; , \nnl[0.4cm]
\varepsilon_{tad}(t) & = & \phantom{-} \frac{1}{4} \, \int \!
\frac{d^{d}p}{(2\pi)^d} \; \; \bar{\Sigma}_{tad}(t) \; \; i \,
G^{<}_{\phi \phi}(\vec{p},t,t) \: + \: \frac{1}{2} \, \int \!
\frac{d^{d}p}{(2\pi)^d} \; \; \delta m_{tad}^2 \; \; i \,
G^{<}_{\phi \phi}(\vec{p},t,t) \; , \nnl[0.4cm]
\varepsilon_{sun}(t) & = &
 -
\frac{1}{4} \, \int \! \frac{d^{d}p}{(2\pi)^d} \; \; i \,
I_1^{<}(\vec{p},t,t) \: + \: \frac{1}{2} \, \int \!
\frac{d^{d}p}{(2\pi)^d} \; \; \delta m_{sun}^2 \; \; i \,
G^{<}_{\phi \phi}(\vec{p},t,t) \; . \nn \eea
The calculation in Fig. \ref{plot_energy01} has been performed for
the initial distribution DT (which represents a free gas of Bose
particles at temperature $T_0 \approx 1.736 \, m$) with a coupling
constant of $\lambda/m = 18$. This state is stationary in the
well-known Boltzmann limit (cf. Section 1.4), but it is not for
the Kadanoff-Baym equation. In the full quantum calculations the
system evolves from the uncorrelated initial state and the
correlation energy density $\varepsilon_{sun}$ decreases rapidly
with time. The decrease of the correlation energy
$\varepsilon_{sun}$ which is -- with exception of the sunset mass
counterterm contribution -- initially zero is approximately
compensated by an increase of the kinetic energy density
$\varepsilon_{kin}$. Since the kinetic energy increases in the
initial phase, the final temperature $T_f$ is slightly higher than
the initial 'temperature' $T_0$. The remaining difference is
compensated by the tadpole energy density $\varepsilon_{tad}$ such
that the total energy density is conserved.

While the sunset energy density and the kinetic energy density
always show a time evolution comparable to Fig.
\ref{plot_energy01} the  change of the tadpole energy density
depends on the initial configuration and may be positive as well.
 Since the self energies are obtained within a
$\Phi$-derivable scheme the fundamental conservation laws, as e.g.
energy conservation, are respected to all orders in the coupling
constant. When neglecting the $\propto \lambda^2$ sunset
contributions and starting with a non-static initial state of
identical energy density  one observes the same compensating
behaviour  between the kinetic and the tadpole terms.

\subsubsection{Time evolution of the spectral function}

Within the Kadanoff-Baym calculations the full quantum information
of the two-point functions is retained. Consequently, one has
access to the spectral properties of the nonequilibrium system
during its time evolution. The spectral function $A(x,y)$ in our
case is given by
\bea A(x,y) \: = \: < \, [ \, \phi(x) \, , \, \phi(y) \, ]_{-} \,
> \: = \: i \,
\left[ \, G^{>}(x,y) \: - \: G^{<}(x,y) \, \right] .
\label{spec_def} \eea
From the dynamical calculations  the spectral function in
Wigner-space for each system time $T=(t_1\!\!+\!t_2)/2$ is
obtained via Fourier transformation with respect to the relative
time coordinate $\Delta t = t_1\!\!-\!t_2$:
\bea A(\vec{p}, p_0, T) \: = \: \int_{-\infty}^{\infty} \!\!\!\!
d\Delta t \; \; e^{i \Delta t \, p_0} \; \; A(\vec{p},
t_1=T\!+\!\Delta t/2, t_2=T\!-\!\Delta t/2) . \label{spec_fourier}
\eea
Note that a damping of the function $A(\vec{p},t_1,t_2)$ in
relative time $\Delta t$ corresponds to  a finite width $\Gamma$
of the spectral function in Wigner-space. This width in turn can
be interpreted as the inverse life time of the interacting scalar
particle. We recall, that the spectral function  -- for all
 times $T \equiv t$ and for all momenta $\vec{p}$ -- obeys the normalization
\bea \int_{-\infty}^{\infty} \frac{dp_0}{2 \pi} \ p_0 \
A(\vec{p},p_0,T) \; = \; 1 \qquad \forall \; \vec{p},\,T
\label{specnorm} \eea
which is nothing but a reformulation of the equal-time commutation
relation (or quantization) for the fields.
\begin{figure}[hbpt]
\begin{center}
\resizebox{0.9\columnwidth}{!}{\includegraphics{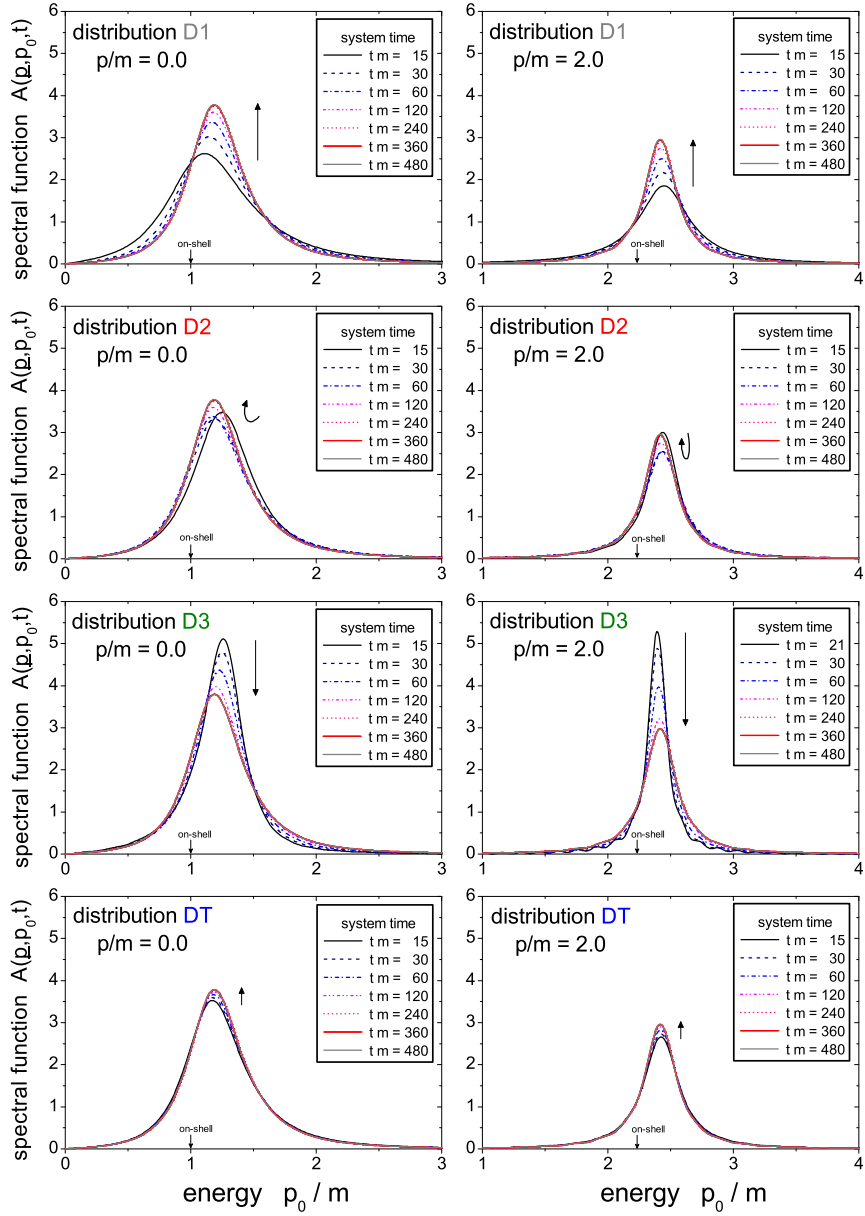}}
\end{center}
\caption[]{Time evolution of the spectral function
$ A(\vec{p}, p_0, T) $ for the initial distributions D1, D2, D3
and DT (from top to bottom) for the two momenta $| \, \vec{p} \, |
/ m = 0.0$ (l.h.s.) and $| \, \vec{p} \, | / m = 2.0$ (r.h.s.).
The spectral function is shown for several times $t \cdot m =$ 15,
30, 60, 120, 240, 360, 480 as indicated by the different line
types. \label{plot_spec01}}
\end{figure}

In Fig. \ref{plot_spec01} we display the time evolution of the
spectral function for the  initial distributions D1, D2 and DT for
two different momentum modes $| \, \vec{p} \, | / m = 0.0$ and $|
\, \vec{p} \, | / m = 2.0$. Since the spectral functions are
antisymmetric in energy for the  momentum symmetric configurations
considered, i.e. $A(\vec{p},-p_0,T) = - A(\vec{p},p_0,T)$,
 we only show the positive energy part. For our initial value problem in
two-times and space the Fourier transformation
(\ref{spec_fourier}) is restricted for system times $T$ to an
interval $\Delta t \in [-2T,2T\,]$. Thus in the very early phase
the spectral function assumes a finite width already due to the
limited support of the Fourier transform in the interval $\Delta t
\in [-2T,2T\,]$ and a Wigner representation is not very
meaningful. We, therefore, present the spectral functions for
various system times $T\equiv t$ starting from $t \cdot m = 15$ up
to $t \cdot m = 480$.

For the free thermal initialization DT the evolution of the
spectral function is very smooth and comparable to the smooth
evolution of the equal-time Green function as discussed in Section
1.2.4. In this case the spectral function is already close to the
equilibrium shape at small times being initially only slightly
broader than for late times. The maximum of the spectral function
 (for all momenta) is higher than the (bare) on-shell value and
nearly keeps its position during the whole time evolution. This
results from a positive tadpole mass shift, which is only partly
compensated by a downward shift originating from the sunset
diagram.
%
%
\begin{figure}[hbpt]
\begin{center}
\resizebox{0.48\columnwidth}{!}{\includegraphics{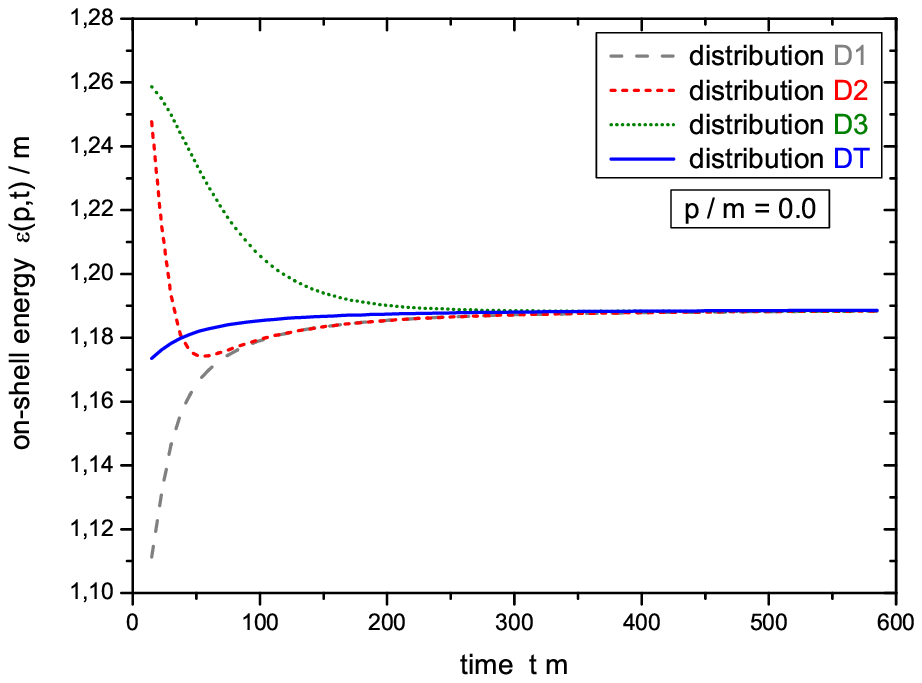} }
\resizebox{0.48\columnwidth}{!}{\includegraphics{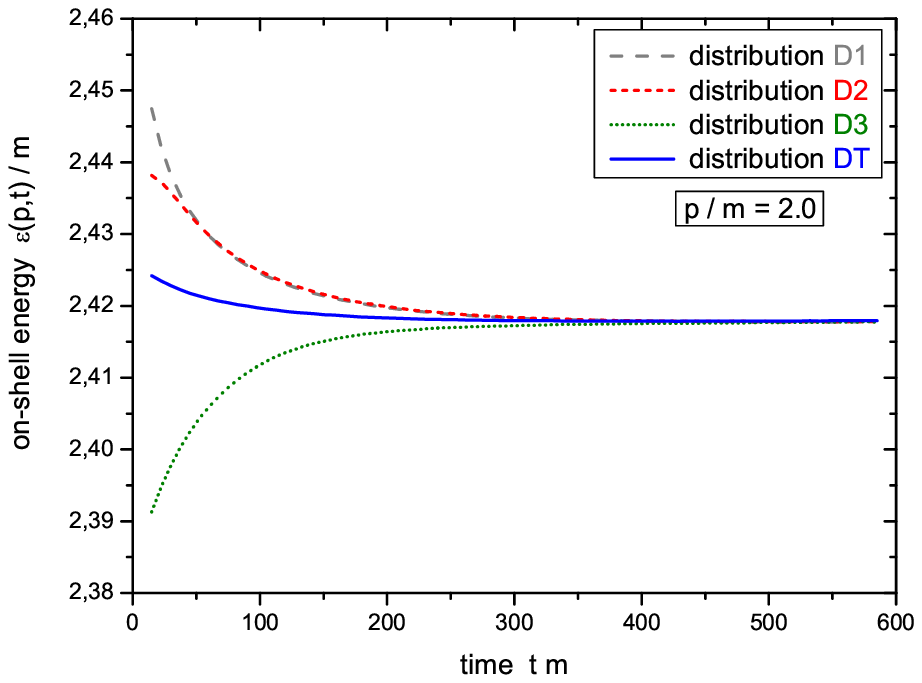} }
\end{center}
\vspace{-0.0cm} \caption[]{Time evolution of the on-shell energies
$\varepsilon(\vec{p},t)$ of the momentum modes $|\,\vec{p}\,|/m =
0.0$ and $|\,\vec{p}\,|/m = 2.0$ for the different initializations
D1, D2, D3 and DT. The on-shell self energies are extracted from
the maxima of the time-dependent spectral functions.
\label{plot_ose01}}
\end{figure}
The time evolution for the initial distributions D1, D2 and D3 has
a richer structure. For the  distribution D1 the spectral function
is broad for small system times (see the line for $t \cdot m =
15$) and becomes a little sharper in the course of the time
evolution (as presented for the momentum mode $|\,\vec{p}\,| / m =
0.0$ as well as for $|\,\vec{p}\,| / m = 2.0$). In line with the
decrease in width the height of the spectral function is
increasing (as demanded by the normalization property
(\ref{specnorm})). This is indicated by the small arrow close to
the peak position. Furthermore, the maximum of the spectral
function (which is approximately the on-shell energy) is shifted
slightly upwards for the zero mode and downwards for the mode with
higher momentum. Although the real part of the (retarded) sunset
self-energy leads (in general) to a lowering of the effective
mass, the on-shell energy of the momentum modes is still higher
than the one for the initial mass $m$ (indicated by the 'on-shell'
arrow) due to the positive mass shift from the tadpole
contribution.
%
\begin{figure}[hbpt]
\begin{center}
\resizebox{0.48\columnwidth}{!}{\includegraphics{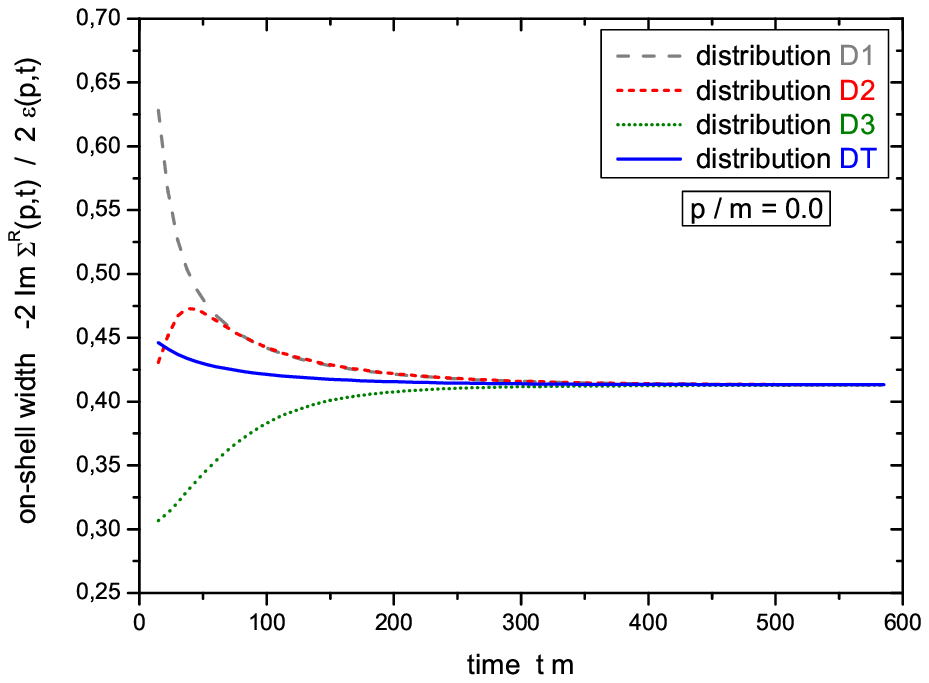} }
\resizebox{0.48\columnwidth}{!}{\includegraphics{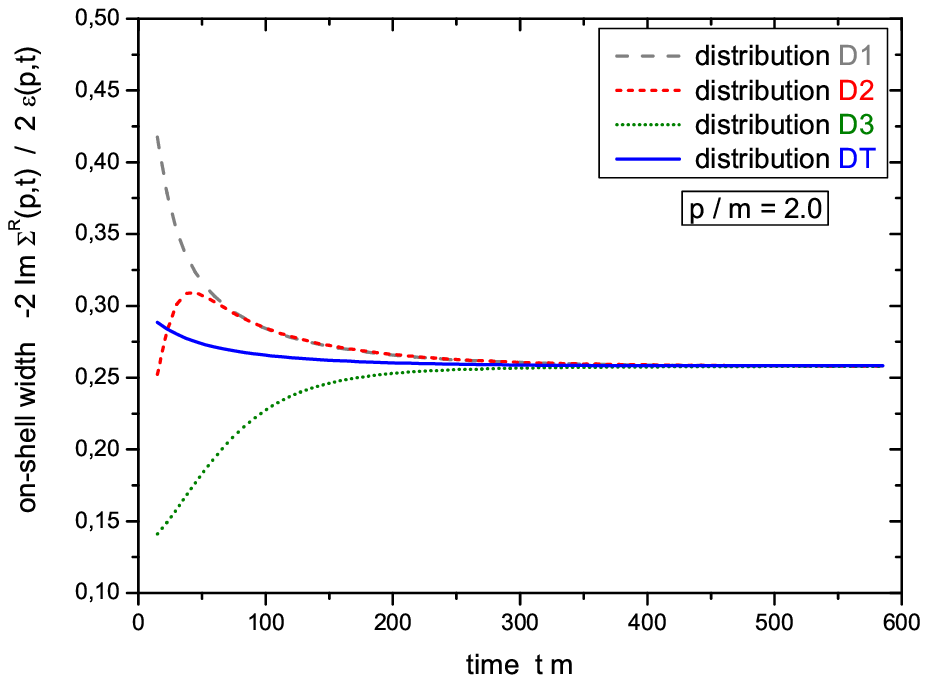} }
\end{center}
\caption[]{Time evolution of the on-shell widths
$-2\,Im\, \Sigma^{R}(\vec{p},\varepsilon(\vec{p},t),t) / 2\,
\varepsilon(\vec{p},t)$ of the momentum modes $|\,\vec{p}\,|/m =
0.0$ and $|\,\vec{p}\,|/m = 2.0$ for the different initializations
D1, D2, D3 and DT. \label{plot_osw01}}
\end{figure}
For the initial distribution D3 we find the opposite behaviour.
Here the spectral function is quite narrow for early times  and
increasing its width during the time evolution. Correspondingly,
the height of the spectral function decreases with time. This
behaviour is observed for the zero momentum mode $|\,\vec{p}\,|/m
= 0.0$ as well as for the finite momentum mode $|\,\vec{p}\,|/m =
2.0$. Especially in the latter case the width for early times
 is so small that the spectral function shows oscillations
 originating from the finite range of the Fourier-transformation
from relative time to energy. Although we have already increased
the system time for the first curve to $t \cdot m = 21$ (for $t
\cdot m = 15$ the oscillations are much stronger) the spectral
function is not fully resolved, i.e. it is not sufficiently damped
in relative time $\Delta t$ in the interval available for the
Fourier transform. For later times the oscillations vanish and the
spectral function tends to the common equilibrium shape.

The time evolution of the spectral function for the initial
distribution D2 is somehow in between the last two cases. Here the
spectral function develops (at intermediate times) a slightly
higher width than in the beginning before it is approaching the
narrower static shape again. The corresponding evolution of the
maximum is again indicated by the (bent) arrow. Finally, all
spectral functions show the (same) equilibrium form represented by
the solid gray line.

As already observed in Section 1.2.4 for the equal-time Green
functions, we emphazise that there is no universal time evolution
for the nonequilibrium systems. In fact, the evolution of the
system during the equilibration process depends on the initial
conditions. On the other hand, the time dependence of the spectral
function is only moderate such that one might also work with some
time-averaged or even the equilibrium spectral function. In order
to investigate this issue in more quantitative detail we
concentrate on the maxima and widths of the spectral functions in
the following.

Since the solution of the Kadanoff-Baym equation provides the full
spectral information for all system times the evolution of the
on-shell energies can be studied as well as the spectral widths.
In Fig. \ref{plot_ose01} we  display the time dependence of the
on-shell energies $\varepsilon(\vec{p},t)$ of the momentum modes
$|\,\vec{p}\,|/m = 0.0$ (l.h.s.) and $|\,\vec{p}\,|/m = 2.0$
(r.h.s.) for the four initial distributions D1, D2, D3 and DT. We
see  that the on-shell energy for the zero momentum mode increases
with time for the initial distribution D1 and to a certain extent
for the free thermal distribution DT (as can be also extracted
form Fig. \ref{plot_spec01}). The on-shell energy of distribution
D3 shows a monotonic decrease during the evolution while it passes
through a minimum for distribution D2 before joining the line for
the initialization D1. For momentum $|\,\vec{p}\,|/m = 2.0$ an
opposite behaviour is observed. Here the on-shell energy for
distribution D1 (and less pronounced for the distribution DT) are
reduced in time whereas it is increased in the case of D3. The
result for the initialization D2 is monotonous for this mode and
matches the one for D1 already for moderate times. Thus we find,
that the time evolution  of the on-shell energies does not only
depend on the initial conditions, but might also be different for
various momentum modes. It turns out -- for the initial
distributions investigated -- that the above described
characteristics change around $|\,\vec{p}\,|/m = 1.5$ and are
retained for larger momenta (not presented here).

Furthermore, we present in Fig. \ref{plot_osw01} the time
evolution of the on-shell width for the usual momentum modes for
the different initial distributions. The on-shell width $\Gamma$
is given by the imaginary part of the retarded sunset self-energy
at the on-shell energy of each respective momentum mode as
\begin{equation}
\label{ggg} \Gamma =
-2\,Im\,\Sigma^{R}(\vec{p},\varepsilon(\vec{p},t),t) \,/\,
2\,\varepsilon(\vec{p},t). \end{equation} As already discussed in
connection with Fig. \ref{plot_spec01} we observe for both
momentum modes a strong decrease of the on-shell width for the
initial distribution D1 associated with a narrowing of the
spectral function. In contrast, the on-shell widths of
distribution D3 increase with time such that the corresponding
spectral functions broaden towards the common static shape. For
the initialization D2 we observe an non-monotonic evolution of the
on-shell widths connected with a broadening of the spectral
function at intermediate times. Similar to the case of the
on-shell energies we find, that the results for the on-shell
widths of the distributions D1 and D2 coincide well above a
certain system time. As expected from the lower plots of Fig.
\ref{plot_spec01} the on-shell width for the free thermal
distribution DT exhibits only a weak time dependence with a slight
decrease in the initial phase of the time evolution.

In summarizing this Subsection we point out, that there is no
universal time evolution of the spectral functions for the initial
distributions considered. Peak positions and widths depend on the
initial configuration and evolve differently in time. However, we
find only effects in the order of $<$10\% for the on-shell
energies in the initial phase of the system evolution and initial
variations of $<$50\% for the widths of the dominant momentum
modes. Thus, depending on the physics problem of interest, one
might discard an explicit time-dependence of the spectral
functions and adopt the equilibrium shape.

\subsubsection{The equilibrium state}

In Section 1.2 we have seen that arbitrary initial momentum
configurations of the same energy density approach a stationary
limit for $t \rightarrow \infty$, which  is the same for all
initial distributions. In this Subsection we will investigate
whether this stationary state is the proper thermal state for
interacting Bose particles.

As shown before, in the present calculations within the three-loop
approximation of the 2PI effective action we describe kinetic
equilibration via the sunset self-energies and also obtain a
finite width for the particle spectral function, since the Green
functions have a non-zero imaginary part (in energy space). It is
not obvious, however, if the stationary state obtained for $t
\rightarrow \infty$ corresponds to the proper equilibrium state.

In order to clarify the nature of the asymptotic stationary state
of our calculations we first change into Wigner space. The Green
function and the spectral function in energy $p_0$ are obtained by
Fourier-transformation with respect to the relative time $\Delta t
= t_1 - t_2$ at every system time $\bar{t} = (t_1 + t_2)/2$,
\bea G^{\gtrless}(\vec{p},p_0,\bar{t}) \; = \;
\int_{-\infty}^{\infty} \!\!\!\!\! d\Delta t \;\;
e^{i\,p_0\,\Delta t} \;\; G^{\gtrless}(\vec{p},t_1 = \bar{t} +
\Delta t / 2, t_2 = \bar{t} - \Delta t /2 ) \, , \eea
\bea A(\vec{p},p_0,\bar{t}) \; = \; \int_{-\infty}^{\infty}
\!\!\!\!\! d\Delta t \;\; e^{i\,p_0\,\Delta t} \;\; A(\vec{p},t_1
= \bar{t} + \Delta t / 2, t_2 = \bar{t} - \Delta t /2 ) \, . \eea
We recall, that the spectral function can also be obtained
directly from the Green functions (cf. Section 1.1) by
\bea A(\vec{p},p_0,\bar{t}) \; = \; i \; \left[ \:
G^{>}(\vec{p},p_0,\bar{t}) \: - \: G^{<}(\vec{p},p_0,\bar{t}) \:
\right] \, . \eea
Now we introduce the energy and momentum dependent distribution
function $N$ at any system time $\bar{t}$ by the definition
\bea i \, G^{<}(\vec{p},p_0,\bar{t}) & = & A(\vec{p},p_0,\bar{t})
\; \; N(\vec{p},p_0,\bar{t}) \, , \nnl[0.4cm]
i \, G^{>}(\vec{p},p_0,\bar{t}) & = & A(\vec{p},p_0,\bar{t}) \; [
\: N(\vec{p},p_0,\bar{t}) + 1 \: ] \, \label{distdef} \eea
\begin{figure}[hbpt]
\begin{center}
\vspace*{-0.5cm}
\resizebox{0.5\columnwidth}{!}{\includegraphics{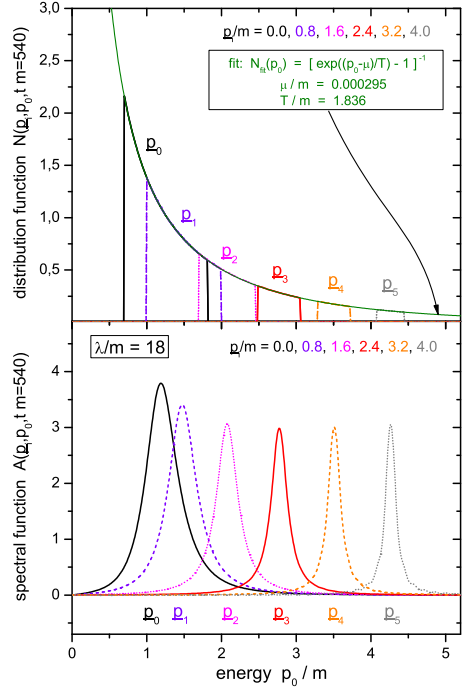} }
\end{center}
\vspace{-0.7cm} \caption[]{Spectral function $A$ for various
momentum modes $|\,\vec{p}\,|/m =
0.0,\,0.8,\,1.6,\,2.4,\,3.2,\,4.0$ as a function of energy for
late times $t \cdot m = 540$ (lower part). Corresponding
distribution function $N$ at the same time for the same momentum
modes (upper part). All momentum modes can be fitted with a single
Bose function of temperature $T_{eq} /m = 1.836$ and a chemical
potential close to zero. \label{plot_kmsna}}
\end{figure}
In equilibrium (at temperature $T$) the Green functions obey the
Kubo-Martin-Schwinger relation (KMS) for all momenta $\vec{p}$
\cite{KMS,oldkaba,oldba}
%
\bea G^{>}_{eq}(\vec{p},p_0) \; = \; e^{p_0/T} \;
G^{<}_{eq}(\vec{p},p_0) \qquad \qquad \forall \;\; \vec{p} \, .
\label{kms} \eea
If there exists a conserved quantum number in the theory we have,
furthermore, a contribution of the corresponding chemical
potential in the exponential function which leads to a shift of
arguments: $p_0/T \rightarrow (p_0 - \mu)/T$. In the present case,
however, there is no conserved quantum number and thus the
equilibrium state has  $\mu = 0$.

From the KMS condition of the Green functions (\ref{kms}) we
obtain the equilibrium form of the distribution function
(\ref{distdef}) at temperature $T$ as
\bea N_{eq}(\vec{p},p_0) \; = \; N_{eq}(p_0) \; = \;
\frac{1}{e^{p_0/T} - 1} \; = \; N_{bose}(p_0/T) \, ,
\label{distequi} \eea which is the well-known Bose distribution.
As is obvious from (\ref{distequi}) the equilibrium distribution
can only be a function of energy and not of the momentum variable
in addition.

In Figure \ref{plot_kmsna} (lower part) we present the spectral
function $A(\vec{p},p_0)$ for the initial distribution D2 at late
times $\bar{t} \cdot m = 540$ for various momentum modes
$|\vec{p}| / m = 0.0, 0.8, 1.6,$ $2.4, 3.2, 4.0$ as a function of
the energy $p_0$. -- We note, that for all other initial
distributions - with equal energy density - the spectral function
looks very similar at this time since the systems proceed to the
same stationary state (cf. Section 1.3). -- We recognize that the
spectral function is quite broad, especially for the low momentum
modes, while for the higher momentum modes its width is slightly
lower.

 The distribution function $N(p_0)$ as extracted from (\ref{distdef})
 is displayed in Fig. \ref{plot_kmsna} (upper part) for the same
momentum modes as a function of the energy $p_0$. We find that $N(p_0)$
for all momentum modes can be fitted by a single Bose-function with
temperature $T/m = 1.836$. Thus the distribution function emerging from
the Kadanoff-Baym time evolution for $t \rightarrow \infty$ approaches
a Bose function in the energy that is independent of the momentum as
demanded by the equilibrium form (\ref{distequi}).

Fig. \ref{plot_kmsna} (upper part) demonstrates, furthermore, that the
 KMS-condition is fulfilled not only for on-shell energies, but for all
 $p_0$.  We, therefore, have obtained the full off-shell equilibrium
state by integrating the Kadanoff-Baym equations in time. In addition,
the limiting stationary state is the correct equilibrium state for all
energies $p_0$, i.e. also away from the quasi-particle energies.

We note in closing this Subsection, that the chemical potential
$\mu$ --  used as a second fit parameter -- is already close to
zero for these late times as expected for the correct equilibrium
state of the neutral $\phi^4$-theory that is characterized by a
vanishing chemical potential $\mu$ in equilibrium which is
actually achieved in the calculations for $t \rightarrow \infty$.
This, at first sight, seems trivial but  is a consequence of our
'exact' treatment. In contrast, the Boltzmann equation (cf.
Section 1.4) in general leads to a stationary state for $t
\rightarrow \infty$ with a finite chemical potential. We will
attribute this failure of the Boltzmann approach to the absence of
particle number nonconserving processes in the quasi-particle
limit (see below).

\subsection{Full versus approximate dynamics}
The Kadanoff-Baym equations studied in the previous Sections
represent the full quantum-field theoretical equations on the
single-particle level. However, its numerical solution is quite
involved and it is of strong interest to investigate, in how far
approximate schemes deviate from the full calculation. Nowadays,
transport models are widely used in the description of quantum
systems out of equilibrium (cf. Introduction). Most of these
models work in the 'quasi-particle' picture, where all particles
obey a fixed energy-momentum relation and the energy  is no
independent degree of freedom anymore; it is determined by the
momentum and the (effective) mass of the particle. Accordingly,
these particles are treated with their $\delta$-function spectral
shape as infinitely long living, i.e. stable objects. This
assumption is very questionable e.g. for high-energy heavy ion
reactions, where the particles achieve a large width due to the
frequent collisions with other particles in the high-density
and/or high-energy regime. Furthermore, this is doubtful for
particles that are unstable even in the vacuum. The question, in
how far the quasiparticle approximation influences the dynamics in
comparison to the full Kadanoff-Baym calculation, is of widespread
interest \cite{koe1}.

\subsubsection{Derivation of the Boltzmann approximation}

In the following we will give a short derivation of the Boltzmann
equation starting directly from the Kadanoff-Baym dynamics in the
two-time and momentum-space representation. This derivation is
briefly reviewed since we want  i) to emphasize the link of the
full Kadanoff-Baym equation with its approximated version and ii)
to clarify the assumptions that enter the Boltzmann equation.

Since the Boltzmann equation describes the time evolution of
distribution functions for quasi-particles we first consider the
quasi-particle Green-functions in two-time representation for
homogeneous systems
\bea \label{qpgreen} \phantom{a} \\[-0.8cm]
\begin{array}{cccrcl}
G^{\gtrless}_{\phi \phi,qp}(\vec{p},t,t^{\prime}) & = &
\displaystyle{\frac{-i}{2 \omega_{\vec{p}}}} \!\!\!\! & \!\!\!\!
\{\; N_{qp}(\mp\vec{p}) \;\; e^{\pm i
\omega_{\vec{p}}(t-t^{\prime})} & + &
 [\, N_{qp}(\pm\vec{p})\!+\!1 \,] \;\;
e^{\mp i \omega_{\vec{p}}(t-t^{\prime})} \;\}
\\[0.6cm]
G^{\gtrless}_{\phi \pi,qp}(\vec{p},t,t^{\prime}) & = &
\displaystyle{\frac{1}{2}} \!\!\!\! & \!\!\!\! \{\; \mp N_{qp}(\mp
\vec{p}) \;\; e^{\pm i \omega_{\vec{p}}(t-t^{\prime})} & \pm & [\,
N_{qp}(\pm \vec{p})\!+\!1 \,] \;\; e^{\mp i
\omega_{\vec{p}}(t-t^{\prime})} \;\} \nnl[0.6cm]
G^{\gtrless}_{\pi \phi,qp}(\vec{p},t,t^{\prime}) & = &
\displaystyle{\frac{1}{2}} \!\!\!\! & \!\!\!\! \{\; \pm N_{qp}(\mp
\vec{p}) \;\; e^{\pm i \omega_{\vec{p}}(t-t^{\prime})} & \mp & [\,
N_{qp}(\pm \vec{p})\!+\!1 \,] \;\; e^{\mp i
\omega_{\vec{p}}(t-t^{\prime})} \;\} \nnl[0.6cm]
G^{\gtrless}_{\pi \pi,qp}(\vec{p},t,t^{\prime}) & = &
\displaystyle{\frac{-i\,\omega_{\vec{p}}}{2}} \!\!\!\! & \!\!\!\!
\{\; N_{qp}(\mp \vec{p}) \;\; e^{\pm i
\omega_{\vec{p}}(t-t^{\prime})} & + & [\, N_{qp}(\pm
\vec{p})\!+\!1 \,] \;\; e^{\mp i \omega_{\vec{p}}(t-t^{\prime})}
\;\} \, , \nn
\end{array}
\eea
where for each momentum $\vec{p}$ the Green functions are freely
oscillating in relative time $t-t^\prime$ with the on-shell energy
$\omega_{p}$. The time-dependent quasi-particle distribution
functions are given with the energy variable fixed to the on-shell
energy as $N_{qp}(\vec{p},\bar{t}) \equiv
N(\vec{p},p_0=\omega_{\vec{p}},\bar{t})$, where the on-shell
energies $\omega_{\vec{p}}$ might depend on  time as well. Such a
time variation e.g. might be due to an effective mass as generated
by the time-dependent tadpole self-energy. In this case the
on-shell energy reads \begin{equation} \omega_{\vec{p}}(\bar{t}) =
\sqrt{\vec{p}^{\,2} + m^2 + \bar{\Sigma}^{\delta}_{ren}(\bar{t})}.
\end{equation}
Vice versa we can define the quasi-particle distribution function
by means of the quasi-particle Green functions at equal times
$\bar{t}$ as
\bea \label{defqpdist} N_{qp}(\vec{p},\bar{t}) & = & \left[ \;
\frac{\omega_{\vec{p}}(\bar{t})}{2}\: i\,G^{<}_{\phi
\phi,qp}(\vec{p},\bar{t},\bar{t}) \; + \; \frac{1}{2
\omega_{\vec{p}}(\bar{t})}\: i\,G^{<}_{\pi
\pi,qp}(\vec{p},\bar{t},\bar{t}) \; \right]
\\[0.5cm]
&& - \; \frac{1}{2} \, \left[ \; G^{<}_{\pi
\phi,qp}(\vec{p},\bar{t},\bar{t}) \; - \; G^{<}_{\phi
\pi,qp}(\vec{p},\bar{t},\bar{t}) \phantom{\frac{1}{2
\omega_{\vec{p}}}} \!\!\!\!\!\!\!\!\! \; \right] . \nn \eea
Using the equations of motions for the Green functions in diagonal
time direction (\ref{eomall}) (exploiting $ G^{<}_{\phi
\pi}(\vec{p},\bar{t},\bar{t}) =
       - [ \, G^{<}_{\pi \phi}(\vec{p},\bar{t},\bar{t}) \, ]^{*} ) $
the time evolution of the distribution function is given by
\bea
\partial_{\bar{t}} \: N_{qp}(\vec{p},\bar{t})
&=& - \, Re \left\{ \; I_{1\,;\,qp}^{<}(\vec{p},\bar{t},\bar{t})
\; \right\} - \, \frac{1}{\omega_{\vec{p}}(\bar{t})} \: \, Im
\left\{ \; I_{1,2\,;\,qp}^{<}(\vec{p},\bar{t},\bar{t}) \; \right\}
\, . \eea
The time derivatives of the on-shell energies cancel out since the
 quasiparticle Green functions obey \begin{equation}
 G^{<}_{\pi
\pi}(\vec{p},\bar{t},\bar{t}) \: = \: \omega^2_{\vec{p}}(\bar{t})
\, G^{<}_{\phi \phi}(\vec{p},\bar{t},\bar{t}) \end{equation}
 as
seen from (\ref{qpgreen}). Furthermore, it is remarkable that
contributions containing the energy $\omega^2_{\vec{p}}$ - as
present in the equation of motion for the Green functions
(\ref{eomall}) -  no longer show up. The time evolution of the
distribution function is entirely determined by (equal-time)
collision integrals containing (time derivatives of the) Green
functions and self-energies.
\bea \label{coll8} I_{1;qp}^{<}(\vec{p},\bar{t},\bar{t}) &=&
\int_{t_0}^{\bar{t}} \!\! dt^{\prime} \;\;\; \left(
\Sigma^{<}_{qp}(\vec{p},\bar{t},t^{\prime}) \; G^{>}_{\phi
\phi,qp}(\vec{p},t^{\prime}\!,\bar{t}) \;-\;
\Sigma^{>}_{qp}(\vec{p},\bar{t},t^{\prime}) \; G^{<}_{\phi
\phi,qp}(\vec{p},t^{\prime}\!,\bar{t}) \right) \, , \\[0.7cm]
I_{1,2;qp}^{<}(\vec{p},\bar{t},\bar{t}) &=& \int_{t_0}^{\bar{t}}
\!\! dt^{\prime} \;\;\; \left(
\Sigma^{<}_{qp}(\vec{p},\bar{t},t^{\prime}) \; G^{>}_{\phi
\pi,qp}(\vec{p},t^{\prime}\!,\bar{t}) \;-\;
\Sigma^{>}_{qp}(\vec{p},\bar{t},t^{\prime}) \; G^{<}_{\phi
\pi,qp}(\vec{p},t^{\prime}\!,\bar{t})\right) \, . \nn \eea
Since we are dealing with a system of on-shell quasi-particles
within the Boltzmann approximation, the Green functions in the
 collision integrals (\ref{coll8}) are given by the respective
quasi-particle quantities of (\ref{qpgreen}). Moreover, the
collisional self-energies (\ref{sems}) are obtained in accordance
with the quasi-particle approximation as
\bea \label{qpsigma} \Sigma^{\gtrless}_{qp}(\vec{p},t,t^{\prime})
\!\!&=&\!\! -i\frac{\lambda^2}{6} \int\!\!\!
\frac{d^{d}q}{(2\pi)^{d}} \! \int\!\!\! \frac{d^{d}r}{(2\pi)^{d}}
\! \int\!\!\! \frac{d^{d}s}{(2\pi)^{d}} \;\: (2\pi)^{d} \:
\delta^{(d)}\!(\vec{p}\!-\!\vec{q}\!-\!\vec{r}\!-\!\vec{s})
\;\:\frac{1}{2\omega_{\vec{q}}\,2\omega_{\vec{r}}\,2\omega_{\vec{s}}}
\\[0.5cm]
&& \!\!\!\!\!\!\!\!\!\!\!\!\!\!\!
\begin{array}{rcccl}
\left\{ \phantom{\frac{1^1}{2}} \!\!\!\! \right. \!&\! N_{qp}(\mp
\vec{q}) \!&\! N_{qp}(\mp \vec{r}) \!&\! N_{qp}(\mp \vec{s}) \!&\!
\; e^{+i\,[\,t-t^{\prime}\,]\,
[\,\pm\omega_{\vec{q}}\pm\omega_{\vec{r}}\pm\omega_{\vec{s}}\,]}
\nnl[0.5cm] +\,3 \!&\! N_{qp}(\mp \vec{q}) \!&\! N_{qp}(\mp
\vec{r}) \!&\! [\,N_{qp}(\pm \vec{s})\!+\!1\,] \!&\! \;
e^{+i\,[\,t-t^{\prime}\,]\,
[\,\pm\omega_{\vec{q}}\pm\omega_{\vec{r}}\mp\omega_{\vec{s}}\,]}
\nnl[0.5cm] +\,3 \!&\! N_{qp}(\mp \vec{q}) \!&\! [\,N_{qp}(\pm
\vec{r})\!+\!1\,] \!&\! [\,N_{qp}(\pm \vec{s})\!+\!1\,] \!&\! \;
e^{+i\,[\,t-t^{\prime}\,]\,
[\,\pm\omega_{\vec{q}}\mp\omega_{\vec{r}}\mp\omega_{\vec{s}}\,]}
\nnl[0.5cm] + \!&\! [\,N_{qp}(\pm \vec{q})\!+\!1\,] \!&\!
[\,N_{qp}(\pm \vec{r})\!+\!1\,] \!&\! [\,N_{qp}(\pm
\vec{s})\!+\!1\,] \!&\! \; e^{+i\,[\,t-t^{\prime}\,]\,
[\,\mp\omega_{\vec{q}}\mp\omega_{\vec{r}}\mp\omega_{\vec{s}}\,]}
\left. \phantom{\frac{1^1}{2}} \!\!\!\! \right\} .
\end{array} \nn
\eea
For a free theory the distribution functions $N_{qp}(\vec{p})$ are
obviously constant in time which, of course, is no longer valid
for an interacting system out of equilibrium. Thus one has to
specify the above expressions for the quasi-particle Green
functions (\ref{qpgreen}) to account for the time dependence of
the distribution functions.

The actual Boltzmann approximation is defined in the limit, that
the distribution functions have to be taken always at {\it the
latest time argument} of the two-time Green function \cite{koe1}.
Accordingly, for the general non-equilibrium case, we introduce
the ansatz for the Green functions in  the collision term
\bea \label{qpgreenboltz} \hspace{-0.3cm} G^{\gtrless}_{\phi
\phi,qp}(\vec{p},t,t^{\prime}) &= &\! \frac{-i}{2
\omega_{\vec{p}}} \: \{\;  N_{qp}(\mp\vec{p},t_{max}) \; e^{\pm i
\omega_{\vec{p}}(t-t^{\prime})} \:+\:
    [\,N_{qp}(\pm\vec{p},t_{max})\!+\!1 \,] \;\,
e^{\mp i \omega_{\vec{p}}(t-t^{\prime})} \;\} \phantom{aaa}
\\[0.2cm]
G^{\gtrless}_{\phi \pi,qp}(\vec{p},t,t^{\prime}) & = & \frac{1}{2}
\: \{\; \mp N_{qp}(\mp\vec{p},t_{max}) \; e^{\pm i
\omega_{\vec{p}}(t-t^{\prime})} \;\pm\; [\,
N_{qp}(\pm\vec{p},t_{max})\!+\!1 \,] \; e^{\mp i
\omega_{\vec{p}}(t-t^{\prime})} \;\} \, . \nn \eea
with the maximum time $t_{max} = max(t,t^{\prime})$. The same
ansatz is employed for the time dependent on-shell energies which
enter the representation of the quasi-particle two-time Green
functions (\ref{qpgreenboltz}) with their value at $t_{max}$, i.e.
$\omega_{\vec{p}} = \omega_{\vec{p}}(t_{max}=max(t,t^{\prime}))$.

The collision term contains a time integration which extends from
an initial time $t_0$ to the current time $\bar{t}$.
 All two-time
Green functions and self-energies depend on the current time
$\bar{t}$ as well as on the integration time $t^{\prime} \le
\bar{t}$. Thus only distribution functions at the current time,
i.e. the maximum time of all appearing two-time functions, enter
the collision integrals and the evolution equation for the
distribution function becomes local in time. Since the
distribution functions are given at fixed  time $\bar{t}$, they
can be taken out of the time integral. When inserting the
expressions for the self-energies and the Green functions in the
collision integrals the evolution equation for the quasi-particle
distribution function reads
\bea \label{boltz_collterm1} && \partial_{\bar{t}} \:
N_{qp}(\vec{p},\bar{t}) \;=\; \frac{\lambda^2}{3} \int\!\!\!
\frac{d^{d}q}{(2\pi)^{d}} \! \int\!\!\! \frac{d^{d}r}{(2\pi)^{d}}
\! \int\!\!\! \frac{d^{d}s}{(2\pi)^{d}} \;\; (2\pi)^{d} \:
\delta^{(d)}\!(\vec{p}\!-\!\vec{q}\!-\!\vec{r}\!-\!\vec{s})
\;\:\frac{1}{2\omega_{\vec{p}}\:2\omega_{\vec{q}}\:
             2\omega_{\vec{r}}\:2\omega_{\vec{s}}}
\\[0.6cm]
&& \!\!\!\!\!\!\!\!\!\!
\begin{array}{rcccccccccccl}
\left\{ \displaystyle{\phantom{int_{t_0}^{\bar{t}} \frac{1^1}{2}}}
\!\!\!\!\!\!\!\!\!\!\!\!\!\!\! \right. \!\!\!\!\!\! &\! [ \!\!&\!
\bar{N}_{ \vec{p},\bar{t}} \!\!&\!\! \bar{N}_{\!-\vec{q},\bar{t}}
\!\!&\!\! \bar{N}_{\!-\vec{r},\bar{t}} \!\!&\!\!
\bar{N}_{\!-\vec{s},\bar{t}} \!\!&\!\! - \!&\!
      N_{ \vec{p},\bar{t}} \!\!&\!\!
      N_{\!-\vec{q},\bar{t}} \!\!&\!\!
      N_{\!-\vec{r},\bar{t}} \!\!&\!\!
      N_{\!-\vec{s},\bar{t}} \!&\!\!\! ] \!&\!
\displaystyle{\int_{t_0}^{\bar{t}}} \!\!\! dt^{\prime} \;
\cos([\,\bar{t}\!-\!t^{\prime}\,]\,
[\,\omega_{\vec{p}}\!+\!\omega_{\vec{q}}
\!+\!\omega_{\vec{r}}\!+\!\omega_{\vec{s}}\,]) \nnl[0.6cm] +3\!&\!
[ \!\!&\! \bar{N}_{ \vec{p},\bar{t}} \!\!&\!\!
\bar{N}_{\!-\vec{q},\bar{t}} \!\!&\!\!
\bar{N}_{\!-\vec{r},\bar{t}} \!\!&\!\!
      N_{ \vec{s},\bar{t}} \!\!&\!\! - \!&\!
      N_{ \vec{p},\bar{t}} \!\!&\!\!
      N_{\!-\vec{q},\bar{t}} \!\!&\!\!
      N_{\!-\vec{r},\bar{t}} \!\!&\!\!
\bar{N}_{ \vec{s},\bar{t}} \!&\!\!\! ] \!&\!
\displaystyle{\int_{t_0}^{\bar{t}}} \!\!\! dt^{\prime} \;
\cos([\,\bar{t}\!-\!t^{\prime}\,]\,
[\,\omega_{\vec{p}}\!+\!\omega_{\vec{q}}
\!+\!\omega_{\vec{r}}\!-\!\omega_{\vec{s}}\,]) \nnl[0.6cm] +3\!&\!
[ \!\!&\! \bar{N}_{ \vec{p},\bar{t}} \!\!&\!\!
\bar{N}_{\!-\vec{q},\bar{t}} \!\!&\!\!
      N_{ \vec{r},\bar{t}} \!\!&\!\!
      N_{ \vec{s},\bar{t}} \!\!&\!\! - \!&\!
      N_{ \vec{p},\bar{t}} \!\!&\!\!
      N_{\!-\vec{q},\bar{t}} \!\!&\!\!
\bar{N}_{ \vec{r},\bar{t}} \!\!&\!\! \bar{N}_{ \vec{s},\bar{t}}
\!&\!\!\! ] \!&\! \displaystyle{\int_{t_0}^{\bar{t}}} \!\!\!
dt^{\prime} \; \cos([\,\bar{t}\!-\!t^{\prime}\,]\,
[\,\omega_{\vec{p}}\!+\!\omega_{\vec{q}}
\!-\!\omega_{\vec{r}}\!-\!\omega_{\vec{s}}\,]) \nnl[0.6cm] +\!&\!
[ \!\!&\! \bar{N}_{ \vec{p},\bar{t}} \!\!&\!\!
      N_{ \vec{q},\bar{t}} \!\!&\!\!
      N_{ \vec{r},\bar{t}} \!\!&\!\!
      N_{ \vec{s},\bar{t}} \!\!&\!\! - \!&\!
      N_{ \vec{p},\bar{t}} \!\!&\!\!
\bar{N}_{ \vec{q},\bar{t}} \!\!&\!\! \bar{N}_{ \vec{r},\bar{t}}
\!\!&\!\! \bar{N}_{ \vec{s},\bar{t}} \!&\!\!\! ] \!&\!
\displaystyle{\int_{t_0}^{\bar{t}}} \!\!\! dt^{\prime} \;
\cos([\,\bar{t}\!-\!t^{\prime}\,]\,
[\,\omega_{\vec{p}}\!-\!\omega_{\vec{q}}
\!-\!\omega_{\vec{r}}\!-\!\omega_{\vec{s}}\,]) \left.
\phantom{\frac{1^1}{2}} \!\!\!\!\!\!\! \right\} ,
\end{array}
\nn \eea
where we have introduced the abbreviation $N_{\vec{p},\bar{t}} =
N_{qp}(\vec{p},\bar{t})$ for the distribution function at current
time $\bar{t}$ and $\bar{N}_{\vec{p},\bar{t}} =
N_{qp}(\vec{p},\bar{t}) + 1$ for the according Bose factor.
Furthermore, a possible time dependence of the on-shell energies
is suppressed in the above notation.

The contributions in the collision term (\ref{boltz_collterm1})
for particles of momentum $\vec{p}$ are ordered as they describe
different types of scattering processes  where, however,  we
always find the typical gain and loss structure. The first line in
(\ref{boltz_collterm1}) corresponds to the production and
annihilation of four on-shell particles ($ 0 \rightarrow 4 $, $ 4
\rightarrow 0 $), where a particle of momentum $\vec{p}$ is
produced or destroyed simultaneous with three other particles with
momenta $\vec{q}, \vec{r}, \vec{s}$. The second line and the forth
line describe ($ 1 \rightarrow 3 $) and ($ 3 \rightarrow 1 $)
processes where the quasi-particle with momentum $\vec{p}$ is the
single one or appears with two other particles. The relevant
 contribution in the Boltzmann limit is the third line which
respresents ($ 2 \rightarrow 2 $) scattering processes;
quasi-particles with momentum $\vec{p}$ can be scattered out of
their momentum cell by collisions with particles of momenta
$\vec{q}$ (second term) or can be produced within a reaction of
on-shell particles with momenta $\vec{r}$, $\vec{s}$ (first term).

The time evolution of the quasi-particle distribution is given as
an initial value problem for the function $N_{qp}(\vec{p})$
prepared at initial time $t_0$. For large system times $\bar{t}$
(compared to the initial time) the time integration over the
trigonometric function results in an energy conserving
$\delta$-function:
\bea \label{energy_delta} \lim_{\bar{t}-t_0 \rightarrow \infty} \:
\int_{t_0}^{\bar{t}} \!\! dt^{\prime} \: \cos((\bar{t}-t^{\prime})
\: \hat{\omega}) \; = \; \lim_{\bar{t}-t_0 \rightarrow \infty} \:
\frac{1}{\hat{\omega}} \: \sin((\bar{t}-t_0) \: \hat{\omega}) \; =
\; \pi \: \delta(\hat{\omega}) \; . \eea\\
Here $\hat{\omega} = \omega_{\vec{p}} \pm \omega_{\vec{q}} \pm
\omega_{\vec{r}} \pm \omega_{\vec{s}}$ represents the energy sum
which is conserved in the limit $\bar{t}-t_0 \rightarrow \infty$
where the initial time $t_0$ is considered as fixed. In this limit
the time evolution of the distribution function amounts to
\bea \label{boltz_collterm2} &&
\partial_{\bar{t}} \: N_{qp}(\vec{p},\bar{t})
\;=\; \frac{\lambda^2}{6} \int\!\!\! \frac{d^{d}q}{(2\pi)^{d}} \!
\int\!\!\! \frac{d^{d}r}{(2\pi)^{d}} \! \int\!\!\!
\frac{d^{d}s}{(2\pi)^{d}} \;\; (2\pi)^{d+1}
\;\:\frac{1}{2\omega_{\vec{p}}\:2\omega_{\vec{q}}\:
             2\omega_{\vec{r}}\:2\omega_{\vec{s}}}
\\[0.6cm]
\left\{ \phantom{\frac{1}{2}} \right. \!\!\!\!\!\!\!\!\!
\!\!\!\!\!\!\!\!\! &\phantom{+}& \!\! [ \,
\bar{N}_{\vec{p},\bar{t}} \; \bar{N}_{\vec{q},\bar{t}} \;
\bar{N}_{\vec{r},\bar{t}} \; \bar{N}_{\vec{s},\bar{t}} \; - \;
      N_{\vec{p},\bar{t}} \;
      N_{\vec{q},\bar{t}} \;
      N_{\vec{r},\bar{t}} \;
      N_{\vec{s},\bar{t}} \, ] \;\;
\delta^{(d)}\!(\vec{p}\!+\!\vec{q}\!+\!\vec{r}\!+\!\vec{s}) \;\;
\delta (\omega_{\vec{p}}\!+\!\omega_{\vec{q}}
   \!+\!\omega_{\vec{r}}\!+\!\omega_{\vec{s}}) \nnl[0.5cm]
&+3& \!\! [ \, \bar{N}_{\vec{p},\bar{t}} \;
\bar{N}_{\vec{q},\bar{t}} \; \bar{N}_{\vec{r},\bar{t}} \;
      N_{\vec{s},\bar{t}} \; - \;
      N_{\vec{p},\bar{t}} \;
      N_{\vec{q},\bar{t}} \;
      N_{\vec{r},\bar{t}} \;
\bar{N}_{\vec{s},\bar{t}} \, ] \;\;
\delta^{(d)}\!(\vec{p}\!+\!\vec{q}\!+\!\vec{r}\!-\!\vec{s}) \;\;
\delta(\omega_{\vec{p}}\!+\!\omega_{\vec{q}}
  \!+\!\omega_{\vec{r}}\!-\!\omega_{\vec{s}}) \nnl[0.6cm]
&+3& \!\! [ \, \bar{N}_{\vec{p},\bar{t}} \;
\bar{N}_{\vec{q},\bar{t}} \;
      N_{\vec{r},\bar{t}} \;
      N_{\vec{s},\bar{t}} \; - \;
      N_{\vec{p},\bar{t}} \;
      N_{\vec{q},\bar{t}} \;
\bar{N}_{\vec{r},\bar{t}} \; \bar{N}_{\vec{s},\bar{t}} \, ] \;\;
\delta^{(d)}\!(\vec{p}\!+\!\vec{q}\!-\!\vec{r}\!-\!\vec{s}) \;\;
\delta(\omega_{\vec{p}}\!+\!\omega_{\vec{q}}
  \!-\!\omega_{\vec{r}}\!-\!\omega_{\vec{s}}) \nnl[0.5cm]
&+& \!\! [ \, \bar{N}_{\vec{p},\bar{t}} \;
      N_{\vec{q},\bar{t}} \;
      N_{\vec{r},\bar{t}} \;
      N_{\vec{s},\bar{t}} \; - \;
      N_{\vec{p},\bar{t}} \;
\bar{N}_{\vec{q},\bar{t}} \; \bar{N}_{\vec{r},\bar{t}} \;
\bar{N}_{\vec{s},\bar{t}} \, ] \;\;
\delta^{(d)}\!(\vec{p}\!-\!\vec{q}\!-\!\vec{r}\!-\!\vec{s}) \;\;
\delta(\omega_{\vec{p}}\!-\!\omega_{\vec{q}}
  \!-\!\omega_{\vec{r}}\!-\!\omega_{\vec{s}})
\left. \phantom{\frac{1}{2}} \!\!\!\!\! \right\} . \nn \eea
In the energy conserving long-time limit (\ref{energy_delta}) only
the $2 \rightarrow 2$ scattering processes are non-vanishing,
because all other terms do not contribute since the energy
$\delta$-functions can not be fulfilled for on-shell
quasi-particles. Furthermore, the system evolution is explicitly
local in time because it depends only on the current
configuration; there are no memory effects from the integration
over past times as present in the full Kadanoff-Baym equation.

 In the following we will solve the
energy conserving Boltzmann equation for on-shell particles:
\bea \label{boltz_collterm3} &&
\partial_{\bar{t}} \: N_{qp}(\vec{p},\bar{t})
\;=\; \frac{\lambda^2}{2} \int\!\!\! \frac{d^{d}q}{(2\pi)^{d}} \!
\int\!\!\! \frac{d^{d}r}{(2\pi)^{d}} \! \int\!\!\!
\frac{d^{d}s}{(2\pi)^{d}} \;\; (2\pi)^{d+1}
\;\:\frac{1}{2\omega_{\vec{p}}\:2\omega_{\vec{q}}\:
             2\omega_{\vec{r}}\:2\omega_{\vec{s}}}
\\[0.6cm]
&\phantom{+}& \!\! [ \, \bar{N}_{\vec{p},\bar{t}} \;
\bar{N}_{\vec{q},\bar{t}} \;
      N_{\vec{r},\bar{t}} \;
      N_{\vec{s},\bar{t}} \; - \;
      N_{\vec{p},\bar{t}} \;
      N_{\vec{q},\bar{t}} \;
\bar{N}_{\vec{r},\bar{t}} \; \bar{N}_{\vec{s},\bar{t}} \, ] \;\;
\delta^{(d)}\!(\vec{p}\!+\!\vec{q}\!-\!\vec{r}\!-\!\vec{s}) \;\;
\delta(\omega_{\vec{p}}\!+\!\omega_{\vec{q}}
  \!-\!\omega_{\vec{r}}\!-\!\omega_{\vec{s}}) \, .
\nn \eea
The numerical algorithm employed for the solution of
(\ref{boltz_collterm3}) is basically the same as for the solution
of the Kadanoff-Baym equation (cf. Section 1.2).
 We explicitly calculate the time integral in
(\ref{boltz_collterm1}). Energy conservation can be assured by a
precalculation including a shift of the lower boundary $t_0$ to
earlier times.  We note, that in contrast to the Kadanoff-Baym
equation no correlation energy is generated in the Boltzmann
limit!

 In addition to the procedure presented  above we calculate the
 actual momentum-dependent on-shell
energy for every momentum mode by a solution of the dispersion
relation including contributions from the tadpole and the real
part of the (retarded) sunset self energy. In this way one can
guarantee that at every time $t$ the particles are treated as
quasi-particles with the correct energy-momentum relation.

Before presenting the actual numerical results we  comment on the
derivation of the  Boltzmann equation within the conventional
scheme that  is different from the one presented above. Here, at
first the Kadanoff-Baym equation (in coordinate space) is
transformed to the Wigner representation by Fourier transformation
with respect to the relative coordinates in space and time. The
problem then is formulated in terms of energy and momentum
variables together with a single system time. For non-homogeneous
systems a mean spatial coordinate is necessary as well. As a next
step the 'semiclassical approximation' is introduced, which
consists of a gradient expansion of the convolution integrals in
coordinate space within the Wigner transformation. For the time
evolution only contributions up to first order in the gradients
are kept. Finally, the quasi-particle assumption is introduced as
follows: The Green functions appearing in the transport equation
-- explicitly or implicitly via the self-energies -- are written
in Wigner representation as a product of a distribution function
$N$ and the spectral function $A$. The quasi-particle assumption
is then realized by employing a $\delta$-like form for the
spectral function which connects the energy variable to the
momentum. By integrating the first order transport equation over
all (positive) energies, furthermore,  the Boltzmann equation for
the time evolution of the on-shell distribution function
(\ref{boltz_collterm3}) is obtained.
\begin{figure}[t]
\resizebox{0.85\columnwidth}{!}{\includegraphics{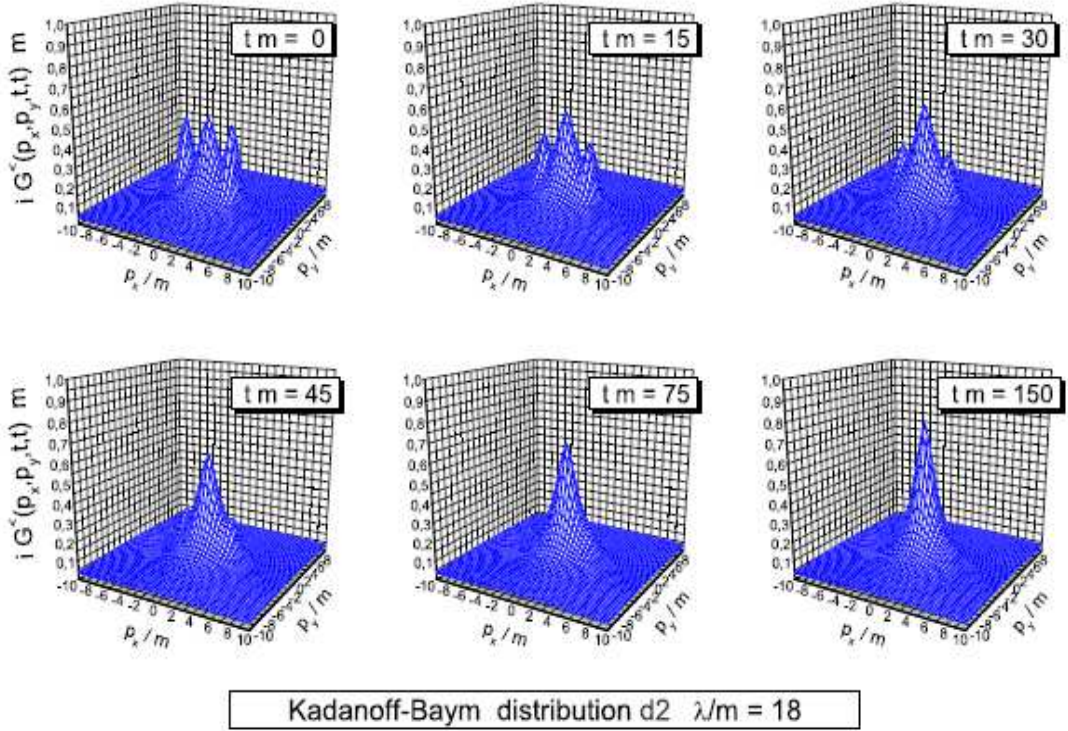} }
\caption{Evolution of the Green function in momentum space. The
equal time Green function is displayed for various times
 $t \cdot m =$ 0, 15, 30, 45, 75, 150. Starting from an initially
non-isotropic shape it develops towards a rotational symmetric
distribution in momentum space.\label{plot_3d1} }
\end{figure}

Inspite of the fact, that the Bolzmann equation
(\ref{boltz_collterm3}) can be obtained in different subsequent
approximation schemes, it is of basic interest, how its actual
solutions compare to those from the full Kadanoff-Baym dynamics.

\subsubsection{Boltzmann vs. Kadanoff-Baym dynamics}

In the following we will compare the solutions of the Boltzmann
equation with the solution of the Kadanoff-Baym theory. We start
with a presentation of the non-equilibrium time evolution of two
colliding particle accumulations (tsunamis) within the full
Kadanoff-Baym calculation (see Figure \ref{plot_3d1}).

During the time evolution the bumps at finite momenta (in $p_x$
direction) slowly disappear, while the one close to zero momentum
-- which initially  stems from the vacuum contribution to the
Green function -- is decreased as seen for different snapshots at
times $t \cdot m =$ 0, 15, 30, 45, 75, 150 in Fig. \ref{plot_3d1}.
The system with initially apparent collision axis slowly merges --
as expected -- into an isotropic final distribution in momentum
space.

For the comparison between the full Kadanoff-Baym dynamics and the
Boltzmann approximation we concentrate on equilibration times. To
this aim we  define a 'quadrupole' moment for a given momentum
distribution $n(\vec{p})$ at time $T$ as
\bea Q(\bar{t}) \: = \: \frac{\displaystyle{\int \!\!
\frac{d^{d}p}{(2 \pi)^d} \;\; ( p_x^2 - p_y^2 ) \;\;
N(\vec{p},\bar{t})}}
     {\displaystyle{\int \!\! \frac{d^{d}p}{(2 \pi)^d} \;\;
N(\vec{p},\bar{t})}} \; ,\eea
which vanishes for  the equilibrium state. For the Kadanoff-Baym
case we employ the actual distribution function by the relation
\bea \label{quad} n(\vec{p},\bar{t}) \: = \: \sqrt{ G^{<}_{\phi
\phi}(\vec{p},\bar{t},\bar{t}) \,
       G^{<}_{\pi  \pi }(\vec{p},\bar{t},\bar{t})}
\: - \: \frac{1}{2} \: . \eea
Note that when  constructing the distribution function by means of
equal-time Green functions the energy variable has been
effectively integrated out. This has the advantage that the
distribution function is given independently of the actual
on-shell energies.

\begin{figure}[t]
\vspace*{0.0 cm} \hspace*{0.0 cm}
\begin{center}
\resizebox{0.7\columnwidth}{!}{\includegraphics{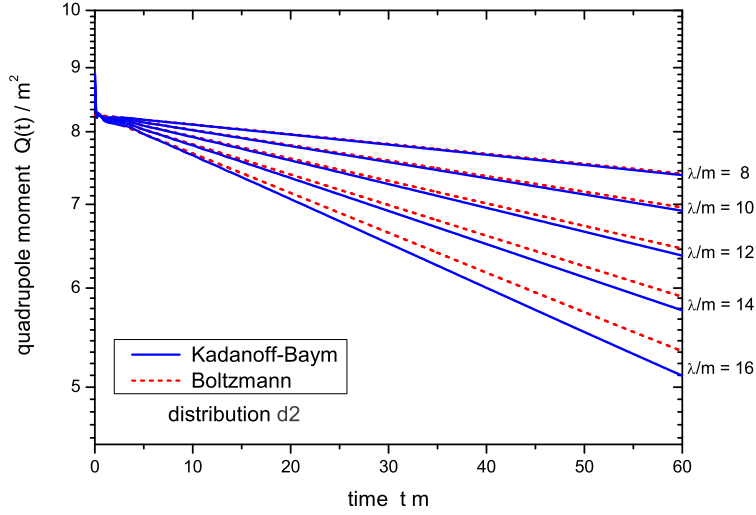} }
\end{center}
\caption{ Decrease of the quadrupole moment in
time for different coupling constants $\lambda / m = 8\,(2)\,16$
for the full Kadanoff-Baym calculation and the Boltzmann
approximation. \label{plot_qmom}}
\end{figure}

The relaxation of the quadrupole moment (\ref{quad}) has been
studied for two different initial distributions: The evolution of
distribution d2 is displayed in Fig. \ref{plot_3d1} while for
distribution d1 the position and the width of the two particle
bumps have been  modified. The calculated quadrupole moment
(\ref{quad}) shows a nearly exponential decrease with time (cf.
Fig. \ref{plot_qmom}) and one can extract a relaxation rate
$\Gamma_Q$ via the relation \begin{equation} Q(\bar{t}) \sim
\exp\left(- \Gamma_Q \bar{t}\right). \end{equation}
\begin{figure}[t]
\vspace*{0.0 cm} \hspace*{0.0 cm}
\begin{center}
\resizebox{0.7\columnwidth}{!}{\includegraphics{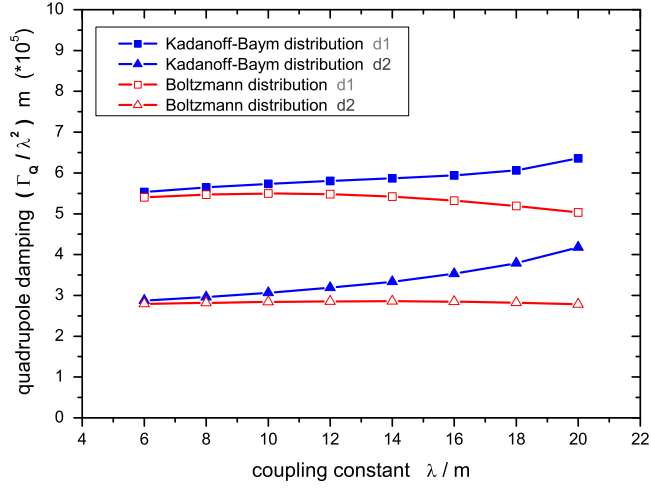} }
\end{center}
\vspace{-0.6cm} \caption{ Relaxation rate (divided by the coupling
$\lambda$ squared) for Kadanoff-Baym and Boltzmann calculations as
a function of the interaction strength. For the two different
initial configurations the full Kadanoff-Baym evolution leads to a
faster equilibration. \label{plot_quad}}
\end{figure}

Fig. \ref{plot_quad} shows for both initializations that the
relaxation in the full quantum calculation occurs faster for large
coupling constants than in the quasi-classical approximation,
whereas for small couplings the equilibration times of the full
and the approximate evolutions are comparable. We find that the
scaled relaxation rate $\Gamma_Q/\lambda^2$ is nearly constant in
the Boltzmann case, but increases with the coupling strength in
the Kadanoff-Baym calculation (especially for initial distribution
d2).

These findings are readily explained: Since the free Green
function -- as used in the Boltzmann calculation -- has only
support on the mass shell, only $(2 \leftrightarrow 2)$ scattering
processes are described in the Boltzmann limit. All other
processes with a different number of incoming and outgoing
particles vanish (as noted before). Within the full Kadanoff-Baym
calculation this is  different, since here the spectral function
-- determined from the self-consistent Green function -- aquires a
finite width. Thus the Green function has support at all energies
although it drops fast far off the mass shell. Especially for
large coupling constants, where the spectral function is
sufficiently broad, the three particle production process gives a
significant contribution to the collision integral. Since the
width of the spectral function increases with the interaction
strength, such processes become more important in the high
coupling regime. As a consequence the difference between both
approaches is larger for stronger interactions as observed in Fig.
\ref{plot_quad}. For small couplings $\lambda / m$ in both
approaches basically the usual $2 \leftrightarrow 2$ scattering
contributes and the results for the thermalization rate $\Gamma_Q$
are quite similar.

In summarizing this Subsection we point out that the full solution
of the Kadanoff-Baym equations does include $0 \leftrightarrow 4$,
$1 \leftrightarrow 3$ and $2 \leftrightarrow 2$ off-shell
collision processes which -- in comparison to the Bolzmann
on-shell $2 \leftrightarrow 2$ collision limit -- become important
when the spectral width of the particles reaches  $\sim$ 1/3 of
the particle mass. On the other hand, the simple Boltzmann limit
works surprisingly well for smaller couplings and those cases,
where the spectral function is sufficiently narrow.

\section{Off-shell relativistic transport theory}

Formal derivations of off-shell transport equations have been
presented about 40 years ago by Kadanoff and Baym \cite{KB} but
actual solutions have barely been addressed. This Section is
devoted to a transparent derivation of generalized transport
equations including numerical illustrations as well as detailed
comparisons to the full theory.

\subsection{\label{sec:general_transport_derivation} Derivation of off-shell transport theory}

The derivation of generalized transport equations starts by
rewriting the Kadanoff-Baym equation for the Wightman functions in
coordinate space ($ x_1\!=\!(t_1,\vec{x}_1),
x_2\!=\!(t_2,\vec{x}_2) $) (\ref{kabaeqcs}) as
\bea \label{eq:kbcs1} [ \, \partial^{\mu}_{x_1}
\partial_{\mu}^{x_1} + m^2 + \Sigma^{\delta}(x_1) \, ] \;\: i
G^{\gtrless}(x_1,x_2) \; = \; i\,I_1^{\gtrless}(x_1,x_2) \, . \eea
The collision terms on the r.h.s. of (\ref{eq:kbcs1}) are given in
$D = d+1$ space-time dimensions by convolution integrals over
coordinate space self-energies and Green functions:
\bea \label{eq:i1cs} I_1^{\gtrless}(x_1,x_2) \: = \: & - & \!\!
\int_{t_0}^{t_1} \!\!\! d^{D}\!z \; \; \left[ \, \Sigma^{>}(x_1,z)
- \Sigma^{<}(x_1,z) \, \right] \;\; G^{\gtrless}(z,x_2) \\[0.3cm]
& + & \!\! \int_{t_0}^{t_2} \!\!\! d^{D}\!z \; \;
\Sigma^{\gtrless}(x_1,z) \; \; \left[ \, G^{>}(z,x_2) -
G^{<}(z,x_2) \, \right] \, . \nn \eea
In the general case of an arbitrary (scalar) quantum field theory
$\Sigma^{\delta}$ is the local (non-dissipative) part of the path
self-energy while $\Sigma^{\gtrless}$ resemble the non-local
collisional self-energy contributions. In the representation
(\ref{eq:i1cs}) the integration boundaries are exclusively given
for the time coordinates, while the integration over the spatial
coordinates extends over the whole spatial volume from $- \infty$
to $+ \infty$ in $d$ dimensions.

Since transport theories   are formulated in phase-space one
changes  to the Wigner representation via Fourier transformation
with respect to the rapidly varying ('intrinsic') relative
coordinate $\Delta x = x_1 - x_2$ and treats the system evolution
in terms of the ('macroscopic') mean space-time coordinate $x =
(x_1 + x_2)/2$ and the four-momentum $p = (p_0,\vec{p})$. The
functions in Wigner space are obtained as
\bea \label{eq:wignertrafo} \bar{F}(p,x) \; = \;
\int_{-\infty}^{\infty} \!\!\! d^{D}\!\Delta x \;\; \;
e^{+i\:\Delta x_{\mu}\:p^{\mu}} \;\; F(x_1=x+\Delta
x/2,\,x_2=x-\Delta x/2) \, . \eea
For the formulation of transport theory in the Wigner
representation we have to focus not only on the transformation
properties of ordinary two-point functions as given in
(\ref{eq:wignertrafo}), but also of convolution integrals as
appearing in Eq. (\ref{eq:i1cs}). A convolution integral in $D$
dimensions (for arbitrary functions $F, G$),
\bea \label{eq:conv} H(x_1,x_2) \; = \; \int_{-\infty}^{\infty}
\!\!\! d^{D}\!z \;\; F(x_1,z) \;\; G(z,x_2) \eea
transforms as
\bea
\bar{H}(p,x)
& = &
\int_{-\infty}^{\infty} \!\!\! d^{D}\!\Delta x \;\;
\; e^{+i\:\Delta x_{\mu}\:p^{\mu}} \;\;
H(x_1,x_2)
\\[0.2cm]
& = & \int_{-\infty}^{\infty} \!\!\! d^{D}\!\Delta x \;\; \;
e^{+i\:\Delta x_{\mu}\:p^{\mu}} \;\; \int_{-\infty}^{\infty}
\!\!\! d^{D}\!z \;\; F(x_1,z) \;\; G(z,x_2) \nnl[0.4cm] & = &
\left. \; e^{+i\,\frac{1}{2} \,
(\partial_{p^{\phantom{\prime}}}^{\mu} \!\! \cdot \:
 \partial^{x^{\prime}}_{\mu}
 \: - \;
 \partial_{x^{\phantom{\prime}}}^{\mu} \!\! \cdot \:
 \partial^{p^{\prime}}_{\mu} ) } \;\:
\left[ \; \bar{F}(p,x) \;\; \bar{G}(p^{\prime},x^{\prime}) \;
\right] \right|_{x^{\prime} = x,\: p^{\prime} = p} \; . \nn \eea
In accordance with the standard assumption of transport theory we
assume that all functions only smoothly evolve in the mean
space-time coordinates and thus restrict to first order
derivatives. All terms proportional to second or higher order
derivatives in the mean space-time coordinates (also mixed ones)
will be dropped. Thus the Wigner transformed convolution integrals
(\ref{eq:conv}) are given in {\it first order gradient
approximation} by,
\bea \label{eq:firstordergrad} \bar{H}(p,x) \; = \; \bar{F}(p,x)
\;\; \bar{G}(p,x) \; + \; i \, \frac{1}{2} \: \{ \, \bar{F}(p,x)
\, , \, \bar{G}(p,x) \, \} \; + \; {\cal O}(\partial^2_x) \, ,
\eea
using the relativistic generalization of the Poisson bracket
\bea
\label{eq:def_poisson}
\{ \, \bar{F}(p,x) \, , \, \bar{G}(p,x) \, \}
\; = \;
\partial^{p}_{\mu} \, \bar{F}(p,x) \cdot
\partial_{x}^{\mu} \, \bar{G}(p,x) \; - \;
\partial_{x}^{\mu} \, \bar{F}(p,x) \cdot
\partial^{p}_{\mu} \, \bar{G}(p,x) \; .
\eea

In order to obtain the dynamics for the spectral functions within
the approximate scheme we start with the Dyson-Schwinger equations
for the retarded and advanced Green functions in coordinate space
(\ref{dseqretcs},\ref{dseqadvcs}). -- Note that the convolution
integrals in (\ref{dseqretcs}) and (\ref{dseqadvcs}) extend over
the whole space and time range in contrast to the equations of
motion for the Wightman functions given in (\ref{eq:kbcs1}) and
(\ref{eq:i1cs})! -- The further procedure consists in the
following steps: First we \\
\\
i) transform the above equations into the Wigner representation
and apply the first order gradient approximation. In this limit
the convolution integrals yield the product terms and the general
Poisson bracket of the self-energies and the Green functions $\{\,
\Sigma^{R/A}, G^{R/A} \,\}$. We, furtheron, represent both
equations in terms of real quantities by the decomposition of the
retarded and advanced Green functions and self-energies as
\bea
\begin{array}{ccccccc}
\bar{G}^{R/A}
&\!\!=\!\!& Re\,\bar{G}^{R} \,\pm\, i\,Im\,\bar{G}^{R}
&\!\!=\!\!& Re\,\bar{G}^{R} \,\mp\, i\,\bar{A} / 2\; ,
\phantom{aaaaaa}
\bar{A} &\!\!=\!\!& \mp \, 2 \, Im\,\bar{G}^{R/A} \, ,
\\[0.7cm]
\bar{\Sigma}^{R/A}
&\!\!=\!\!& Re\,\bar{\Sigma}^{R} \,\pm\, i\,Im\,\bar{\Sigma}^{R}
&\!\!=\!\!& Re\,\bar{\Sigma}^{R} \,\mp\, i\,\bar{\Gamma} / 2 \; ,
\phantom{aaaaaa}
\bar{\Gamma} &\!\!=\!\!& \mp \, 2\, Im\,\bar{\Sigma}^{R/A} \, .
\end{array}
\phantom{aa} \eea
We find that in Wigner space the real parts of the retarded and
advanced Green functions and self-energies are equal, while the
imaginary parts have opposite sign and are proportional to the
spectral function $\bar{A}$ and the width $\bar{\Gamma}$,
respectively. The next step consists in \\ \\ ii) the separation
of the real part and the imaginary part of the two equations for
the retarded and advanced Green functions, that have to be
fulfilled independently. Thus we obtain four real-valued equations
for the self-consistent retarded and advanced Green functions. In
the last step \\ \\
iii) we get simple relations by linear
combination of these equations, i.e. by adding/subtrac\-ting the
relevant equations.

This finally leads to two algebraic relations for the spectral
function $\bar{A}$ and the real part of the retarded Green
function $Re\,\bar{G}^{R}$ in terms of the width $\bar{\Gamma}$
and the real part of the retarded self-energy
$Re\,\bar{\Sigma}^{R}$ as \cite{caju1,caju2}:
\bea
\label{eq:specrel1}
[ \, p_0^2 - \vec{p}^{\,2} - m^2 -
\bar{\Sigma}^{\delta} + Re\,\bar{\Sigma}^{R} \, ] \; Re\,\bar{G}^{R}
& = &
1 \: + \: \frac{1}{4} \: \bar{\Gamma} \; \bar{A} \, ,
\\[0.6cm]
\label{eq:specrel2} [ \, p_0^2 - \vec{p}^{\,2} - m^2 -
\bar{\Sigma}^{\delta} + Re\,\bar{\Sigma}^{R} \, ] \; \bar{A} & = &
\bar{\Gamma} \; Re\,\bar{G}^{R} \, . \eea
Note that all terms with first order gradients have disappeared in
(\ref{eq:specrel1}) and (\ref{eq:specrel2}). A first consequence
of (\ref{eq:specrel2}) is a direct relation between the real and
the imaginary parts of the retarded/advanced Green function, which
reads (for $\bar{\Gamma} \neq 0$):
\bea \label{ins1}
 Re\,\bar{G}^{R} \; = \; \frac{p_0^2 -
\vec{p}^{\,2} - m^2 - \bar{\Sigma}^{\delta} - Re\,\bar{\Sigma}^{R}
}{\bar{\Gamma}} \; \bar{A} \; . \eea
Inserting (\ref{ins1}) in (\ref{eq:specrel1})  we end up with the
following result for the spectral function and the real part of
the retarded Green function
\bea
\label{eq:specorder0}
\bar{A} \; = \;
\frac{\bar{\Gamma}}{[ \, p_0^2 - \vec{p}^{\,2} - m^2
- \bar{\Sigma}^{\delta} - Re\,\bar{\Sigma}^{R} \, ]^2 + \bar{\Gamma}^2/4}
& = &
\frac{\bar{\Gamma}}{\bar{M}^2 + \bar{\Gamma}^2/4} \, , \phantom{aaa}
\\[0.6cm]
\label{eq:regretorder0} Re\,\bar{G}^{R} \; = \; \frac{[ \, p_0^2 -
\vec{p}^{\,2} - m^2 - \bar{\Sigma}^{\delta} - Re\,\bar{\Sigma}^{R}
\, ]} {[ \, p_0^2 - \vec{p}^{\,2} - m^2 - \bar{\Sigma}^{\delta} -
Re\,\bar{\Sigma}^{R} \, ]^2 + \bar{\Gamma}^2/4} & = &
\frac{\bar{M}}{\bar{M}^2 + \bar{\Gamma}^2/4} \, , \eea
where we have introduced the mass-function $\bar{M}(p,x)$ in
Wigner space:
\bea \label{eq:massfunction} \bar{M}(p,x) & = & p_0^2 -
\vec{p}^{\,2} - m^2 - \bar{\Sigma}^{\delta}(x) -
Re\,\bar{\Sigma}^{R}(p,x) \; . \eea
The  spectral function (\ref{eq:specorder0}) shows a typical
Breit-Wigner shape with energy- and momentum-dependent self-energy
terms. Although the above equations are purely algebraic solutions
and contain no derivative terms, they are valid up to the first
order in the gradients!

In addition, subtraction of the real parts and adding up the
imaginary parts lead to the time evolution equations
\bea
\label{eq:specorder1}
p^{\mu} \, \partial_{\mu}^x \, \bar{A}
& = &
\frac{1}{2} \,
\{ \, \bar{\Sigma}^{\delta} + Re\,\bar{\Sigma}^{R} \, ,
   \, \bar{A} \, \}
\: + \: \frac{1}{2} \,
\{ \, \bar{\Gamma} \, , \, Re\,\bar{G}^{R} \, \} \, ,
\\[0.6cm]
p^{\mu} \, \partial_{\mu}^x \, Re\,\bar{G}^{R}
& = &
\label{eq:regretorder1}
\frac{1}{2} \,
\{ \, \bar{\Sigma}^{\delta} + Re\,\bar{\Sigma}^{R} \, ,
   \, Re\,\bar{G}^{R} \, \}
\: - \: \frac{1}{8} \, \{ \, \bar{\Gamma} \, , \, \bar{A} \, \} \,
. \eea
The Poisson bracket containing the mass-function $\bar{M}$ leads
to the well-known drift operator
$p^{\mu}\,\partial^{x}_{\mu}\,\bar{F}$ (for an arbitrary function
$\bar{F}$), i.e.
\bea \label{eq:mass_poisson} \{\, \bar{M} \, , \, \bar{F} \,\} & =
& \{\, p_0^2 - \vec{p}^{\,2} - m^2 - \bar{\Sigma}^{\delta} -
Re\,\bar{\Sigma}^{R} \, , \, \bar{F} \,\} \\[0.6cm] & = & 2 \,
p^{\mu} \, \partial_{\mu}^x \: \bar{F} \: - \:
\{\,\bar{\Sigma}^{\delta} + Re\,\bar{\Sigma}^{R} \, , \, \bar{F}
\,\} \; , \eea
such that the first order equations (\ref{eq:specorder1}) and
(\ref{eq:regretorder1}) can be written in a more comprehensive
form as
\bea
\label{eq:specorder1final}
\{ \, \bar{M} \, , \, \bar{A} \, \}
& = &
\{ \, \bar{\Gamma} \, , \, Re\,\bar{G}^{R} \, \} \, ,
\\[0.6cm]
\{ \, \bar{M} \, , \, Re\,\bar{G}^{R} \, \} & = &
\label{eq:regretorder1final} - \, \frac{1}{4} \, \{ \,
\bar{\Gamma} \, , \, \bar{A} \, \} \, . \eea
When inserting (\ref{eq:specorder0}) and (\ref{eq:regretorder0})
we find that these first order time evolution equations are {\em
solved} by the algebraic expressions. In this case the following
relations hold:
\bea
\{\, \bar{M} \, , \, \bar{A} \,\}
\; = \;
\{\, \bar{\Gamma} \, , \, Re\,\bar{G}^{R} \,\}
& = &
\{\, \bar{M} \, , \, \bar{\Gamma} \,\} \;\;
\frac{\bar{M}^2 - \bar{\Gamma}^2/4}
{[\, \bar{M}^2 + \bar{\Gamma}^2/4 \,]^2} \, ,
\\[0.6cm]
\{\, \bar{M} \, , \, Re\,\bar{G}^{R} \,\} \; = \; - \, \frac{1}{4}
\, \{\, \bar{\Gamma} \, , \, \bar{A} \,\} & = & \{\, \bar{M} \, ,
\, \bar{\Gamma} \,\} \;\; \frac{ \bar{M} \, \bar{\Gamma} / 2 }
{[\, \bar{M}^2 + \bar{\Gamma}^2/4 \,]^2} \, . \eea
Thus we have derived the proper structure of the spectral function
(\ref{eq:specorder0}) within the first-order gradient (or
semiclassical) approximation. Together with the explicit form for
the real part of the retarded Green function
(\ref{eq:regretorder0}) we now have fixed the dynamics of the
spectral properties, which is consistent up to first order in the
gradients.

\subsubsection{Kadanoff-Baym transport}
As a next step we rewrite the memory terms in the collision
integrals (\ref{eq:i1cs}) such that the time integrations extend
from $- \infty$ to $+ \infty$. In this respect we consider the
initial time $t_0 = - \infty$ whereas the upper time boundaries
$t_1, t_2$ are taken into account by $\Theta$-functions, i.e.
\bea \label{eq:i1csnew} I_1^{\gtrless}(x_1,x_2) \: = \: & - & \!\!
\int_{-\infty}^{\infty} \!\!\! d^{D}x^{\prime} \;\;\;
\Theta(t_1-t^{\prime}) \: \left[ \, \Sigma^{>}(x_1,x^{\prime}) -
\Sigma^{<}(x_1,x^{\prime}) \, \right] \;\;
G^{\gtrless}(x^{\prime},x_2) \nnl[0.2cm] & + & \!\!
\int_{-\infty}^{\infty} \!\!\! d^{D}x^{\prime} \;\;\;
\Sigma^{\gtrless}(x_1,x^{\prime}) \; \; \Theta(t_2-t^{\prime}) \:
\left[ \, G^{>}(x^{\prime},x_2) - G^{<}(x^{\prime},x_2) \, \right]
\nnl[0.5cm] \: = \: & - & \!\! \int_{-\infty}^{\infty} \!\!\!
d^{D}x^{\prime} \;\;\;\; \Sigma^{R}(x_1,x^{\prime}) \;
G^{\gtrless}(x^{\prime},x_2) \: + \:
\Sigma^{\gtrless}(x_1,x^{\prime}) \; G^{A}(x^{\prime},x_2) \; .
\phantom{aaaa} \eea
We now perform the analogous steps as invoked before for the
retarded and advanced Dyson-Schwinger equations. We start with a
first order gradient expansion of the Wigner transformed
Kadanoff-Baym equation using (\ref{eq:i1csnew}) for the memory
integrals. Again we separate the real and the imaginary parts in
the resulting equation, which have to be satisfied independently.
At the end of this procedure we obtain a generalized transport
equation \cite{KB,GL98a,botmal,caju1,caju2,Leupold,knoll3,knoll6}:
\\
\bea
\label{eq:general_transport}
\underbrace{
\phantom{\frac{1}{1}} \!\!\!
2\,p^{\mu}\:\partial^{x}_{\!\mu} \: i\bar{G}^{\gtrless}
\, - \,
\{ \, \bar{\Sigma}^{\delta} \!+\! Re\,\bar{\Sigma}^{R} ,
   \, i \bar{G}^{\gtrless} \, \} }
\, - \,
\{ \, i\bar{\Sigma}^{\gtrless} \, , \, Re\,\bar{G}^{R} \, \}
& = &
i\bar{\Sigma}^{<} \; i\bar{G}^{>}
\, - \,
i\bar{\Sigma}^{>} \; i\bar{G}^{<}
\nnl[0.4cm]
\{ \, \bar{M} \, , \, i \bar{G}^{\gtrless} \, \}  \;\; \qquad \qquad
\, - \,
\{ \, i\bar{\Sigma}^{\gtrless} \, , \, Re\,\bar{G}^{R} \, \}
& = &
i\bar{\Sigma}^{<} \; i\bar{G}^{>}
\, - \,
i\bar{\Sigma}^{>} \; i\bar{G}^{<} \phantom{aaaaaaa}
\eea\\
as well as a generalized mass-shell equation \\
\bea
\label{eq:general_mass}
\underbrace{
\phantom{\frac{1}{1}} \!\!\!\!
[ \, p^2 - m^2 - \bar{\Sigma}^{\delta} - Re\,\bar{\Sigma}^{R} \, ]}_{\bar{M}}
\;\,
i \bar{G}^{\gtrless}
\: = \:
i\bar{\Sigma}^{\gtrless} \; Re\,\bar{G}^{R}
\, + \, \frac{1}{4} \, \{ \,
i\bar{\Sigma}^{>} , \, i\bar{G}^{<} \, \}
\, - \, \frac{1}{4} \, \{ \,
i\bar{\Sigma}^{<} , \, i\bar{G}^{>} \, \}  \phantom{aaa}
\eea\\
with the mass-function $\bar{M}$ specified in
(\ref{eq:massfunction}). Since the Green function
$G^{\gtrless}(x_1,x_2)$ consists of an antisymmetric real part and
a symmetric imaginary part with respect to the relative coordinate
$x_1-x_2$, the Wigner transform of this function is purely
imaginary. It is thus convenient to represent the Wightman
functions in Wigner space  by the real-valued quantities $i
\bar{G}^{\gtrless}(p,x)$. Since the collisional self-energies obey
the same symmetry relations in coordinate space and in
phase-space, they will be kept also as $i
\bar{\Sigma}^{\gtrless}(p,x)$ furtheron.

In the transport equation (\ref{eq:general_transport}) one
recognizes on the l.h.s. the drift term
$p^{\mu}\:\partial^{x}_{\!\mu} \: i\bar{G}^{\gtrless}$, as well as
the Vlasov term with the local self-energy $\bar{\Sigma}^{\delta}$
and the real part of the retarded self-energy
$Re\,\bar{\Sigma}^{R}$. On the other hand the r.h.s. represents
the collision term with its typical `gain and loss' structure. The
loss term $i\bar{\Sigma}^{>} \; i\bar{G}^{<}$ (proportional to the
Green function itself) describes the scattering out of a
respective phase-space cell whereas the gain term
$i\bar{\Sigma}^{<} \; i\bar{G}^{>}$ takes into account scatterings
into the actual cell. The last term on the l.h.s. $\{\,
i\bar{\Sigma}^{\gtrless} , Re\,\bar{G}^{R} \, \}$ is very {\em
peculiar} since it does not contain directly the distribution
function $i\bar{G}^{<}$. This second Poisson bracket vanishes in
the quasiparticle approximation and thus does not appear in the
on-shell Boltzmann limit. As demonstrated in detail in Refs.
\cite{KB,caju1,caju2,Leupold,knoll3,knoll6}
 the second Poisson bracket $\{ \,
i\bar{\Sigma}^{\gtrless} , Re\,\bar{G}^{R} \, \}$ governs the
evolution of the off-shell dynamics for nonequilibrium systems.

Although the generalized transport equation
(\ref{eq:general_transport}) and the generalized mass-shell
equation (\ref{eq:general_mass}) have been derived from the same
Kadanoff-Baym equation in a first order gradient expansion, both
equations are not exactly equivalent \cite{botmal,caju1,knoll6}.
Instead, they deviate from each other by contributions of second
gradient order, which are hidden in the term $\{\,
i\bar{\Sigma}^{\gtrless} , Re\,\bar{G}^{R} \, \}$ (see below or
Refs. \cite{botmal,knoll6} for extended discussions). This raises
the question: {\it which one of these two equations has to be
considered of higher priority?} The question is answered in
practical applications by the prescription of solving the
generalized transport equation (\ref{eq:general_transport}) for
$i\bar{G}^{<}$ in order to study the dynamics of the
nonequilibrium system in phase-space. Since the dynamical
evolution of the spectral properties is taken into account by the
equations derived in first order gradient expansion from the
retarded and advanced Dyson-Schwinger equations, one can neglect
the generalized mass-shell equation (\ref{eq:general_mass}). Thus
for our actual numerical studies in Section 2.2 we will  use the
generalized transport equation (\ref{eq:general_transport})
supported by the algebraic relations (\ref{eq:specorder0}) and
(\ref{eq:regretorder0}).

\subsubsection{Transport in the Botermans-Malfliet scheme}
Furthermore, one recognizes by subtraction of the $i\bar{G}^{>}$ and
$i\bar{G}^{<}$ mass-shell and transport equations, that the dynamics
of the spectral function $\bar{A} = i\bar{G}^{>} - i\bar{G}^{<}$ is
determined in the same way as derived from the retarded and advanced
Dyson-Schwinger equations (\ref{eq:specorder0}) and
(\ref{eq:specorder1final}).
The inconsistency between the two equations
(\ref{eq:general_transport}) and (\ref{eq:general_mass})
vanishes  since the differences are
contained in the collisional contributions on the r.h.s.
of (\ref{eq:general_transport}).

In order to evaluate the $\{\, i\bar{\Sigma}^{<} , Re\,\bar{G}^{R}
\,\}$-term on the l.h.s. of (\ref{eq:general_transport}) and to
explore the differences between the KB- and BM-form of the
transport equations (see below) it is useful to introduce
distribution functions for the Green functions and self-energies
as \\
\bea
\label{eq:ansatz_dist}
i\bar{G}^{<}(p,x) \; = \;
\bar{N}(p,x) \; \bar{A}(p,x) \: ,
\quad &&
i\bar{G}^{>}(p,x) \; = \;
[ \, 1 \, + \, \bar{N}(p,x) \, ] \; \bar{A}(p,x) \: ,
\\[0.6cm]
i\bar{\Sigma}^{<}(p,x) \; = \;
\bar{N}^{\Sigma}(p,x) \; \bar{\Gamma}(p,x) \: ,
\quad &&
i\bar{\Sigma}^{>}(p,x) \; = \;
[ \, 1 \, + \, \bar{N}^{\Sigma}(p,x) \, ] \; \bar{\Gamma}(p,x) \: .
\phantom{aaaa}
\eea\\
In equilibrium the distribution function with respect to the Green
functions $\bar{N}$ and the self-energies $\bar{N}^{\Sigma}$ are
given as Bose functions in the energy $p_0$ at given temperature;
they thus are equal in equilibrium but in general might differ
out-of-equilibrium. Following the argumentation of Botermans and
Malfliet \cite{botmal} the distribution functions $\bar{N}$ and
$\bar{N}^{\Sigma}$ in (\ref{eq:ansatz_dist}) should be identical
within the second term of the l.h.s. of
(\ref{eq:general_transport}) in order to obtain a consistent first
order gradient expansion (without hidden higher order gradient
terms). In order to demonstrate their argument we write
\bea i\bar{\Sigma}^{<} \; = \; \bar{\Gamma} \; \bar{N}^{\Sigma} \;
= \; \bar{\Gamma} \; \bar{N} \; + \; \bar{K} \: . \eea
The `correction' term
\bea
\bar{K} \; = \; \bar{\Gamma} \;
( \, \bar{N}^{\Sigma} \; - \; \bar{N} \, )
\; = \;
( \, i\bar{\Sigma}^{<} \; i\bar{G}^{>} \; - \;
     i\bar{\Sigma}^{>} \; i\bar{G}^{<} \,) \; \bar{A}^{-1} \: ,
\eea
is proportional to the collision term  of the generalized
transport equation (\ref{eq:general_transport}), which itself is
already  of first order in the gradients. Thus, whenever a
distribution function $\bar{N}^{\Sigma}$ appears within a Poisson
bracket, the difference term $(\bar{N}^{\Sigma} - \bar{N}$)
becomes of second order in the gradients and should be omitted for
consistency. As a consequence $\bar{N}^{\Sigma}$ can be replaced
by $\bar{N}$ and thus the self-energy $\bar{\Sigma}^{<}$ by
$\bar{G}^< \cdot \bar{\Gamma} / \bar{A}$ in the Poisson bracket
term $\{ \bar{\Sigma}^{<} , Re\,\bar{G}^{R} \}$. The generalized
transport equation (\ref{eq:general_transport}) then can be
written in short-hand notation
\bea \label{eq:general_transport_bm} \frac{1}{2} \, \bar{A} \;
\bar{\Gamma} \; \left[ \, \{ \, \bar{M} \, , \, \, i\bar{G}^{<} \,
\} \: - \: \frac{1}{\bar{\Gamma}} \; \{ \, \bar{\Gamma} \, , \,
\bar{M} \cdot i\bar{G}^{<} \, \} \, \right] \; = \;
i\bar{\Sigma}^{<} \; i\bar{G}^{>} \, - \, i\bar{\Sigma}^{>} \;
i\bar{G}^{<} \;\;\;\;\;\; \eea
with the mass-function $\bar{M}$ (\ref{eq:massfunction}). The
transport equation (\ref{eq:general_transport_bm}) within the
Botermans-Malfliet (BM) form resolves the discrepancy between the
generalized mass-shell equation (\ref{eq:general_mass}) and the
generalized transport equation in its original Kadanoff-Baym (KB)
form (\ref{eq:general_transport}).

 However, it is presently not clear in how far the
generalized transport equations in KB-form
(\ref{eq:general_transport}) or in BM-form
(\ref{eq:general_transport_bm}) reproduce the same dynamics as for
the full Kadanoff-Baym theory (\ref{kabaeqcs}). Moreover, the
differences in the time evolution of nonequilibrium systems
between the KB-form (\ref{eq:general_transport}) and BM-form
(\ref{eq:general_transport_bm}) are not known either. We will thus
perform an explicit comparison between the different limits for
$\phi^4$-theory in 2+1 dimensions in the following Section.

\subsection{\label{sec:gradient_dynamics} Numerical studies on off-shell transport}

In this Section we will perform numerical studies of the dynamics
inherent in the generalized transport equations  in comparison to
solutions of the full Kadanoff-Baym equations (\ref{kabaeqcs}) for
the $\phi^{4}$-theory in 2+1 space-time dimensions within the
three-loop approximation for the effective action. This fixes the
self-energies $\bar{\Sigma}^{\delta}$ and
$i\bar{\Sigma}^{\gtrless}$ in (\ref{eq:general_transport}) and
(\ref{eq:general_mass}) to be the same as in the case of the full
Kadanoff-Baym theory given by the tadpole (\ref{tadpole_cs}) and
the sunset (\ref{sunset_cs}) contributions, respectively.

For the first investigation we concentrate on the dynamics of the
generalized transport equation (\ref{eq:general_transport}). As in
Section 1.2 we restrict ourselves to homogeneous systems in space.
Consequently the derivatives with respect to the mean spatial
coordinate $\vec{x}$ vanish, such that the generalized transport
equation (\ref{eq:general_transport}) reduces to
\bea
\label{eq:homo_general_transport}
2\,p_0\:\partial_{t} \: i\bar{G}^{<}
\, - \,
\{ \, \bar{\Sigma}^{\delta} \!+\! Re\,\bar{\Sigma}^{R} \, ,
   \, i \bar{G}^{<} \, \}
\, - \, \{ \, i\bar{\Sigma}^{<} \, , \, Re\,\bar{G}^{R} \, \} \: =
\: i\bar{\Sigma}^{<} \; i\bar{G}^{>} \, - \, i\bar{\Sigma}^{>} \;
i\bar{G}^{<} \;\;\;\;\;\; \eea
with the simplified Poisson brackets (for arbitrary functions
$\bar{F}, \bar{G}$)
\bea
\label{eq:def_homo_poisson}
&&
\{ \, \bar{F}(\vec{p},p_0,t) \, , \, \bar{G}(\vec{p},p_0,t) \, \}
\\[0.4cm]
&& \qquad \qquad \; = \;
\partial_{p_0} \,\bar{F}(\vec{p},p_0,t) \;
\partial_{t}   \,\bar{G}(\vec{p},p_0,t) \; - \;
\partial_{t}   \,\bar{F}(\vec{p},p_0,t) \;
\partial_{p_0} \,\bar{G}(\vec{p},p_0,t) \; . \phantom{aaa}
\nn \eea
In order to obtain a numerical solution the Kadanoff-Baym equation
(\ref{kabaeqcs}) as well as the generalized transport equations
(\ref{eq:general_transport}) are transformed to momentum space.
 For the actual numerical integration
of the generalized transport equation we will use the same grid in
momentum space as employed for the full KB-theory in Section 1.2.

\subsubsection{\label{sec:gradient_equilibration} Phases of
Equilibration}

In order to specify the problem for the first order transport
equation in time, the initial state has to be fixed, i.e., the
initial Green function $i\bar{G}^{<}(\vec{p},p_0,t=0)$ has to be
specified for all momenta $\vec{p}$ and all energies $p_0$.
According to (\ref{eq:ansatz_dist})
$i\bar{G}^{<}(\vec{p},p_0,t=0)$ can be written as a product of a
distribution function $\bar{N}(\vec{p},p_0,t=0)$ and a spectral
function $\bar{A}(\vec{p},p_0,t=0)$. In order to determine
$\bar{N}$ and $\bar{A}$  we employ an iterative scheme: We first
assume initial momentum-distribution functions
$n(\vec{p},t\!=\!0)$  equivalent to those used in the
investigation of the full Kadanoff-Baym theory in Section 1.2 in
order to allow for a comparison of the two schemes. These initial
momentum distributions $n(\vec{p},t\!=\!0)$ then fix
$i\bar{G}^{<}(\vec{p},t=0,t=0)$ on the time-diagonal axis by
\cite{Juchem03}
\begin{equation}
\label{init0} 2 \omega_{\bf p}\ i\bar{G}^{<}(\vec{p},t=0,t=0) =
n(\vec{p},t\!=\!0)+ n(-\vec{p},t\!=\!0) + 1.
\end{equation}
Next, the complete initial phase-space distribution function
$\bar{N}(\vec{p},p_0,t\!=\!0)$ in Wigner representation as well as
the spectral function $\bar{A}(\vec{p},p_0,t=0)$ is specified as a
function of the energy $p_0$ in an iterative way. As a starting
point we assume $\bar{N}({\vec p},p_0)$ to be constant in the
energy variable $p_0$. In order to determine the initial Green
function $i\bar{G}^{<}$ we employ the self-consistent iteration
procedure for the full spectral function as described in Appendix
D of Ref. \cite{Juchem03} using the specified nonequilibrium
distribution function $\bar{N}$. The iteration process then yields
the fully self-consistent spectral function
$\bar{A}(\vec{p},p_0,t\!=\!0)$ for this initial distribution and
thus determines the initial Green function via \\
\bea
i\bar{G}^{<}(\vec{p},p_0,t\!=\!0) \; = \;
\bar{N}(\vec{p},p_0,t\!=\!0) \; \bar{A}(\vec{p},p_0,t\!=\!0) \: .
\eea\\
The advantage of the initialization prescription introduced above
is that the actual spectral function -- directly obtained by
$\bar{A} = i\bar{G}^{>} - i\bar{G}^{<}$ from the Green functions
-- complies with the one determined from the self-energies
(\ref{eq:specorder0}) in accordance with the first order gradient
expansion scheme. During the nonequilibrium time evolution this
correspondence is maintained since the {\em analytic} expression
for the spectral function already is a solution of the generalized
transport equation itself. Furthermore, the real part of the
retarded Green function, that enters the peculiar second Poisson
bracket on the l.h.s. in (\ref{eq:general_transport}), can be
taken in the first order scheme (\ref{eq:regretorder0}) which
simplifies the calculations considerably.

\begin{figure}[tbh]
\resizebox{0.48\columnwidth}{!}{\includegraphics{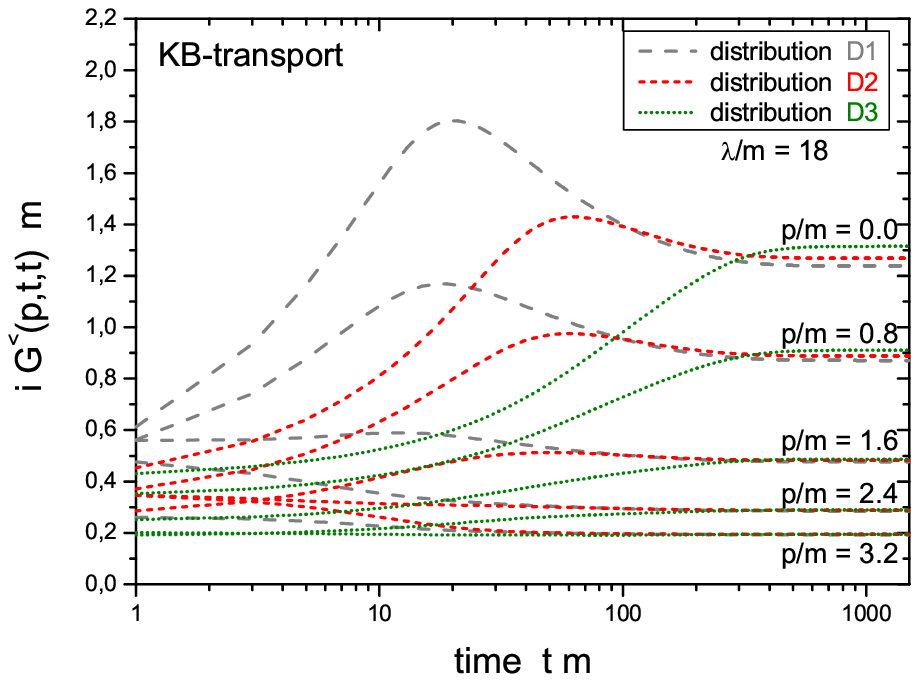} }
\resizebox{0.48\columnwidth}{!}{\includegraphics{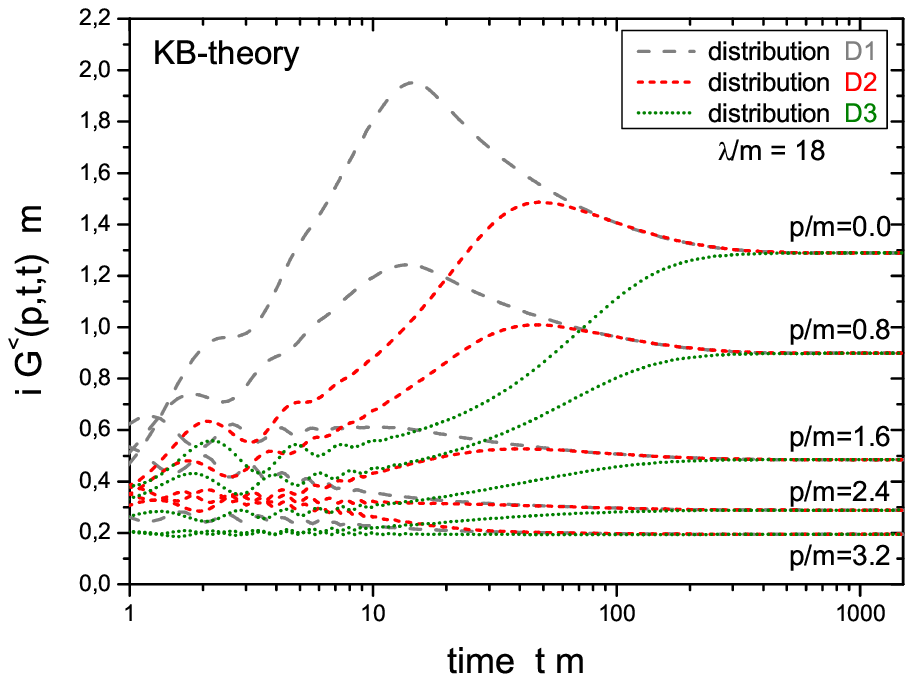} }
\caption{\label{fig:4phd_gradient_modes} Left part: Evolution of
several momentum modes $|\,\vec{p}\,|/m =$ 0.0, 0.8, 1.6, 2.4,
3.2, 4.0 of the equal-time Green function on a logarithmic time
scale for the different initializations D1, D2 and D3 for the
generalized transport equation
(\protect\ref{eq:general_transport}). Right part: same as above
but for the full Kadanoff-Baym equation (\protect\ref{kabaeqcs}).}
\end{figure}

Now we turn to the actual solutions of the generalized transport
equation in the KB form (\ref{eq:general_transport}). In Fig.
\ref{fig:4phd_gradient_modes} (left part) we show the time
evolution of the equal-time Green function
$iG^{<}(|\,\vec{p}\,|,t,t)$ for the polar symmetric initial states
D1, D2 and D3. Displayed are several momentum modes
$|\,\vec{p}\,|/m =$ 0.0, 0.8, 1.6, 2.4, 3.2, 4.0 of the equal-time
Green function on a logarithmic time scale. As in the full
Kadanoff-Baym theory (right part) one finds that for all
initializations the quantum system approaches a stationary state
for $t \rightarrow \infty$, i.e. all momentum modes approach a
constant.

However, the respective momentum modes of the different
initializations do not achieve identical values for
$t \rightarrow \infty$, as seen
in particular for the low momenta $|\,\vec{p}\,|/m =$ 0.0, 0.8 in
 Fig. \ref{fig:4phd_gradient_modes} (left part).
This is not surprising since the various initializations --
obtained within the self-consistent scheme described above -- do
not correspond to exactly the same energy. This is why the
respective long-time limits differ slightly. The small difference
in energy is, of course, most prominently seen in the low momentum
(energy) modes. Moreover, the dynamics within the generalized
transport equation (\ref{eq:general_transport}) is in general very
similar to the full Kadanoff-Baym theory (right part). For all
three initial states we find (apart from the very initial phase $t
\cdot m <$ 5) the same structures during the equilibration
process. In particular for the initializations D1 and D2 the
characteristic overshooting for the low momentum modes is seen as
in the full quantum evolution, which does not show up in solutions
of the corresponding on-shell Boltzmann limit. Since in the
Boltzmann limit a strictly monotonous evolution of the momentum
modes is seen  (cf. Ref. \cite{Juchem03}) this overshooting has to
be attributed to an off-shell quantum effect. Even the positions
of the maxima are in a comparable range: For the initialization D1
they are shifted to slightly larger times and are a little bit
lower than in the full calculation; the same holds for the initial
state D2. The initial distribution D3 yields a monotonous
behaviour for all momentum modes within the generalized transport
formulation which is again in a good agreement with the full
dynamics.

Some comments are worthwhile with respect to the comparison
performed above: The spectral function in the Kadanoff-Baym
calculation is completely undetermined in the initial state; it
develops during the very early phase to an approximate form (which
in the following still evolves in time). In contrast to this, the
spectral function in the generalized transport formulation
(\ref{eq:general_transport}) has a well-defined structure already
from the beginning. This principle difference results from the
fact, that in the Kadanoff-Baym case we deal with a true initial
value problem in the two time directions ($t_1,t_2$).
Additionally, the relative time integral in ($t_1-t_2$) -- to
obtain the spectral function in energy $p_0$ by Wigner
transformation -- is very small.  Consequently, the spectral shape
in Wigner space is determined  by the finite integration interval
in time rather than by the interactions itself. On the other hand,
we have used an infinite relative time range in deriving the
generalized transport equation within the first order gradient
expansion. Thus in this case we deal with a completely resolved
spectral function already at the initial time! This demonstrates
why both approaches can only be compared to a limited extent for
the very early times!

Finally, concentrating on the very early time behaviour, we find a
significant difference between the full and the approximate
dynamics in the gradient scheme (\ref{eq:general_transport}). For
the genera\-lized transport equation we observe a monotonous
evolution of the equal-time Green function momentum modes whereas
strong oscillations are seen in the initial phase for the solution
of the full Kadanoff-Baym theory. Thus -- with respect to the
early time behaviour -- the generalized transport equation behaves
much more like the Boltzmann approximation, which is a first order
differential equation in time as well. However, the Kadanoff-Baym
evolution is given by an integro-differential equation of second
order in time. In this case the phase correlations between the
Green functions $G^{<}_{\phi \phi},\,G^{<}_{\pi \phi},\,G^{<}_{\pi
\pi}$ are kept and the instantaneous switching-on of the
interaction results in an oscillatory behaviour of the single
momentum modes.

\subsubsection{\label{sec:gradient_spectralevolution} Evolution of the
Spectral Function}
\begin{figure}[t]
\centerline{\resizebox{0.95\columnwidth}{!}{\includegraphics{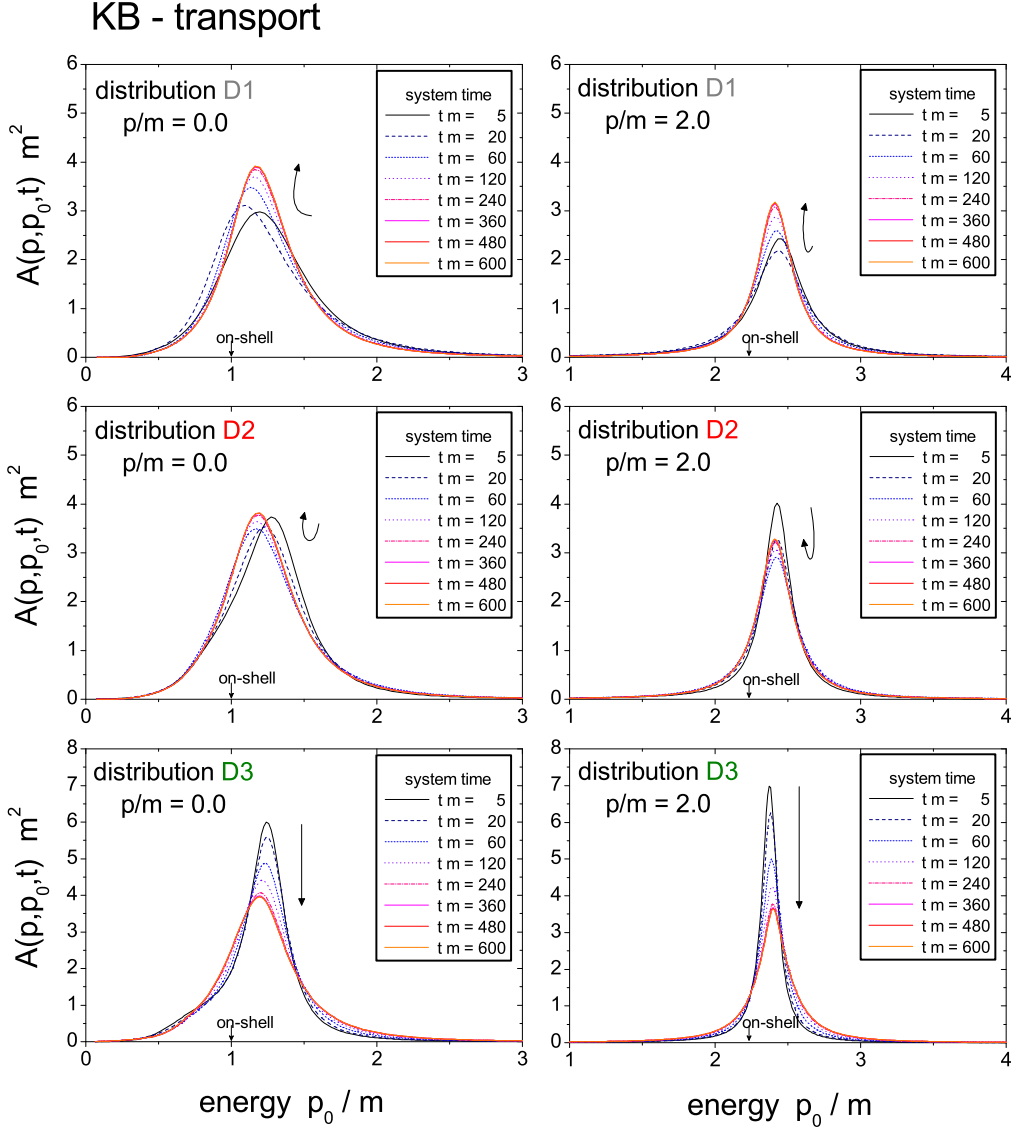} }}
\caption{\label{fig:4phd_gradient_spec} Time evolution of the
spectral function $\bar{A}(\vec{p},p_0,t)$ for the initial
distributions D1, D2 and D3 (from top to bottom) with coupling
constant $\lambda / m =$ 18 for the two momentum modes
$|\,\vec{p}\,| / m =$ 0.0 (l.h.s.) and $|\,\vec{p}\,| / m =$ 0.0
(r.h.s.). The spectral function from the transport eq.
(\ref{eq:general_transport}) is shown for times $t \cdot m =$ 5,
20, 60, 120, 240, 360, 480, 600 as indicated by the different line
types.} \vspace*{-0.1cm}
\end{figure}

Since the Green functions develop in time also the spectral properties
of the system change as well. In Fig.  \ref{fig:4phd_gradient_spec} the
time evolution of the spectral functions for the initializations D1, D2
and D3 within the gradient scheme are displayed for two particular
momentum modes $|\,\vec{p}\,| / m =$ 0.0 (l.h.s.) and $|\,\vec{p}\,| /
m =$ 2.0 (r.h.s.) for various system times $t \cdot m =$ 5, 20, 60,
120, 240, 360, 480, 600 up to the long-time limit. This representation
corresponds to Fig. \ref{plot_spec01}, where the respective evolution
of the spectral function is studied for the full Kadanoff-Baym theory.
We find that the time evolution of the spectral functions obtained from
the generalized transport equation (\ref{eq:general_transport}) is very
similar to the one from the full quantum calculation (see below). The
zero-mode spectral function for the initial distribution D1 becomes
sharper with time and is moving to slightly higher energies. The
opposite characteristics is observed for the zero-mode spectral
function for the initialization D3, which broadens with time (reducing
the peak correspondingly) and slowly shifts to smaller energies.
Together with the weak evolution for the distribution D2 (which only
slightly broadens at intermediate times and returns to a narrower shape
at smaller energies in the long-time limit) the evolution of all three
initializations in the semiclassical approximation is well comparable
to the full Kadanoff-Baym dynamics (cf. Fig. \ref{plot_spec01}).
Furthermore, the maxima of the zero-mode spectral functions are located
above the bare mass (as indicated by the on-shell arrow) for all
initial states during the time evolution.

\begin{figure}[t]
\centerline{\resizebox{0.7\columnwidth}{!}{\includegraphics{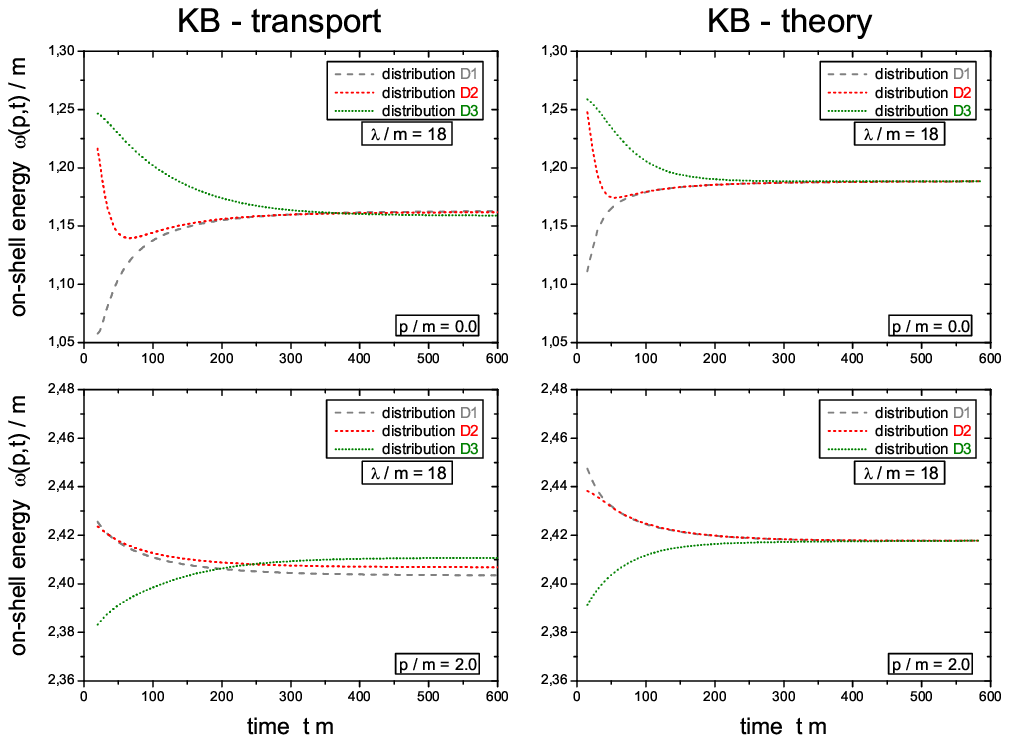} }}
\caption{\label{fig:4phd_gradient_ose01} Time evolution of the
on-shell energies $\omega(\vec{p},t)$ of the momentum modes
$|\,\vec{p}\,|/m = 0.0$ and $|\,\vec{p}\,|/m = 2.0$ for the
different initializations D1, D2 and D3 with $\lambda / m = 18$ in
the semiclassical KB limit (\ref{eq:general_transport}) (l.h.s.).
The on-shell self-energies are extracted from the maxima of the
time-dependent spectral functions. The respective results
 from the full Kadanof-Baym theory are displayed on the r.h.s.}
\end{figure}

\begin{figure}[ht]
\centerline{\resizebox{0.7\columnwidth}{!}{\includegraphics{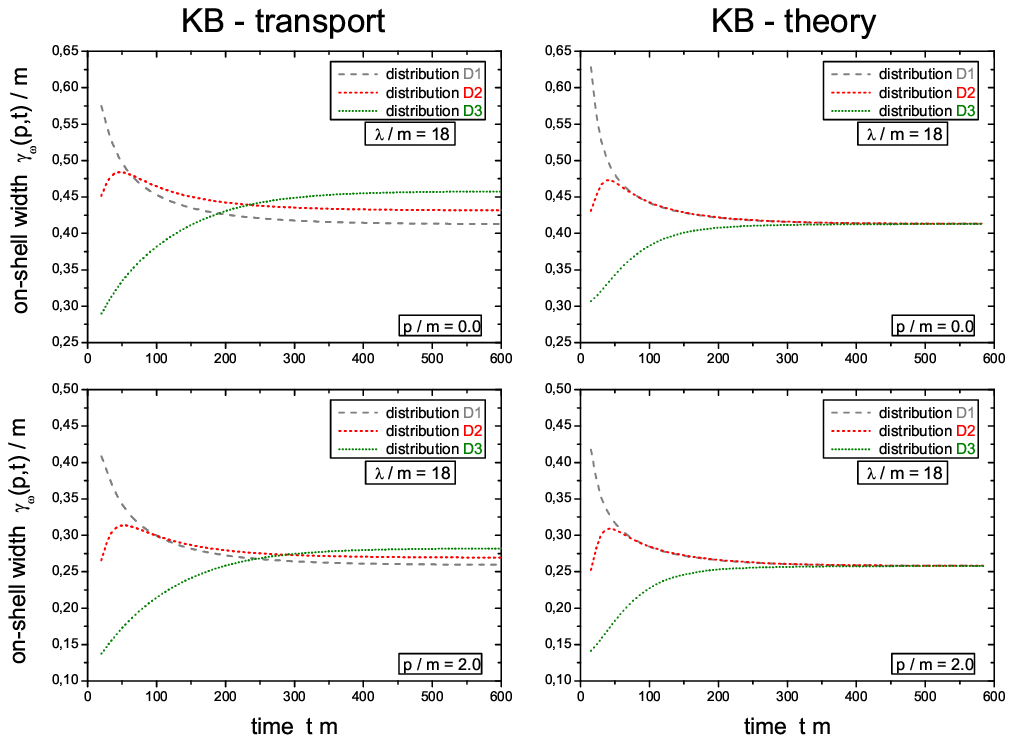} }}
\caption{\label{fig:4phd_gradient_osw01} Time evolution of the
on-shell widths $-
Im\,\bar{\Sigma}^{R}(\vec{p},\omega(\vec{p},t),t) /
\omega(\vec{p},t)$ of the momentum modes $|\,\vec{p}\,|/m = 0.0$
and $|\,\vec{p}\,|/m = 2.0$ for the different initializations D1,
D2 and D3 with $\lambda / m = 18$ in the semiclassical transport
eq. (\ref{eq:general_transport}) (l.h.s.). The respective results
 from the full Kadanof-Baym theory are displayed on the r.h.s.}
\vspace{-0.5cm}
\end{figure}

The spectral functions for the momentum mode
$|\,\vec{p}\,| / m =$ 2.0 are in a good agreement with
the Kadanoff-Baym dynamics as well.
Again we observe -- for the initial distribution D1 --
a narrowing of the spectral function,
while for D3 the spectral function broadens with time.
Moreover, the width of the spectral function starting
from distribution D2 shows a non-monotonous behaviour
with a maximum at intermediate times.

In order to study the dynamics of the spectral function
in a more quantitative manner we display in
Fig. \ref{fig:4phd_gradient_ose01}
the time evolution of the on-shell energies (as derived from the
maxima of the spectral function) for the
momentum modes $|\,\vec{p}\,| / m =$ 0.0 (upper plot)
and 2.0 (lower plot) for the initializations D1, D2 and D3
with $\lambda / m =$ 18 (l.h.s.).
By comparison with the corresponding results from the Kadanoff-Baym
theory (r.h.s.) we observe a close similarity of the
evolutions within the full and the semiclassical KB scheme.
The effective mass of the zero momentum
mode decreases for initialization D3, passes a minimum for D2
and increases for the initial state D1.

As familiar from the Kadanoff-Baym calculations in Section 1.2 the
behaviour of the on-shell energies is different for higher
momentum modes. We find for the momentum mode $|\,\vec{p}\,| / m
=$ 2.0 a monotonous decrease of the on-shell energy for the
initializations D1 and D2 and an increase for distribution D3.
Altogether, the evolution of the on-shell energies for the higher
modes is rather moderate compared to the lower ones in accordance
with the dominant momentum contribution and the weakening of the
retarded self-energy for higher energy modes.

Finally, the on-shell energies approach a stationary state
for all modes and all initializations.
However, the long-time limit of the equal momentum modes is
not exactly the same for all initial distributions D1, D2 and D3.
As discussed above this small difference can be traced back to the specific initial
state generation from the given momentum distribution.

Next we consider the time evolution of the on-shell width
as determined by the imaginary part of the retarded self-energy
at the maximum position of the spectral function.
In Fig. \ref{fig:4phd_gradient_osw01} (l.h.s.) the on-shell width
is displayed for the two momentum modes
$|\,\vec{p}\,| / m =$ 0.0 and
$|\,\vec{p}\,| / m =$ 2.0 for all three initial distributions
D1, D2 and D3 with $\lambda / m = $18 as a function of time.
For both momentum modes the on-shell width increases for the
distribution D3, while it has a maximum at intermediate times
($t \cdot m \approx$ 40) for the initialization D2.
Thus the results -- together with the reduction of the on-shell
width for both momentum modes for the initialization D1 -- is in
good agreement with the results obtained for the full Kadanoff-Baym
theory (r.h.s.).
However, the stationary values for the on-shell widths
deviate again slightly in accordance with the different preparation of the
initial state in the gradient scheme.

In summarizing  we find that the main characteristics
of the full quantum evolution of the spectral function are
maintained in the semiclassical transport equation
(\ref{eq:general_transport}) as well.
This includes the evolution of the on-shell energies as well
as the width of the spectral function.
Since the generalized transport equation is formulated directly
in Wigner space one has access to the spectral properties
at all times, whereas the very early times in
the Kadanoff-Baym case have to be excluded
due to the very limited support in the relative time
interval ($t_1 - t_2$) for the Wigner transformation.

\subsubsection{\label{sec:gradient_stationary} Stationary State of the
Semiclassical Evolution}

\begin{figure}[b!]
\hspace*{1mm}
\resizebox{0.45\columnwidth}{!}{\includegraphics{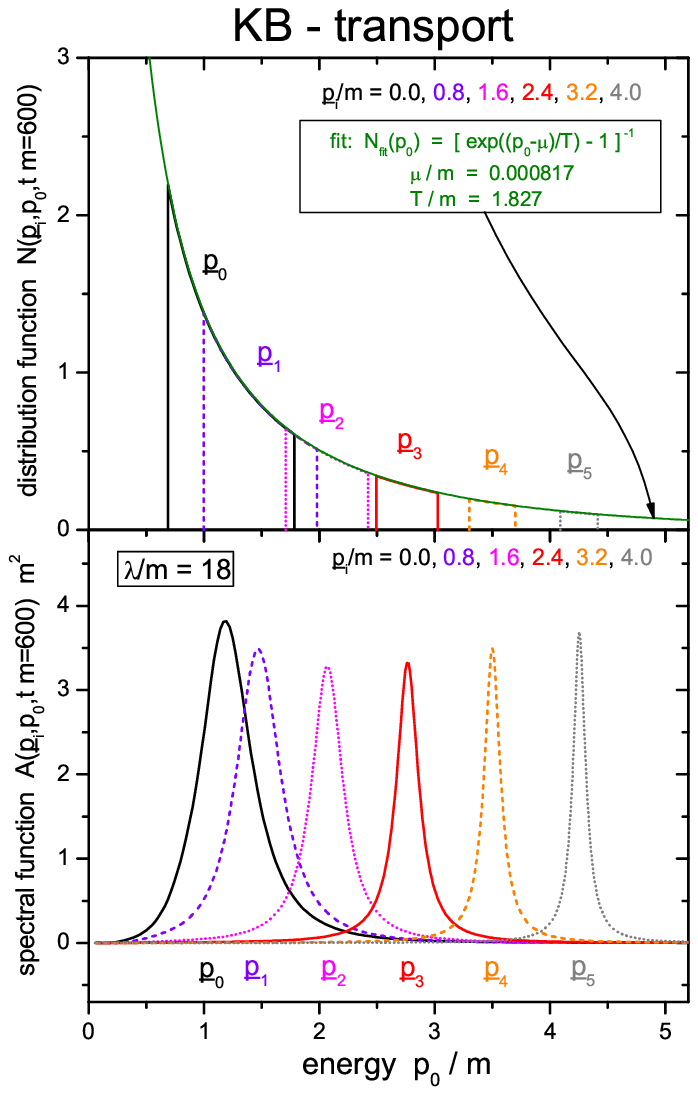} }
\hspace*{7mm}
\resizebox{0.45\columnwidth}{!}{\includegraphics{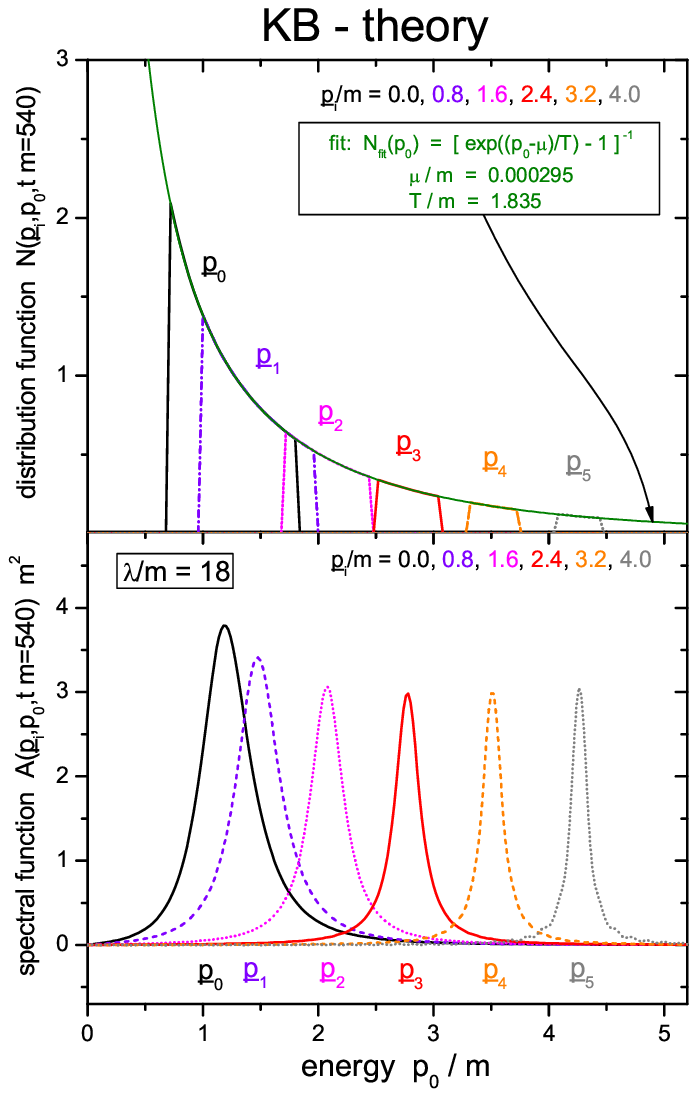} }
\caption{\label{fig:4phd_gradient_kmsna} Spectral function
$\bar{A}(p_0)$ for various momentum modes as a function of energy
$p_0 / m$ at the late time $t \cdot m =$ 600 (lower part) for
initial distribution D2 with coupling constant $\lambda/m = 18$ in
the semiclassical transport limit (\ref{eq:general_transport})
(l.h.s.). The corresponding distribution function $\bar{N}$ (at
the same time for the same momentum modes) is shown in the upper
part. All momentum modes can be fitted for all energies by a
single Bose function of temperature $T / m =$ 1.827 and a chemical
potential $\mu / m$ close to zero. The respective results
 from the full Kadanof-Baym theory are displayed on the r.h.s.}
\end{figure}

As we have observed in the previous Subsections the evolution
within the generalized transport equation
(\ref{eq:general_transport}) leads to a stationary state
for all three different initializations D1, D2 and D3.
Thus we turn to the investigation of this long-time limit itself, here
in particular for the initialization D2.
In Fig. \ref{fig:4phd_gradient_kmsna} (l.h.s.) we show the distribution
function $\bar{N}$ of various momentum modes
$|\,\vec{p}\,| / m =$ 0.0, 0.8, 1.6, 2.4, 3.2, 4.0 for large times
($t \cdot m$ = 600) as derived from the Green function itself and the
spectral function via the relation $\bar{N} = i\bar{G}^{<} / \bar{A}$.
The distribution function for a given momentum mode is calculated
for all energies $p_0$ where the corresponding spectral function
-- as displayed in the lower part of Fig.
\ref{fig:4phd_gradient_kmsna} -- exceeds a value of 0.5. Since the
width of the late time spectral function decreases with increasing
momentum, the energy range (for which the distribution function is
shown), is smaller for larger momentum modes. We find, that all
momentum modes of $\bar{N}$ can be fitted at all energies by a
single Bose function with a temperature $T / m =$ 1.827 and a very
small chemical potential $\mu / m =$ 0.000817. Thus the
generalized transport formulation (\ref{eq:general_transport})
leads to a complete (off-shell) equilibration of the system very
similar to the solution of the full Kadanoff-Baym equation (r.h.s.
of Fig. \ref{fig:4phd_gradient_kmsna}). Furthermore, the long-time
limit of the semiclassical time evolution exhibits a vanishing
chemical potential $\mu / m$ in accordance with the properties of
the neutral $\phi^4$-theory. This might have been expected since
in the generalized transport equation particle number
non-conserving processes of the type $1 \leftrightarrow 3$ --
which lead to the decrease of the chemical potential -- are
included by means of the dynamical spectral function. {\it Thus
the semiclassical approximation} (\ref{eq:general_transport}) {\it
solves the problems within the Boltzmann limit}, which does not
yield a relaxation of the chemical potential, since only on-shell
$2 \leftrightarrow 2$ transitions of quasiparticles are taken into
account as demonstrated in Section 1.4.

After observing, that the chemical potential decreases to zero in
the long-time limit, it is interesting to study the relaxation
process itself. The relaxation of the chemical potential $\mu / m$
 for the three different initializations D1, D2 and D3
(with coupling constant $\lambda / m =$ 18) shows an approximately
exponential decrease in time. The relaxation rates -- as
determined from the slope of the exponential decline -- are also
approximately the same for all distributions. They are given
explicitely by $\Gamma^{D1}_{\mu} \approx 0.98 \cdot 10^{-2}$ for
distribution D1, $\Gamma^{D2}_{\mu} \approx 1.01 \cdot 10^{-2}$
for distribution D2 and $\Gamma^{D3}_{\mu} \approx 1.07 \cdot
10^{-2}$ for distribution D3. Thus the relaxation rates are in the
same range as those found within the full Kadanoff-Baym theory
(Section 1.3).

We conclude that -- although there is a small relative shift of
the different time scales of kinetic and chemical equilibration as
a function of the coupling strength $\lambda$ with respect to the
full KB solutions -- the results of the generalized KB transport
equations are very similar. The differences we attribute to
higher-order multi-particle effects in off-shell transitions.
While the kinetic equilibration proceeds approximately with the
coupling constant squared (as indicated by the calculations for
non-polar-symmetric systems), the chemical relaxation rate is a
higher order process in the coupling constant $\lambda$.

\subsubsection{\label{sec:gradient_quadrupole} Quadrupole Relaxation}

\begin{figure}[tbh]
\centerline{\resizebox{0.7\columnwidth}{!}{\includegraphics{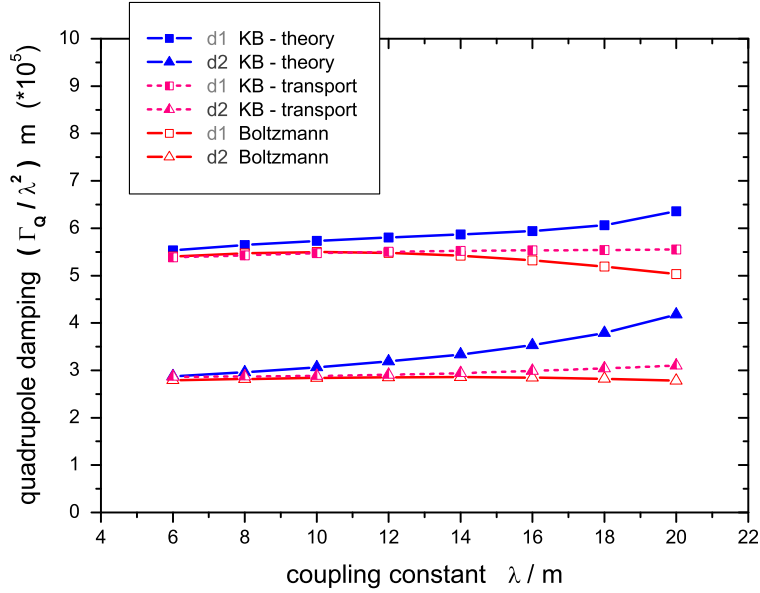} }}
\caption{\label{fig:4phd_gradient_quad} Scaled Relaxation rate for
the generalized transport equation as a function of the coupling
strength (half-filled symbols) for the initial distributions d1=D1
(squares) and d2=D2 (triangles). Additionally the results obtained
within the Kadanoff-Baym (full symbols) and the Boltzmann
calculation (open symbols) are shown for comparison. }
\end{figure}

In this Subsection we no longer restrict to polar symmetric
systems and discuss the time evolution of more general initial
distributions within the generalized transport approximation
(\ref{eq:general_transport}). We start with conditions similar to
those employed in Section 1.4 but combined with the initialization
scheme for the semiclassical limit. Again the decrease of the
quadrupole moment of the distribution
\bea
\label{quadpole}
Q(t) \: = \:
\frac{\displaystyle{\int \!\! \frac{d^{2}p}{(2 \pi)^2} \;\;
[\, p_x^2 - p_y^2 \,] \;\; N(\vec{p},t)}}
     {\displaystyle{\int \!\! \frac{d^{2}p}{(2 \pi)^2} \;\;
N(\vec{p},t)}} \; , \eea
is approximately exponential in time ($\propto \exp(-\Gamma_Q
\cdot t)$) and thus allows for the extraction of a quadrupole
damping rate $\Gamma_Q$. The scaled quadrupole damping rates -- as
obtained for the two initial distributions D1 and D2 -- are
displayed in Fig. \ref{fig:4phd_gradient_quad} as a function of
the coupling strength $\lambda / m$.

The calculations show that the quadrupole relaxation rates within
the semiclassical approxi\-mation (\ref{eq:general_transport}) for
both initial distributions D1 and D2 is well within in the range
of the full Kadanoff-Baym and the on-shell Boltzmann case.
Additionally, the quadrupole relaxation rate is rather flat in the
coupling $\lambda$ when divided by the coupling constant squared
($\lambda^2$) as already observed for the other two evolution
schemes in Section 1.4. Nevertheless, the relaxation in the full
KB-theory (\ref{kabaeqcs}) proceeds slightly faster than in the
transport limits for large couplings. The latter effect is again
attributed to higher order off-shell transition effects which are
no longer incorporated in the generalized KB transport equation.

\subsubsection{\label{sec:gradient_boma} Kadanoff-Baym versus
Botermans-Malfliet transport}

In this Subection we will perform a comparison of the generalized
transport equation in the original {Kadanoff-Baym} (KB) form
(\ref{eq:general_transport}) with the modified
{Botermans-Malfliet} (BM) form (\ref{eq:general_transport_bm}). As
discussed in detail in Section 2.1.2 the latter form results from
the replacement of the collisional self-energy by
\begin{equation} \label{replace}
i\bar{\Sigma}^{<} \rightarrow i\bar{G}^{<} \cdot \bar{\Gamma} / \bar{A}
\end{equation}
in the second Poisson bracket on the l.h.s. of the original
kinetic equation (\ref{eq:general_transport}). This replacement
leads to a {\it consistent first order equation in the gradients
and achieves consistency of the resulting transport equation with
the corresponding generalized mass-shell relation}
(\ref{eq:general_mass}). As noted before, the replacement
(\ref{replace}) might be questionable in the general case; thus
explicit numerical investigations are needed.

\begin{figure}[th]
\centerline{\resizebox{0.7\columnwidth}{!}{\includegraphics{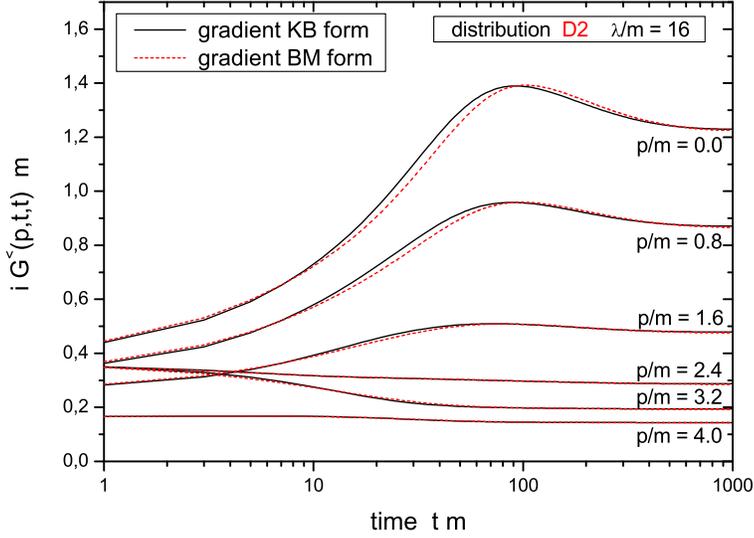} }}
\caption{\label{fig:4phd_gradient_bm} Time evolution of various
equal-time Green function momentum modes within the generalized
transport equation (original KB form, solid black lines) and
within the Botermans-Malfiet (BM) form (dashed red lines) for
initial distribution D2 with coupling constant $\lambda/m = 16$.}
\end{figure}

In Fig. \ref{fig:4phd_gradient_bm} we compare the time evolution
within the generalized transport equation in the KB form to the
consistent equation in BM form. In this respect several momentum
modes of the equal-time Green function are displayed evolving in
time from an initial distribution D2 for a coupling constant
$\lambda / m =$ 16. We find that the deviations between both
approximations (KB and BM) are rather moderate. Only for very
small momentum modes $|\,\vec{p}\,| / m \leq$ 1.6 deviations
between both modes are visible. For the very low momentum modes
the range of difference starts at $t \cdot m \approx$ 10 and
extends to $t \cdot m \approx$ 100 for the non-zero modes. For the
zero momentum mode the deviation lasts even longer. In this region
the semiclassical transport in the BM form is slightly 'slower'
than in the original KB choice. Nevertheless, also the BM form
exhibits the typical overshooting behaviour of the low momentum
modes beyond the stationary limit as observed for the KB form.
However, the maxima are shifted slightly to later times. Finally,
both gradient approximations converge in the long-time limit to
very similar configurations.

In summarizing we point out that the approximation of the full
Kadanoff-Baym dynamics by the generalized transport equations in
Kadanoff-Baym (\ref{eq:general_transport}) or Botermans-Malfliet
form (\ref{eq:general_transport_bm}) holds very well for the
different momentum modes of the Green function $i\bar{G}^{<}$
itself. Slight deviations are only visible for the zero momentum
mode at early to intermediate times (Figs.
 \ref{fig:4phd_gradient_modes} and
\ref{fig:4phd_gradient_bm}) for a logarithmic representation of
the time axis. Consequently, {\it the characteristic features of
quantum equilibration obtained for the full Kadanoff-Baym theory
are retained in the generalized transport limits}.

\subsection{Testparticle representation}

The generalized transport equation (\ref{eq:general_transport_bm})
allows to extend the traditional on-shell transport approaches for
which efficient numerical recepies have been set up
\cite{Ca90,CMO,CB99,CBS,URQMD1,URQMD2,Stoecker,Bertsch,T3,CAB,Falter,Koreview,Aich,ART,Ehehalt}
(and Refs. therein). In order to obtain a practical solution to
the transport equation (\ref{eq:general_transport_bm}) we use a
testparticle ansatz for the Green function $G^{<}$, more
specifically for the real and positive semidefinite quantity
\bea F_{XP} \; =  i \, G^{<}(X,P) \; \sim \; \sum_{i=1}^{N} \;
\delta^{(3)} ({\vec{X}} \, - \, {\vec{X}}_i (t)) \; \;
\delta^{(3)} ({\vec{P}} \, - \, {\vec{P}}_i (t)) \; \; \delta(P_0
- \, \epsilon_i(t)) \: . \label{testparticle} \eea
 In the
most general case (where the self energies depend on four-momentum
$P$, time $t$ and the spatial coordinates $\vec{X}$) the equations
of motion for the testparticles  read \cite{caju2}
\hspace{-0.9cm} \bea \label{eomr} \frac{d {\vec X}_i}{dt} \! & = &
\!  \frac{1}{1 - C_{(i)}} \, \frac{1}{2 \epsilon_i} \: \left[ \, 2
\, {\vec P}_i \, + \, {\vec \nabla}_{P_i} \, Re \Sigma^{ret}_{(i)}
\, + \, \frac{ \epsilon_i^2 - {\vec P}_i^2 - M_0^2 - Re
\Sigma^{ret}_{(i)}}{\Gamma_{(i)}} \: {\vec \nabla}_{P_i} \,
\Gamma_{(i)} \: \right],
\\[0.3cm]
\label{eomp}
\frac{d {\vec P}_i}{d t} \! & = & \!
- \frac{1}{1-C_{(i)}} \,
\frac{1}{2 \epsilon_{i}} \:
\left[ {\vec \nabla}_{X_i} \, Re \Sigma^{ret}_i
\: + \: \frac{\epsilon_i^2 - {\vec P}_i^2 - M_0^{2}
- Re \Sigma^{ret}_{(i)}}{\Gamma_{(i)}}
\: {\vec \nabla}_{X_i} \, \Gamma_{(i)} \:
\right],
\\[0.3cm]
\label{eome} \frac{d \epsilon_i}{d t} \!  & = & \! \phantom{- }
\frac{1}{1 - C_{(i)}} \, \frac{1}{2 \epsilon_i} \: \left[
\frac{\partial Re \Sigma^{ret}_{(i)}}{\partial t} \: + \:
\frac{\epsilon_i^2 - {\vec P}_i^2 - M_0^{2} - Re
\Sigma^{ret}_{(i)}}{\Gamma_{(i)}} \: \frac{\partial
\Gamma_{(i)}}{\partial t} \right], \eea
where the notation $F_{(i)}$ implies that the function is taken at
the coordinates of the testparticle, i.e.
$F_{(i)} \equiv F(t,\vec{X}_{i}(t),\vec{P}_{i}(t),\epsilon_{i}(t))$.

In (\ref{eomr}-\ref{eome}) a common multiplication factor
$(1-C_{(i)})^{-1}$ appears, which contains the energy derivatives
of the retarded self energy
\bea \label{correc} C_{(i)} \: = \: \frac{1}{2 \epsilon_i} \left[
\frac{\partial}{\partial \epsilon_i} \, Re \Sigma^{ret}_{(i)} \: +
\: \frac{\epsilon_i^2 - {\vec P}_i^2 - M_0^2 - Re
\Sigma^{ret}_{(i)}}{\Gamma_{(i)}} \: \frac{\partial }{\partial
\epsilon_i} \, \Gamma_{(i)} \right] \: . \eea
It yields a shift of the system time $t$ to the 'eigentime' of
particle $i$ defined by $\tilde{t}_{i} = t /(1-C_{(i)})$. As the
reader immediately verifies, the derivatives with respect to the
'eigentime', i.e. $d \vec{X}_i / d \tilde{t}_i$, $d \vec{P}_i / d
\tilde{t}_i$ and $d \epsilon_i / d \tilde{t}_i$ then emerge
without this renormalization factor for each testparticle
$i$ when neglecting higher order time derivatives in line with the
semiclassical approximation scheme.

Some limiting cases should be mentioned explicitly: In case of a
momentum-independent 'width' $\Gamma(X)$ we take $M^{2} = P^2 - Re
\Sigma^{ret}$ as an independent variable instead of $P_0$, which
then fixes the energy (for given $\vec{P}$ and $M^{2}$) to
\bea
P_{0}^{2} \; = \; \vec{P}^{2} \: + \: M^{2} \: + \:
Re \Sigma_{X\vec{P}M^2}^{ret} \, .
\label{energyfix}
\eea
Eq. (\ref{eome}) then turns to ($\Delta M_i^2 = M_i^2 - M_0^2$)
\bea \label{eomm} \frac{d \Delta M_i^2}{dt} \; = \; \frac{\Delta
M_i^2}{\Gamma_{(i)}} \; \frac{d \Gamma_{(i)}}{dt} \hspace{1cm}
\leftrightarrow \hspace{1cm} \frac{d}{dt} \ln \left( \frac{\Delta
M_i^2}{\Gamma_{(i)}} \right) = 0 \eea
for the time evolution of the test-particle $i$ in the invariant
mass squared. In case of $\Gamma = const.$ the familiar equations
of motion for testparticles in on-shell transport approaches are
regained.

\subsubsection{Model simulations for the momentum-dependent transport
equations in testparticle representation} \label{trialpot} To
demonstrate the physical content of the equations of motion for
testparticles (\ref{eomr}-\ref{eome}) we perform an exploratory
study with a momentum-dependent trial potential. The potential is
chosen of the type:
\bea
Re \Sigma^{ret} - \frac{i}{2} \Gamma
\: = \:
\frac{V(P_0,\vec{P})}{1+\exp\{(|\vec{r}|-R)/a_0\}}
\; - \; i \left(
\frac{W(P_0,\vec{P})}{1+\exp\{(|\vec{r}|-R)/a_0\}}
\: + \: \frac{\Gamma_V}{2} \right)
\eea
with a constant (but finite) vacuum width $\Gamma_V$. While the
spatial extension of the potential is given by a Woods-Saxon shape
(with parameters $R = 5$ fm and $a_0 = 0.6$ fm) its momentum
dependence for the real as well as for the imaginary part is
introduced by
\bea
V(P_0,\vec{P}) \; \; = C_V \:
\frac{\Lambda_V^2}{\Lambda_V^2 - (P_0^2 - \vec{P}^2)} \: , \qquad
W(P_0,\vec{P}) \; \; = C_W \:
\frac{\Lambda_W^2}{\Lambda_W^2 - (P_0^2 - \vec{P}^2)}
\label{momdeppart} \: .
\eea
Here the constants $C_V$ ($C_W$) give the 'strength' of the
complex potential while $\Lambda_V$ ($\Lambda_W$) play the role of
cutoff-parameters. Due to the structure in the denominator of
(\ref{momdeppart}) the momentum-dependent part of this potential
is explicitly Lorentz-covariant but should be considered for $M^2
\ll \Lambda_N^2$.

In the simulations the testparticles are propagated with different
initial mass parameters $M_i$, which are shifted relative to each
other by $\Gamma_V/20$ [GeV] around a mean mass of 1.0 GeV. To
each testparticle a momentum in positive $z$-direction is
attributed so that all of them have initially the same energy
$P_0$ = 2.0 GeV. All testparticles are initialized on the negative
$z$-axis with ($|\vec{X}_i(t=0)| \approx 15$ fm) and then evolved
in time according to the equations of motion
(\ref{eomr}-\ref{eome}).

\begin{figure}[h]
\resizebox{0.5\columnwidth}{!}{\includegraphics{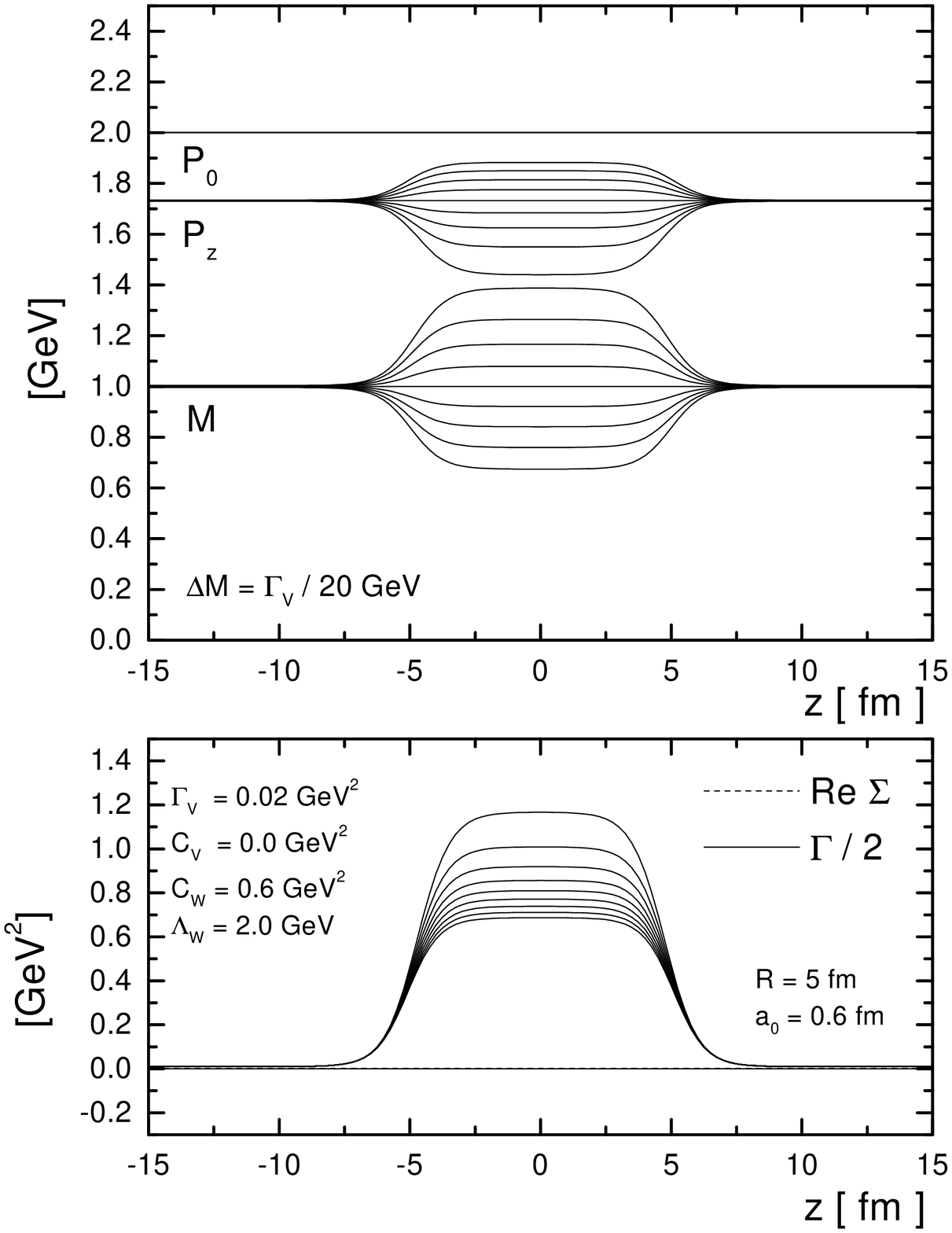} }
\resizebox{0.5\columnwidth}{!}{\includegraphics{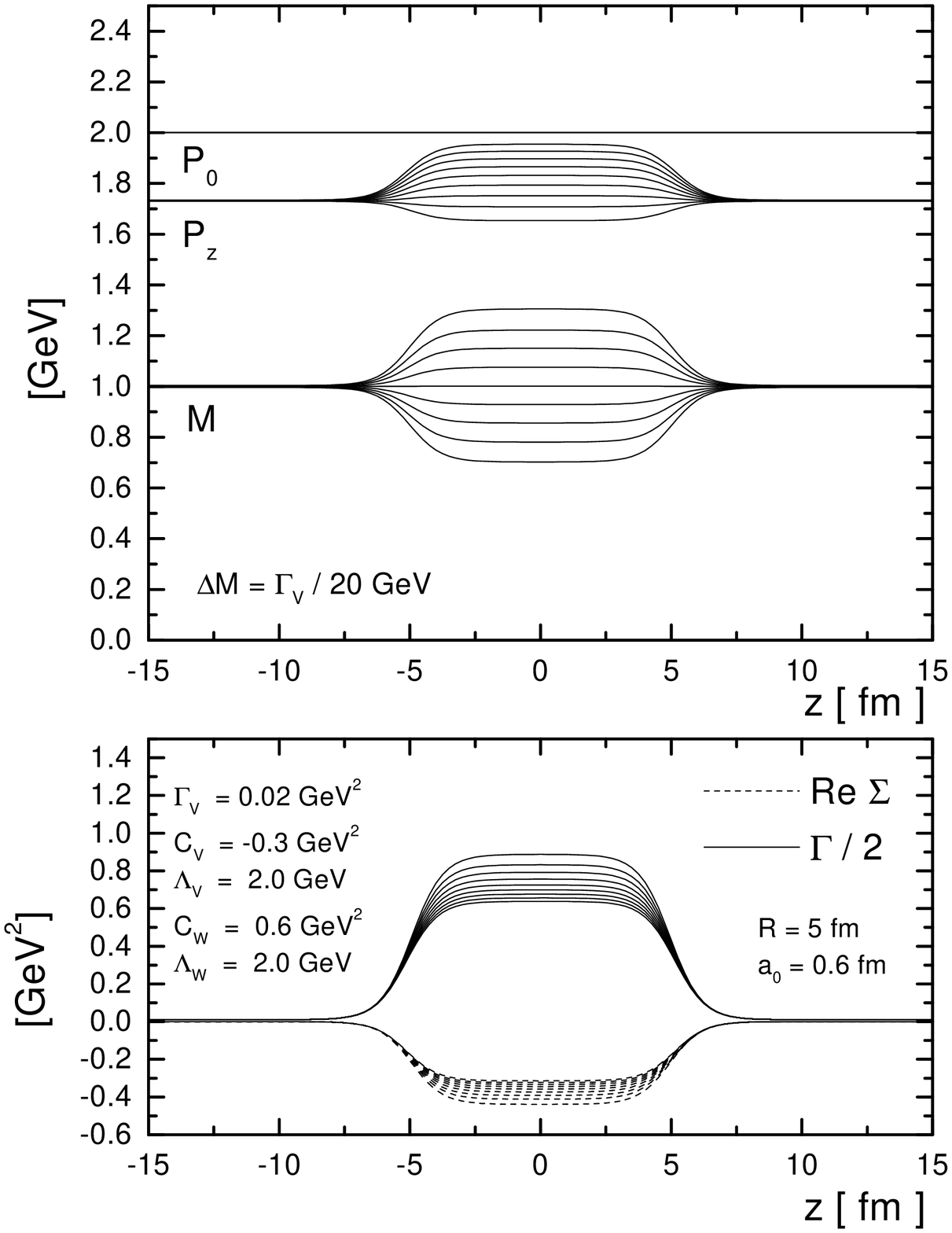} }
\caption{L.h.s.: upper part: $P_{i0}$, $P_{iz}$ and $M_i$ as a
function of $z(t)$ for a momentum-dependent imaginary potential
with $C_W = 0.6$ GeV$^2$ and $\Lambda_W = 2.0$ GeV (lower part).
The vacuum width is chosen as $\Gamma_V = 0.02$ GeV$^2$ and the
initial separation in the mass parameter of the testparticles is
$\Delta M = \Gamma_V / 20$. R.h.s.: upper part: $P_{i0}$, $P_{iz}$
and $M_i$ as a function of $z(t)$ for a momentum-dependent complex
potential with $C_V = -0.3$ GeV$^2$, $\Lambda_V = 2.0$ GeV, $C_W =
0.6$ GeV$^2$ and $\Lambda_W = 2.0$ GeV (lower part).}
\label{fig1m}
\end{figure}

In our first simulation we consider a purely imaginary potential
with a strength of $C_W = 0.6$ GeV$^2$ and a cutoff-parameter
$\Lambda_W = 2.0$ GeV. The evolution in energy $P_{0i}$, momentum
$P_{zi}$ and in the mass parameter $M_i$ for all testparticles is
shown in Fig. \ref{fig1m} (l.h.s., upper part) as a function of
$z(t)$. When the testparticles enter the potential region, their
momenta and mass parameters are modified.  The imaginary potential
leads to a spreading of the trajectories in the mass parameter
$M_i$ which in turn reflects a broadening of the spectral
function. The relation between the imaginary self energy and the
spreading in mass is fully determined by relation (\ref{eomm}).
Since we have chosen a potential with no explicit time dependence
the energy of each testparticle is a constant of time. According
to the explicit momentum dependence of our 'trial' potential each
single testparticle is affected with different strength. Since the
imaginary potential is strongest for small momenta (which
correspond to the highest lines in the lower graph of Fig.
\ref{fig1m}) (l.h.s.) the momentum and mass coordinates of those
testparticles are changed predominantly that are initialized with
the lowest momenta (i.e. with the largest masses). As a result one
observes a rather asymmetric distribution in the mass parameters
(and in the momenta) in the potential zone.  For $z(t) \gg R$ the
mass and momentum coordinates of the testparticles regain the
proper asymptotic value.

In the second example we allow for an addititonal real part of the
self energy. The calculation is performed with the parameters $C_V
= -0.3$ GeV$^2$, $C_W = 0.6$ GeV$^2$ and $\Lambda_V = \Lambda_W =
2.0$ GeV. The momentum-dependent real part (r.h.s., lower part of
Fig. \ref{fig1m}) causes  an additional shift of the testparticle
momenta. Since the real part of the potential is larger for small
initial momenta, these testparticle momenta are shifted up
somewhat more than for particles with larger momenta. This gives
rise to a reduction of the asymmetry which was introduced by the
momentum-dependent imaginary part of the self energy (r.h.s.,
upper part of Fig. \ref{fig1m}).  The mass parameters of the
testparticles are only weakly influenced by the real part of the
potential.

%
\begin{figure}[h]
\vspace*{0.8cm}
\centerline{\resizebox{0.5\columnwidth}{!}{\includegraphics{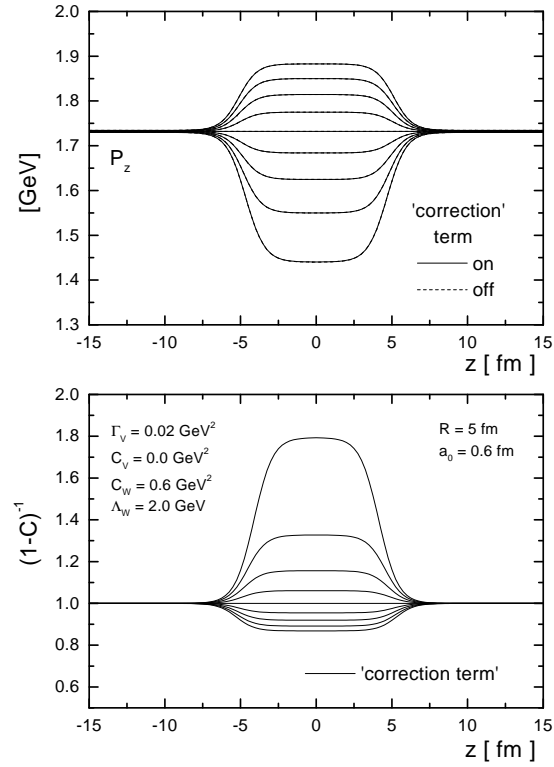}
}}
\caption{Lower part: Correction term $(1-C_{(i)})^{-1}$ as a
function of $z(t)$ for the same imaginary potential as in Fig.
\ref{fig1m}. Upper part: $P_{iz}$ as a function of $z(t)$ with and
without including the correction term. Both curves are identical.}
\label{fig3m}
\end{figure}

We, furthermore, study the implications of the correction term
$(1-C_{(i)})^{-1}$ using the same imaginary potential as in Fig.
\ref{fig1m}. The time evolution of the correction factor for each
testparticle $i$ is displayed in the lower part of Fig
\ref{fig3m}. While it is $> 1$ for large initial mass parameters
$M_i > M_0$, it is $ < 1$ for mass parameters $M_i < M_0$. In the
upper part of Fig. \ref{fig3m} the momenta of the testparticles
are shown for two calculational limits: in the first one the
correction term is taken into account (as in the previous
calculations), while in the second one the corrections due to the
energy dependence of the retarded self energy are neglected.
However, the calculations with and without the correction factor
exhibit no difference in the testparticle momenta as a function of
$z$.
 The same holds for the mass parameters $M_i$ which
are not displayed here since they provide no new information due
to energy conservation. We thus conclude that the particle
trajectory (in phase space) is independent of these correction
factors, since the correction term -- when displayed as $P_{z}(z)$
in phase-space -- leads only to a rescaling of the 'eigentime' of
the testparticles as pointed out above.

\subsubsection{Collision terms}
The collision term of the Kadanoff-Baym equation can only be
worked out in more detail by giving explicit approximations for
$\Sigma^{<}$ and $\Sigma^{>}$. A corresponding collision term can
be formulated in full analogy to Refs. \cite{CB99,CaWa}, e.g. from
Dirac-Brueckner theory (cf. also Section 1.2), following detailed
balance as $$ I_{coll}(X,\vec{P},M^2) = Tr_2 Tr_3 Tr_4 A(X,{\vec
P},M^2) A(X,{\vec P}_2, M_2 ^2) A(X,{\vec P}_3, M_3 ^2) A(X,{\vec
P}_4, M_4 ^2) $$ $$
|T(({\vec P},M^2) + ({\vec P}_2,M_2^2) \rightarrow ({\vec
P}_3,M_3^2) + ({\vec P}_4,M_4^2))|_{{\cal A,S}}^2 \; \;
\delta^{(4)}({P} + {P}_2 - {P}_3 - {P}_4) $$ \begin{equation}
\label{Icoll} \left[
 N_{X{\vec P}_3 M_3^2} \  N_{X {\vec P}_4 M_4^2} \  {\bar f}_{X
{\vec P} M^2} \  {\bar f}_{X {\vec P}_2 M_2^2} \  - \  N_{X{\vec
P} M^2} \  N_{X {\vec P}_2 M_2^2} \  {\bar f}_{X {\vec P}_3 M_3^2}
\  {\bar f}_{X {\vec P}_4 M_4^2} \right]
\end{equation} with
\begin{equation}
\label{pauli} {\bar f}_{X {\vec P} M^2} = 1 + \eta \, N_{X {\vec
P} M^2} \end{equation}
 and $\eta = \pm 1$ for bosons/fermions,
respectively. The indices ${\cal A,S}$ stand for the
antisymme\-tric/symmetric matrix element of the in-medium
scattering amplitude $T$ in case of fermions/bo\-sons. In Eq.
(\ref{Icoll}) the trace over particles 2,3,4 reads explicitly for
fermions
\begin{equation} \label{trace} Tr_2 = \sum_{\sigma_2, \tau_2}
\frac{1}{(2 \pi)^4} \int d^3 P_2 \frac{d M^2_2}{2
\sqrt{\vec{P}^2_2+M^2_2}}, \end{equation}
 where $\sigma_2, \tau_2$
denote the spin and isospin of particle 2. In case of bosons we
have
\begin{equation}
\label{trace2} Tr_2 = \sum_{\sigma_2, \tau_2} \frac{1}{(2 \pi)^4}
\int d^3 P_2 \frac{d P_{0,2}^2}{2}, \end{equation} since here the
spectral function $A_B$ is normalized as
\begin{equation}
\label{sb} \int \frac{d P_0^2}{4 \pi} A_B(X,P) = 1
\end{equation} whereas for fermions we have
\begin{equation}
\label{sb1}
\int \frac{d P_0}{2 \pi} A_F(X,P) = 1.
\end{equation}
We mention that the spectral function $A_F$ in case of fermions in
(\ref{Icoll}) is obtained
by considering only particles of positive energy and assuming the spectral
function to be identical for spin 'up' and 'down' states. In general, the
spectral function for fermions ${\hat A}_{\alpha \beta}(X,P)$ is a Dirac-tensor
with $\alpha \beta$ denoting the Dirac indices. It is normalized as
\begin{equation}
\label{specF}
\int \frac{d P_0}{2 \pi} {\hat A}_{\alpha \beta} (X,P) =
(\gamma^0)_{\alpha \beta},
\end{equation}
which implies
\begin{equation}
\frac{1}{4} \sum_{\alpha} \int \frac{d P_0}{2 \pi} (\gamma^0 {\hat
A}(X,P))_{\alpha \alpha} = 1. \end{equation}
Now expanding ${\hat
A}$ in terms of free spinors $u_s(P)$ ($s$=1,2) and $v_s(P)$ as
\begin{equation}
(\hat{A})_{\alpha \beta} \: = \:
\sum_{r,s=1}^{2}
\bar{u}_s(\vec{P}_,M)_{\beta} \; A_{rs}^{p} \; u_r(\vec{P},M)_{\alpha}
\: + \:
\bar{v}_s(\vec{P}_,M)_{\beta} \; A_{rs}^{ap} \; v_r(\vec{P},M)_{\alpha}
\end{equation}
one can separate particles and antiparticles. By neglection of the
antiparticle contributions (i.e. $A_{rs}^{ap} \equiv 0$) and within the
assumption that the spectral function for the particles is diagonal
in spin-space (i.e. $A_{rs}^{p} = \delta_{rs} A_s^{p}$) as well as
spin symmetric, one can define $A_F$ as
\begin{equation}
A_F \equiv A_1^{p} = A_2^{p} \: .
\end{equation}
Neglecting the 'gain-term' in eq. (\ref{Icoll}) one recognizes
that the collisional width of the particle in the rest frame is
given by
\begin{equation}
\label{gcoll}
\Gamma_{coll}(X,\vec{P},M^2) = Tr_2 Tr_3 Tr_4 \;
|T(({\vec P},M^2) + ({\vec P}_2,M_2^2) \rightarrow ({\vec
P}_3,M_3^2) + ({\vec P}_4,M_4^2))|_{{\cal A,S}}^2 \end{equation}
$$ A(X,{\vec P}_2,M_2^2) A(X,{\vec P}_3,M_3^2) A(X,{\vec P}_4,
M_4^2) \; \; \delta^4(P + P_2 - P_3-P_4) \ N_{X {\vec P}_2 M_2^2}
\, {\bar f}_{X {\vec P}_3 M^2_3} \, {\bar f}_{X {\vec P}_4 M^2_4}
\,
, $$ \\ where -- as in eq. (\ref{Icoll}) -- local on-shell
scattering processes are assumed for the transitions $P + P_2
\rightarrow P_3 + P_4$. We note that the extension of eq.
(\ref{Icoll}) to inelastic scattering processes (e.g. $NN
\rightarrow N\Delta$) or ($\pi N \rightarrow \Delta$ etc.) is
straightforward when exchanging the elastic transition amplitude
$T$ by the corresponding inelastic one and taking care of
Pauli-blocking or Bose-enhancement for the particles in the final
state.

For particles of infinite life time in vacuum -- such as protons
-- the collisional width (\ref{gcoll}) has to be identified with
twice the imaginary part of the self energy. Thus the transport
approach determines the particle spectral function dynamically via
(\ref{gcoll}) for all hadrons if the in-medium transition
amplitudes $T$ are known {\it in their full off-shell dependence}.
Since this information is not available for configurations of hot
and dense matter, which is the major subject of future
development, a couple of assumptions and numerical approximation
schemes have to be invoked in actual applications.

\section{The dynamical quasiparticle approach to hot QCD}

This Section addresses the question how to achieve a suitable
approximation for the spectral functions of partons (quarks and
gluons) in hot QCD, i.e. at temperatures above the deconfinement
transition $T_c$. Since the spectral functions are fully
determined by complex selfenergies $\Sigma^{ret}$ for the partons,
it is sufficient to determine effective selfenergies, e.g., from
lattice QCD.

\subsection{Application to parton dynamics}
The 'Big Bang' scenario implies that in the first micro-seconds of
the universe the entire system has emerged from a partonic system
of quarks, antiquarks and gluons -- a quark-gluon plasma (QGP) --
to color neutral hadronic matter consisting of interacting
hadronic states (and resonances) in which the partonic degrees of
freedom are confined. The nature of confinement and the dynamics
of this phase transition has motivated a large community for
several decades (cf.\ \cite{Jacobs,QM01} and Refs.\
therein). Early concepts of the QGP were guided by the idea of a
weakly interacting system of partons since the entropy density $s$
and energy density $\epsilon$ were found in lattice QCD to be
close to the Stefan Boltzmann (SB) limit for a relativistic
noninteracting system \cite{Karsch}. However, experimental
observations at the Relativistic Heavy Ion Collider (RHIC)
indicated that the new medium created in ultrarelativistic Au+Au
collisions was interacting more strongly than hadronic matter and
consequently this notion had to be given up. Moreover, in line
with earlier theoretical studies in Refs.
\cite{Thoma,Andre,Shuryak} the medium showed phenomena of an
almost perfect liquid of partons \cite{STARS,Miklos3} as extracted
from the strong radial expansion and elliptic flow of hadrons as
well the scaling of the elliptic flow with parton number {\it
etc}. All the latter collective observables have been severely
underestimated in conventional string/hadron transport models
\cite{Cassing03,Brat04,Cassing04} whereas hydrodynamical
approaches did quite well in describing (at midrapidity) the
collective properties of the medium generated during the early
times for low and moderate transverse momenta \cite{Heinz,Bass2}.
The question about the constituents and properties of this QGP
liquid is discussed controversely in the literature (cf. Refs.
\cite{GerryEd,GerryRho,Eddi}) and practically no dynamical
concepts are available to describe the dynamical freezeout of
partons to color neutral hadrons that are finally observed
experimentally. Since the partonic system appears to interact more
strongly than even hadronic systems the notation strong QGP (sQGP)
has been introduced in order to distinguish from the dynamics
known from perturbative QCD (pQCD).

Lattice QCD (lQCD) calculations provide some guidance to the
thermodynamic properties of the partonic medium close to the
transition at a critical temperature $T_c$ up to a few times
$T_c$, but lQCD calculations for transport coefficients presently
are not accurate enough \cite{lattice2} to allow for firm
conclusions. Furthermore, it is not clear whether the partonic
system really reaches thermal and chemical equilibrium in
ultrarelativistic nucleus-nucleus collisions \cite{ZhangKo} such
that nonequilibrium models are needed to trace the entire
collision history.  The available string/hadron transport models
\cite{URQMD1,URQMD2,Cass99} partly fail - as pointed out above -
nor do partonic cascade simulations
\cite{Geiger,Zhang,Molnar,Bass} (propagating massless partons)
sufficiently describe the reaction dynamics when employing cross
sections from perturbative QCD. Some models, e.g. the Multiphase
Transport Model AMPT \cite{AMPT}, employ strong enhancement
factors for the cross sections, however, use only on-shell
massless partons in the partonic phase as in Ref. \cite{Zhang}.
The same problem comes about in the parton cascade model of Ref.
\cite{Carsten} where additional 2$ \leftrightarrow$ 3 processes
like $gg \leftrightarrow ggg$ are incorporated but massless
partons are involved.

On the other hand it is well known that strongly interacting
quantum systems require descriptions in terms of propagators with
sizeable selfenergies  for the relevant degrees of freedom (cf.
Sections 1 and 2). While the real part of the selfenergies can be
related to  mean-field potentials, the imaginary parts  provide
information about the lifetime and/or reaction rate of time-like
'particles' \cite{Sascha1}.  The studies of Peshier
\cite{Andre04,Andre05} indicate that the effective degrees of
freedom in a partonic phase should have a width $\gamma$  in the
order of the pole mass $M$ already slightly above $T_c$. This
opens up the problem how to interpret/deal with the space-like
part of the distribution functions and how to 'pro\-pagate'
effective degrees in space-time in equilibrium as well as out of
equilibrium without violating 'microcausality'. These questions
will now be addressed in the Dynamical QuasiParticle Model (DQPM)
\cite{Cassing06}.

\subsection{Off-shell elements in the DQPM}

The dynamical quasiparticle model  goes back to
Peshier \cite{Andre04,Andre05} and starts with the entropy density
 in the quasiparticle limit ~\cite{Andre05,R38,R39,R40},
\begin{equation}   \label{sdqp} \hspace{0.5cm}
  s^{dqp}
  =
  - d_g \!\int\!\!\frac{d \omega}{2 \pi} \frac{d^3p}{(2 \pi)^3}
  \frac{\partial n_B}{\partial T}
   \l( \Im\ln(-\Delta^{-1}) + \Im\Pi\,\Re\Delta \r) \end{equation} $$
   - d_q \!\int\!\!\frac{d \omega}{2 \pi} \frac{d^3p}{(2 \pi)^3}
  \frac{\partial n_F((\omega-\mu_q)/T)}{\partial T}
   \l( \Im\ln(-S_q^{-1}) + \Im\Sigma_q\,\Re S_q \r)
   \!, $$
$$
   - d_{\bar q} \!\int\!\!\frac{d \omega}{2 \pi} \frac{d^3p}{(2 \pi)^3}
  \frac{\partial n_F((\omega+\mu_q)/T)}{\partial T}
   \l( \Im\ln(-S_{\bar q}^{-1}) + \Im\Sigma_{\bar q}\,\Re S_{\bar q} \r)
   \!, $$

\noindent where $n_B(\omega/T) = (\exp(\omega/T)-1)^{-1}$ and
$n_F((\omega-\mu_q)/T) = (\exp((\omega-\mu_q)/T)+1)^{-1}$ denote
the Bose and Fermi distribution functions, respectively, while
$\Delta =(P^2-\Pi)^{-1}$, $S_q = (P^2-\Sigma_q)^{-1}$ and $S_{\bar
q} = (P^2-\Sigma_{\bar q})^{-1}$ stand for the full (scalar)
quasiparticle propagators of gluons $g$, quarks $q$ and antiquarks
${\bar q}$. The degeneracy factor for gluons is $d_g= 2 (N_c^2-1)$
= 16 while for quarks $q$ and antiquarks ${\bar q}$ we get $d_q =
d_{\bar q} = 2 N_c N_f $ = 18 for three flavors $N_f$. In Eq.
(\ref{sdqp}) $\Pi$ and $\Sigma = \Sigma_q \approx \Sigma_{\bar q}$
denote the (retarded) quasiparticle selfenergies. In principle,
$\Pi$ as well as $\Delta$ are Lorentz tensors and should be
evaluated in a nonperturbative framework. The DQPM treats these
degrees of freedom as independent scalar fields with scalar
selfenergies. In case of the fermions $S_q, S_{\bar q}$ and
$\Sigma_q, \Sigma_{\bar q}$ (for $q$ and ${\bar q}$) have Lorentz
scalar and vector contributions but only scalar terms are kept in
(\ref{sdqp}) for simplicity which are assumed to be identical for
quarks and antiquarks. Note that one has to treat quarks and
antiquarks separately in (\ref{sdqp}) as their abundance differs
at finite quark chemical potential $\mu_q$.

Since the nonperturbative evaluation of the propagators and
selfenergies in QCD is a formidable task (and addressed in
Dyson-Schwinger (DS) Bethe-Salpeter (BS)
approaches \cite{DS1,DS2,DS3,DS4,DS5}) an alternative and
practical procedure is to use physically motivated {\em
Ans\"atze} with Lorentzian spectral functions for quarks\footnote{In the following
the abbreviation is used that 'quarks' denote quarks and antiquarks
if not specified explicitly.}
and gluons,
\begin{equation}
 \rho(\omega)
 =
 \frac\gamma{ E} \l(
   \frac1{(\omega-E)^2+\gamma^2} - \frac1{(\omega+E)^2+\gamma^2}
 \r) ,
 \label{eq:rho}
\end{equation} and to fit the few parameters to results from lQCD. With the
convention $E^2(\bm p) = \bm p^2+M^2-\gamma^2$, the parameters
$M^2$ and $\gamma$ are directly related to the real and imaginary
parts of the corresponding (retarded) self-energy, e.g. $\Pi =
M^2-2i\gamma\omega$ in case of the 'scalar' gluons.

\begin{figure}[htb!]
  \begin{center}
  \vspace{0.1cm}
\resizebox{0.8\columnwidth}{!}{\includegraphics{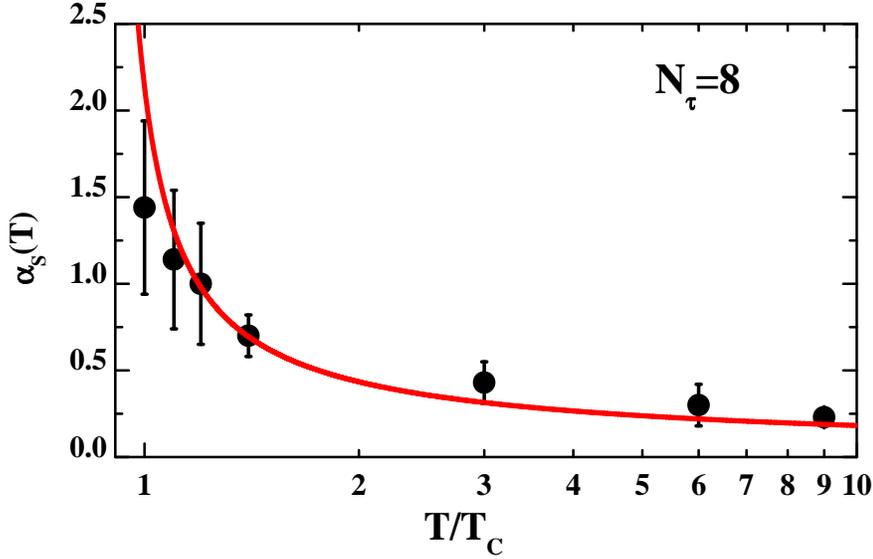} }
    \caption{The coupling $\alpha_s(T) = g^2(T)/(4\pi)$ (solid red
    line) as a function of $T/T_c$ in comparison to the long range part of the
strong coupling as extracted from Ref. \protect\cite{Bielefeld} from the free energy
of a quark-antiquark pair in quenched lQCD (for $N_\tau$ = 8).  }
    \label{ffig0}
  \end{center}
\end{figure}

Following  \cite{Andre05} the quasiparticle mass (squared) for
gluons is assumed to be given by the thermal mass in the
asymptotic high-momentum regime, i.e.
\begin{equation}
 M^2(T) = \frac{g^2}{6} \left( (N_c + \frac{1}{2}N_f)\, T^2
 + \frac{N_c}{2} \sum_q \frac{\mu_q^2}{\pi^2}
 \right) \, ,
 \label{eq:M2} \end{equation}
and for quarks (assuming vanishing constituent masses here) as,
\begin{equation}
m^2(T) = \frac{N_c^2-1}{8 N_c}\, g^2 \left( T^2 +
\frac{\mu_q^2}{\pi^2} \right) \, ,\label{eq:M2b} \end{equation}
with a running coupling (squared),
\begin{equation}
 g^2(T/T_c) = \frac{48\pi^2}{(11N_c - 2 N_f)  \ln(\lambda^2(T/T_c-T_s/T_c)^2}\ ,
 \label{eq:g2}
\end{equation} which permits for an enhancement near $T_c$
\cite{pQP,Rossend,Rafelski}. Here $N_c = 3$ stands for the number
of colors while $N_f$ denotes the number of flavors and $\mu_q$
the quark chemical potentials. The parameters $\lambda = 2.42$ and
$T_s/T_c = 0.46$ have been fixed in Ref. \cite{Andre05}. As demonstrated
in Fig. \ref{ffig0} this functional form for the
strong coupling $\alpha_s = g^2/(4\pi)$ is in accordance with the
lQCD calculations of Ref. \cite{Bielefeld} for the long range part
of the $q - \bar{q}$ potential.

\noindent The width  for gluons and quarks (for $\mu_q = 0$) is adopted in
the form \cite{Pisar89LebedS}
\begin{equation}
  \gamma_g(T)
  =
  N_c \frac{g^2 T}{8 \pi} \,  \ln\frac{2c}{g^2} \, , \hspace{2cm}
    \gamma_q(T)
  =
  \frac{N_c^2-1}{2 N_c} \frac{g^2 T}{8 \pi} \,  \ln\frac{2c}{g^2}
  \,.
 \label{eq:gamma}
\end{equation} where $c=14.4$ (from Ref. \cite{Andre}) is related to a
magnetic cut-off.

The physical processes contributing to the width $\gamma_g$ are
both $gg \leftrightarrow gg$, $gq \leftrightarrow gq$ scattering
as well as splitting and fusion reactions $gg \leftrightarrow g$,
$gg \leftrightarrow ggg$, $ggg \leftrightarrow gggg$ or $g
\leftrightarrow q \bar{q}$ etc. On the fermion side elastic
fermion-fermion scattering $pp \leftrightarrow pp$, where $p$
stands for a quark $q$ or antiquark $\bar{q}$, fermion-gluon
scattering $pg \leftrightarrow pg$, gluon bremsstrahlung $pp
\leftrightarrow pp+g$ or quark-antiquark fusion $q \bar{q}
\leftrightarrow g$ etc. emerge. Note, however, that the explicit
form of (\ref{eq:gamma}) is derived for hard two-body scatterings
only. It is worth to point out that the ratio of the masses to
their widths $ \sim g \ln(2c/g^2)$ approaches zero only
asymptotically for $T \rightarrow \infty$ such that the width of
the quasiparticles is comparable to the pole mass slightly above
$T_c$ up to all terrestrial energy scales.

Within the DQPM the real and imaginary parts of the propagators
$\Delta$ and $S$ now are fixed and the entropy density
(\ref{sdqp}) can be evaluated numerically once the free parameters
in (\ref{eq:g2}) are determined. In the following we will assume 3
light quark flavors $N_f = 3$. Since the presently available
unquenched lQCD calculations (for three flavors) for the entropy
density are still accompanied with rather large error bars the
parameters of the DQPM are taken the same as in the pure
Yang-Mills sector:  $\lambda = 2.42$, $T_s/T_c= 0.46$ as
determined in Refs. \cite{Andre,Andre05}. This is legitimate since
an approximate scaling of thermodynamic quantities from lQCD is
observed when dividing by the number of degrees of freedom and
scaling by the individual critical temperature $T_c$ which is a
function of the different number of parton species \cite{Karsch5}.
However, these parameters will have to be refitted once more
accurate 'lattice data' become available.

\begin{figure}[htb!]
  \vspace{0.5cm}
\resizebox{0.75\columnwidth}{!}{\includegraphics{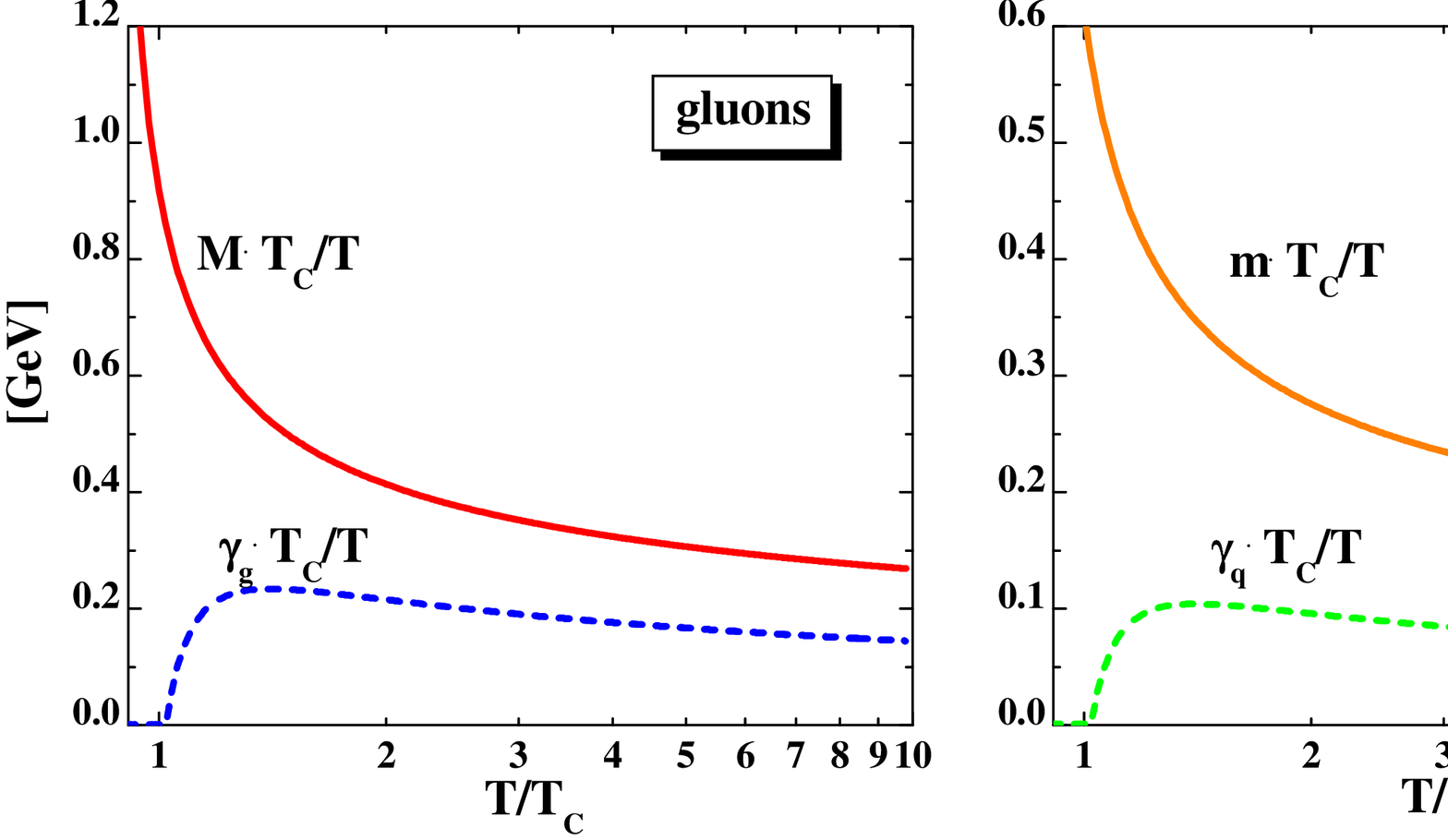} }
    \caption{ The mass $M$ (solid red line) and width $\gamma_g$ (dashed blue line)
    for gluons (l.h.s.) and the mass $m$ (solid orange line) and width
    $\gamma_q$ (dashed green line) for
     quarks (r.h.s.) as a function of $T/T_c$ in the DQPM
     for $\lambda = 2.42$, $T_s/T_c= 0.46$, and $c=$ 14.4 (for $\mu_q = 0$).
     All quantities have been multiplied
     by the dimensionless factor $T_c/T$.}
    \label{ffig1}
\end{figure}

The resulting values for the gluon and quark masses - multiplied
by $T_c/T$ - are displayed in Fig. \ref{ffig1} (for $\mu_q=0$) by
the solid lines while the gluon and quark width ($\gamma_g,
\gamma_q)$ - multiplied by $T_c/T$ - are displayed in terms of the
dashed lines as a function of $T/T_c$.  The actual numbers
imply very 'broad' quasiparticles
already slightly above $T_c$. For $\mu_q = 0$ the ratio
$\gamma_q/\gamma_g$ =4/9 is the same as for the ratio of the
squared masses $m^2/M^2 = 4/9$ and reflects the ratio of the
Casimir eigenvalues in color space. Consequently the ratio of the
width to the pole mass is smaller for quarks (antiquarks) than for
gluons in the whole temperature range.

In order to fix the scale $T_c$, which is not specified so far,
one may directly address unquenched lQCD calculations (for 3 light
flavors). However, here the situation is presently controversal
between different groups (cf. Refs. \cite{xx1,xx2} and the
discussion therein). An alternative way is to calculate the
pressure $P$ from the thermodynamical relation (at $\mu_q$ = 0),
\begin{equation}
\label{pressure} s =\frac{\partial P}{\partial T} \ ,
\end{equation} by integration of the entropy density $s$ over $T$,
where one may tacitly identify the 'full' entropy density $s$ with
the quasiparticle entropy density $s^{dqp}$ (\ref{sdqp}). Since
for $T < T_c$ the DQPM entropy density drops to zero (with
decreasing $T$) due to the high quasiparticle masses and the width
$\gamma$ vanishes as well (cf. Fig. \ref{ffig1}) the integration
constant may be assumed to be zero in the DQPM which focusses on
the quasiparticle properties above $T_c$.

\begin{figure}[htb!]
  \vspace{0.5cm} \hspace{1cm}
\resizebox{0.65\columnwidth}{!}{\includegraphics{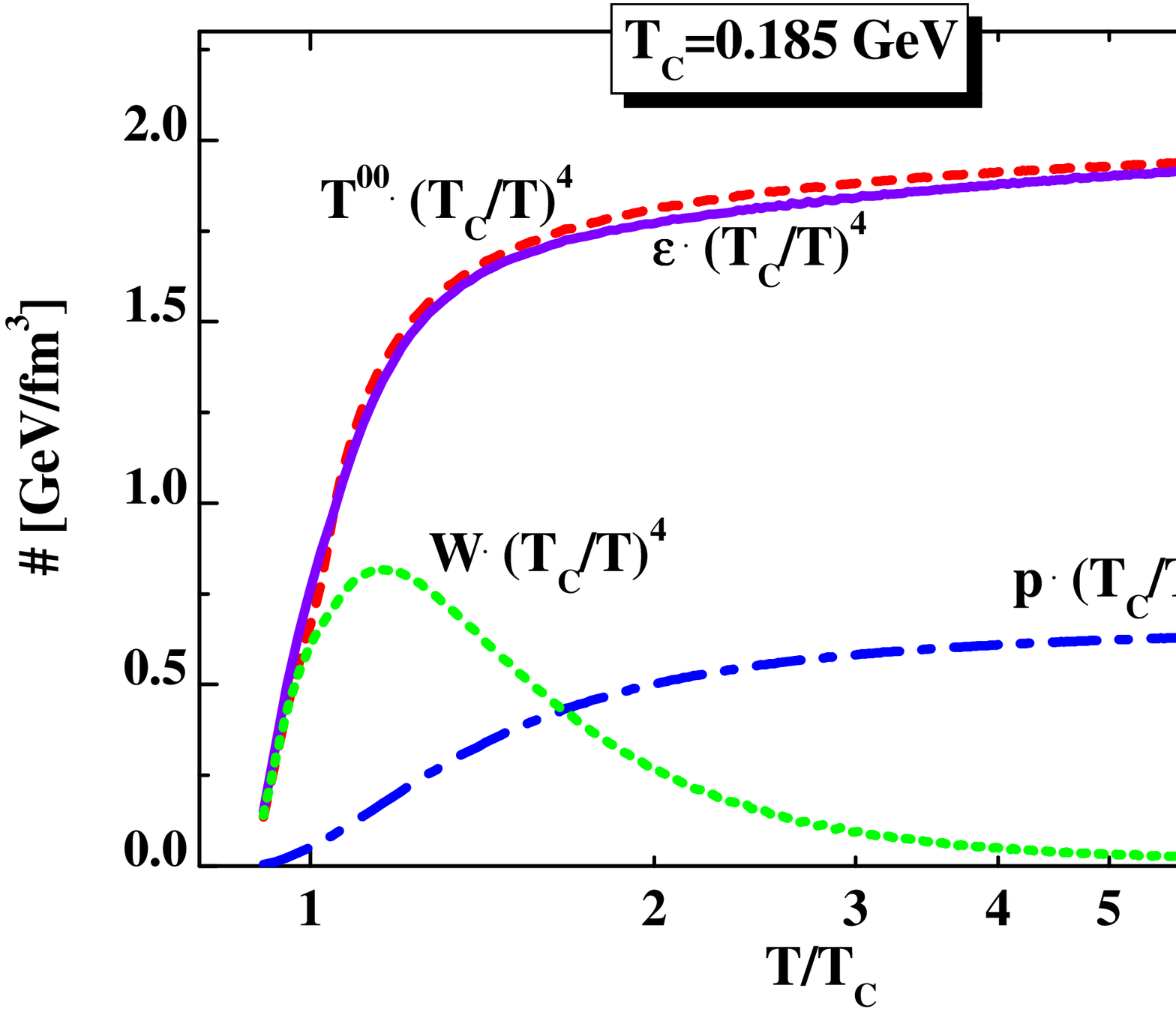} }
    \caption{The DQPM results for $\epsilon (T/T_c) (T_c/T)^4$  (solid violet line),
    $P(T/T_c) (T_c/T)^4$ (blue
dash-dotted line) and the interaction measure $W(T/T_c) (T_c/T)^4$
(\ref{wint}) (green dotted line). Note the logarithmic scale in
$T/T_c$. The energy density $\epsilon$ (\ref{eps}) practically
coincides with the quasiparticle energy density $T^{00}$ from
(\ref{ent}) (dashed red line). }
    \label{ffig2}
\end{figure}

\vspace{1.0cm}

\noindent The energy density $\epsilon$ then follows from the
thermodynamical relation \cite{pQP,Peshi} \begin{equation}
\label{eps} \epsilon = T s -P \end{equation} (for $\mu_q$ = 0) and
thus is also fixed by the entropy $s(T)$ as well as the
interaction measure
\begin{equation} \label{wint} W(T): = \epsilon(T) - 3P(T) = Ts - 4
P \end{equation} that vanishes for massless and noninteracting
degrees of freedom.

The actual results for  $\epsilon \cdot (T_c/T)^4$ are displayed
in Fig. \ref{ffig2} (solid violet line), $P \cdot (T_c/T)^4$ (blue
dash-dotted line) as well as the interaction measure $W \cdot
(T_c/T)^4$ (\ref{wint}) (green dotted line) and show the typical
pattern from lQCD calculations \cite{Karsch5}. The scale $T_c$ may
now be fixed (estimated) by requiring that the critical energy
density $\epsilon(T_c)$ is roughly the same for the pure
Yang-Mills case as for the full theory with dynamical quarks in
line with the approximate scaling of lQCD \cite{Karsch5}. Since
$\epsilon (T_c)$ in the Yang Mills sector is about 1 GeV/fm$^3$
(cf. \cite{Cassing06}) the scaled energy density $\epsilon(T/T_c)$
from Fig. \ref{ffig2} can be employed to fix the critical
temperature $T_c \approx$ 0.185 GeV. This leads to the 'thumb
rule' $\epsilon \approx 2\  (T/T_c)^4$ [GeV/fm$^3$] for $T > 1.2
T_c$ which is roughly fulfilled according to Fig. \ref{ffig2}.

\begin{figure}[htb!]
\hspace{1cm}
\resizebox{0.65\columnwidth}{!}{\includegraphics{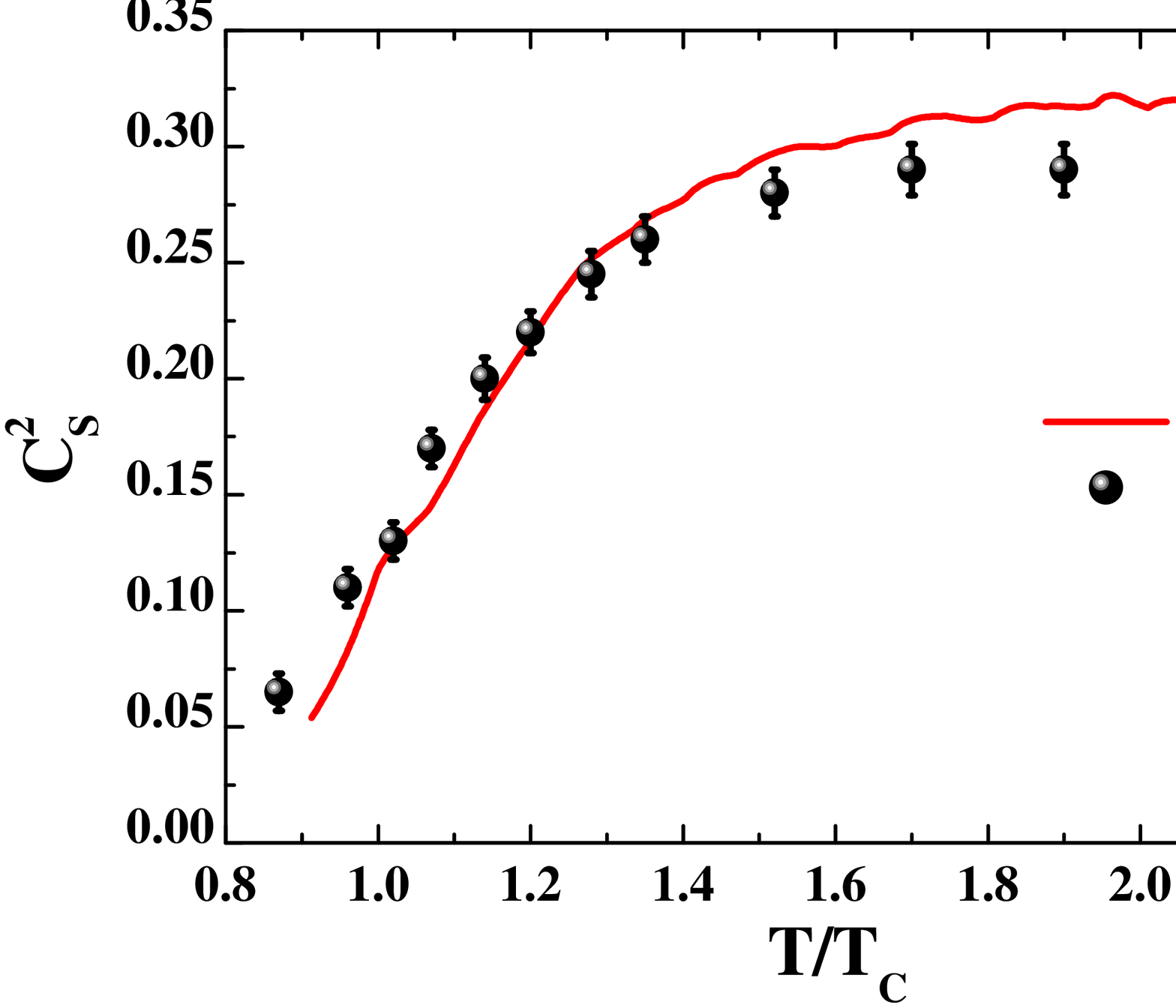} }
    \caption{The DQPM results for the sound velocity (squared) (\ref{sound})
    as a function of $T/T_c$ in comparison to the lQCD results from Ref.
    \cite{Fodor5}.}
    \label{ffig2b}
\end{figure}

\vspace{1.0cm}

A first test of the DQPM in comparison to lQCD calculations is
given for the sound velocity (squared), \begin{equation}
\label{sound} c_s^2 = \frac{d P}{d \epsilon} \ , \end{equation}
which does not depend on the absolute (uncertain) scale of $T_c$.
This comparison is shown in Fig. \ref{ffig2b} by the solid line
where the lQCD results have been taken from Ref. \cite{Fodor5} and
correspond to $N_f = 2+1$ with $N_t = 6$. The DQPM is seen to
reproduce the drop in $c_s^2$ close to $T_c$ within errorbars and
to reach the asymptotic value $c_s^2$ = 1/3 approximately for $T >
2 \ T_c$. This comparison, however, has to be taken with some care
since the present DQPM assumes massless current quarks whereas the lQCD
calculations employ finite quark masses.

\subsubsection{Time-like and space-like quantities}
For the further analysis of the DQPM  it is useful to introduce
the shorthand notations
\begin{equation} \label{conv} \hspace{1.5cm}
 {\rm \tilde Tr}^{\pm}_g \cdots
 =
 d_g\!\int\!\!\frac{d \omega}{2 \pi} \frac{d^3p}{(2 \pi)^3}\,
 2\omega\, \rho_g(\omega)\, \Theta(\omega) \, n_B(\omega/T) \ \Theta(\pm P^2) \, \cdots
 \,\end{equation}
$$   {\rm \tilde Tr}^{\pm}_q \cdots
 =
 d_q\!\int\!\!\frac{d \omega}{2 \pi} \frac{d^3p}{(2 \pi)^3}\,
 2\omega\, \rho_q(\omega)\, \Theta(\omega) \, n_F((\omega-\mu_q)/T) \ \Theta(\pm P^2) \, \cdots
 \,$$ $$  {\rm \tilde Tr}^{\pm}_{\bar q} \cdots =
 d_{\bar q}\!\int\!\!\frac{d \omega}{2 \pi} \frac{d^3p}{(2 \pi)^3}\,
 2\omega\, \rho_{\bar q}(\omega)\, \Theta(\omega) \, n_F((\omega+\mu_q)/T) \ \Theta(\pm P^2) \, \cdots
$$

\noindent
 with $P^2= \omega^2-{\bf p}^2$ denoting the invariant mass
squared.  The quark and
antiquark degrees of freedom, i.e. 2$d_q$ = 36, are
by roughly a factor of two more abundant than the gluonic degrees
of freedom. The $\Theta(\pm P^2)$ function in (\ref{conv})
separates time-like quantities from space-like quantities and can
be inserted for any observable of interest. Note, however, that
not all space-like quantities have a direct physical
interpretation!

We note in passing that the entropy density (\ref{sdqp}) is
dominated by the time-like contributions for quarks and gluons and
shows only minor space-like parts (cf. Refs.
\cite{Andre05,Cassing06}). Furthermore, the entropy density from
the DQPM is only 10 - 15 $\%$ smaller than the Stefan Boltzmann
entropy density $s_{SB}$ for $T > 2\  T_c$ as in the case of lQCD
\cite{Karsch}.

Further quantities of interest are the 'quasiparticle densities'
\begin{equation}
   N^{\pm}_g (T) = {\rm {\tilde Tr}^{\pm}_g }\ 1, \hspace{1cm}
   N^{\pm}_q (T) = {\rm {\tilde Tr^{\pm}_q }}\ 1, \hspace{1cm}
   N^{\pm}_{\bar q} (T) = {\rm {\tilde Tr^{\pm}_{\bar q} }}\ 1,
   \label{eq: N+}
\end{equation} that correspond to the time-like (+) and space-like (-) parts
of the integrated distribution functions. Note that only the
time-like integrals over space have a particle number
interpretation. In QED this corresponds e.g. to time-like photons
($\gamma^*$) which are virtual in intermediate processes but may
also be seen asymptotically by dileptons (e.g. $e^+ e^-$ pairs)
due to the decay $\gamma^* \rightarrow e^+ + e^- (\mu^+ + \mu^-)$
\cite{Cass99,RW00} (cf. Section 4).

Scalar densities for quarks and gluons - only defined in the
time-like sector - are given by \begin{equation} \label{scalar}
N^s_g(T) = {\rm {\tilde Tr}^+_g }\ \left(
\frac{\sqrt{P^2}}{\omega} \right), \, \hspace{0.3cm}  N^s_q(T) =
{\rm{\tilde Tr^+_q }}\ \left( \frac{\sqrt{P^2}}{\omega} \right),
\, \hspace{0.3cm}  N^s_{\bar q}(T) = {\rm{\tilde Tr^+_{\bar q} }}\
\left( \frac{\sqrt{P^2}}{\omega} \right) \, \end{equation} and
have the virtue of being Lorentz invariant.

\begin{figure}[htb!]
\resizebox{0.75\columnwidth}{!}{\includegraphics{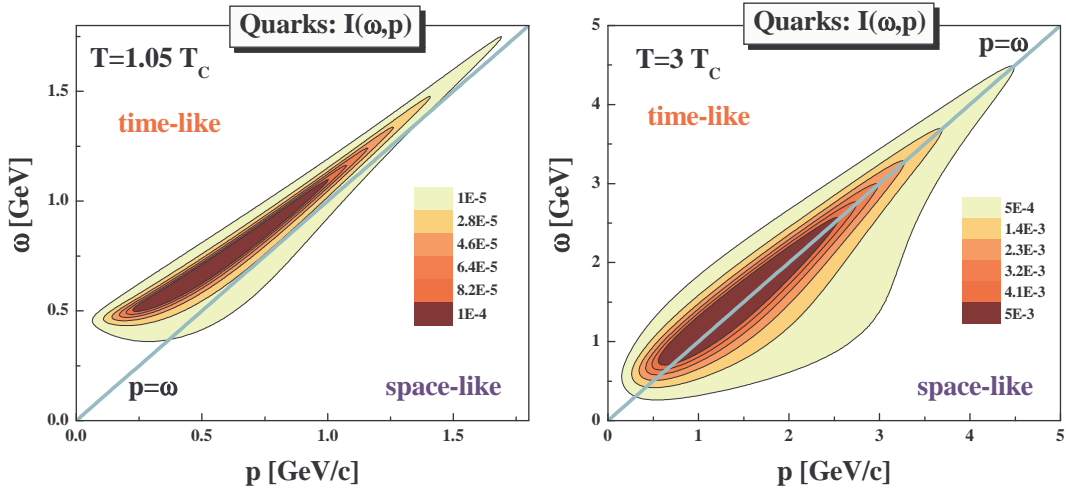} }
    \caption{The integrand $I(\omega,p)$ (\ref{explain}) as a function
    of $\omega$ and $p$ for quarks at temperatures $T=1.05\  T_c$ (l.h.s.)
    and $T= 3\  T_c$ (r.h.s.). At $T=1.05\  T_c$ (l.h.s.) the
    quasiparticle mass amounts to $m \approx 0.55$ GeV and the width to
    $\gamma_q \approx$ 0.034  GeV while at $T=3\  T_c$ (r.h.s.) $m \approx$
    0.7 GeV and $\gamma_q \approx $ 0.25 GeV. The contour lines in both
    figures extend over one order of magnitude.  Note that for a convergence
    of the integrals (\ref{eq: N+}) the upper limits for
    $\omega$ and $p$ have to be increased by
    roughly an order of magnitude compared to the area shown in the figure.}
    \label{ffig3}
\end{figure}

Before coming to the actual results for the quantities (\ref{eq:
N+}) and (\ref{scalar}) it is instructive to have a look at the
integrand in the quark density (\ref{eq: N+}) which reads as (in
spherical momentum coordinates with angular degrees of freedom
integrated out)
\begin{equation}
\label{explain} I(\omega, p) =  \frac{d_q}{2 \pi^3}\ p^2 \ \omega
\, \rho_q(\omega,p^2)\, n_F((\omega-\mu_q)/T)  \, . \end{equation}
Here the integration is to be taken over $\omega$ and $p$ from $0$
to $\infty$. The integrand $I(\omega, p)$ is shown in Fig.
\ref{ffig3} for $T=1.05 T_c$ (l.h.s.) and $T=3 T_c$ (r.h.s.)
($\mu_q=0$) in terms of contour lines spanning both one order of
magnitude. For the lower temperature the quark mass is about 0.55
GeV and the width $\gamma \approx $ 0.034 GeV such that the
quasiparticle properties are close to an on-shell particle. In
this case the integrand $I(\omega,p)$ is essentially located in
the time-like sector and the integral over the space-like sector
is almost negligible. This situation changes for $T = 3 T_c$ where
the mass is about 0.7 GeV while the width increases to $\gamma
\approx $ 0.25 GeV. This situation is close to the systems studied
in Sections 1 and 2 in case of $\phi^4$-theory and a strong
coupling ($\lambda/m \approx$ 18). As one observes from the r.h.s.
of Fig. \ref{ffig3} the maximum of the integrand is shifted
towards the line $\omega = p$ and higher momentum due to the
increase in temperature by about a factor of three; furthermore,
the distribution reaches far out in the space-like sector due to
the Fermi factor $n_F(\omega/T)$ which favors small $\omega$. Thus
the relative importance of the time-like (+) part to the
space-like (-) part is dominantly controlled by the width $\gamma$
- relative to the pole mass - which determines the fraction of
$N_q^-$ with negative invariant mass squared $(P^2 < 0)$ relative
to the time-like part $N_q^+$ ($P^2 > 0$).

\begin{figure}[htb!]
\resizebox{0.75\columnwidth}{!}{\includegraphics{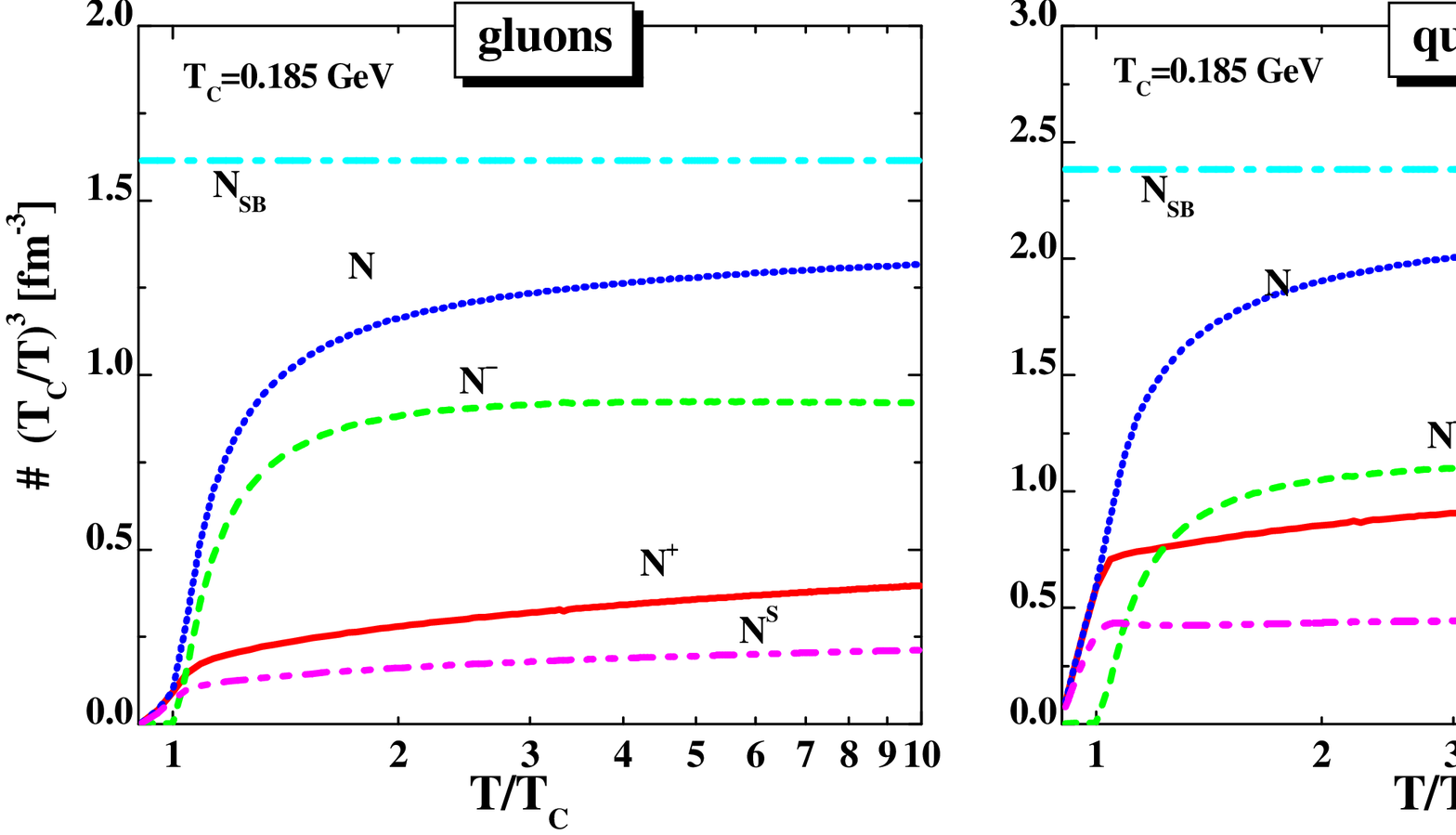} }
    \caption{The various 'densities' (\ref{eq: N+}) for gluons (l.h.s.) and quarks (r.h.s.)
    including antiquarks (at $\mu_q=0$). The lower magenta
lines represent the scalar densities $N^s$, the red solid lines
the time-like densities $N^+$, the green lines the quantities
$N^-$ while the thick solid blue lines are the sum $N=N^+ + N^-$
as a function of $T/T_c$.  The upper dash-dotted lines display the
Stefan Boltzmann limits $N_{SB}$ for reference. All densities are
multiplied by the dimensionless factor $(T_c/T)^3$ to divide out
the leading scaling with temperature. }
    \label{ffig4}
\end{figure}

The actual results for the different 'densities' (multiplied by
$(T_c/T)^3$) are displayed in  Fig. \ref{ffig4} for gluons (l.h.s.)
and quarks (r.h.s.) including the antiquarks. The lower (magenta)
lines represent the scalar densities $N^s$, the red solid lines
the time-like densities $N^+$, the green lines the quantities
$N^-$ while the thick solid blue lines are the sum $N=N^+ + N^-$
as a function of $T/T_c$ (assuming  $T_c$ = 0.185 GeV). It is
seen that $N^+$ is substantially smaller than $N^-$ in case of
gluons in the temperature range 1.1 $\leq T/T_c \leq$ 10.
The quantity $N$ follows closely the Stefan
Boltzmann limit $N_{SB}$ for a massless noninteracting system of
bosons which is given in Fig. \ref{ffig4} (l.h.s.) by the upper
dash-dotted line. Though  $N$ differs by less than 20\% from the
Stefan Boltzmann (SB) limit for $T > 2 T_C$ the physical
interpretation is essentially different! Whereas in the SB limit
all gluons move on the light cone without interactions only a
small fraction of gluons can be attributed to quasiparticles with
density $N^+$ within the DQPM that propagate within the lightcone.
The space-like part $N^-$ corresponds to 'gluons' exchanged in
$t$-channel scattering processes and thus cannot be propagated
explicitly in off-shell transport approaches without violating
causality and/or Lorentz invariance. In case of quarks (or
antiquarks) the results are qualitatively similar but now the
time-like part $N^+$ comes closer to the space-like part $N^-$
since the ratio of the width to the pole mass ($\gamma_q/m$) is
smaller than the corresponding ratio for gluons as stated above.
Furthermore, the quantity $N$ is closer to the respective SB limit
(for massless fermions) due to the lower effective mass of the
quarks. In this respect the quarks and antiquarks are closer to
(but still far from) the massless on-shell quasiparticle limit.

The scalar densities $N^s$ (lower magenta lines) follow smoothly the time-like densities
$N^+$ (for gluons as well as quarks+antiquarks) as a function of
temperature and uniquely relate to the corresponding time-like
densities $N^+$ or the temperature $T$ in thermal equilibrium.

The separation of $N^+$ and $N^-$ so far has no direct dynamical
implications except for the fact that only the fraction $N^+$ can
explicitly be propagated in transport models as argued above. Following
Ref. \cite{Cassing06} we, furthermore, consider the energy densities,
 \begin{equation} \label{energy} T_{00,x}^\pm(T) = {\rm {\tilde Tr^\pm_x
}}\ \omega  \ , \end{equation} that specify time-like and
space-like contributions to the quasiparticle energy densities
($x= g, q , {\bar q}$).

\begin{figure}[htb!]
\resizebox{0.75\columnwidth}{!}{\includegraphics{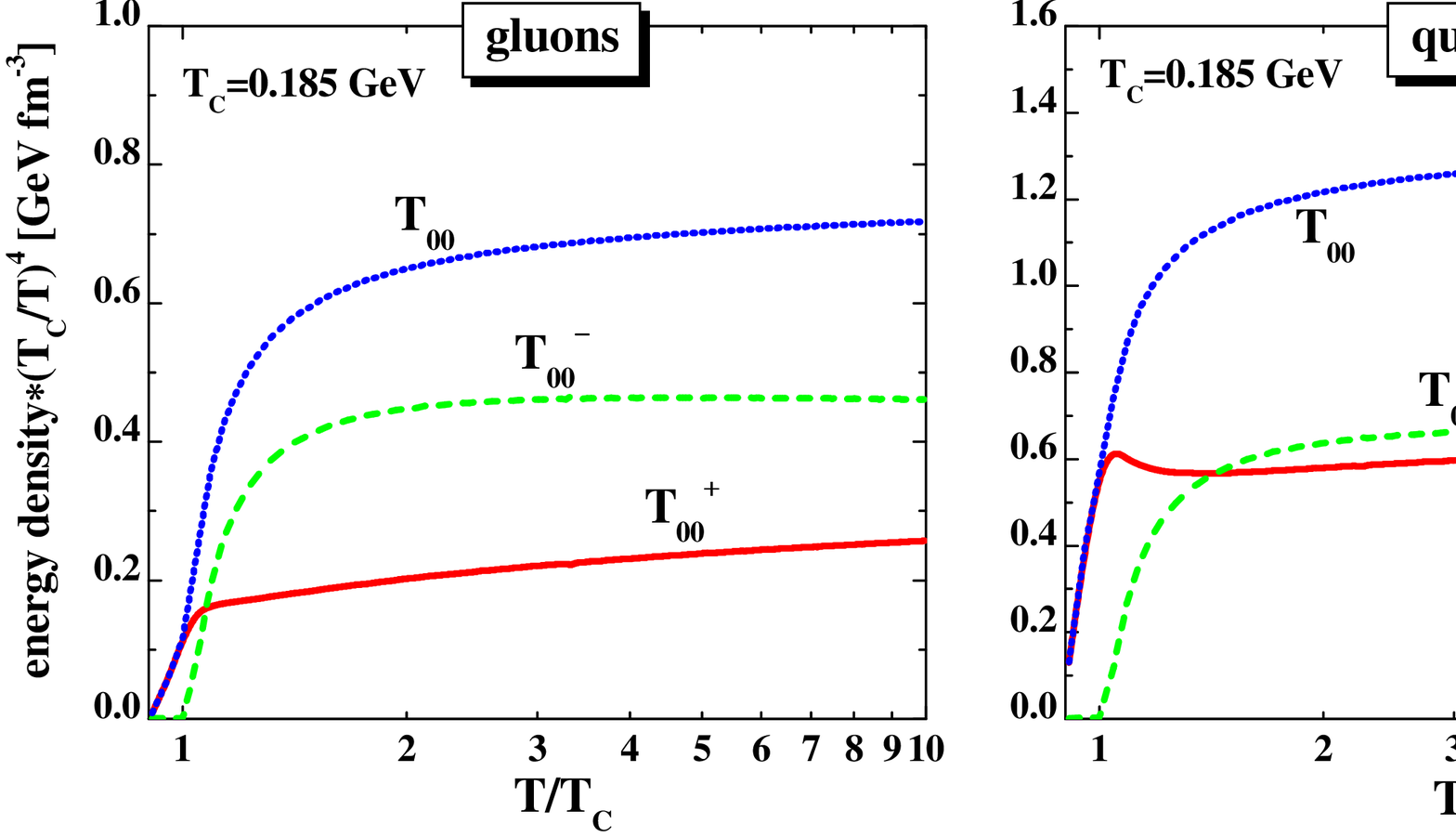} }
    \caption{The time-like energy density $T_{00}^+$ (red solid line),
 the space-like energy density $T_{00}^-$  (dashed green line)
and their sum $T_{00}=T_{00}^+ + T_{00}^-$ (upper blue line)
as a function of $T/T_c$ for gluons (l.h.s.) and quarks (+ antiquarks)
(r.h.s.).  All densities are multiplied
by the dimensionless factor $(T_c/T)^4$. }
    \label{ffig5}
\end{figure}

The result for the quasiparticle energy densities $T^+_{00}$ and
$T^-_{00}$ are displayed in Fig. \ref{ffig5} for gluons (l.h.s.)
and quarks+antiquarks (r.h.s.) as a function of $T/T_c$. All
quantities have been multiplied by the dimensionless factor
$(T_c/T)^4$ in order to divide out the leading temperature
dependence. The lower red solid lines show the time-like
components $T_{00}^+$ while the dashed green lines display the
space-like parts $T_{00}^-$ which dominate over the time-like
parts except in the vicinity of $T_c$. The general behaviour of
the scaled energy densities $T_{00}^\pm$ is similar as for the
'densities' $N^\pm$ given in Fig. \ref{ffig4} since the extra
factor $\omega$ in the integrand does not change significantly the
time-like and space-like parts. As in Ref. \cite{Cassing06} the
space-like parts are interpreted as potential energy densities
while the time-like fractions are the gluon and quark
quasiparticle contributions which propagate within the lightcone.

Summing up the time-like and space-like contributions for gluons,
quarks and antiquarks we obtain the total energy density $T^{00}$,
\begin{equation} \label{ent} T^{00} = T_{00,g}^+ + T_{00,g}^- + T_{00,q}^+ +
T_{00,q}^- + T_{00,{\bar q}}^+ + T_{00,{\bar q}}^- \ ,
\end{equation} which is displayed in Fig. \ref{ffig2} by the dashed red line
(scaled by $T_c/T)^4$). As in the case of the pure Yang-Mills
system in Ref. \cite{Cassing06} the quantity $T^{00}$ practically
coincides with the energy density $\epsilon$ (\ref{eps}) obtained
from the thermodynamical relations. Small differences of less than
5\% show up  which indicates that the DQPM in its present
formulation is not fully consistent in the thermodynamical sense.
Since these differences are small on an absolute scale and
significantly smaller than differences between present independent
lQCD calculations for 3 quark flavors one may consider $T^{00}(T)
\approx \epsilon(T)$ and separate the kinetic energy densities
$T^+_{00}$ from the potential energy densities $T^-_{00}$ as a
function of the temperature $T$ or - in equilibrium - as a
function of the scalar densities $N^s$ or time-like densities
$N^+$, respectively.

\begin{figure}[htb!]
\resizebox{0.75\columnwidth}{!}{\includegraphics{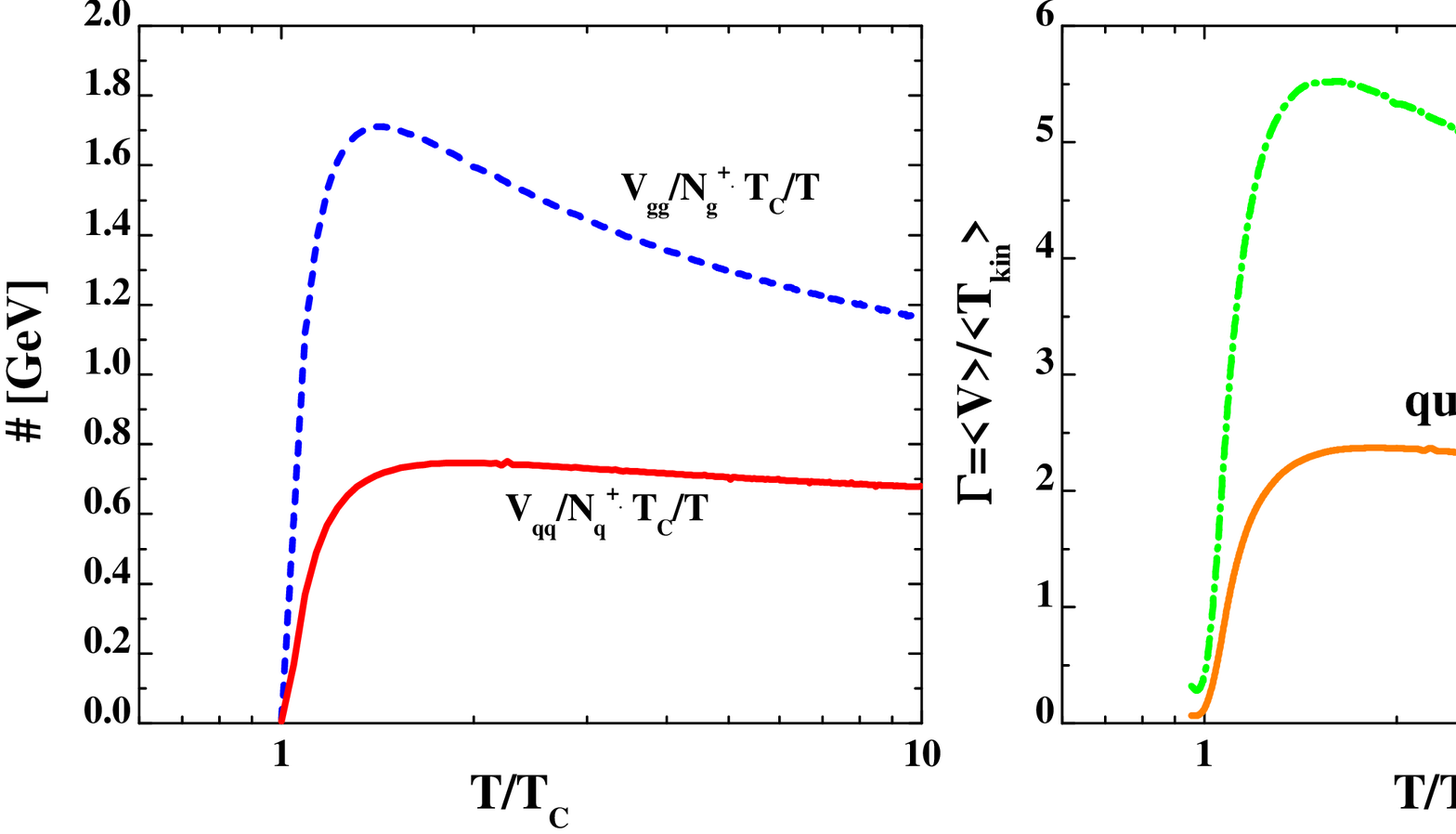} }
    \caption{l.h.s.: The  potential energies per degree of
freedom  for gluons $V_{gg}/N^+_g = T_{00,g}^-/N^+_g$ (dashed blue
line) and for quarks (antiquarks) $V_{qq}/N^+_q =
T_{00,q}^-/N^+_q$ (solid red line) as a function of $T/T_c$. All
energies are multiplied by the dimensionless factor $(T_c/T)$.
r.h.s.: The plasma parameter $\Gamma$ (\ref{Plasma}) for gluons
(dot-dashed green line)  and quarks (solid orange line) as a
function of $T/T_c$. Note that $\Gamma \approx 1-2$ separates a
gas phase from a liquid phase in case of Lenard-Jones type of
interactions.}
    \label{ffig6}
\end{figure}

It is instructive to show the  potential energies per degree of
freedom  $V_{gg}/N^+_g = T_{00,g}^-/N^+_g$ and $V_{qq}/N^+_q =
T_{00,q}^-/N^+_q$ as a function of  $T/T_c$. The corresponding
quantities are displayed in Fig. \ref{ffig6} (l.h.s.) multiplied by
$T_c/T$ in terms of the solid red line and the dot-dashed blue
line.  It is seen that the potential energies per degree of
freedom steeply rise in the vicinity of $T_c$ and then increase
approximately linear with temperature $T$. As expected from the
larger width of the gluons the latter also show a potential energy
per degree of freedom which is roughly a factor of two larger than
the corresponding quantity for quarks (antiquarks). Consequently
rapid changes in the temperature (or density) - as in the
expansion of the fireball in ultrarelativistic nucleus-nucleus
collisions - are accompanied by a dramatic change in the potential
energy density and thus to a violent acceleration of the
quasiparticles. It is speculated here that the large collective
flow of practically all hadrons seen at RHIC \cite{STARS} might be
attributed to the early strong partonic forces expected from the
DQPM.

Furthermore, the plasma parameter $\Gamma$  defined by the ratio
of the average potential energy per particle to the average
kinetic energy per particle, \begin{equation} \label{Plasma}
\Gamma_g = \frac{V_{gg}}{T_{kin,g}} \ , \hspace{2cm} \Gamma_q =
\frac{V_{qq}}{T_{kin,q}} \ , \end{equation} is displayed in the
r.h.s. of Fig. \ref{ffig6} for gluons (dot-dashed green line) and
quarks (solid orange line) as a function of $T/T_c$. Here the
kinetic energy densities are evaluated as \cite{Andre}
\begin{equation} \label{kinetic} T_{kin,g} = {\rm {\tilde Tr}}^+_g
(\omega - \sqrt{P^2}), \hspace{2cm} T_{kin,q} = {\rm {\tilde
Tr}}^+_q (\omega - \sqrt{P^2}). \end{equation} The present results
clearly indicate that the plasma parameters $\Gamma_g, \Gamma_q$
are larger than unity for both quarks and gluons up to 10 $T_c$
(except for the vicinity of $T_c$) such that the system should be
in a liquid phase provided that some attractive interaction
between the constituents persists. Note that the present
evaluation of the plasma parameter $\Gamma$ is entirely carried
out within the DQPM and no longer based on estimates for the
potential energy as in Refs. \cite{Thoma,Andre}. The present
results indicate that the sQGP should persist for a  large range
in temperature (or energy density) and thus also show up in
nucleus-nucleus collisions at Large Hadron Collider (LHC)
energies. Consequently, a partonic liquid is expected to be seen
also at LHC energies and as a consequence the observed scaling of
elliptic flow of practically all hadrons with the number of
constituent quarks (as seen at RHIC) should persist also at LHC.

\subsubsection{Selfenergies and effective interactions of time-like quasiparticles}
Since in transport dynamical approaches there are no
thermodynamical Lagrange parameters like the inverse temperature
$\beta = T^{-1}$ or the quark chemical potential $\mu_q$, which
have to be introduced in thermodynamics in order to specify the
average values of conserved quantities (or currents in the
relativistic sense), derivatives of physical quantities with
respect to the scalar densities $ N^s_x$  (or time-like
densities $ N^+_x$) ($x=g,q,{\bar q}$) are considered in the following (cf.
Refs. \cite{Cassing06,Toneev}).

\begin{figure}[htb!]
\resizebox{0.72\columnwidth}{!}{\includegraphics{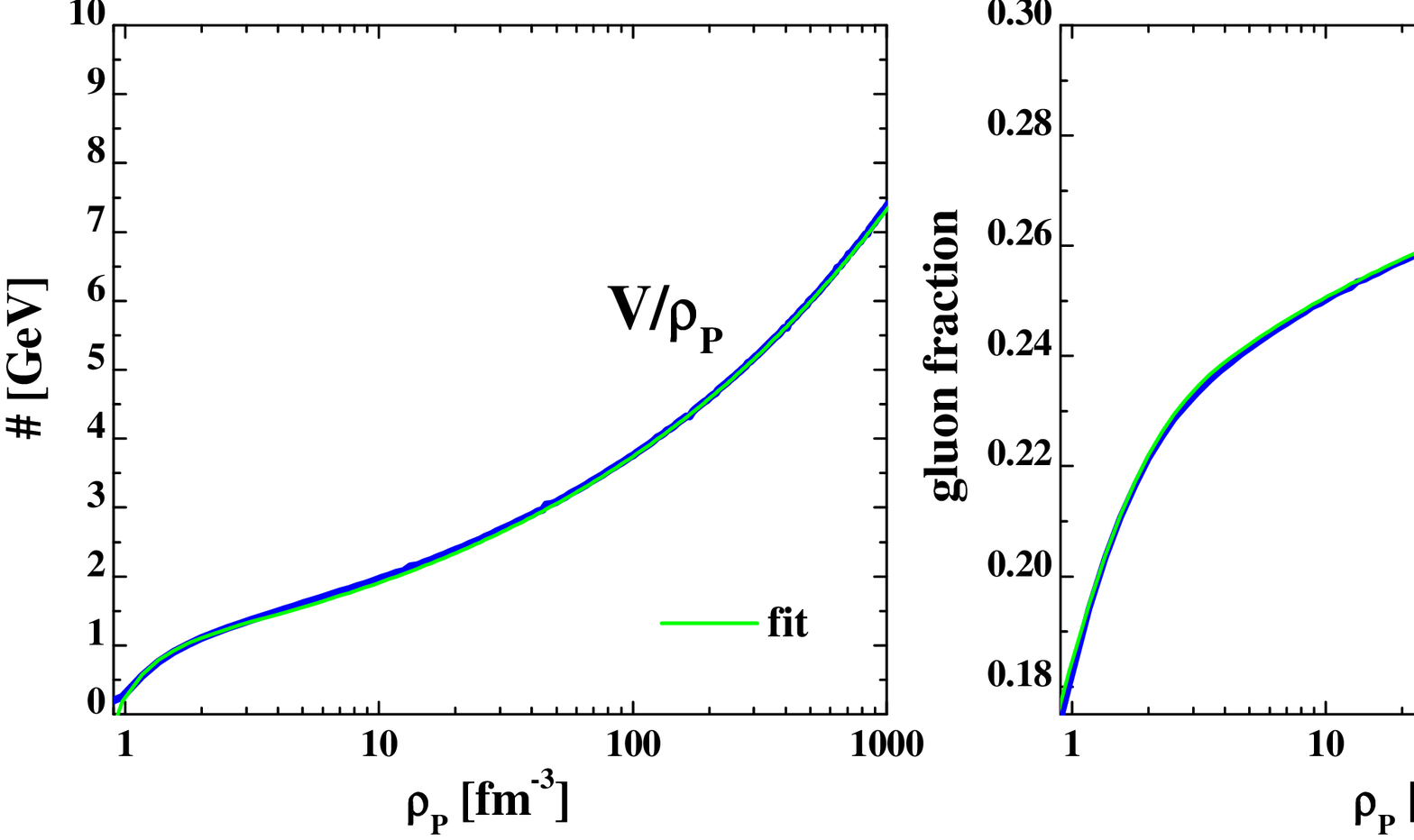} }
    \caption{The parton
potential energy density $V$ (\ref{potp})  - divided by the parton
density $\rho_p$ -  from the DQPM (l.h.s., blue line)  as a
function of the parton density  $\rho_p$ (\ref{partond}). The
functional form of $V(\rho_p)$ is well reproduced by the
expression (\ref{fit1}) when divided by $\rho_p$ (green line) such
that the lines cannot be separated by eye. The r.h.s. shows the
gluon fraction $\alpha$ (\ref{partond}) from the DQPM (dark blue
line) as a function of $\rho_p$ together with the fit (\ref{fit2})
(green line). Again both lines cannot be distinguished by
eye.}    \label{ffig7}
\end{figure}

The independent  potential energy densities  $V_x : = T_{00,x}^-$
now can be expressed as functions of the scalar densities $N^s_x$
(or $N^+_x$) instead of the temperature $T$ (and/or quark chemical
potential $\mu_q$). For a determination of mean-field potentials
for gluons and quarks (antiquarks) it is useful to consider the
partonic potential energy density \begin{equation} \label{potp} V
:= T_{00,g}^- + T_{00,q}^- + T_{00,{\bar q}}^- = {\tilde V}_{gg} +
{\tilde V}_{qq} + {\tilde V}_{qg} \end{equation} and to separate a
pure gluonic interaction density ${\tilde V}_{gg}$ from a pure
fermionic interaction density ${\tilde V}_{qq}$ as well as a
gluon-fermion interaction density ${\tilde V}_{qg}$.
Correspondingly, a parton density $\rho_p$ and gluon fraction
$\alpha$ is defined via \begin{equation} \label{partond} \rho_p =
N^+_g + N^+_q + N^+_{\bar q} \ , \hspace{2cm} \alpha =
\frac{N^+_g}{N^+_g + N^+_q + N^+_{\bar q}} \ . \end{equation} In
the present DQPM (for $T_c$ = 0.185 GeV) the parton density
$\rho_p$ (\ref{partond}) turns out to be a simple function of
temperature,
\begin{equation} \label{pde} \rho_p \left(\frac{T}{T_c} \right) \approx \left( \frac{T}{T_c}
\right)^{3.15} \hspace{1.0cm} {\rm [fm^{-3}] } \ , \end{equation}
such that the average distance between the partons is given by
$d(T/T_c) = \rho_p^{-1/3} \approx (T_c/T)^{1.05}$ [fm] $\approx
T_c/T$ [fm]. These relations allow to convert temperatures scales
to geometrical scales in a simple fashion.

In Fig. \ref{ffig7} the parton potential energy density $V$
(\ref{potp}) - divided by the parton density $\rho_p$ -  is shown
as a function of $\rho_p$ (l.h.s.). The functional dependence of
$V$ on $\rho_p$ can be well approximated by the expression (cf.
l.h.s. of Fig. \ref{ffig7}) \begin{equation} \label{fit1} V(\rho_p)
\approx 0.975 \rho_p^{1.292} - 0.71 \rho_p^{-2.1} \ \ [{\rm
GeV/fm^3}] \ , \end{equation} where the numbers in front carry a
dimension in order to match the units in GeV/fm$^3$. The gluon
fraction $\alpha$ (\ref{partond}) is shown on the r.h.s. of Fig.
\ref{ffig7} and is well approximated by \begin{equation}
\label{fit2} \alpha(\rho_p) = 0.29-0.075 \rho_p^{-0.28} -0.15
\exp(-1.6 \rho_p). \end{equation}

Adding half of the interaction density ${\tilde V}_{qp}$ to the
gluon part and fermion part separately, we have $T_{00,g}^- =
{\tilde V}_{gg} + 0.5 {\tilde V}_{qg}$ and $T_{00,q}^- +
T_{00,{\bar q}}^- = {\tilde V}_{qq} + 0.5 {\tilde V}_{qg}$ such
that ${\tilde V}_{qq} - {\tilde V}_{gg} =: \Delta {\tilde V} =
T_{00,q}^- + T_{00,{\bar q}}^- - T_{00,g}^-$. The relative
fraction of this quantity to the total potential energy density is
evaluated as \begin{equation} \label{frac5} \kappa(\rho_p) =
\frac{\Delta {\tilde V}}{V} = \frac{\Delta {\tilde V}}{T_{00,q}^-
+ T_{00,{\bar q}}^- + T_{00,g}^-} . \end{equation} Using the {\it
Ansatz}: \begin{equation} \label{ansatz2} {\tilde V}_{gg}+{\tilde
V}_{qq} = (1-\xi) V \end{equation} then gives
\begin{equation}  {\tilde V}_{gg} = 0.5 (1-\xi - \kappa) V \ ,
\hspace{1cm} {\tilde V}_{qq} = 0.5 (1-\xi + \kappa) V \ ,
\hspace{1cm}  {\tilde V}_{qg} = \xi V \ , \end{equation} with
still unknown fraction $\xi$ for the interaction density ${\tilde
V}_{qg}$.

In order to determine mean-field potentials $U_g (\rho_p)$ for
gluons or $U_q(\rho_p)$ for quarks (in the rest frame of the
system) one has to consider the derivatives  \begin{equation} \label{mfields}
U_g(\rho_p) : = \frac{\partial ({\tilde V}_{gg} + {\tilde
V}_{qg})}{\partial N_g^+} \  , \hspace{1cm} U_q(\rho_p) : =
\frac{\partial ({\tilde V}_{qq} + {\tilde V}_{qg})}{\partial
(N_q^+ + N_{\bar q}^+)} \ , \end{equation} which by virtue of
(\ref{ansatz2}) can be computed as \begin{equation}
\label{mfields2} \hspace{-0.3cm} U_g(\rho_p)  = \frac{1}{2}
\frac{\partial (1-\kappa+\xi) V }{\partial \rho_p} \frac{\partial
\rho_p}{\partial N_g^+} \  , \hspace{0.4cm} U_q(\rho_p) =
\frac{1}{2} \frac{\partial (1+\kappa+\xi) V }{\partial \rho_p}
\frac{\partial \rho_p}{\partial (N_q^+ + N_{\bar q}^+)} \ .
\end{equation} The fraction $\xi$ of the interaction density - in
principle a function of $\rho_p$ but here taken to be a constant -
now can be fixed in comparison to the gluon mean-field from Ref.
\cite{Cassing06} where the pure Yang-Mills sector has be
investigated in the same way. This leads to $\xi \approx 0.3$ and
separates the total potential energy density $V$ into $\approx 26
\%$ for the gluon-gluon interaction part, 30\% for the quark-gluon
interaction part (including the antiquarks) and $\approx$ 44\% for
the fermionic interaction part.

\begin{figure}[htb!]
\resizebox{0.72\columnwidth}{!}{\includegraphics{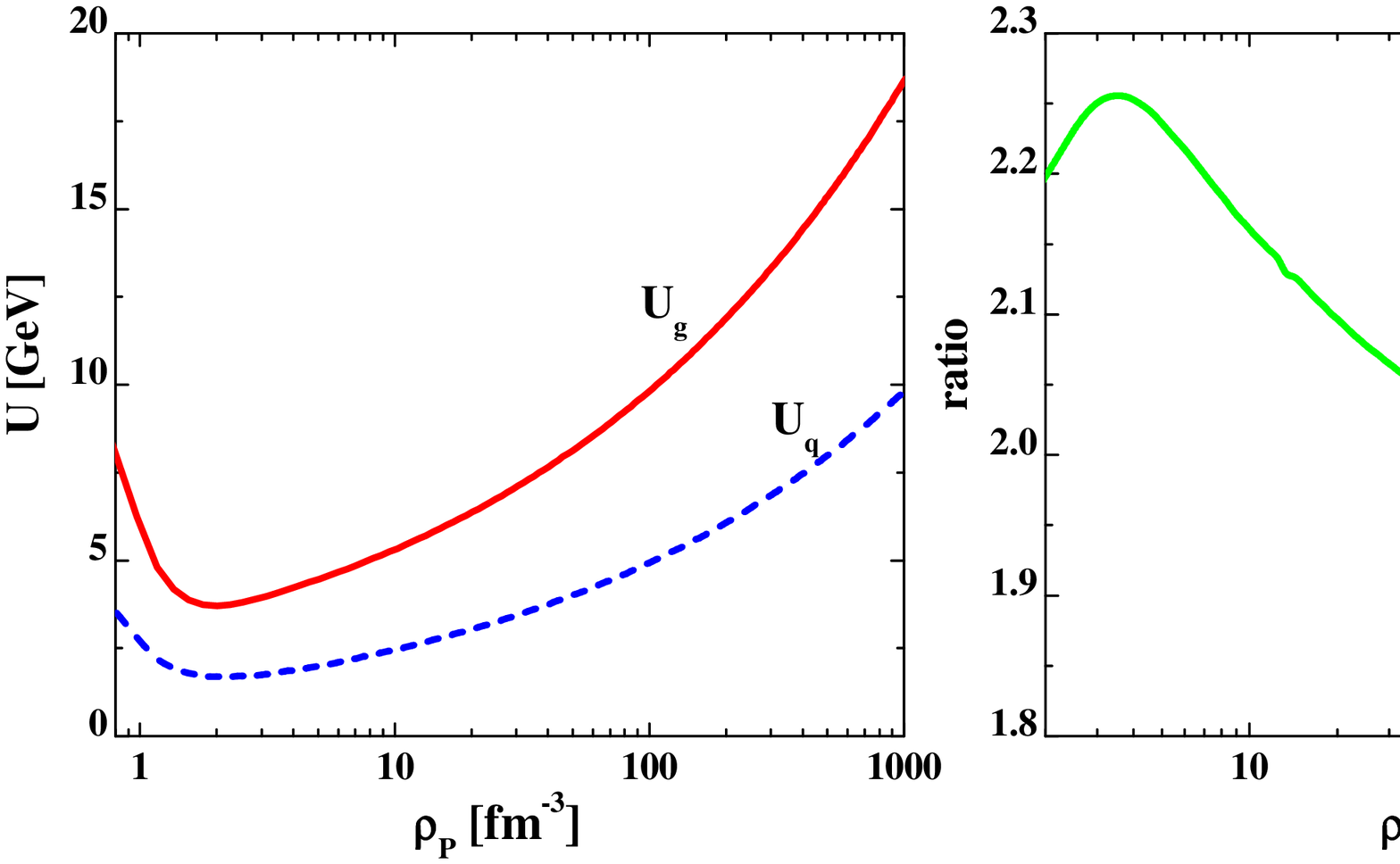} }
    \caption{The mean-field potentials $U_g(\rho_p)$ for gluons
    (solid red line) and $U_q(\rho_p)$ for quarks (dashed blue
    line) as a function of the parton density $\rho_p$ (l.h.s.).
    The r.h.s. displays their ratio as a function of $\rho_p$ which is $\approx 2.05$
    within 10\% accuracy. }
    \label{ffig8}
\end{figure}

The corresponding results for $U_g(\rho_p)$ and $U_q(\rho_p)$ are
displayed in the l.h.s. of Fig. \ref{ffig8} in terms of the solid
red line and dashed blue line, respectively, and show distinct
minima at $\rho_p \approx$ 2.2 fm$^{-3}$ which corresponds to an
average partonic distance of $\approx$ 0.77 fm. The actual
numerical results for the mean-fields can be fitted by the
expressions,
\begin{equation}
\label{pott} \hspace{1.7cm} U_g(\rho_p) \approx 70 \
e^{-\rho_p/0.31} + 2.65\ \rho_p^{0.21} + 0.45\  \rho_p^{0.4} \, \
\ [{\rm GeV}] \ , \end{equation} $$ U_q(\rho_p) \approx 32 \
e^{-\rho_p/0.31} + 1.1\ \rho_p^{0.21} + 0.3\  \rho_p^{0.41} \, \ \
[{\rm GeV}] \ , $$

\noindent where $\rho_p$ is given in fm$^{-3}$ and the actual
numbers in front carry a dimension in order to match to the proper
units of GeV for the mean-fields. The ratio $U_g/U_q \approx 2.05$
as can be seen in the r.h.s. of Fig. \ref{ffig8} for a very wide
range of parton densities $\rho_p$ within 10\% accuracy.

Some comments on the Lorentz structure of the mean fields $U_g$
and $U_q$ appear appropriate. Note that by taking the derivatives
with respect to the time-like densities one implicitly assumes
that a 4-vector current is the physical source of the selfenergies
and that $U_g, U_q$ are the 0'th components of  vector fields
$U^\nu_g$ and $U^\nu_q$ ($\nu = 0,1,2,3$). The spatial components
are assumed to vanish in the rest frame of the system and can be
evaluated by a proper Lorentz boost to the frame of interest. This
implies that the dynamical forces (as space-time derivatives of a
Lorentz vector) are Lorentz tensors as in case of QED (or vector
selfenergies as in the nuclear physics context \cite{SIGMAM}). On
the other hand one might consider the notion of purely scalar
selfenergies where derivatives of the potential energy density
with respect to the scalar density (e.g. $\partial
T_{00,x}^-/\partial N^s_x$) define an effective mass $M_x^*$
\cite{SIGMAM}. In a selfconsistent framework then the
quasiparticles masses (\ref{eq:M2b}) should be given by the
derivatives with respect to the scalar densities. However, this
relation is not fulfilled at all since a numerical evaluation of
the scalar derivatives gives effective masses that are larger by
more than an order of magnitude than the quasiparticle masses
introduced in (\ref{eq:M2b})! This result might have been
anticipated since the effective forces for a gauge (vector) field
theory should be dominated by Lorentz forces as in case of QED.

 Some information on the
properties of the effective gluon-gluon, quark-gluon and
quark-quark interaction may be extracted from the second
derivatives of the potential energy density $V$, i.e.
\begin{equation} \label{interaction}  \hspace{1cm} v_{gg}(\rho_p):
= \frac{\partial^2 {\tilde V}_{gg}}{\partial N_g^{+2}} \approx
\frac{1}{2} \frac{\partial^2 (1-\xi-\kappa) V}{\partial \rho_p^2}
\left( \frac{\partial \rho_p}{\partial N_g^+} \right)^2\ ,
\end{equation} $$ v_{qq}(\rho_p): = \frac{\partial^2 {\tilde
V}_{qq}}{\partial (N_q^{+}+N_{\bar q}^+)^2}
 \approx
\frac{1}{2} \frac{\partial^2 (1-\xi+\kappa) V}{\partial \rho_p^2}
\left( \frac{\partial \rho_p}{\partial( N_q^+ + N_{\bar q}^+)}
\right)^2\ ,$$ $$ v_{qg}(\rho_p): = \frac{\partial^2 {\tilde
V}_{qg}}{\partial (N_q^{+}+N_{\bar q}^+) \partial N_g^+}
 \approx
 \frac{\partial^2 (\xi V)}{\partial
\rho_p^2} \left( \frac{\partial \rho_p}{\partial( N_q^+ + N_{\bar
q}^+)} \right)   \left( \frac{\partial \rho_p}{\partial N_g^+}
\right)  \ .$$

\begin{figure}[htb!]
\resizebox{0.72\columnwidth}{!}{\includegraphics{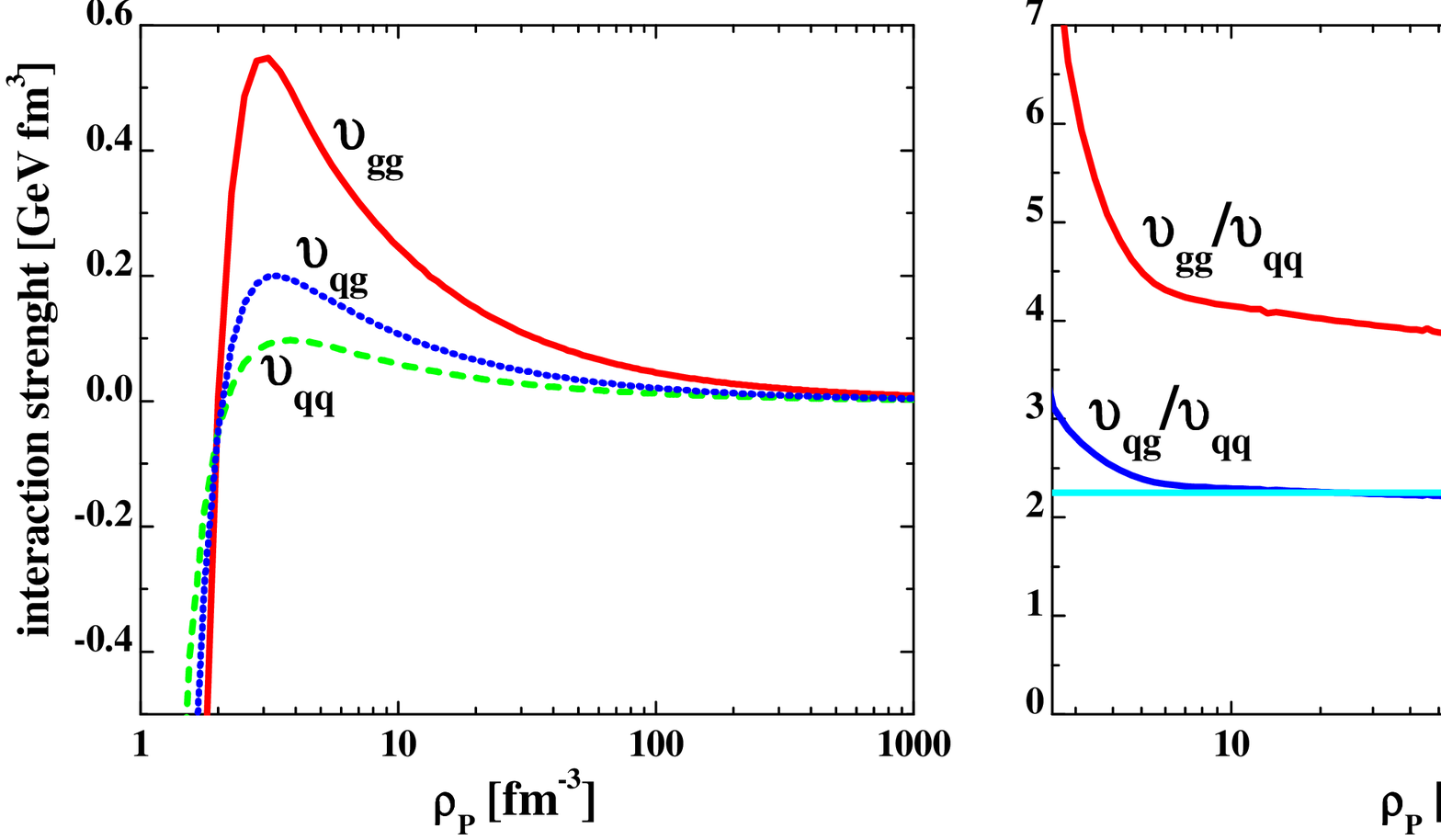} }
    \caption{l.h.s. The effective gluon-gluon (solid red line),
    gluon-quark (solid blue line)
and quark-quark interaction (dashed green line) from the DQPM for
$\xi$ = 0.3 (see text). The r.h.s. displays the ratios
$v_{gg}/v_{qq}$ (solid red line) and $v_{qg}/v_{qq}$ (solid blue
line). The straight light blue line is the ratio of the Casimir
eigenvalues, i.e. $C_g/C_q$= 9/4.}
    \label{ffig10}
\end{figure}

\noindent The numerical results for the interactions
(\ref{interaction}) are displayed in Fig. \ref{ffig10} (l.h.s.) for
the effective gluon-gluon (solid red line), gluon-quark (solid
blue line) and quark-quark interaction (dashed green line). All
interactions show up to become strongly attractive at low parton
density $\rho_p < $ 2.2 fm$^{-3}$, change sign and become
repulsive for all higher parton densities.  Note that the
change of quasiparticle momenta (apart from collisions) will be
essentially driven by the (negative) space-derivatives $-\nabla
U_j(x) = - d U_j(\rho_p)/d \rho_p \ \nabla \rho_p(x)$ which
implies that the partonic quasiparticles (at low parton density)
will bind with decreasing density, i.e. form 'glueballs', mesons,
baryons or antibaryons dynamically close to the phase boundary and
repel each other for $\rho_p >$ 2.2 fm$^{-3}$. Note that color
neutrality is imposed by color-current conservation and only acts
as a boundary condition for the quantum numbers of the
bound/resonant states in color space.

The r.h.s. of Fig. \ref{ffig10} displays the ratios $v_{gg}/v_{qq}$
(solid red line) and $v_{qg}/v_{qq}$ (solid blue line) as a
function of the parton density $\rho_p$ and demonstrates that
$v_{qg}/v_{qq} \approx 9/4$, which is the ratio of the Casimir
eigenvalues.

A straight forward way to model the parton condensation or
clustering to confined glueballs or hadrons dynamically (close to
the phase transition) is to adopt screened Coulomb-like potentials
$v_c(r,\Lambda)$ with the strength $\int d^3r \ v_c(r,\Lambda)$
fixed by the interactions $v_{gg}(\rho_p), v_{qg}(\rho_p),
v_{qq}(\rho_p)$ from (\ref{interaction}) and a screening length
$\Lambda$ from lQCD studies. For the 'dilute parton regime'
($\rho_p < $ 2.2 fm$^{-3}$), where two-body interactions should
dominate, one may solve a Schr\"odinger (Dirac or Klein-Gordon)
equation for the bound and/or resonant states. This task is not
addressed further  since for the actual applications in the PHSD
approach (cf. Section 4) the formation of glueballs is discarded
and the formation of resonant hadronic states close to the phase
boundary is described by density-dependent transition matrix
elements
 ($\sim |v_{qq}(\rho_p)|^2$)
between partons of 'opposite' color with fixed flavor content. In
this way the energy-momentum conservation, the flavor current
conservation as well as 'color neutrality' are explicitly
fulfilled in the (PHSD) transport calculations for interacting
particles with spectral functions of finite width \cite{PHSDnew}.

\subsection{Finite quark chemical potential $\mu_q$}
The extension of the DQPM to finite quark chemical potential
$\mu_q$ is more delicate since a guidance by lQCD is presently
very limited. In the simple quasiparticle model one may use the
stationarity of the thermodynamic potential with respect to
self-energies and (by employing Maxwell relations) derive a
partial differential equation for the coupling $g^2(T,\mu_q)$
which may be solved with a suitable boundary condition for
$g^2(T,\mu_q=0)$ \cite{pQP}. Once $g^2(T,\mu_q)$ is known one can
evaluate the changes in the quasiparticle masses with respect to
$T$ and $\mu_q$, i.e. $\partial M_x^2/\partial \mu_q$ and
$\partial M_x^2/\partial T$ (for $x=g,q,\bar{q}$) and calculate
the change in the 'bag pressure' $\Delta B$ (cf. Refs.
\cite{pQP,Reb} for details). However, such a strategy cannot be
taken over directly since additionally the quasiparticle widths
$\gamma_x(T,\mu_q)$ have to be known in the $(T,\mu_q)$ plane in
case of the DQPM.

In hard-thermal-loop (HTL) approaches \cite{BraP2,BlaJ} the
damping of a hard quark (or gluon) does not depend on the quark
chemical potential explicitly \cite{Vija} and one might  employ
(\ref{eq:gamma}) also at finite $\mu$. This, however, has to  be
considered with care since  HTL approaches assume small couplings
$g^2$ and should be applied at sufficiently high temperature,
only. Present lQCD calculations suggest that the ratio of pressure
to energy density, $P/\epsilon$, is approximately independent of
$\mu_q$ as a function of the energy density $\epsilon$
 \cite{Fodorx}. Accordingly, the functional dependence of the
quasiparticle width $\gamma$ on $\mu_q$ and $T$ has to be modeled
in line with 'lattice phenomenology' (see below).

\subsubsection{A scaling hypothesis}
Assuming three light flavors ($q= u,d,s)$ and all chemical
potentials to be equal ($\mu_u = \mu_d = \mu_s = \mu$) equations
(\ref{eq:M2}) and (\ref{eq:M2b}) demonstrate that the effective
gluon and quark masses are  a function of \begin{equation}
\label{Tstar} T^{*2} = T^2+\frac{\mu^2}{\pi^2} . \end{equation}
Since the coupling (squared) (\ref{eq:g2}) is a function of
$T/T_c$ a straight forward extension of the DQPM to finite $\mu$
is to consider the coupling as a function of $T^*/T_c(\mu)$ with a
$\mu$-dependent critical temperature,   \begin{equation}
\label{Tstar2} T_c(\mu) \approx T_c(\mu=0)(1 - \frac{1}{2\pi^2}
\frac{\mu^2}{T_c(0)^2}) \approx T_c(0)(1 - 0.05
\frac{\mu^2}{T_c(0)^2}) . \end{equation} The coefficient in front
of the $\mu^2$-dependent part can be compared to lQCD calculations
at finite (but small) $\mu$ which gives $0.07(3)$ \cite{Karsch9}
instead of 0.05 in (\ref{Tstar2}). Consequently one has to expect
an approximate scaling of the DQPM results if the partonic width
is assumed to have the form (\ref{eq:gamma}), \begin{equation}
\label{gammamu} \hspace{2cm} \gamma_g(T,\mu)   =
  N_c\  \frac{g^2(T^*/T_c(\mu))}{4 \pi} \, T \
  \ln\frac{2c}{g^2(T^*/T_c(\mu))} \, , \end{equation}
  $$     \gamma_q(T,\mu)
  =
  \frac{N_c^2-1}{2 N_c} \frac{g^2(T^*/T_c(\mu))}{4 \pi} \,  T \
   \ln\frac{2c}{g^2(T^*/T_c(\mu))}
  \ ,
$$

\noindent where $g^2(T/T_c)$ has been replaced by $g^2(T^*/T_c(\mu))$.
 In fact, as will be demonstrated below, this choice leads to an approximate
independence of the potential energies per degree of freedom as a
function of $\mu_q$.  Nevertheless, the conjecture (\ref{gammamu})
should be explicitly controlled by lQCD studies for $N_f$=3 at
finite quark chemical potential. Unfortunately, this task is
presently out of reach and one has to live with the uncertainty in
(\ref{gammamu}) which is assumed in the following investigations.

\begin{figure}[htb!]
\vspace{0.5cm}
\resizebox{0.72\columnwidth}{!}{\includegraphics{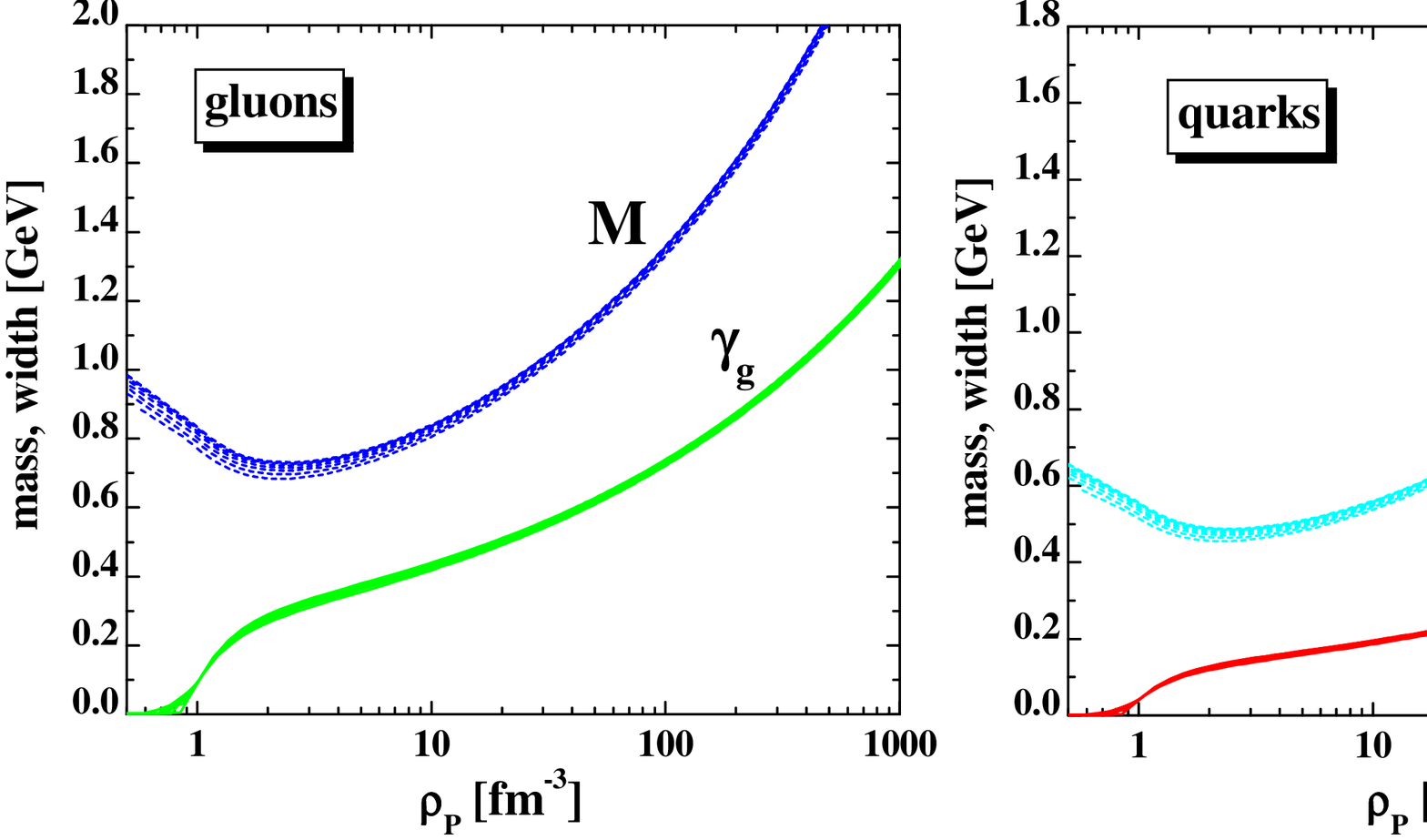} }
    \caption{The gluon mass $M$ and width $\gamma_g$ (l.h.s.) as a function of the parton
    density $\rho_p$ for various chemical potentials $\mu_q = \mu$ from $\mu =0$ to $\mu=$ 0.21 GeV
    in steps of 0.03 GeV. The r.h.s. displays the mass $m$ for quarks and width $\gamma_q$
    for the same quark chemical potentials. }
    \label{ffig18}
\end{figure}

Within the scaling hypothesis (\ref{Tstar2}), (\ref{gammamu}) the
results for the masses and widths in Section 3.2 stay about the
same as a function of $T^*/T_c(\mu)$ when dividing by the
temperature $T$. This also holds approximately when displaying the
masses and widths as a function of the parton density $\rho_p$ for
different chemical potentials $\mu$ as demonstrated in Fig.
\ref{ffig18}. The latter quantities can well be fitted by the
expressions \begin{equation} \label{fitmass} \hspace{3cm}
M_g(\rho_p) \approx 0.41 \ \rho_p^{0.255} + 0.38 \ \rho_p^{-0.7}
\hspace{1cm} {\rm [GeV]} \ , \end{equation}  $$ \gamma_g(\rho_p)
\approx 0.235 \ \rho_p^{0.245} - 0.14 \ \rho_p^{-2} \hspace{1cm}
{\rm [GeV]} , $$ $$ m_q(\rho_p) \approx \frac{2}{3} M_g(\rho_p) ,
\hspace{1cm} \gamma_q(\rho_p) \approx \frac{4}{9} \
\gamma_g(\rho_p)
 , $$

\noindent with $\rho_p$ given in units of fm$^{-3}$. Note that
according to the parametrization (\ref{fitmass}) the width might
become negative for very small $\rho_p$; in actual transport
applications it will be set to zero (cf. Section 4).

\subsubsection{Time-like and space-like quantities}
\begin{figure}[htb!]
\resizebox{0.72\columnwidth}{!}{\includegraphics{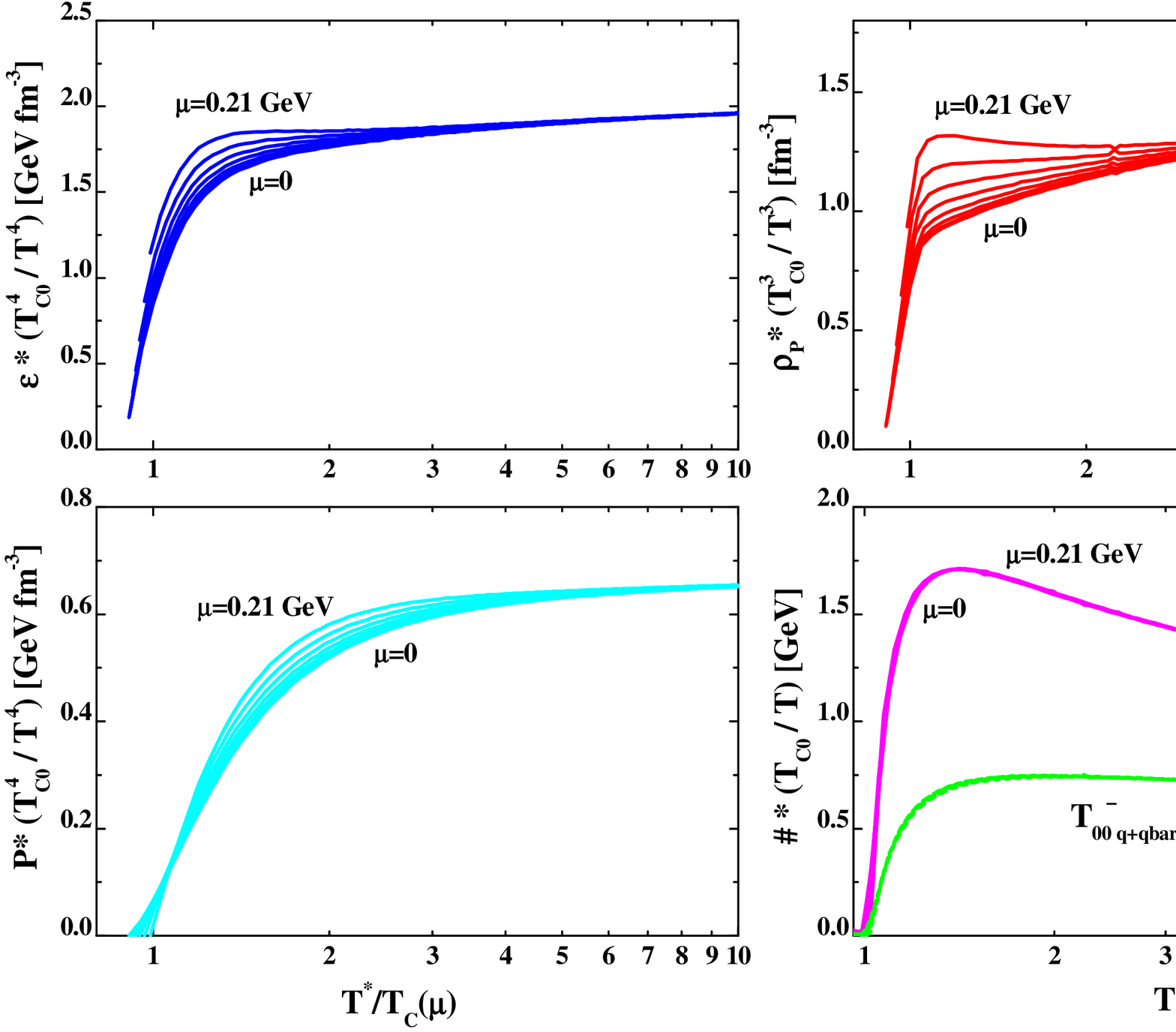} }
    \caption{The energy density $\epsilon(T^*/T_c(\mu))$ (\ref{eps})
    for quark chemical potentials $\mu$ from 0 to 0.21 GeV in steps
    of 0.03 GeV as a function of the scaled temperature
    $T^*/T_c(\mu)$ for $N_f$ =3 (upper l.h.s.). The pressure $P(T^*/T_c(\mu))$
    from the thermodynamical relation (\ref{pressure})
    for quark chemical potentials $\mu$ from 0 to 0.21 GeV in steps
    of 0.03 GeV as a function of the scaled temperature
    $T^*/T_c(\mu)$ for $N_f$ =3 (lower l.h.s.). The parton density
    $\rho_p$ (for quark chemical potentials $\mu$ from 0 to 0.21 GeV) is shown in
    the upper r.h.s. The potential energy
    per time-like 'gluon' (upper magenta lines) and the potential
    energy per time-like 'quark+antiquark' (lower green lines) for
    the same     quark chemical potentials  are displayed  in the
    lower r.h.s. Note that $\epsilon$ and $P$ are
    scaled by the dimensionless factor $(T_{c0}/T)^4$ (with
    $T_{c0}$ = 0.185 GeV) while the parton density $\rho_p$ is scaled by the
    factor $(T_{c0}/T)^3$.}
    \label{ffig14}
\end{figure}

The more interesting question is how the energy density $\epsilon$
(\ref{eps}) and the pressure $P$ (from (\ref{pressure})) change
with quark chemical potential $\mu=\mu_q$ in the DQPM. This
information is provided in Fig. \ref{ffig14} where the upper l.h.s.
shows the energy density $\epsilon$ (\ref{eps}) (scaled in terms
of $T_{c0}=T_c(\mu=0)=$ 0.185 GeV ) as a function of $T^*/T_c(\mu)$.
Here a scaling of the 'temperature' $T^*$ with $T_c(\mu)$ (\ref{Tstar2})
is used since the phase boundary changes with the quark chemical
potential $\mu$. The energy density $\epsilon$ is seen to scale
well with $(T/T_{c0})^4$ as a function of temperature for
$T^*/T_c(\mu)
> 3$, however, increases slightly with $\mu$ close to the phase
boundary where the scaling is violated on the level of 20\%. This
violation in the scaling (seen in the upper left part of the
figure) is essentially due to an increase of the pressure $P$
which is displayed in the lower left part of the figure as a
function of $T^*/T_c(\mu)$ for the same chemical potentials $\mu$
from $\mu = 0$ to 0.21 GeV in steps of 0.03 GeV. Note that a quark
chemical potential of 0.21 GeV corresponds to a baryon chemical
potential of $\mu_B= 3 \mu = $0.63 GeV which is already substantial
and the validity of (\ref{Tstar2}) becomes questionable.

Since the pressure $P$ is obtained from an integration of the
entropy density $s$ over temperature (\ref{pressure}) the increase
in $P$ with $\mu$ can directly be traced back to a corresponding
increase in entropy density. The latter is dominated by the
time-like quasiparticle contributions thus 'counting' the
effective degrees of freedom, \begin{equation} \label{rhog} \rho_p
= N_g^+ + N^+_{q+{\bar q}} \ , \hspace{2cm} N^+_{q+{\bar q}} =
N_q^+ + N_{\bar q}^+ \ . \end{equation}

The upper r.h.s. of Fig. \ref{ffig14} shows $\rho_p$ versus
$T^*/T_c(\mu)$ (multiplied by $T_{c0}^3/T^3$) for chemical
potentials from $\mu=0$ to $\mu$ = 0.21 GeV in steps of 0.03 GeV.
Indeed the scaled  parton density increases with $\mu$; this
increase is most pronounced for lower temperatures and becomes
substantial for $\mu$ = 0.21 GeV. One should recall, however, that
a tri-critical endpoint (in the $T,\mu$ plane) is expected for
$\mu_B = 3 \mu \approx$ 0.4 GeV \cite{FodorKatz} which corresponds
to $\mu \approx$  0.13 GeV. When restricting to the interval $0
\leq \mu \leq $ 0.13 GeV the explicit change in the parton density
$\rho_p$ with $\mu$ stays very moderate. This also holds for the
energy density $\epsilon$ and the pressure $P$.

Quite remarkably the potential energy per time-like fermion
$T^-_{00, q+{\bar q}}/N^+_{q+{\bar q}}$ changes very little with
$\mu$ as can be seen from the lower right part of Fig. \ref{ffig14}
where the latter quantity is displayed for the same chemical
potentials as before (lower green lines) as a function of
$T^*/T_c(\mu)$. This also holds for the potential energy per time-like
gluon $T^-_{00,g}/N^+_g$ (upper magenta lines). Accordingly the potential
energy per time-like degree of  freedom is essentially a function of
$T^*/T_c(\mu)$ alone.

\begin{figure}[htb!]
\vspace{0.2cm}
\resizebox{0.75\columnwidth}{!}{\includegraphics{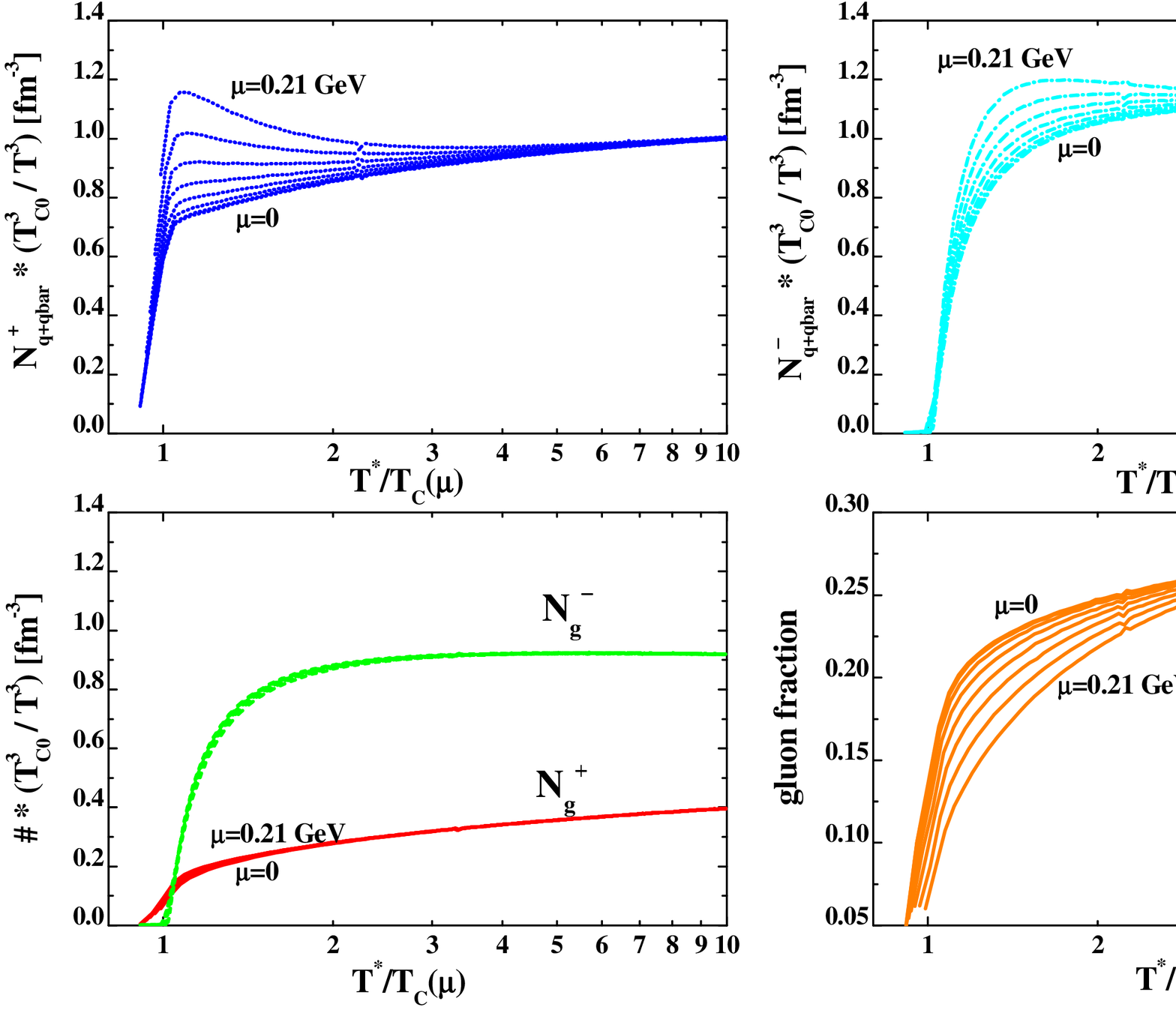} }
    \caption{Upper parts: The time-like fermion density $N_{q+{\bar q}}^+$ (l.h.s., dark blue lines)
    and the space-like quantity $N^-_{q+{\bar q}}$ (r.h.s., light blue lines) for quark chemical
     potentials $\mu$ from 0 to 0.21 GeV in steps
    of 0.03 GeV as a function of the scaled temperature
    $T^*/T_c(\mu)$. Lower parts: the time-like gluon density
    $N_g^+$ (l.h.s., lower red lines) and the space-like quantity $N_g^-$
    (l.h.s., dashed green lines) for the same quark
    chemical potentials as a function of $T^*/T_c(\mu)$.
    Both, $N_g^-$ and $N_g^+$ change only very little with $\mu$
    in the whole temperature range; the relative changes are of
    the order of the line width.
    The gluon fraction (\ref{glufrac})
    is shown on the r.h.s. as a function of the scaled temperature  $T^*/T_c(\mu)$
    for quark chemical potentials $\mu$ from 0 to 0.21 GeV in steps
    of 0.03 GeV. Note that all 'densities' have been multiplied by the dimensionless
    factor $(T_{c0}/T)^3$.}
    \label{ffig13}
\end{figure}

The time-like densities for the fermions $N^+_{q+{\bar q}}$
(\ref{rhog}) (upper l.h.s., dark blue lines) and the space-like
quantities $N^-_{q+{\bar q}}$ (upper r.h.s., light blue lines) are
shown in Fig. \ref{ffig13} as a function of $T^*/T_c(\mu)$ for
chemical potentials $\mu$ from $\mu$ = 0 to $\mu=$ 0.21 GeV in
steps of 0.03 GeV (scaled by $T_{c0}^3/T^3$). Both quantities
increase with $\mu$ in a comparable fashion such that their ratio
stays approximately constant for $T^* > 1.5 \ T_c(\mu)$. The
space-like quantity $N_g^-$ (lower l.h.s., dashed green lines)
practically is independent from $\mu$ as well as the time-like
gluon density $N_g^+$ (lower l.h.s.,  red lines). Accordingly the
gluon fraction \begin{equation} \label{glufrac} \alpha(T,\mu) =
\frac{N_g^+}{N^+_g + N^+_{q + {\bar q}}} \ ,\end{equation} which
is displayed on the lower r.h.s. of Fig. \ref{ffig13} for the same
set of quark chemical potentials as a function of $T^*/T_c(\mu)$,
decreases with $\mu$. It drops to zero below the phase boundary
because the difference between the gluon effective mass and the
fermion effective mass becomes large below $T_c$.

\begin{figure}[htb!]
\vspace{0.2cm}
\resizebox{0.75\columnwidth}{!}{\includegraphics{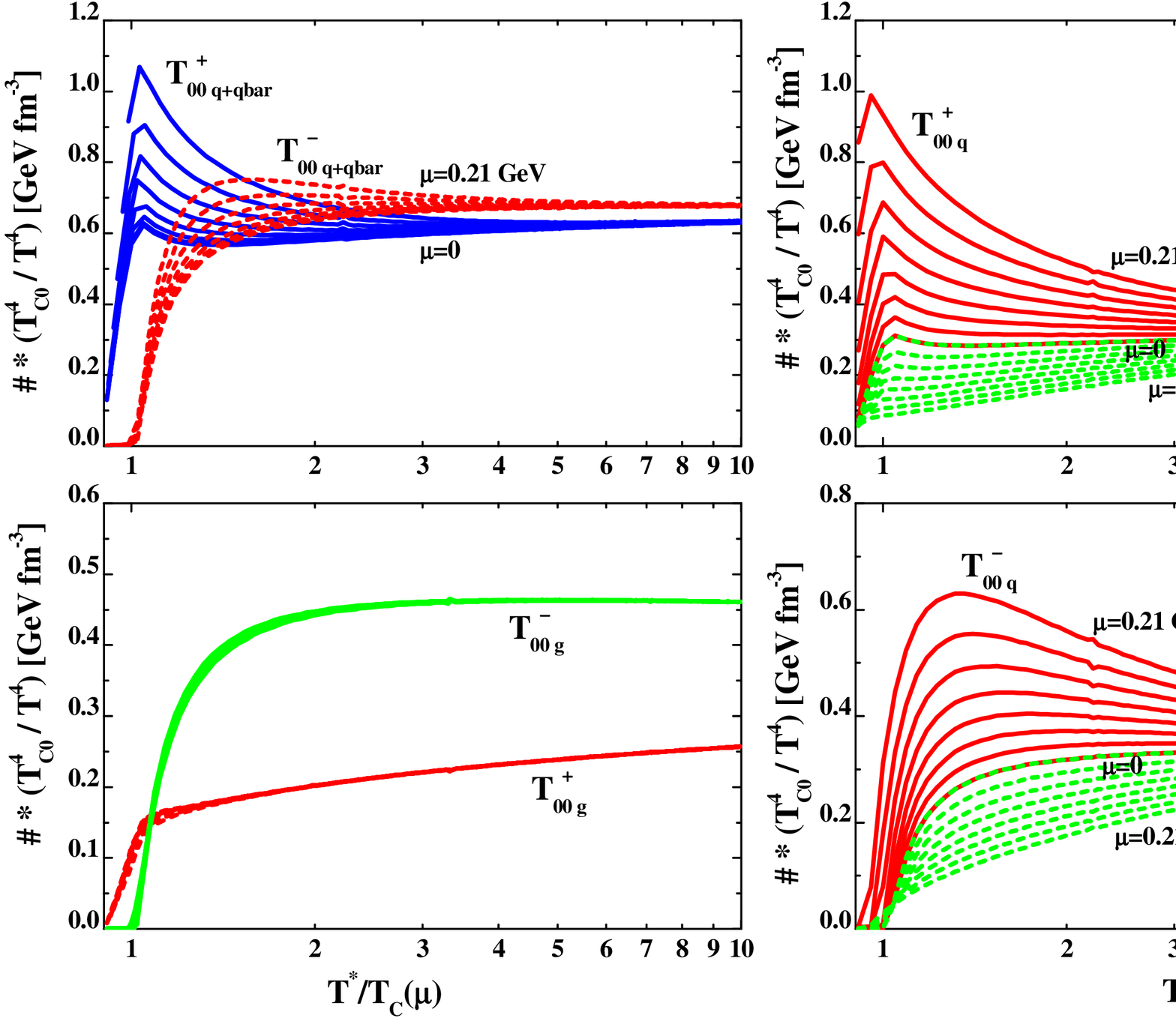} }
    \caption{l.h.s.: The time-like fermion energy density $T^+_{00, q+{\bar q}}$ (blue lines)
    and the space-like  fermion energy density $T^-_{00, q+{\bar q}}$ (dashed red
    lines) for quark chemical
     potentials $\mu$ from 0 to 0.21 GeV in steps
    of 0.03 GeV as a function of the scaled temperature
    $T^*/T_c(\mu)$ (upper l.h.s.). The lower l.h.s. shows the
    time-like (dashed red lines) and space-like gluon energy
    density (green lines) for the same chemical potentials as a function of the
    scaled temperature. Both,  $T_{00, g}^-$ and  $T_{00, g}^+$ are roughly
    independent from $\mu$ in the whole temperature range.  The time-like (upper r.h.s) and
    space-like energy densities (lower r.h.s.) for quarks (upper
    red lines) and antiquarks (lower green lines) are separately displayed
     for quark chemical
     potentials $\mu$ from 0 to 0.21 GeV in steps
    of 0.03 GeV  for $N_f = 3$. All quantities have been multiplied by the dimensionless
    factor $T_{c0}^4/T^4$.}
    \label{ffig15}
\end{figure}

We continue with the time-like and space-like energy densities for
the fermions and gluons as a function of $\mu$ and $T^*/T_c(\mu)$
which are displayed on the l.h.s. of Fig. \ref{ffig15}. The upper
part shows that the space-like and time-like energy densities for
the fermions increase with $\mu$ roughly in a similar fashion as the fermion
'densities' such
that their ratio is approximately independent on $\mu$. The lower
part of Fig. \ref{ffig15} (l.h.s.) indicates that the space-like
energy density for gluons (green lines) as well as the time-like
energy density for gluons (red lines)
are approximately independent from $\mu$ within line width.
 When separating the time-like fermion energy
density into contributions from quarks $q$ and antiquarks ${\bar
q}$ (r.h.s. of Fig. \ref{ffig15}) we find an increase of $T_{00,
q}^+$ with $\mu$ which is not fully compensated by a decrease of
$T_{00, {\bar q}}^+$ with quark chemical potential. Similar
dependences on $\mu$ and $T^*$ are found for the space-like sector
(lower part, r.h.s.). Thus when summing up all space-like and
time-like energy densities from the fermions and the gluons a
small net increase in the total energy density with $\mu$
survives (see also upper left part of Fig. \ref{ffig14}).

We note in passing that derivatives of the various energy
densities w.r.t. the time-like gluon or fermion densities (as
investigated in detail in Section 3.3) are approximately independent
of $\mu$ such that the effective potentials $U_g(\rho_p)$,
$U_q(\rho_p)$ and $U_q(\rho_p)$ (\ref{mfields2}) stay practically
the same. Since this result may be inferred already from the
$\mu$-(in)-dependence of the potential energy per time-like degree of freedom
(displayed in the lower r.h.s. of Fig. \ref{ffig14}) an
explicit representation is discarded. This implies that the
mean-fields (\ref{mfields2}) or parametrizations (\ref{pott})  may
be employed also at finite (moderate) net quark density $N_q^+ -
N_{\bar q}^+$ which simplifies an implementation in parton
transport models (as e.g. PHSD).

\begin{figure}[htb!]
\vspace{0.5cm}
\resizebox{0.75\columnwidth}{!}{\includegraphics{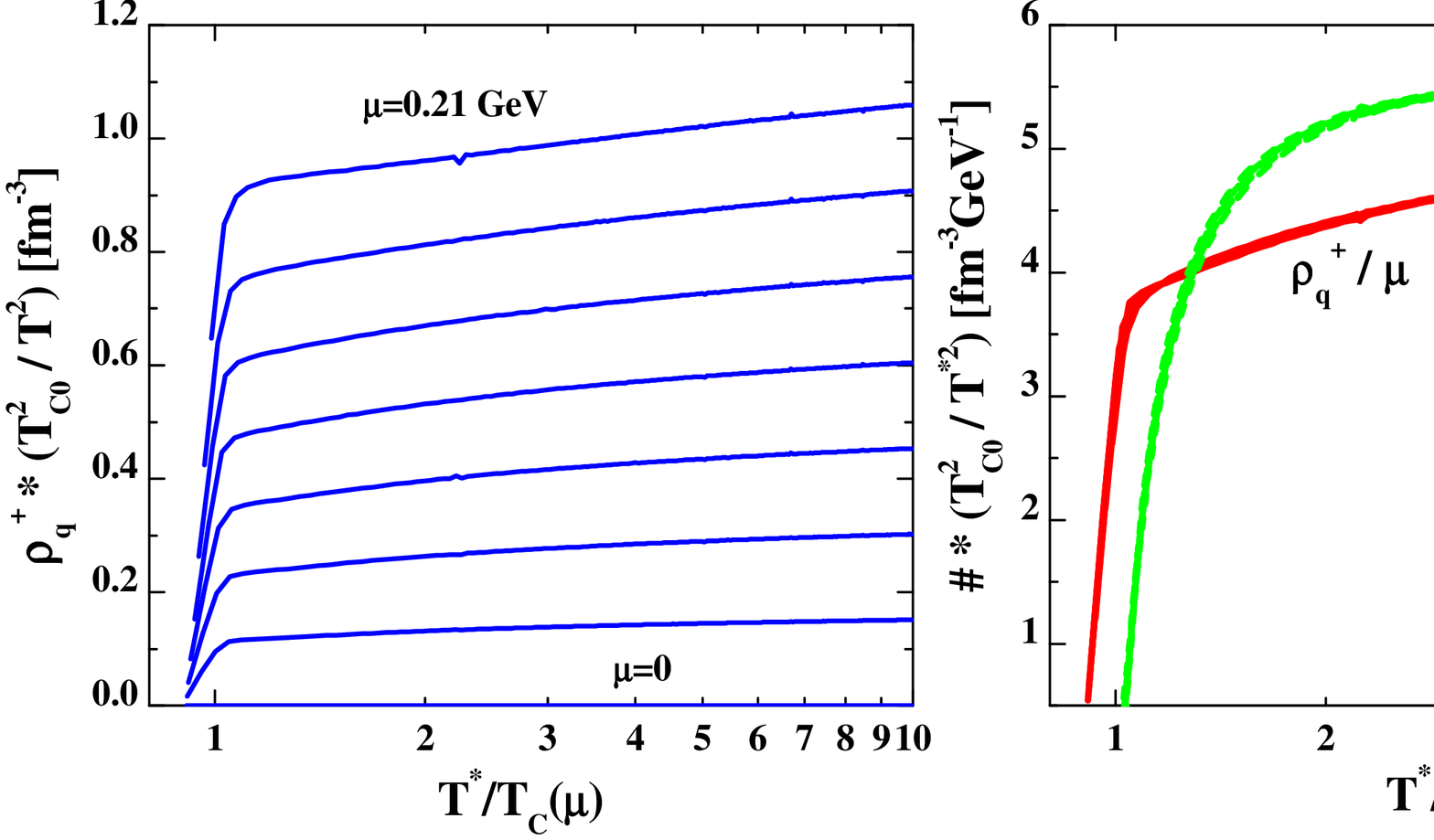} }
    \caption{l.h.s.: The time-like quark density $\rho_q^+$ (\ref{qdens}) (scaled by
    the dimensionless factor $T^2_{c0}/T^2$)
    for quark chemical
     potentials $\mu$ from 0 to 0.21 GeV in steps
    of 0.03 GeV as a function of the scaled temperature
    $T^*/T_c(\mu)$. r.h.s.: The time-like quark density $\rho_q^+$ (\ref{qdens})
    (red solid lines) and the space-like quantity $\rho_q^-$
    (\ref{qdens})(dashed green lines)     for the same quark chemical
     potentials $\mu$. In this part of the figure all quantities are scaled by
    the factor $T^2_{c0}/T^{*2}/\mu$.  }
   \label{ffig16}
\end{figure}

Whereas for vanishing quark chemical potential $\mu$=0 the quark
and antiquark densities are the same this no longer holds for
finite $\mu$ where the differences \begin{equation} \label{qdens}
\rho_q^{\pm} = N_q^{\pm} - N_{\bar q}^{\pm} \end{equation} are of
separate interest since $\rho_q = \rho_q^+ + \rho_q^-$ is the
zero'th component of a conserved flavor current (separately for
each flavor $u,d,s,..$). In order to obtain some idea about
space-like and time-like contributions of the net quark density
$\rho_q$ in the DQPM we first plot the time-like component
$\rho_q^+$ in the l.h.s. of Fig. \ref{ffig16} as a function of
$T^*/T_c(\mu)$ for different $\mu$ from 0 to 0.3 GeV (as before).
In order to divide out the leading scaling with temperature  the
time-like densities have been multiplied here by $T_{c0}^2/T^2$ on
the r.h.s. of Fig. \ref{ffig16} (as known from Fermi systems in the
nuclear physics context). The actual results show an approximately
linear increase in $\mu$ (l.h.s.) which suggests to study the
scaled quantities $\rho_q^{\pm} \cdot T_{c0}^2/(\mu T^{2})$.
However, a rather good scaling is obtained when multiplying
$\rho_q^{\pm}$ by $T_{c0}^2/(\mu T^{*2})$. The latter quantities
are displayed in the r.h.s. of Fig. \ref{ffig16} for the same set
of chemical potentials as before. This approximate scaling allows
to estimate the net quark density as \begin{equation}
\label{guess} \rho_q \approx 10 \ \mu \ T^{*2}/T_{c0}^2 \ \ [{\rm
fm}^{-3}] \end{equation}
 with $\mu$ given in units of GeV in case of $N_f$ =3.

\subsubsection{Comparison to lattice QCD}
\begin{figure}[htb!]
  \begin{center}
\vspace{0.5cm}
\resizebox{0.7\columnwidth}{!}{\includegraphics{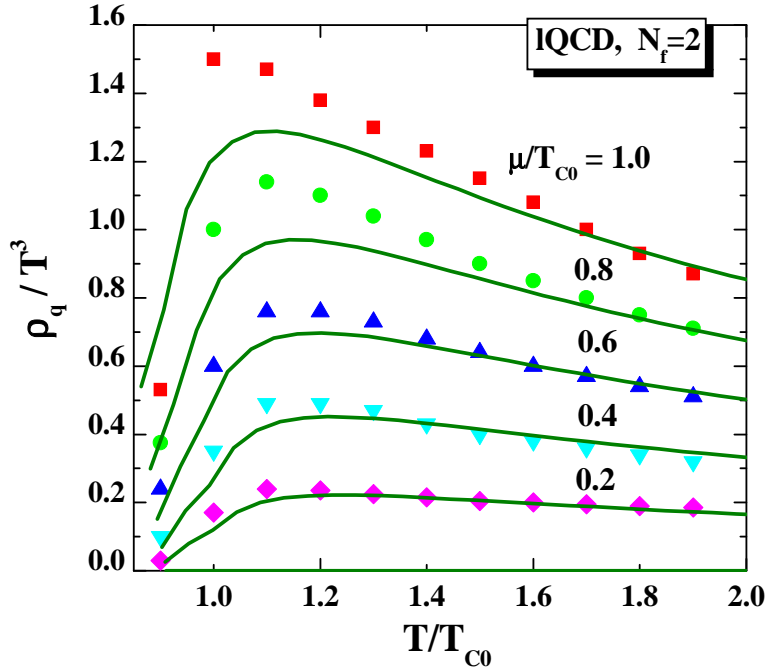} }
    \caption{The quark density $\rho_q$ (\ref{qdens}) (divided by the factor
    $T^3$)    for quark chemical
     potentials $\mu/T_{c0}$ from 0 to 1.0 in steps of 0.2
     as a function of the scaled temperature
    $T/T_{c0}$ from the DQPM (green lines) for $N_f$ =2.
    The different symbols represent the lQCD results from Ref.
    \cite{Karsch8} for the same quark chemical potentials as a
    function of $T/T_{c0}$ (also for $N_f$ =2). Note that the temperature
    axis here is given by $T/T_{c0}= T/T_c(\mu=0)$ and not by $T^*/T_c(\mu)$!}
   \label{ffig17}
  \end{center}
\end{figure}

The approximate scaling depicted in Fig. \ref{ffig16} is a
prediction of the DQPM in case of 3 quark flavors and should be
controlled  by lQCD calculations. For two light flavors ($N_f =
2$) some comparison can be made in the low temperature (and low
$\mu$) range. Unfortunately the simple scaling relations (\ref{guess}) do not
hold at small $T$ such that an explicit comparison has to be
presented between the DQPM for $N_f$ = 2 and the lQCD calculations
from Ref. \cite{Karsch8}. The latter calculations have been
carried out on a 16$^3 \times 4$ lattice with two continuum
flavors (of p4-improved staggered fermions) with mass $m= 0.4 T$.
Though the fermion masses are not really 'light' in the lQCD
calculations the actual lQCD results may serve a test of the
present DQPM in the 2-flavor sector (for low $\mu$ and $T$). The
explicit lQCD results for the net quark density $\rho_q$ (divided
by $T^3$) are displayed in Fig. \ref{ffig17} as a function of
$T/T_{c0}$ in terms of the various symbols. The symbols of equal
color correspond to $\mu/T_{c0}$ = 1.0, 0.8, 0.6, 0.4, and 0.2
from top to bottom and by eye show an approximately linear
dependence on $\mu$. The explicit results from the DQPM for $N_f$
= 2 are presented in terms of the dark green lines and
approximately follow the lQCD results (at least for smaller $\mu$).
Since the systematic errors
of the lQCD calculations are not known to the author an explicit
refitting of the parameters $\lambda, T_s/T_c$ and $c$ for $N_f$
=2  is discarded here. Nevertheless, the
qualitative (and partly quantitative) agreement between the DQPM
and lQCD provides a test for the basic concepts of the DQPM.

\subsection{Dilepton radiation from the sQGP}
The properties of the sQGP so far have been fixed in the DQPM by
specifying the (vector) selfenergies/potentials as well as
effective interactions for the time-like partons. As shown in Ref.
\cite{Andre} this leads to a strongly interacting partonic system
with a shear viscosity to entropy density ratio close to $\eta/s \approx$
0.2. However, the predictions from the DQPM should be controlled
by independent lQCD studies to get some idea about the reliability
of the approach. As mentioned before transport coefficients like
the shear viscosity $\eta$ are available from lQCD \cite{lattice2}
but the present accuracy is not satisfactory. On the other hand
some lQCD information is available from the Bielefeld group on the
electromagnetic correlator which is initimately related to the
dilepton emission rate \cite{RW00,Karsch6,Karsch7}.

\begin{figure}[htb!]
  \begin{center}
\resizebox{0.7\columnwidth}{!}{\includegraphics{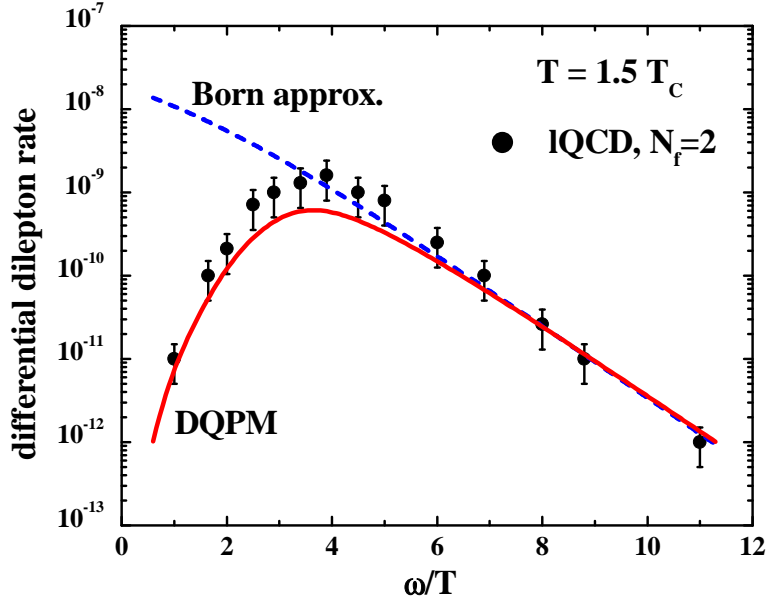} }
    \caption{The 'back-to-back' dilepton emission rate (\ref{leptons2}) from the
    DQPM (solid red line) in comparison to the Born approximation
    (\ref{leptons1}) for massless partons (dashed blue line) and the results
    from the lQCD analysis in Ref. \cite{Karsch7} at $T= 1.5\
    T_c$ in case of $N_f=2$ (full dots).}
    \label{ffig12}
  \end{center}
\end{figure}

In order to provide some further comparison to lQCD results we
consider the dilepton production rate in thermal equilibrium (at
temperature $T$) in a two-flavor QGP as in Ref. \cite{Karsch7}.
For massless quarks and antiquarks the emission rate of
'back-to-back' leptons ($q {\bar q} \rightarrow e^+ e^-$) is given by
\cite{Karsch7,Braten} \begin{equation} \label{leptons1} \frac{d
W}{d \omega d^3 p}({\bf p}=0) = \frac{5 \alpha^2}{36 \pi^4} \
n_F(\frac{\omega}{2T}) \ n_F(\frac{\omega}{2T}) \ , \end{equation}
where $\omega$ is the invariant mass of the lepton pair and $n_F$
denotes the Fermi distribution function. In (\ref{leptons1})
$\alpha$ is the electromagnetic coupling constant.  For massless
partons, furthermore, the magnitude of the lepton momenta is given
by $|{\bf p}| = \omega/2$ while their direction is opposite in the
dilepton rest frame. Neglecting the rest mass of leptons their
energy is $\omega/2$ in the dilepton rest frame, too. The
expression (\ref{leptons1}) changes in case of spectral functions
with finite width to
\begin{equation}
\label{leptons2} \frac{d W}{d \omega d^3 p} = \frac{5 \alpha^2}{36
\pi^4} \int_{0}^\infty d \omega_1 \ \int_{0}^\infty d \omega_2 \
\int_0^\infty dp \ \frac{\omega_1}{\pi} \frac{\omega_2}{\pi} \
\rho_q(\omega_1,p;T) \rho_{\bar q}(\omega_2,p;T) \end{equation} $$
\cdot {\sqrt{{\tilde
\lambda}(\omega^2,P_1^2,P_2^2)}}/{\omega^2} \ \delta(\omega -
\omega_1 - \omega_2) \  n_F(\frac{\omega_1}{T}) \
n_F(\frac{\omega_2}{T}) \  $$

\noindent
 with $P_j^2 = \omega_j^2-p^2$ denoting the
invariant mass of the annihilating partons $j=1,2$. In
(\ref{leptons2}) the factor $\sqrt{{\tilde \lambda}}/\omega^2$
gives a flux correction in case of massive quasiparticles with
\begin{equation} {\tilde \lambda}(x,y,z) = (x - y -z)^2 - 4 yz.
\end{equation}

Since the spectral functions in the DQPM are fixed the lepton
emission rate (\ref{leptons2}) can be evaluated without
introducing any further assumption (or parameter). The results for
the differential emission rate for $N_f = 2$ are shown in Fig.
\ref{ffig12} for $T= 1.5 \ T_c$ (solid red line)  in comparison to
the limit (\ref{leptons1}) (dashed blue line) and the lQCD results
from \cite{Karsch7} (full dots). The lQCD dilepton rate has been
obtained from the temporal correlators in lQCD employing the
'maximum entropy method' which has a systematic error in the order
of 30 to 50 \% \cite{Karsch7} depending on the energy scale
considered. The results from Fig. \ref{ffig12} demonstrate a
drastic suppression of low mass lepton pairs due to the finite
mass of the partons. On the other hand the spectra from the DQPM
are in qualitative agreement with the lQCD results from Ref.
\cite{Karsch7} when including the systematic uncertainties of the
latter approach. Thus the DQPM  passes a further test in
comparison to lQCD.

Some note of caution with respect to the present DQPM appears
appropriate: the parameters in the effective coupling
(\ref{eq:g2}) and the width (\ref{eq:gamma}) have been fixed in
the DQPM by the entropy (\ref{sdqp}) to lQCD results for $N_f$=0
assuming the form (\ref{eq:rho}) for the spectral function
$\rho(\omega)$. Alternative assumptions for $\rho(\omega)$ will
lead to slightly different results for the time-like and
space-like densities, energy densities {\it etc.} but not to a
qualitatively different picture. Also it is presently unclear if
the three parameters ($\lambda, T_s/T_c, c$) employed in Section
3.2 are approximately the same in case of two and three dynamical
flavors. More precise calculations from lQCD should allow to put
further constraints on the form of the spectral function
$\rho(\omega)$ and to fix the basic model parameters in the
effective coupling. Also the 'modified' HTL expression
(\ref{gammamu}) for the quasiparticle width has to be controlled
by lQCD at finite $\mu_q$ as well as transport coefficients like
the shear viscosity $\eta$ or related correlators.

\section{Dilepton production, parton propagation and hadronization}

The dynamics of strongly interacting systems are reflected in the
spectral functions, i.e., in the imaginary part of the
retarded/advanced Green functions. The question comes about how to
measure spectral functions experimentally? A promising way to
obtain information about the spectral properties of hadrons is to
measure their electromagnetic decay to $e^+e^-$ or $\mu^+ \mu^-$
pairs. This is particularly relevant for short lived particles in
a hot and/or dense hadronic environment since this allows to look
at 'in-medium' properties of time-like quanta because the decay
products suffer only from a very small (electromagnetic)
final-state-interaction. In fact, a variety of experimental
collaborations have been set up to investigate this particular
question (cf. the lecture by J. Stroth in this volume
\cite{Stroth}).

\subsection{Dilepton production}

The theory of quantum-chromo-dynamics (QCD) describes hadrons as
many-body bound or resonant states of partonic constituents. While
the  properties of hadrons are rather well known in free space
(embedded in a nonperturbative QCD vacuum) the masses and
lifetimes of hadrons in a baryonic and/or mesonic environment are
subject of current research in order to achieve a better
understanding of the strong interaction and the nature of
confinement. In this context the modification of hadron properties
(or spectral functions) in nuclear matter is of fundamental
interest. Related research has focussed in the past on $K^\pm$
properties in the medium \cite{CAB,Sibi} as well as on vector
mesons ($\rho$, $\omega$ and $\phi$).  QCD sum rules \cite{H&L92}
as well as QCD inspired effective Lagrangian models
\cite{BrownRho} have predicted significant changes, e.g., of the
$\rho$, $\omega$ and $\phi$ with the nuclear density $\rho_N$
and/or temperture $T$.

A direct evidence for the modification of the spectral properties
of vector mesons has been obtained from the enhanced production of
lepton pairs above known sources in nucleus-nucleus collisions at
SPS energies \cite{CERES,HELIOS}. As proposed by Li, Ko, and Brown
\cite{Li} and Ko et al. \cite{Li96} the observed enhancement in
the invariant mass range $0.3 \leq M \leq 0.7$ GeV might be due to
a shift of the $\rho$-meson mass following Brown/Rho scaling
\cite{BrownRho} or the Hatsuda and Lee sum rule
prediction~\cite{H&L92}.  On the other hand also more conventional
approaches that describe a melting of the $\rho$-meson in the
medium due to the strong hadronic coupling (along the lines of
Refs.~\cite{Rapp1}) have been found to be compatible with the
early CERES data \cite{CBRW97}. This ambiguous situation has been
clarified to some extent in 2005 by the NA60 Collaboration since
the invariant mass spectra for $\mu^+\mu^-$ pairs from In+In
collisions at 158 A$\cdot$GeV clearly favored the 'melting $\rho$'
scenario \cite{NA60}. Also the more recent data from the CERES
Collaboration (with enhanced mass resolution) \cite{ceres2} show a
preference for the 'melting $\rho$' scenario.

Dileptons have also been measured in heavy-ion collisions at the
BEVALAC by the DLS Collaboration \cite{DLSold,DLSnew} at incident
energies that are two orders-of-magnitude lower than that at the
SPS. The first published spectra at 1 A$\cdot$GeV \cite{DLSold}
(based on a limited data set) have been consistent with the
results from transport model calculations \cite{Wolf90} that
include $pn$ bremsstrahlung, $\pi^0$, $\eta$ and $\Delta$ Dalitz
decay and pion-pion annihilation.  However, in 1997 the DLS
Collaboration released a new set of data \cite{DLSnew} based on
the full data sample and an improved analysis, which showed a
considerable increase in the dilepton yield:  more than a factor
of five above the early DLS data \cite{DLSold} and the
corresponding theoretical results \cite{Wolf90}. With an in-medium
$\rho$ spectral function, as that used in Ref. \cite{CBRW97} for
dilepton production from heavy-ion collisions at SPS energies,
dileptons from the decay of both directly produced $\rho$'s and
pion-pion annihilation have been considered, and a factor of about
two enhancement has been obtained in the theoretical studies
compared to the case of a free $\rho$-spectral function.
Furthermore, in Ref.  \cite{BratKo99} the alternative scenario of
a dropping $\rho$-meson mass and its influence on the properties
of the $N(1520)$ resonance has been investigated. Indeed, an
incorporation of such medium effects lead to an enhancement of the
$\rho$-meson yield, however, was not sufficient to explain the DLS
data. Since  independent transport calculations by HSD (BUU) and
UrQMD  underestimated the DLS data for C+C and Ca+Ca at 1 A GeV by
roughly the same amount these findings have lead to the denotation
'DLS-puzzle' in 1999 \cite{BratKo99} which persisted by about a
decade.

In order to address the in-medium modifications of vector-mesons
from the experimental side the HADES spectrometer has been built
\cite{HADES06,Stroth} which allows to study $e^+e^-$ pair
production in a much wider acceptance region for elementary $pp$,
$pd$ reactions as well as $\pi A$, $p A$  or even $A A$ collisions
up to about 8 A$\cdot$GeV. Meanwhile the HADES Collaboration has
presented first spectra \cite{HADES06,HADES07pt,HADES07} and the
question is: what do these data tell us? The answer to the
questions raised above is nontrivial due to the nonequilibrium
nature of these reactions, and transport models have to be
incorporated to disentangle the various sources that contribute to
the final dilepton spectra seen experimentally. This task requires
an off-equilibrium transport approach which includes the dynamical
evolution of spectral functions.  In the following we briefly
present results from an up-to-date relativistic transport model
(HSD) that incorporates the relevant off-shell dynamics of the
vector mesons (in line with Section 2) \cite{Brat08}.

In the off-shell HSD approach the dilepton production by a
(baryonic or mesonic) resonance $R$ decay can be schematically
presented in the following way:
\begin{eqnarray}
 BB &\to&R X   \label{chBBR} \\
 mB &\to&R X \label{chmBR} \\
      && R \to  e^+e^- X, \label{chRd} \\
      && R \to  m X, \ m\to e^+e^- X, \label{chRMd} \\
      && R \to  R^\prime X, \ R^\prime \to e^+e^- X, \label{chRprd}
\end{eqnarray}
i.e. in a first step a resonance $R$ might be produced in
baryon-baryon ($BB$) or meson-baryon ($mB$) collisions
(\ref{chBBR}), (\ref{chmBR}). Then this resonance can couple to
dileptons directly (\ref{chRd}) (e.g., Dalitz decay of the
$\Delta$ resonance: $\Delta \to e^+e^-N$) or decays to a meson $m$
(+ baryon) or in (\ref{chRMd})  produce dileptons via direct
decays ($\rho, \omega$) or Dalitz decays ($\pi^0, \eta, \omega$).
The resonance $R$ might also decay into another resonance
$R^\prime$  (\ref{chRprd}) which later produces dileptons via
Dalitz decay.  Note, that in the combined model the final
particles -- which couple to dileptons -- can be produced also via
non-resonant mechanisms, i.e. 'background' channels at low and
intermediate energies or string decay at high energies
\cite{Falter}.

The electromagnetic part of all conventional dilepton sources  --
$\pi^0, \eta, \omega$  Dalitz decays, direct decay of vector
mesons $\rho, \omega$ and $\phi$ -- are treated as described in
detail in Ref.~\cite{CB99}. Modifications -- relative to
Ref.~\cite{CB99} -- are related to the Dalitz decay of baryonic
resonances and especially the strength of the $pp$ and $pn$
bremsstrahlung since more recent calculations by Kaptari and
K\"ampfer \cite{Kaptari} indicated that the latter channels might
have been severely underestimated in previous studies on dilepton
production at SIS energies. In detail:  For the bremsstrahlung
channels in $pp$ and $pn$ reactions we adjust the previous
expressions in order to match the recent results from the OBE
model calculations in Ref. \cite{Kaptari}.

\subsubsection{Vector-meson spectral functions}
In order to explore the influence of in-medium effects on the
vector-meson spectral functions we incorporate the effect of
collisional broadening, i.e. the vector meson width has been
implemented as:
\begin{eqnarray}
\Gamma^*_V(M,|\vec p|,\rho_N)=\Gamma_V(M) + \Gamma_{coll}(M,|\vec
p|,\rho_N) . \label{gammas}
\end{eqnarray}
Here $\Gamma_V(M)$ is the total width of the vector mesons
($V=\rho,\omega$) in the vacuum. For the $\rho$ meson we use
\begin{eqnarray}
\Gamma_\rho(M) &\simeq& \Gamma_{\rho\to\pi\pi}(M) =  \Gamma_0
\left(\frac{M_0}{M} \right)^2 \left(\frac{q}{q_0} \right)^3 \ F(M)
\label{Widthrho} \\ && q = {\frac{(M^2-4m_\pi^2)^{1/2}}{2}},
\
  \ q_0 = {\frac{(M_0^2-4m_\pi^2)^{1/2}}{2}}. \nonumber
\end{eqnarray}
In (\ref{Widthrho}) $M_0$ is the vacuum pole mass of the vector
meson spectral function, $F(M)$ is a formfactor taken  as
\begin{eqnarray}
F(M)={\left(\frac{2\Lambda^2 +M_0^2}{2 \Lambda^2 + M^2} \right)^2}
\label{Frapp}\end{eqnarray} with a cut-off parameter
$\Lambda=3.1$~GeV. For the $\omega$ meson a constant total vacuum
width is used: $\Gamma_\omega\equiv \Gamma_\omega(M_0)$, since the
$\omega$ is a narrow resonance in vacuum. The collisional width in
(\ref{gammas}) is approximated as
\begin{eqnarray}
\Gamma_{coll}(M,|\vec p|,\rho_N) = \gamma \ \rho_N \ \langle v \
\sigma_{VN}^{tot} \rangle \approx  \ \alpha_{coll} \
\frac{\rho_N}{\rho_0} . \label{dgamma}
\end{eqnarray}
Here $v=|{\vec p}|/E; \ {\vec p}, \ E$ are the velocity,
3-momentum and energy of the vector meson in the rest frame of the
nucleon current, $\gamma$ is the Lorentz factor for the boost from
the nucleon rest frame to the center-mass-system of heavy-ions;
$\rho_N$ denotes the nuclear density and $\sigma_{VN}^{tot}$ is
the meson-nucleon total cross section.

In order to simplify the actual calculations for dilepton
production the coefficient $\alpha_{coll}$ has been extracted in
the HSD transport calculations from the vector-meson collision
rate in C+C and Ca+Ca reactions (at 1 and 2 A$\cdot$ GeV) as a
function of the density $\rho_N$. The numerical results for
$\Gamma_{coll}(\rho_N)$ then have been divided by $\rho_N/\rho_0$
to fix the coefficient $\alpha_{coll}$ in (\ref{dgamma}).  We
obtain  $\alpha_{coll} \approx 150$~MeV for the $\rho$ and
$\alpha_{coll} \approx 70$~MeV for $\omega$ mesons.  In this way
the average effects of collisional broadening are incorporated and
allow for an explicit representation of the vector-meson spectral
functions versus the nuclear density (see below).

In order to explore the observable consequences of vector-meson
mass shifts at finite nuclear density -- as indicated by the
CBELSA-TAPS data \cite{tapselsa} for the $\omega$ meson -- the
in-medium vector-meson pole masses are modeled (optionally)
according to the Hatsuda and Lee \cite{H&L92} or Brown/Rho scaling
\cite{BrownRho} as
\begin{eqnarray}
\label{Brown} M_0^*(\rho_N)= \frac{M_0} {\left(1 + \alpha {\rho_N
/ \rho_0}\right)},
\end{eqnarray}
where $\rho_N$ is the nuclear density at the resonance decay
position $\vec r$, $\rho_0 = 0.16 \ {\rm fm}^{-3}$ is the normal
nuclear density and $\alpha \simeq 0.16$ for the $\rho$ and
$\alpha \simeq 0.12$ for the $\omega$ meson. The parametrization
(\ref{Brown}) may be employed also at much higher collision
energies (e.g. FAIR and SPS) and one does not have to introduce a
cut-off density in order to avoid negative pole masses. Note that
(\ref{Brown}) is uniquely fixed by the 'customary' expression
$M_0^*(\rho_N) \approx M_0 (1 - \alpha \rho_N/\rho_0)$ in the low
density regime.

\begin{figure}[t]
\resizebox{0.72\columnwidth}{!}{\includegraphics{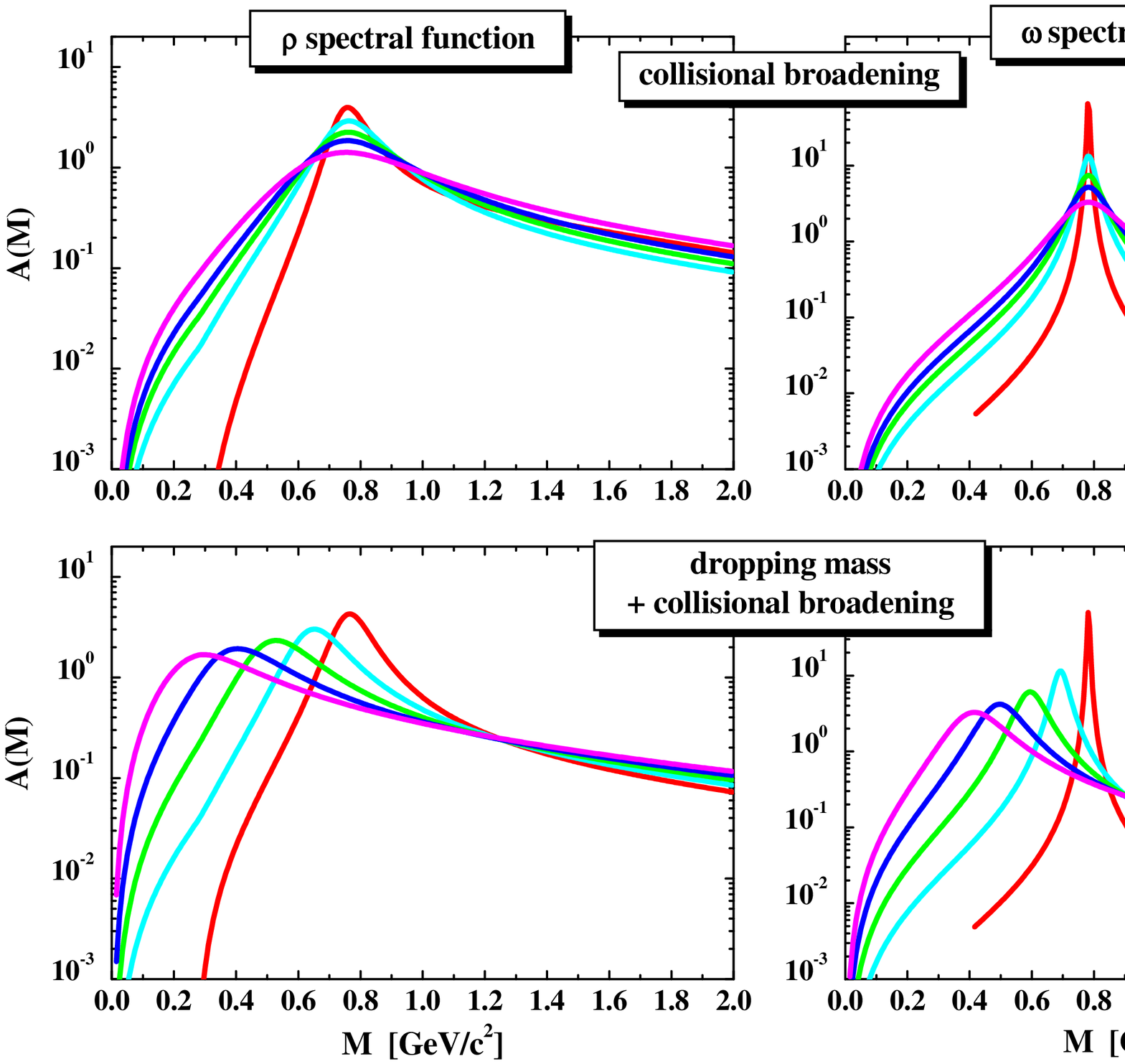} }
\caption{The spectral functions for the $\rho$ and $\omega$ meson
 in the case of the 'collisional broadening' scenario (upper part)
and the 'dropping mass + collisional broadening' scenario (lower
part) for nuclear densities of 0,1,2,3,5$\times\rho_0$ as employed
in the transport calculations (see text for details). }
\label{OFig0}
\end{figure}

The spectral function of the vector meson $V$ with the mass $M$ at
baryon density $\rho_N$ is taken in the Breit-Wigner form:
\begin{eqnarray}
A_V(M,\rho_N) = C_1\cdot {{\frac{2}{\pi}} \
{M^2\Gamma_V^*(M,\rho_N) \over (M^2-M_{0}^{*^2}(\rho_N))^2 + (M
{\Gamma_V^*(M,\rho_N)})^2}},
\label{spfunV}
\end{eqnarray}
with the normalization condition for any $\rho_N$:
\begin{eqnarray}
\int_{M_{min}}^{M_{lim}} A_V(M,\rho_N) dM =1,
\label{SFnorma}\end{eqnarray} where $M_{lim}=2$~GeV is chosen as
an upper limit for the numerical integration. The lower limit of
the vacuum spectral function corresponds to the $2\pi$ decay
$M_{min}=2 m_\pi$, whereas for the in-medium collisional
broadening case $M_{min}=2 m_e \to 0$ with $m_e$ denoting the
electron mass. $M_0^*$ is the pole mass of the vector meson
spectral function which is $M_0^*(\rho_N=0)=M_0$ in vacuum,
however, shifted in the medium for the dropping mass scenario
according to Eq. (\ref{Brown}).

The resulting spectral functions for the $\rho$ and $\omega$ meson
are displayed in Fig. \ref{OFig0} for the case of 'collisional
broadening' (upper part) as well as for the 'dropping mass +
collisional broadening' scenario (lower part) for densities of
0,1,2,3,5 $\times \rho_0$.  Note that in vacuum the hadronic
widths vanish for the $\rho$ below the two-pion mass and for the
$\omega$ below the three-pion mass. With increasing nuclear
density $\rho_N$ elastic and inleastic interactions of the vector
mesons shift strength to low invariant masses. In the 'collisional
broadening' scenario we find a dominant enhancement of strength
below the pole mass for the $\rho$-meson while the $\omega$-meson
spectral function is drastically enhanced in the low- and
high-mass region with density (on expense of the pole-mass
regime). In the 'dropping mass + collisional broadening' scenario
both vector mesons dominantly show a shift of strength to low
invariant masses with increasing $\rho_N$. Qualitatively similar
pictures are obtained for the $\phi$-meson but quantitatively
smaller effects are seen due to the lower effect of mass shifts
and a substantially reduced $\phi N$ cross section which is a
consequence of the $s\bar{s}$ substructure of the $\phi$-meson.

\begin{figure}[t]
\resizebox{0.85\columnwidth}{!}{\includegraphics{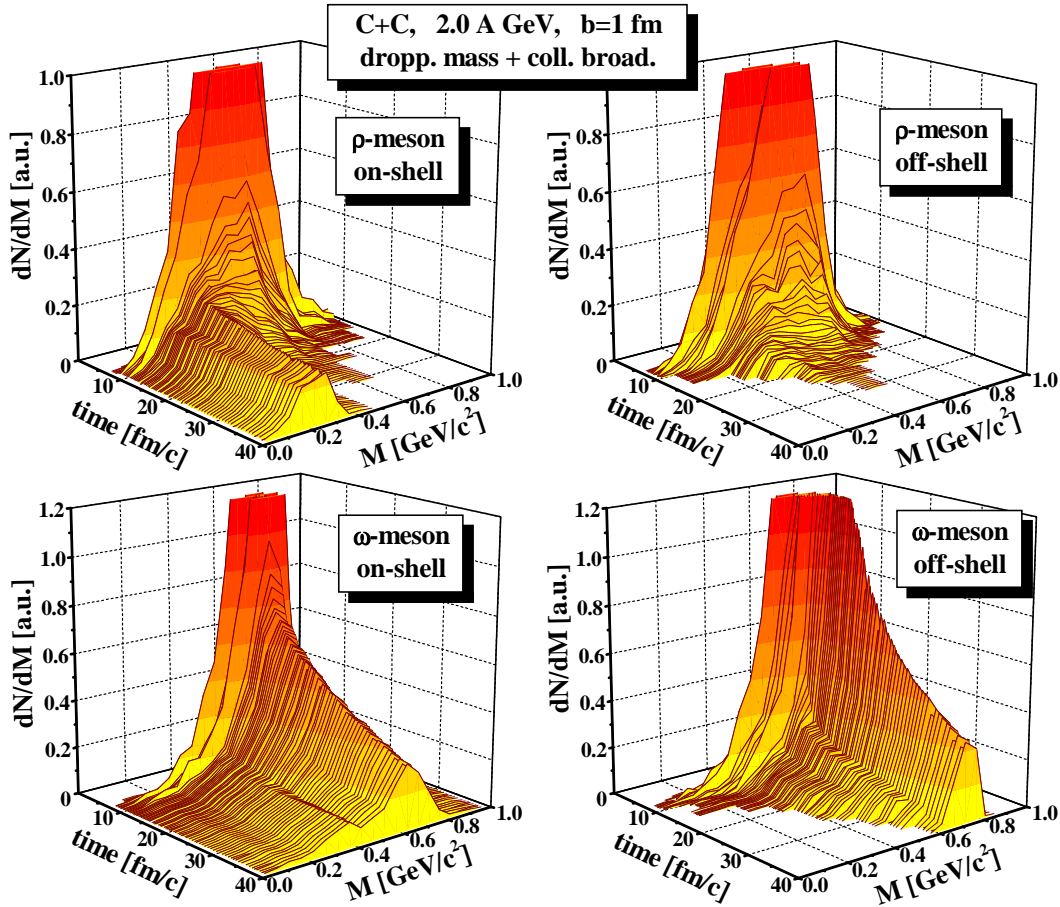} }
\caption{Time evolution
of the mass distribution of $\rho$ (upper part) and $\omega$
(lower part) mesons for  central C+C collisions (b=1 fm) at 2 A
GeV for the dropping mass + collisional broadening scenario. The
l.h.s. of Fig. \protect\ref{Fig3Donoff} corresponds to the
calculations with on-shell dynamics, whereas the r.h.s. shows the
off-shell HSD results.} \label{Fig3Donoff} \vspace*{1mm}
\end{figure}

In order to demonstrate the importance of off-shell transport
dynamics for dilepton production we present in Fig.
\ref{Fig3Donoff} the time evolution of the mass distribution of
$\rho$ (upper part) and $\omega$ (lower part) mesons for  central
C+C collisions (b=1 fm) at 2 A GeV for the dropping mass +
collisional broadening scenario (as an example). The l.h.s. of
Fig. \ref{Fig3Donoff} corresponds to the calculations with
on-shell propagation whereas the r.h.s. stand for the off-shell
dynamics. As seen in Fig. \ref{Fig3Donoff} the initial $\rho$ and
$\omega$ mass distributions are quite broad even for a small
system such as C+C where, however, the baryon density at 2 A GeV
can reach (in some local cells) $\sim 2 \rho_0$. The number of
vector mesons decreases with time due to their decays and the
absorption by baryons ($\rho n \rightarrow \pi N$ or $2 \pi N$).
Most of the $\rho$ mesons decay/disappear already inside the
'fireball' for density  $\rho_N > 0$. Due to the expansion of the
'fireball' the baryon density drops quite fast, so some amount of
$\rho$ mesons reach the very low density zone or even the
'vacuum'. Since for the off-shell case (r.h.s. of Fig.
\ref{Fig3Donoff}) the $\rho$ spectral function changes dynamically
by propagation in the dense medium, it regains the vacuum shape
for $\rho_N \to 0$. This does not happen for the on-shell
treatment (l.h.s. of Fig. \ref{Fig3Donoff}) - the $\rho$ spectral
function does not change its shape by propagation but only by
explicit collisions with other particles. Indeed, there is a
number of $\rho$'s which survive the decay or absorption and leave
the 'fireball' with masses below $2m_\pi$. Thus, the on-shell
treatment leads to the appearance of the $\rho$ mesons in the
vacuum with $M\le 2m_\pi$, which can not decay to $2 \pi$; thus
they live practically 'forever' since the probability to decay to
other channels is very small. Indeed, such $\rho$'s will
continuously shine  low mass dileptons which leads to an
unphysical 'enhancement/divergence' of the dilepton yield at low
masses (note, that the dilepton yield is additionally enhanced by
a factor $1/M^3$).

The same statements are valid for the $\omega$ mesons (cf. the
lower part of Fig.  \ref{Fig3Donoff}): since the $\omega$ is a
long living resonance, a larger amount of $\omega$'s survives with
an in-medium spectral function shape in the vacuum (in case of
on-shell dynamics). Such $\omega$'s with $M < 3m_\pi$ can decay
only to $\pi \gamma$ or electromagnetically (if $M < m_\pi $).
Since such unphysical phenomena appear in on-shell transport
descriptions including an explicit vector-meson propagation an
off-shell treatment is mandatory!

\subsubsection{Dilepton production in C+C and Ca+Ca collisions at SIS energies}

A major aim of the HADES Collaboration was to clarify the
'DLS-puzzle', i.e. to verify/falsify the DLS data \cite{DLSnew}
from the experimental side \cite{Stroth}. Accordingly, the same
systems have been reinvestigated at 1 A$\cdot$GeV with the HADES
detector in order to clarify the issue. The situation, however, is
not as easy since the DLS and HADES acceptances differ
significantly
 \cite{DLSnew,HADES06}. Recently the HADES Collaboration has performed a direct comparison
with the DLS data for C+C at 1 A GeV \cite{DLSnew} by filtering
the HADES data with the DLS acceptance \cite{HADES07}. Both
measurements were found to agree very well \cite{Stroth,HADES07}!
Thus, the 'DLS puzzle' has been solved from the experimental side;
there is no obvious contradiction between the DLS and HADES data!

The major goal for the transport models now was to verify/explain
the solution of the 'DLS-puzzle' from the theoretical side. Note
that transport calculations allow for a direct comparison between
the DLS and HADES measurements by employing the different FILTER
routines to the same set of calculated events.

\subsubsection*{Comparison to the DLS data}

\begin{figure}[tbh]
\phantom{a}\vspace*{2mm}
\resizebox{0.48\columnwidth}{!}{\includegraphics{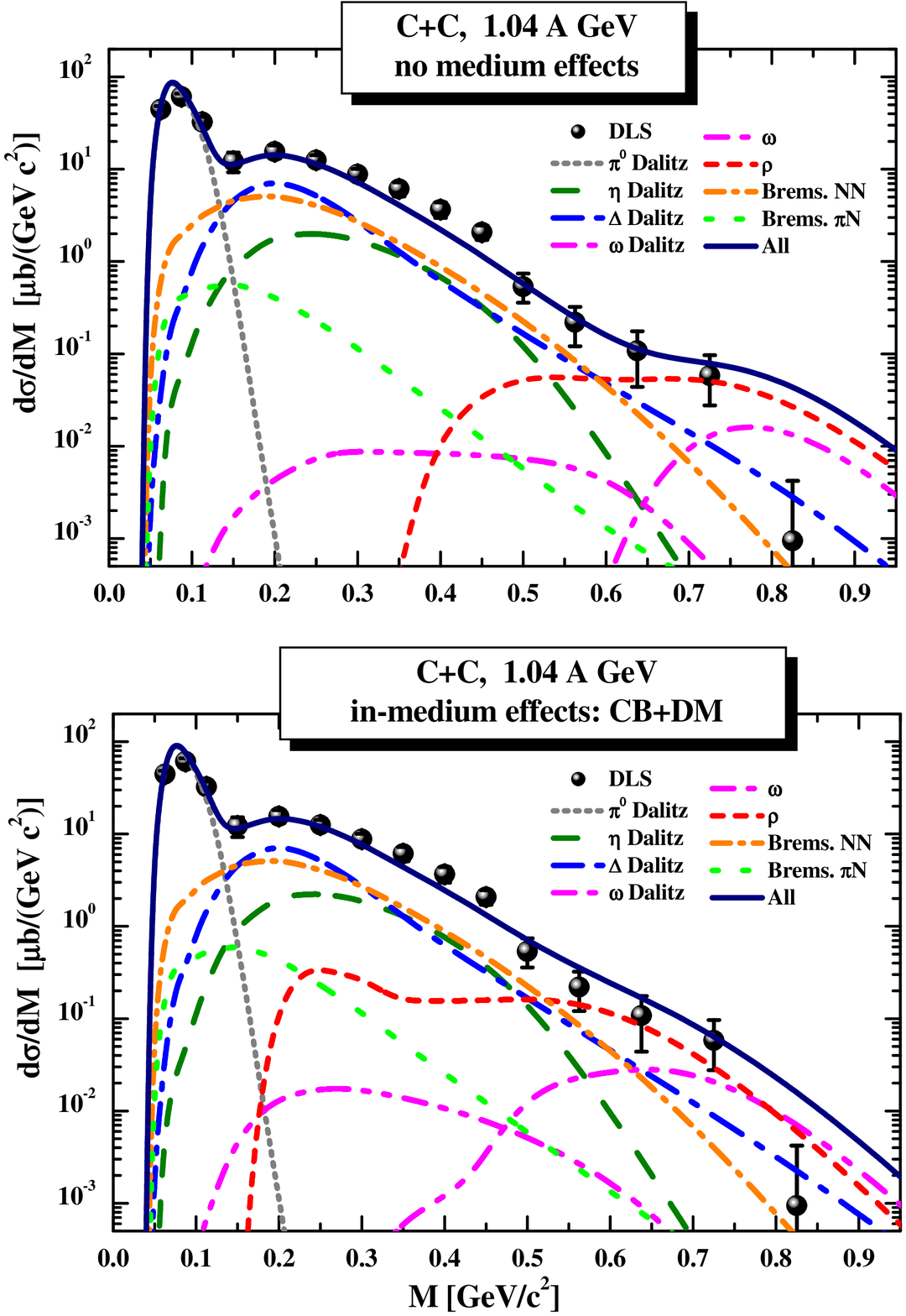} }
\resizebox{0.48\columnwidth}{!}{\includegraphics{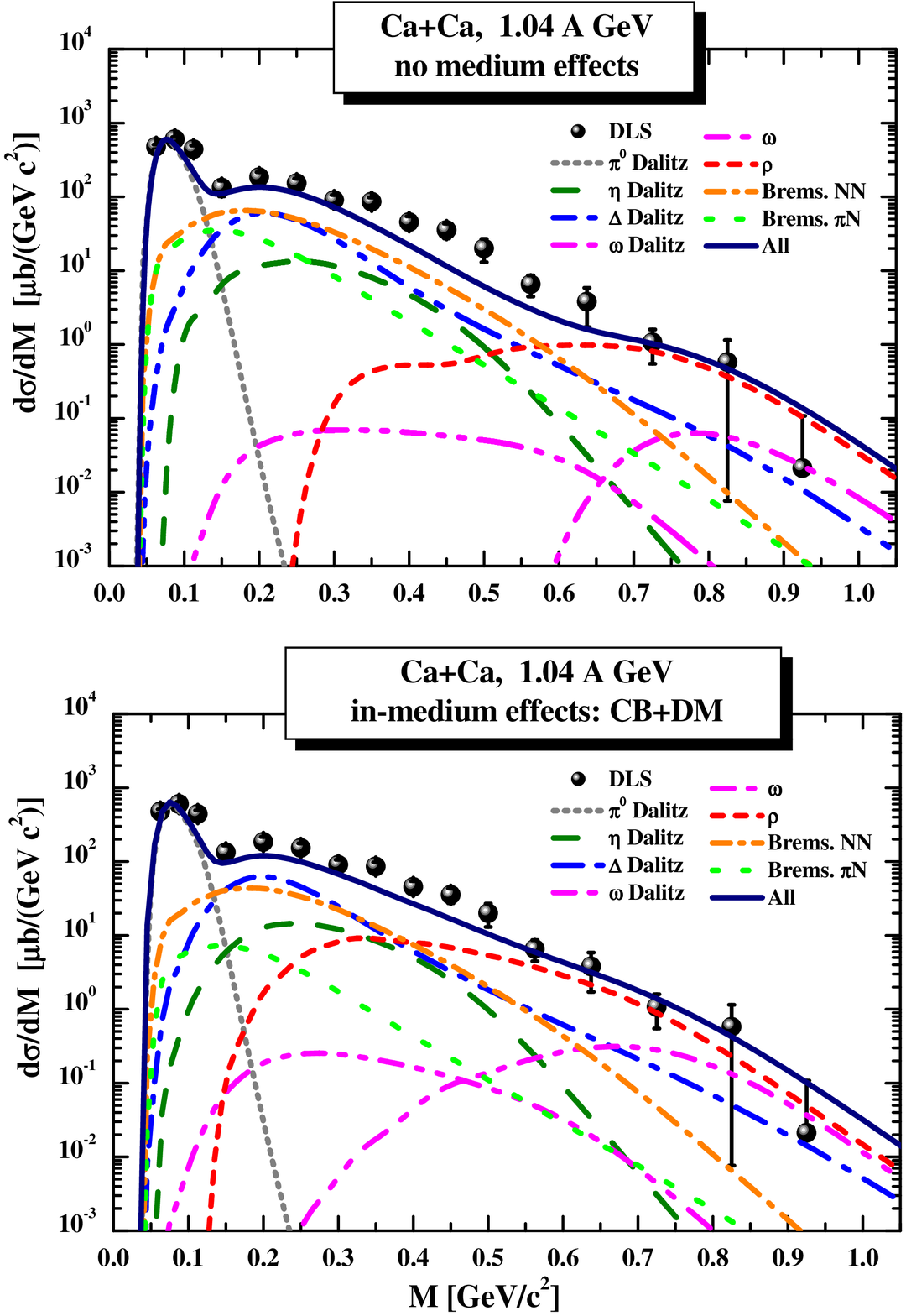} }
\caption{Results of the
HSD transport calculation  for the mass differential dilepton
spectra in case of  $^{12}C + ^{12}C$ (l.h.s.) and $Ca+Ca$
(r.h.s.) at 1.04 A$\cdot$GeV in comparison to the DLS data
\cite{DLSnew}. The upper parts show the case of 'free'
vector-meson spectral functions while the lower parts give the
result for the 'dropping mass + collisional broadening' scenario.
In both scenarios the DLS acceptance filter and mass resolution
have been incorporated. The different color lines display
individual channels in the transport calculation (see legend). }
\label{OFig11}
\end{figure}

We start with a reinvestigation of the DLS data employing  the
additional (and enhanced) bremsstrahlung channels as mentioned
above \cite{Brat08}. The results of the HSD transport calculation
are displayed in Fig. \ref{OFig11} for $^{12}$C+$^{12}$C (l.h.s.)
at 1.04 A$\cdot$GeV in case of 'free' vector-meson spectral
functions (upper part) and in case of the 'dropping mass +
collisional broadening' scenario (lower part) employing the DLS
acceptance filter and mass resolution.

In fact, the situation has substantially improved compared to the
early studies, and the missing yield in the 'free' scenario is now
reduced to a factor of about 1.5 in the mass region from 0.25 to
0.5 GeV. This is due to slightly higher contributions from
$\Delta$ and $\eta$ Dalitz decays and a significantly larger yield
from bremsstrahlung channels. As seen from Fig. \ref{OFig11}, the
bremsstrahlung yield now is similar in shape and magnitude as the
$\Delta$-Dalitz decay contribution and enhanced by factor of up to
5 due to the novel $NN$ bremsstrahlung cross section from Kaptari
\cite{Kaptari} and accounting for the contribution from $pp$
bremsstrahlung additionally to $pn$.  The contribution of the
pion-nucleon bremsstrahlung is quite small in the C+C system due
to the limited energy available in meson-baryon collisions and the
moderate rescattering rate of pions.

Medium modifications for the vector mesons turn out to yield an
enhancement in the region of 0.4 $< M <$ 0.5 GeV but are very
moderate due to the light system C+C. Though a slightly better
description of the DLS data is achieved it would be premature to
claim the presence of in-medium effects on the vector mesons from
these data.

\begin{figure}[!t]
\resizebox{0.48\columnwidth}{!}{\includegraphics{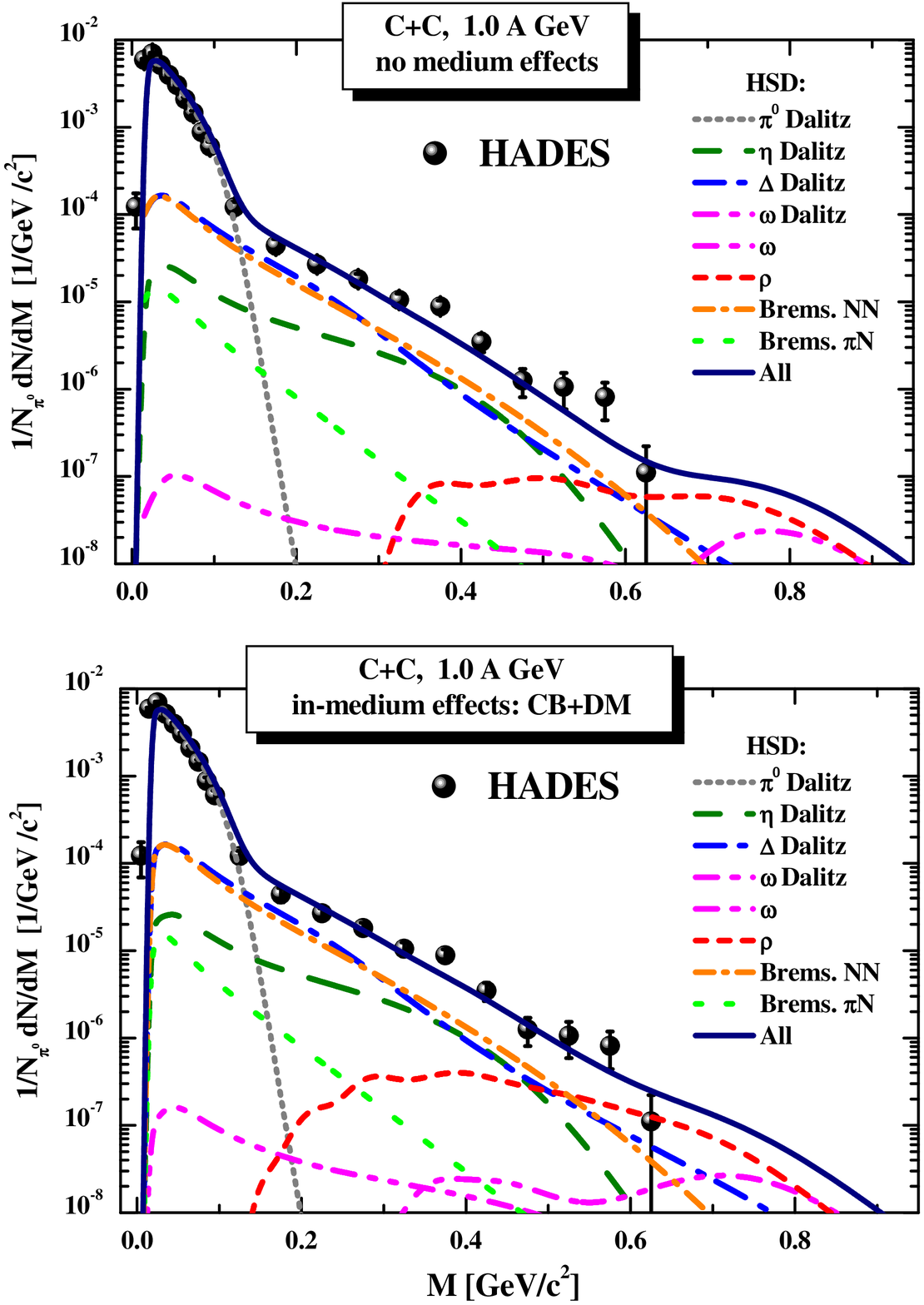} }
\resizebox{0.48\columnwidth}{!}{\includegraphics{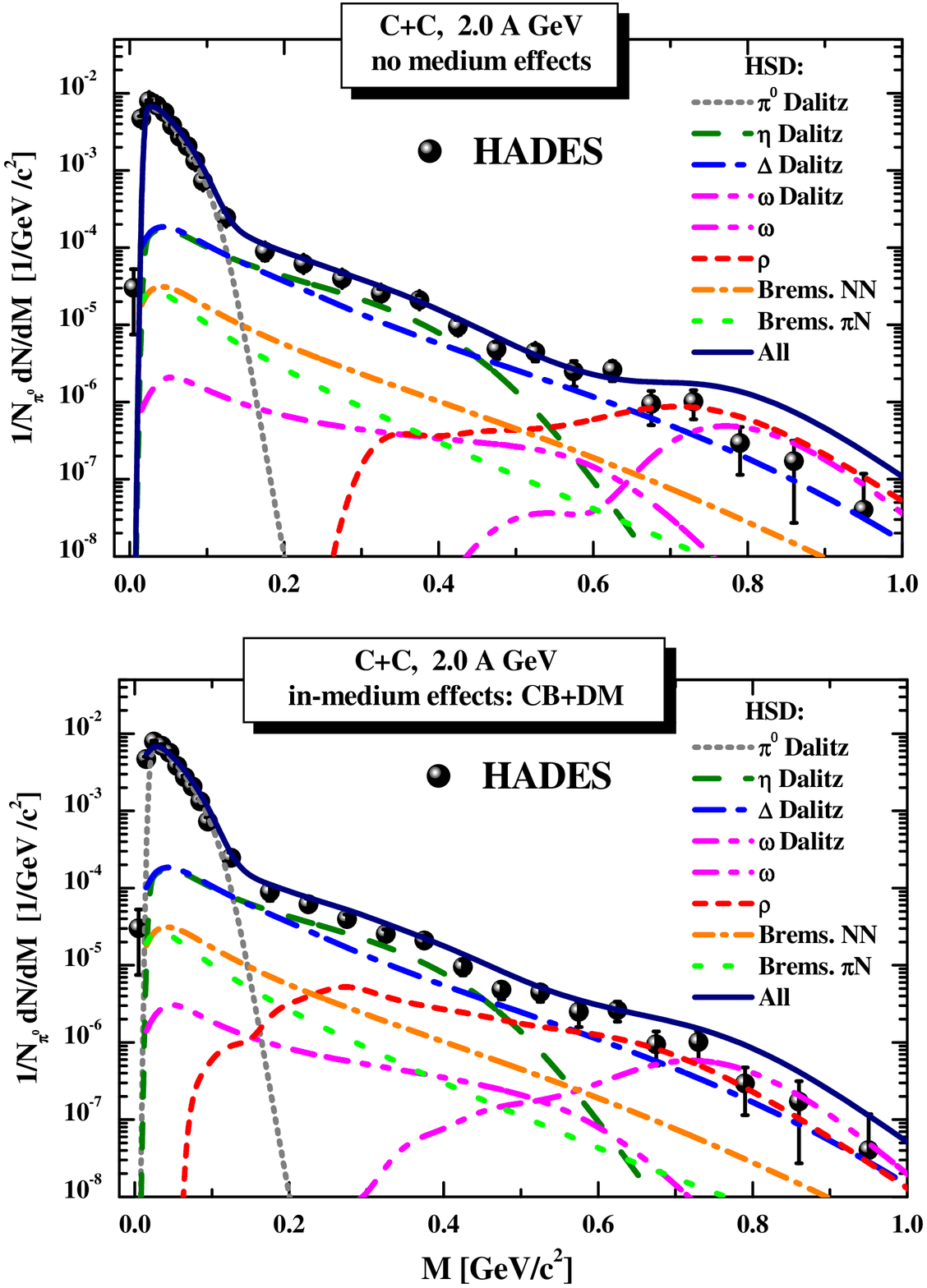} }
\caption{Results of the HSD transport calculation  for the mass
differential dilepton spectra - divided by the average number of
$\pi^0$'s - in case of $^{12}$C + $^{12}$C at 1.0 A$\cdot$GeV in
comparison to the HADES data \cite{HADES07}. The upper part shows
the case of 'free' vector-meson spectral functions while the lower
part gives the result for the 'dropping mass + collisional
broadening' scenario. In both scenarios the HADES acceptance
filter and mass resolution have been incorporated. The different
color lines display individual channels in the transport
calculation (see legend). } \label{OFig13}
\end{figure}

Additional information is provided by the  $^{40}$Ca+ $^{40}$Ca
system at 1.04 A$\cdot$GeV where the measured DLS spectra are
reproduced in a qualitatively and quantitatively similar manner by
the transport calculations (cf. r.h.s. Fig. \ref{OFig11}) as in
case of the lighter system. Again the combined bremsstrahlung
channels provide a contribution in the same order as the $\Delta$
Dalitz decay; the $\eta$ Dalitz decay remains subdominant but the
$\pi N$ bremsstrahlung increases compared to the $C+C$ system due
to more frequent pion rescattering on nucleons. In-medium effects
for the vector mesons can be indentified in the transport
calculations but are hard to see in the total dilepton spectra.

Is the 'DLS puzzle' solved? This question can only be answered in
a convincing manner by comparison with the recent HADES data.

\subsubsection*{Comparison to HADES data}

The same (HSD) multi-differential dilepton spectra - used for
comparison with the DLS data - are now filtered by the HADES
acceptance routines and smeared with the HADES mass resolution
\cite{HADES06}. A comparison of the HSD calculations for the
$^{12}$C + $^{12}$C system at 1.0 A$\cdot$GeV with the HADES data
from \cite{HADES07} is presented in Fig. \ref{OFig13} and
demonstrates that the agreement between data and transport
calculations is reasonable for the HADES data, too. In this case
the  differential dilepton spectrum is divided by the average
number of $\pi^0$'s which in experiment is determined by half of
the average number of ($\pi^+ +\pi^-$). The spectra look slightly
different in shape due to the much higher acceptance at low
invariant mass where the $\pi^0$ Dalitz decay practically exhausts
the dilepton spectrum. As in case of the DLS data the $\eta$
Dalitz decay turns out to be subdominant and the $\Delta$ Dalitz
decay to be of similar magnitude as the combined bremsstrahlung
contribution. Again the effect of in-medium vector-meson spectral
functions is hard to see in the total spectra.

In summary: The 'DLS puzzle' appears to be solved from the
theoretical side, too, if enhanced Bremsstrahlung channels are
accounted for \cite{Brat08}. Note, however, that the problem of
radiative corrections  especially in inelastic collisions of
charged hadrons is presently barely understood from the
theoretical side!

\subsubsection{Dilepton production  at SPS energies}
\begin{figure}[t]
\phantom{a}\vspace*{5mm}
\resizebox{0.7\columnwidth}{!}{\includegraphics{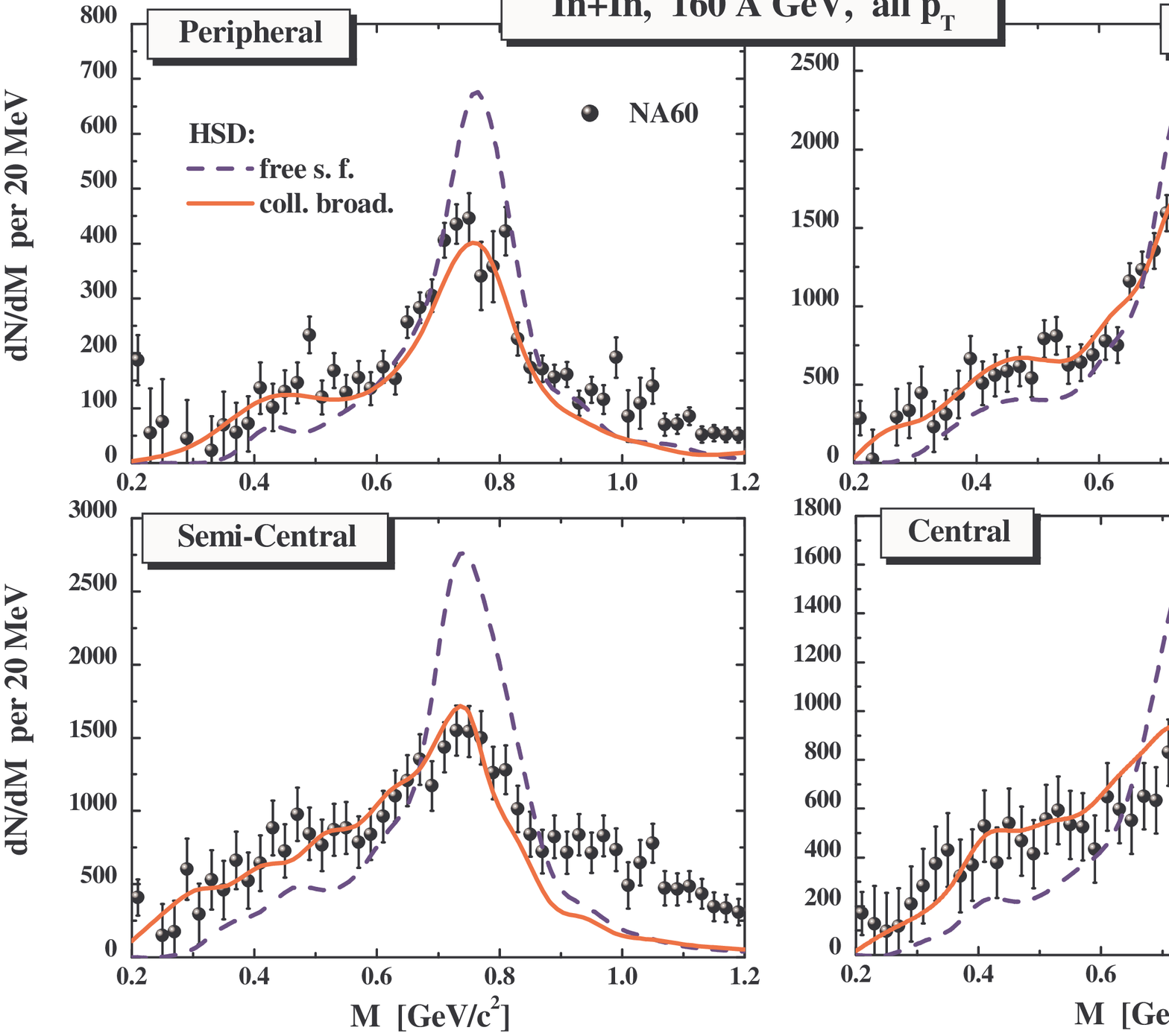} }
\caption{The HSD results  for the mass differential dilepton
spectra in case of $In + In$ at 158 A$\cdot$GeV for peripheral,
semi-peripheral, semi-central and central collisions in comparison
to the data from NA60 \cite{NA60}. The actual NA60 acceptance
filter and mass resolution have been incorporated. The solid red
lines show the HSD results for a scenario including the
collisional bradening of the $\rho$-meson whereas the dashed blue
lines correspond to calculations with  'free' $\rho$ spectral
functions for reference. } \label{Wig1}
\end{figure}

Let's continue with the actual results from HSD for In+In
collisions at 160 A$\cdot$GeV \cite{Olena} which are displayed in
Fig. \ref{Wig1} for peripheral, semi-peripheral, semi-central and
central collisions following the centrality definition of the NA60
Collaboration \cite{NA60}. The dashed (blue) lines give the
dilepton mass spectra when incorporating only the vacuum $\rho$-
spectral function in all hadronic reaction processes. These
reference spectra overestimate the data in the region of the
$\rho$-meson pole mass and underestimate the experimental spectra
in the region below (and above) the pole mass such that in-medium
modifications can clearly be identified. The solid red lines show
the result from HSD in the 'collisional broadening' scenario where
no shift of the $\rho$ pole mass is incorporated but an increase
of the $\rho$-meson width due to hadronic collisions proportional
to the baryon density $\rho_B$ (calculated as $\rho_B^2(x) =
j^{\mu}(x) j_{\mu}(x)$ with $j^{\mu}(x)$ denoting the baryon
four-current).  As can be seen from Fig. \ref{Wig1} the dilepton
mass spectrum is rather well described up to invariant masses of
0.9 GeV. For higher invariant masses the data signal additional
contributions. This result is practically identical to the
calculations of van Hees and Rapp \cite{rapp3} in the expanding
fireball model - when incorporating the spectral function from
Ref. \cite{RappNPA} - and demonstrates that the dominant in-medium
effect seen in the $\mu^+\mu^-$ spectra from NA60 is a broadening
of the $\rho$ meson.

\begin{figure}[t]
\centerline{\resizebox{0.55\columnwidth}{!}{\includegraphics{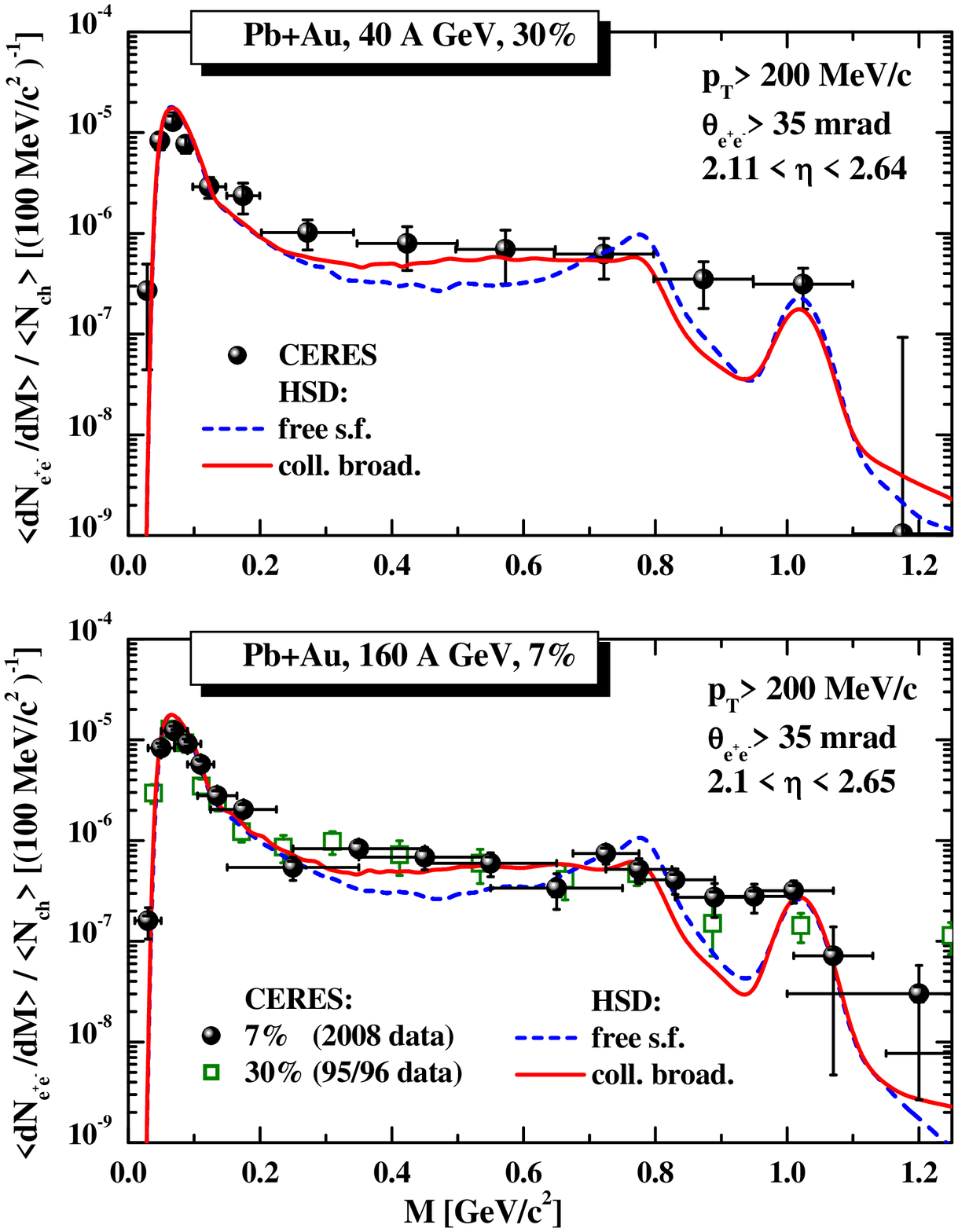}
}}
\caption{The HSD results  for the mass differential dilepton
spectra in 30\% central $Pb + Au$ collisions at 40 A$\cdot$GeV
(upper part) and 7\% central 158 A GeV  (lower part) in comparison
to the data from CERES \cite{CERES2}.  The dashed lines show the
results  for  vacuum spectral functions (for $\rho, \omega, \phi$)
whereas the solid lines correspond to the 'collisional broadening'
scenario.} \label{FigCERES}
\end{figure}

In Ref. \cite{Hees08} a very detailed analysis of the NA60 data
has been performed and shown that this additional yield above 0.9
GeV partly is due to open charm decays, four pion collisions or
'quark-antiquark' annihilation. We mention that our HSD
calculations give only a small contribution from $\pi + a_1$
collisions in this invariant mass range but a preliminary study
within the Parton-Hadron-String-Dynamics (PHSD) model
\cite{PHSDdil} suggests that - apart from open charm decays - the
extra yield seen experimentally by NA60 should be due to massive
'quark-antiquark' annihilations. Since this question is presently
open and discussed controversely we concentrate on low mass
dilepton pairs in the following.

The next step in our study is related to an update of the HSD
calculations in comparison to the recent data from the CERES
Collaboration \cite{CERES2} (with enhanced mass resolution). In
Fig.~\ref{FigCERES}  we present the HSD results  for the
mass-differential dilepton spectra in 30\% central Pb + Au
collisions at 40 A GeV (upper part) and 7\% central 158 A GeV
(lower part) in comparison to the data from CERES \cite{CERES2}.
The dashed lines show the results in case of vacuum spectral
functions (for $\rho, \omega, \phi$) whereas the solid lines
correspond to the 'collisional broadening' scenario. Similar to
the In+In case the experimental data agree better with the
calculations employing the 'collisional broadening' scenario. This
is also in line with  Ref. \cite{Hees08}. On the other hand, the
HSD model underpredicts the yield between the $\omega$ and $\phi$
peaks which might again be attributed to possible contributions
from 'quark-antiquark' annihilations.

\begin{figure}[htb]
\centerline{\resizebox{0.55\columnwidth}{!}{\includegraphics{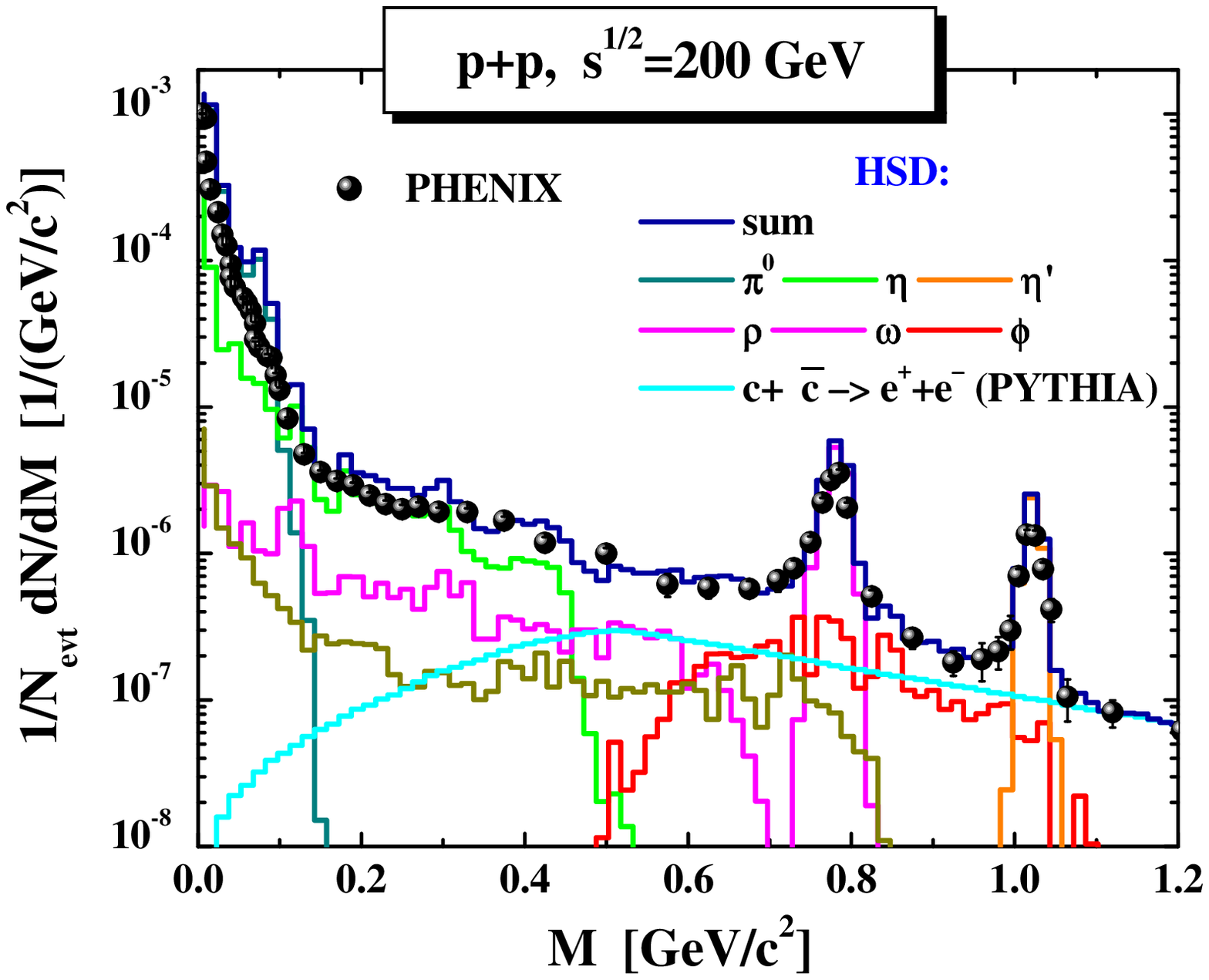}
}}
\caption{The HSD results  for the mass differential dilepton
spectra in case of $pp$ collisions at $\sqrt{s}$ = 200 GeV  in
comparison to the data from PHENIX \cite{PHENIXpp}.  }
\label{WFig2}
\end{figure}

\subsubsection{Dilepton production at RHIC energies}
We step on to RHIC energies and first compare the HSD results
\cite{Olena} for the dilepton invariant-mass spectrum from $pp$
collisions at $\sqrt{s}$ = 200 GeV with the data from PHENIX
(incorporating the acceptance cuts and mass resolution from
PHENIX) \cite{PHENIXpp} in Fig. \ref{WFig2}. Actually the
calculations well reproduce the experimental spectrum which can
entirely be described by meson Dalitz and direct decays as well as
some contribution from open charm decays (light blue solid line as
calculated by PYTHIA). This comparison demonstrates that the
hadron production channels in HSD for elementary $pp$ collisions
are well under control.

\begin{figure}[htb]
\centerline{\resizebox{0.55\columnwidth}{!}{\includegraphics{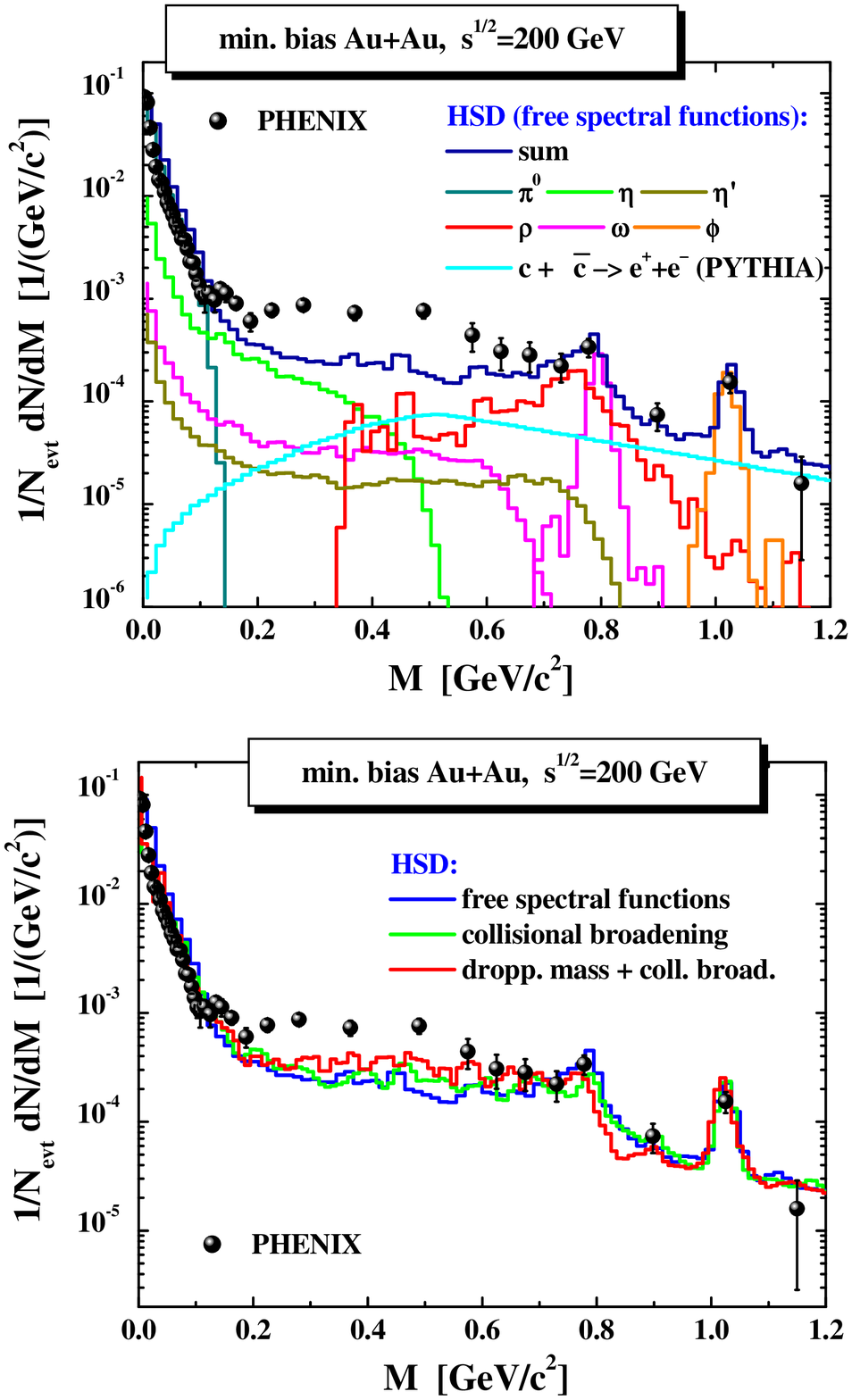}
}}
\caption{The HSD results  for the mass differential dilepton
spectra in case of inclusive Au + Au collisions at $\sqrt{s}$ =
200 GeV  in comparison to the data from PHENIX \cite{PHENIX}.  In
the upper part the results are shown for  vacuum spectral
functions (for $\rho, \omega, \phi$) including the channel
decompositions (see legend for the different color coding of the
individual channels). The lower part shows a comparison for the
total $e^+e^-$ mass spectrum in case of the 'free' scenario (blue
line), the 'collisional broadening' picture (green line) as well
as the 'dropping mass + collisional broadening' model (red line).
} \label{WFig3}
\end{figure}

We recall that HSD also provides a reasonable description of
hadron production in Au+Au collisions at $\sqrt{s}$ = 200 GeV
\cite{Brat03} such that we can directly continue with the results
for $e^+e^-$ pairs which are shown in Fig. \ref{WFig3}  in case of
inclusive Au + Au collisions  in comparison to the data from
PHENIX \cite{PHENIX}. In the upper part of Fig. \ref{WFig3} the
results are shown for vacuum spectral functions (for $\rho,
\omega, \phi$) including the channel decompositions (see legend
for the different color coding of individual channels). Whereas
the total yield (upper blue line) is quite well described in the
region of the pion Dalitz decay as well as the $\omega$- and
$\phi$-mass regime we clearly underestimate  the measured spectra
in the regime from 0.2 to 0.6 GeV by an average factor of 3.

When  including the 'collisional broadening' scenario for the
vector mesons  we achieve the sum spectrum shown by the green line
in the lower part of Fig. \ref{WFig3} which is only slightly
enhanced compared to the 'free' scenario (blue line). Thus the
question emerges if the PHENIX data might signal dropping vector
meson masses contrary to the NA60 data (in Fig. \ref{Wig1})? To
answer this question we have performed additional calculations in
the 'dropping mass + collisional broadening' model where the
$\rho$ and $\omega$ masses have been dropped with baryon density
in accordance with Eq. (\ref{Brown}). The respective HSD results
are displayed in the lower part of Fig.~\ref{WFig3} by the red
line and indeed show a further enhancement of the dilepton yield
which, however, is only small in the mass range 0.2 GeV $< M <$
0.4 GeV such that also this possibility has to be excluded in
comparison to the PHENIX data.

In summary  the presently available dilepton data from In+In
collisions at 158 A$\cdot$GeV (from the NA60 Collaboration) as
well as the low mass dilepton spectra for Pb+Au collisions at 40
and 158 A$\cdot$GeV (from the CERES Collaboration) are well
described in the 'collisional broadening scenario'. However, the
low mass dilepton spectra from Au+Au collisions at RHIC (from the
PHENIX Collaboration) are clearly underestimated in the invariant
mass range from 0.2 to 0.6 GeV. This also holds for the 'dropping
mass + collisional broadening' scenario, i.e., when tentatively
assuming a shift of the vector meson mass poles with the baryon
density.  We attribute this additional low mass enhancement seen
by PHENIX to non-hadronic sources, possibly to virtual
gluon-Compton scattering.


\subsection{Off-shell parton dynamics}
The Parton-Hadron-String Dynamics (PHSD) approach is an extension
of the more familiar HSD model \cite{CB99,Ehehalt} with respect to
the off-shell dynamics of partons (gluons, quarks, antiquarks).
This extension is demanded by experiment since the HSD approach -
without explicit partonic degrees of freedom - failed in a couple
of aspects to describe the rich phenomena observed experimentally
at Relativistic-Heavy-Ion-Collider (RHIC) energies of $\sqrt{s}$ =
200 GeV (cf. Refs. \cite{Cassing03,Brat04,Cassing04}).

The off-shell dynamics of the partons is entirely described by the
generalized transport equations in BM form within the testparticle
representation for $iG^<$ (Section 2) incorporating the effective
selfenergies from the DQPM (Section 3).  In PHSD the following
elastic and inelastic interactions are included $qq
\leftrightarrow qq$, $\bar{q} \bar{q} \leftrightarrow
\bar{q}\bar{q}$, $gg \leftrightarrow gg$, $gg \leftrightarrow g$,
$q\bar{q} \leftrightarrow g$ etc. on the partonic side exploiting
'detailed-balance' with cross sections extracted in Refs.
\cite{Andre,Cassing06}. In this way the partonic evolution is
fixed, however, the effective parton quasi-particles are ill
defined as asymptotic states in vacuum since the 'masses' diverge
for $T/T_c = 0.46$ (cf. equations (3.3) - (3.5)). Consequently the
partons (in PHSD) live only at sufficiently high temperature (or
energy density) and have to recombine to color neutral hadrons in
the expansion of the system.

In PHSD the transition from partonic to hadronic degrees of
freedom is described by local covariant transition rates e.g. for
$q+\bar{q}$ fusion to a meson $m$ of four-momentum $p= (\omega,
{\bf p})$ at space time point $x=(t,{\bf x})$ \cite{PHSDnew}:
\begin{eqnarray}
&&\phantom{a}\hspace*{-5mm} \frac{d N_m(x,p)}{d^4x d^4p}= Tr_q Tr_{\bar q} \
  \delta^4(p-p_q-p_{\bar q}) \
  \delta^4\left(\frac{x_q+x_{\bar q}}{2}-x\right) \nonumber\\
&& \times \omega_q \ \rho_{q}(p_q)
   \  \omega_{\bar q} \ \rho_{{\bar q}}(p_{\bar q})
   \ |v_{q\bar{q}}|^2 \ W(x_q-x_{\bar q},p_q-p_{\bar q}) \nonumber \\
&& \times N_q(x_q, p_q) \
  N_{\bar q}(x_{\bar q},p_{\bar q}) \ \delta({\rm flavor,\, color}).
\label{trans}
\end{eqnarray}
In (\ref{trans}) we have introduced the shorthand notation
\begin{equation} Tr_j = \sum_j \int d^4x_j \ \frac{d^4p_j}{(2\pi)^4} \end{equation} where $\sum_j$
denotes a summation over discrete quantum numbers (spin, flavor,
color); $N_j(x,p)$ is the phase-space density of parton $j$ at
space-time position $x$ and four-momentum $p$.  $\delta($
flavor,\, color) stands symbolically for the conservation of
flavor quantum numbers as well as color neutrality of the formed
hadron $m$. Furthermore, $v_{q{\bar q}}(\rho_p)$ is the effective
quark-antiquark interaction as extracted from the DQPM in
 Section 3.3 as a function of the local
parton ($q + \bar{q} +g$) density $\rho_p$ (or energy density
$\varepsilon$) while $W(x,p)$ is the phase-space distribution of
the formed hadron. It is taken as a Gaussian in coordinate and
momentum space with width $\sqrt{\langle r^2 \rangle}$ = 0.66 fm
for a meson and 1 fm for a baryon/antibaryon (in the rest frame).
The width in momentum space then is fixed by the minimal
uncertainty principle. We note that the final hadron formation
rates are approximately independent on these parameters (within
reasonable variations). Related transition rates are defined for
the fusion of three off-shell quarks ($q+q+q \leftrightarrow B$)
to  color neutral baryonic ($B$) resonances (or $\bar{B}$) of
finite width (or strings) with the help of Jacobi coordinates. The
transition rates (\ref{trans}) fulfill energy and momentum
conservation as well as flavor current conservation since the
parton flavors fix the flavor content of the formed meson $m$ or
baryon $B$.

According to the DQPM the effective interaction turns to be
strongly attractive below $\rho_c \approx$ 2.2 fm$^{-3}$ (cf.
Section 3.3) which implies a dynamical binding of partons to
composite (color neutral) hadrons once the parton density drops
below $\rho_c$. Thus the hadronization starts in PHSD when the
local parton density falls below 2.2 fm$^{-3}$, i.e., the
transition rate (\ref{trans}) becomes nonzero. For comparison we
note that the average parton density in a baryon is about 1.5
fm$^{-3}$ while it is about 1.25 fm$^{-3}$ in case of a meson.
Gluons in PHSD decay to color octet $q\bar{q}$ pairs below
$\rho_c$; their reformation rate is strongly suppressed at low
parton densities due to the large gluon masses from the DQPM (cf.
Fig. \ref{ffig1}).

On the hadronic side PHSD includes explicitly the  baryon octet
and decouplet, the $0^-$ and $1^-$ meson nonets as well as
selected higher resonances. Hadrons of higher masses ($>$ 1.5 GeV
in case of baryons and $>$ 1.3 GeV in case of mesons) are treated
as 'strings' that reflect the continuum excitation spectrum of
mesons or baryons and decay to the known (low mass) hadrons  within HSD
\cite{CB99} according to the JETSET algorithm \cite{JETSET}.

\subsubsection{The dynamics of an expanding partonic fireball}
Due to the complexity of the PHSD approach it is illustrative to
explore the parton and hadronization dynamics in a transparent
model case \cite{PHSDnew} rather than directly step to
relativistic nucleus-nucleus collisions at SPS or RHIC energies.
Accordingly the following study addresses the expansion dynamics
of a partonic fireball at initial temperature $T=1.7\ T_c$ ($T_c$=
0.185 GeV) with quasiparticle properties and four-momentum
distributions determined by the DQPM at temperature $T/T_c$ = 1.7
for quark chemical potential $\mu_q$=0. The initial distribution
in coordinate space is taken as a Gaussian ellipsoid with a
spatial eccentricity \begin{equation} \label{eccen} \epsilon
=\langle y^2-x^2\rangle/\langle y^2 + x^2\rangle \end{equation} in
order to allow for the build-up of elliptic flow (as in
semi-central nucleus-nucleus collisions at relativistic energies).
In order to match the initial off-equilibrium strange-quark
content in relativistic $pp$ collisions the number of $s$ (and
$\bar{s}$ quarks) is assumed to be suppressed by a factor of 3
relative to the abundance of $u,d$ quarks and antiquarks.  In this
way we will be able to investigate additionally the question of
strangeness equilibration. Recall that the dynamical evolution of
the system now is entirely described by the transport dynamics in
PHSD incorporating the off-shell propagation of the partonic
quasiparticles according to Ref. \cite{caju2} (cf. Section 2) as
well as the transition to resonant hadronic states (or strings) in
line with (\ref{trans}) which is solved by Monte Carlo sampling on
a time-depending  expanding grid with local cells of volume
$dV(t)$= 0.25 (1+0.7 $t)^3$ fm$^3$. This choice approximately
corresponds to a comoving grid for the expanding system.

\begin{figure}[thb]
\centerline{\resizebox{0.7\columnwidth}{!}{\includegraphics{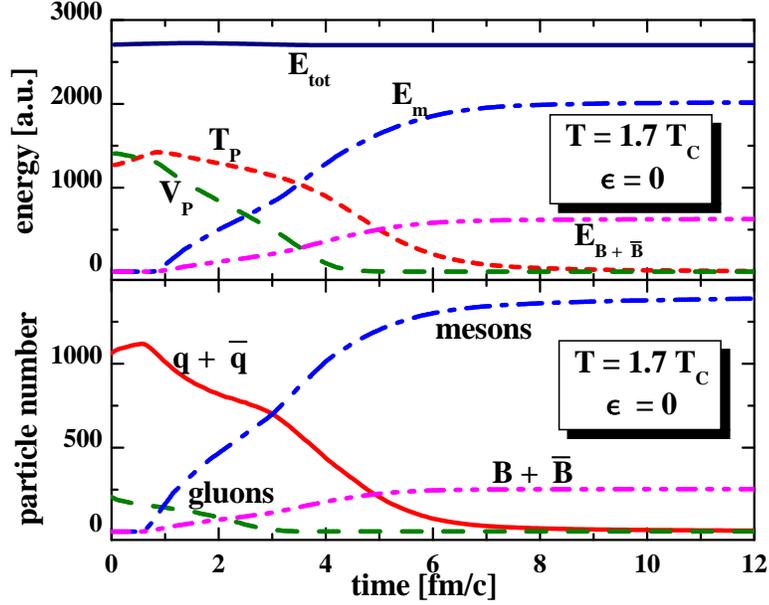} }}
\caption{Upper part: Time evolution of the total energy $E_{tot}$ (upper line),
the partonic contributions from the interaction energy $V_p$ and
the energy of time-like partons $T_p$ in comparison to the energy
contribution from formed mesons $E_m$ and baryons (+ antibaryons)
$E_{B+{\bar B}}$. Lower part:  Time evolution in the parton, meson
and baryon number for an exploding partonic fireball at initial
temperature $T=1.7\  T_c$ with initial eccentricity $\epsilon =
0$.} \label{Kfig1}
\end{figure}

In Fig. \ref{Kfig1} (upper part) the energy balance for the
expanding system at initial temperature $T= 1.7 T_c$, $\mu_q$ = 0
and eccentricity $\epsilon$ =0 is shown. The total energy
$E_{tot}$ (upper line) is conserved within 3\% throughout the
partonic expansion and hadronization phase such that for $t >$
10~fm/c it is given essentially by the energy contribution from
mesons and baryons (+antibaryons). The initial energy splits into
the partonic interaction energy $V_p$ and the energy of the
time-like partons \begin{equation} T_p = \sum_i \sqrt{p^2_i +
M^2_i(\rho_p)} \end{equation} with fractions determined by the
DQPM (Section 3). The hadronization mainly proceeds during the
time interval 1 fm/c $< t < $ 7 fm/c as can be extracted also from
the lower part of Fig. \ref{Kfig1} where the time evolution of the
quark (+antiquark), gluon, meson and baryon (+antibaryon) number
is displayed.

The hadronic decomposition is dominantly determined via parton
fusion to mesonic or baryonic resonances and strings and their
hadronic decays. Since on average the number of hadrons from the
resonance or string decays is larger than the initial number of
fusing partons, the hadronization process leads to an increase of
the total entropy and not to a decrease as in case of coalescence
models \cite{Koal1,Koal2}. This is a direct consequence of the
finite quark and antiquark masses which - by energy conservation -
lead to hadron masses above 0.8 GeV (1.3 GeV) for meson and
baryonic states, respectively; {\it this solves the entropy
problem in hadronization in a natural way}!
\begin{table}[b]
\begin{center}
\begin{tabular}{|c|c|c|c|} \hline
\hspace*{3.0cm} &~~~~$p/\pi^+$~~~  & ~~~$\Lambda/K^+$~~~   & ~~~$K^+/\pi^+$~~~~   \\
 \hline
PHSD  &    0.086    & 0.28            &   0.157         \\ \hline
SM $T =160$~MeV &   0.073   & 0.22   &  0.179 \\          \hline
SM $T =170$~MeV &   0.086  & 0.26   &  0.180 \\
 \hline
  \end{tabular}
\end{center}
\caption{Comparison of particle ratios from PHSD with the
statistical model (SM) \cite{Anton} for $T$= 160 MeV and 170 MeV.}
\end{table}

\subsubsection{Hadronization}
It is, furthermore, interesting to have a look at the final
particle ratios $K^+/\pi^+$, $p/\pi^+$, $\Lambda/K^+$ etc. (after
hadronic decays) which are shown in Table 1. The latter ratios may
be compared to the grandcanonical statistical hadronization model
(SM) at baryon chemical potential $\mu_B = 0$
\cite{PBM,PBM2,Anton}. In this particular case the particle ratios
only depend on temperature $T$ and one may fix a freeze-out
temperature, e.g., by the proton to $\pi^+$ ratio. A respective
comparison is given  in Table 1 for $T$ = 160 MeV and 170 MeV for
the SM which demonstrates that the results from PHSD are close to
those from the SM for $T \approx$ 170 MeV. This also holds roughly
for the $\Lambda/K^+$ ratio. On the other hand the $K^+/\pi^+$
ratio only smoothly depends on the temperature $T$ and measures
the amount of strangeness equilibration. Recall that the system
has been initialized with a relative strangeness suppression
factor of 1/3. The deviation from the SM ratio by about 13\%
indicates that strangeness equilibration is not fully achieved in
the calculations. This is expected since the partons in the
surface of the fireball hadronize before chemical equilibration
may occur.

The agreement between the PHSD results for the baryon to meson
ratio in the strangeness S=0 and S=1 sector may be explained as
follows:  Since the final hadron formation dominantly proceeds via
resonance and string formation and decay - which is approximately
a microcanonical statistical process \cite{beccatini} - the
average over many hadronization events with different energy/mass
and particle number (in the initial and final state) leads to a
grandcanonical ensemble. The latter (for $\mu_B = 0$) is only
characterized by the average energy or an associated Lagrange
parameter $\beta=1/T$ as well known from quantum statistical
mechanics.

\begin{figure}[t]
\centerline{\resizebox{0.7\columnwidth}{!}{\includegraphics{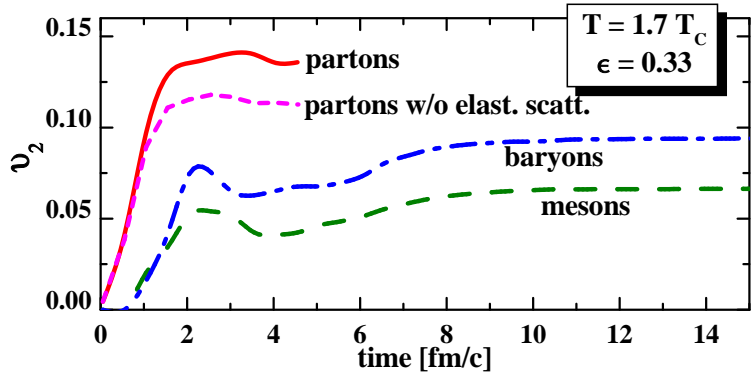} }}
\caption{Time
evolution of the elliptic flow $v_2$ for partons
 and hadrons for the initial spatial eccentricity
$\epsilon =0.33$ for an expanding partonic fireball at initial temperature
$T=1.7\  T_c$.}
 \label{Kfig2}
\end{figure}

\subsubsection{Elliptic flow}
 Of additional interest are the collective
properties of the partonic system during the early time evolution.
In order to demonstrate the build-up of elliptic flow we show in
Fig. \ref{Kfig2} the time evolution of \begin{equation}
\label{vv2} v_2 = \left\langle (p_x^2 - p_y^2)/(p_x^2 + p_y^2)
\right\rangle \end{equation} for partons (solid line), mesons
(long dashed line) and baryons (dot-dashed line) for an initial
eccentricity $\epsilon = 0.33$. As seen from Fig. \ref{Kfig2} the
partonic flow develops within 2 fm/c and the hadrons produced in
time essentially pick up the collective flow from the accelerated
partons. In total the hadron $v_2$ is smaller than the maximal
parton $v_2$ since by parton fusion the average $v_2$ reduces and
a fraction of hadrons is formed early at the surface of the
fireball without a strong acceleration before hadronization. Is is
worth to point out that in PHSD the elliptic flow of partons
predominantly stems from the gradients of the repulsive parton
mean-fields at high parton (energy) density. This is quite
analoguous to $A+A$ collisions at SIS energies where the
collective flow is dominated by the nucleon mean-field potentials
\cite{Lari}. To demonstrate this statement we display in Fig.
\ref{Kfig2} the result for a simulation without elastic partonic
rescattering processes by the short dashed line.

\begin{figure}[t]
\centerline{\resizebox{0.7\columnwidth}{!}{\includegraphics{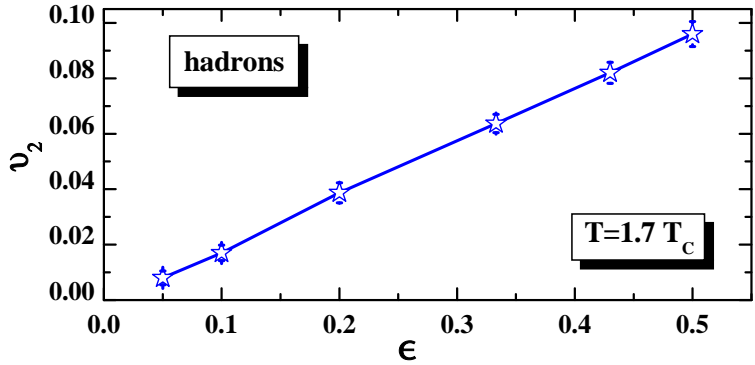} }}
\caption{The elliptic flow $v_2$ versus the initial spatial eccentricity
$\epsilon$  for an expanding partonic fireball at initial
temperature $T=1.7\  T_c$.} \label{Kfig3}
\end{figure}

Fig. \ref{Kfig3} shows the final hadron $v_2$ versus the initial
eccentricity $\epsilon$ and indicates that the ratio
$v_2/\epsilon$ is practically constant ($\approx 0.2$) as in ideal
hydrodynamics (cf. Fig. 3 in Ref. \cite{Voloshin}). Accordingly
the parton dynamics in PHSD are close to ideal hydrodynamics. This
result is expected since the ratio of the shear viscosity $\eta$
to the entropy density $s$ in the DQPM is on the level of $\eta/s
\approx $ 0.2 \cite{Andre} and thus rather close to the suggested
 lower bound of $\eta/s = 1/(4 \pi)$ \cite{Son}.

In summarizing this section we like to point out that the
expansion dynamics of an anisotropic partonic fireball within the
PHSD approach - including dynamical local transition rates from
partons to hadrons (and vice versa) - shows collective features as
expected from ideal hydrodynamics in case of strongly interacting
systems. The hadronization process conserves four-momentum and all
flavor currents and slightly increases the total entropy (by about
15\% in the model case investigated here) since the 'fusion' of
massive partons dominantly leads to the formation of color neutral
strings or resonanaces that decay microcanonically to lower mass
hadrons. This solves the entropy problem associated with the
simple coalescence model in case of massless partons!

Furthermore, the hadron abundancies and baryon to meson ratios are
found to be  compatible with those from the statistical
hadronization model \cite{PBM,PBM2} - which describes well
particle ratios from AGS to RHIC energies - at a freeze-out
temperature of about 170 MeV. Strangeness equilibration is
approximately achieved in the dynamical expansion and driven by
the processes $q\bar{q} \leftrightarrow g \leftrightarrow
s\bar{s}$, which is a resonant process in the DQPM. Nevertheless,
some note of caution has to be added here: Although the final
hadron ratios are compatible with a fixed freeze-out temperature
($\sim$ 170 MeV) we find that the actual hadronization occurs at
very different energy densities (or temperatures) (cf. also Ref.
\cite{Knoll08}) such that the microscopic studies do not support
the sudden freeze-out picture. \\
\\

In closing these lectures the author likes to point out that the
field of quantum physics out-of-equilibrium and especially for
strongly interacting systems is far from being closed. Though
substantial progress has been achieved in the last decade many
fundamental questions still lack a satisfactory answer. This
subject is challenging but also exciting!

\subsubsection*{Acknowlegements}
The author likes to thank E. L. Bratkovskaya, C. Greiner, S.
Juchem, O. Linnyk and A. Peshier for the many substantial
contributions reported in these lectures. Numerous discussions
with further collegues have contributed as well but cannot be
listed in detail. He is also indebted to  H. van Hees for a
careful reading of the manuscript and valuable comments.
\\ The author finally likes to thank the organizers of the Schladming
Winter-School 2008 on 'Non-equilibrium Aspects of Quantum Field
Theory', i.e. R. Alkhofer, H. Gies and B.-J. Sch\"afer, for the
exiting week, the well set up program and the very friendly
atmosphere.

\end{document}